\documentclass{aa}

\usepackage{natbib}
\bibpunct{(}{)}{;}{a}{}{,} 
\usepackage{graphics}   
\usepackage{graphicx}   
\usepackage{epstopdf}
\usepackage[varg]{txfonts}
\usepackage{pdflscape}
\usepackage{hyperref}   
\usepackage[flushleft]{threeparttable}
\usepackage{longtable}
\usepackage{caption}
\hypersetup{
    bookmarks=true,         
    unicode=false,          
    colorlinks=true,       
    linkcolor=blue,          
    citecolor=blue,        
    filecolor=blue,      
    urlcolor=blue           
}

\begin{document} 

   \title{Inferring the velocity of early massive stars from the abundances of extremely metal-poor stars.
}

   \author{
   Arthur Choplin\inst{1,2}, Nozomu Tominaga\inst{1,3} \and Miho N. Ishigaki\inst{4}
                    }

 \authorrunning{Choplin et al.}

      \institute{
Department of Physics, Faculty of Science and Engineering, Konan University, 8-9-1 Okamoto, Kobe, Hyogo 658-8501, Japan
e-mail: arthur.choplin@konan-u.ac.jp
\and Geneva Observatory, University of Geneva, Maillettes 51, CH-1290 Sauverny, Switzerland
\and Kavli Institute for the Physics and Mathematics of the
Universe (WPI), The University of Tokyo, 5-1-5 Kashiwanoha, Kashiwa, Chiba
277-8583, Japan
\and Astronomical Institute, Tohoku University, Aoba, Sendai 980-8578, Japan
                }
  
   \date{Received / Accepted}

  \abstract
   {The nature of the early generation of massive stars can be inferred by investigating the origin of the extremely metal-poor (EMP) stars, likely formed from the ejecta of one or a few previous massive stars.} 
   {We investigate the rotational properties of early massive stars by comparing the abundance patterns of EMP stars with massive stellar models including rotation.}
   {Low metallicity 20 $M_{\odot}$ massive stellar models with eight initial rotation rates between 0 and $70~\%$ of the critical velocity are computed. Explosions with strong fallback are assumed. The ejected material is considered to fit individually the abundance patterns of 272 EMP stars with $-4<$ [Fe/H] $<-3$.}
   {With increasing initial rotation, the [C/H], [N/H], [O/H], [Na/H], [Mg/H], and [Al/H] ratios in the massive star ejecta are gradually increased (up to $\sim 4$~dex) while the $^{12}$C/$^{13}$C ratio is decreased.
   Among the 272 EMP stars considered, $\sim 40-50$~\% are consistent with our models. About $60 - 70~\%$ of the carbon-enhanced EMP star sample can be reproduced against $\sim 20 - 30~\%$ for the carbon-normal EMP star sample. The abundance patterns of carbon-enhanced EMP stars are preferentially reproduced with a material coming from mid to fast rotating massive stars. 
   The overall velocity distribution derived from the best massive star models increases from no rotation to fast rotation. The maximum is reached for massive stars having initial equatorial velocities of $\sim 550 - 640$ km~s$^{-1}$.}
   {Although subject to significant uncertainties, these results suggest that the rotational mixing operating in between the H-burning shell and the He-burning core of early massive stars played an important role in the early chemical enrichment of the Universe. The comparison of the velocity distribution derived from the best massive star models with velocity distributions of nearby OB stars suggests that a greater number of massive fast rotators were present in the early Universe. This may have important consequences for reionization, the first supernovae, or integrated light from high redshift galaxies.}


\titlerunning{The velocity of early massive stars}
\authorrunning{A. Choplin et al.}

   \maketitle

\section{Introduction}\label{intro}

The long-dead early massive stars are the key objects that synthesized the first metals, contributed to  cosmic reionization, and produced the first cosmic explosions \citep[e.g.][and references therein]{tumlinson04, karlsson13, nomoto13}.
Their initial masses, rotational velocities, multiplicity, or magnetic fields were likely the most important parameters that controlled their evolution and death \citep[e.g.][]{yoon05, heger10, maeder12, langer12}. 

Different cosmological simulations have suggested that the first stars were predominantly massive\footnote{Here massive refers to stars with initial mass greater than $8~M_{\odot}$} \citep[generally $\gtrsim 20$~$M_{\odot}$, e.g.][]{abel02, bromm02, bromm04, hirano14, hosokawa16}, but  with  a mass distribution possibly extending towards much lower masses \citep[e.g.][]{clark11, susa13, susa14, stacy16}. 
By recording the angular momentum of the sink particles falling into the growing protostar, \cite{stacy11} have shown that initial velocities of 1000~km~s$^{-1}$ or higher can be reached for Population III (Pop III) stars with $M_{\rm ini}$~$\geq 30$~$M_{\odot}$. 
\cite{hirano18} studied the angular momentum transfer in primordial discs including magnetic fields and suggested that the final rotational state of Pop III protostars may exhibit a net bimodality: either the protostar does not rotate at all or it is a fast rotator that is  close to break-up speed.

Observations can greatly help to infer the nature of early massive stars, either by trying to catch the most distant galaxies or transients \citep[e.g.][]{whalen13a, salvaterra15, oesch16, moriya19} or by observing the still alive nearby extremely metal-poor (EMP) small mass stars, likely formed $\sim 10-14$ Gyr ago \citep[e.g.][]{hill02, sneden03, cayrel04, beers05, norris13, frebel15, starkenburg18}. 
The numerous surveys of the past few decades progressively revealed a population of metal-poor stars, now containing about 500 objects with\footnote{[X/Y]~$= \log_{10}(N_{\rm X} / N_{\rm Y})_{\star} - \log_{10}(N_{\rm X} / N_{\rm Y})_{\odot}$ with $N_{\rm X}$ and $N_{\rm Y}$ the number density of elements X and Y in the observed star and in the Sun.} [Fe/H] $<-3$ \citep[from the SAGA database and JINAbase,][]{suda08, suda17, abohalima18}. Several stars with [Fe/H] $\lesssim -5$ have been observed \citep{christlieb02, keller14, bonifacio15, frebel05, frebel15b, frebel19, aguado18a}, but no metal-free stars were found.
Among the metal-poor stars, many were found to have a super-solar carbon-to-iron ratio \citep[e.g.][]{beers05, aoki07}.
They are called carbon-enhanced metal-poor (CEMP) stars and are generally defined as having [C/Fe] $>0.7$. Their frequency increases with decreasing [Fe/H] to reach $\sim 40~\%$ ($\sim 80~\%$) for [Fe/H] $<-3$ \citep[\text{[}Fe/H\text{]}~$<-4$,][]{placco14c}. Even so,  3D/NLTE effects may  lead to an overestimation of the [C/Fe] ratio and may therefore  significantly affect the CEMP fraction \citep{norris19}. 

At [Fe/H] $\gtrsim -3$, numerous stars show overabundances in both light elements (e.g. carbon) and heavy elements (e.g. barium), which are generally expected to have been acquired from a now extinct AGB companion during a mass-transfer (or wind-mass-transfer) episode \citep[e.g.][]{stancliffe08, bisterzo10, bisterzo12, lugaro12, abate13, abate15a}. This scenario is supported by a large binary fraction among these stars 
\citep{hansen16a}. The enhanced s-process operating in rotating massive stars may also be at the origin of some metal-poor stars enriched in s-elements \citep{choplin17letter, banerjee19}.

Extremely
metal-poor stars ([Fe/H] $<-3$), which are often considered to be the most pristine objects, generally do not show strong overabundances in heavy elements. 
They likely formed from a gas cloud that was enriched by one or a few previous massive stars \citep[e.g.][]{umeda02, limongi03, meynet06, hirschi07, tominaga14, chiaki19}. The surface chemical composition of EMP stars provides a window on the physical processes and nature of the first generation of massive stars. 
Among the CEMP stars with [Fe/H] $<-3$, many are CEMP-no stars (CEMP not strongly enriched in s- or r-process elements), generally defined as having [Ba/Fe] $<0$ \citep{beers05}.

Comparisons between the chemical composition of EMP stars and ejecta from massive star models lead to various studies that investigated the nature of early massive stars and their supernovae (SNe), including  studies considering mixing and fallback in massive Pop III stars \citep[e.g.][]{umeda02,umeda03, nomoto03, iwamoto05, ishigaki14}, rotating massive stars \citep[e.g.][]{meynet06, meynet10, hirschi07, maeder14, takahashi14}, jet-induced SNe from Pop III $25-40$ $M_{\odot}$ stars \citep{maeda03, tominaga09, ezzeddine19}, proton ingestion events in the He-shell during late evolutionary stages \citep{banerjee18, clarkson18}, and enrichment by more than one source \citep{limongi03}. 
No consensus has been reached yet. The need for multiple kinds of progenitors has been raised several times \citep[e.g.][]{yoon16, placco16b}.
In particular, by inspecting the CEMP-no sample morphology in the A(C)-[Fe/H] diagram, \cite{yoon16} divided the CEMP-no sample into two groups and suggested that their existence may indicate the need for more than one kind of stellar progenitor. 
The two groups may also be explained because of two different  dust-cooling regimes during the EMP star formation \citep{chiaki17}. 
With the increasing number of metal-poor stars available, it is now beginning to be possible to perform extensive abundance fitting studies between models and metal-poor stars and to infer the characteristics of the early massive star populations. In this way \cite{ishigaki18} derived an initial mass function for Pop III stars, peaking at $\sim 25$~$M_{\odot}$. 

In massive stars, rotation may considerably affect the evolution and nucleosynthesis \citep[e.g.][]{heger00, meynet00b, brott11, ekstrom12, georgy13, langer12, chieffi13}, as well as the final fate \citep[e.g.][]{woosley93, macfadyen99}. 
Nucleosynthesis may be particularly impacted during the He-burning stage: the rotational mixing operating between the He-burning core and H-burning shell triggers exchanges of material, leading a rich nucleosynthesis \citep[e.g.][see also Sect.~\ref{mixnuc}]{maeder15a, choplin16}. Light elements (mainly from C to Al) and heavy elements (mainly from Fe to Ba, possibly to Pb) are overproduced \citep[e.g.][]{takahashi14, frischknecht16, limongi18, choplin18}. In particular, if  different initial rotation rates
are considered, this process is likely able to cover a wide variety of [C/Fe], [N/Fe], [O/Fe], [Na/Fe], [Mg/Fe], and [Al/Fe] ratios.
\cite{maeder15a} proposed that this back and forth mixing process could be responsible for the peculiar  abundances of CEMP-no stars with [Fe/H]~$<-2.5$. 
It can be motivated by the fact that the [X/Ca] scatter (or [X/Fe] scatter, which is similar) of EMP stars is of the order of $3-4$ dex for C, N, and O and about $1.5-2.5$ dex for Na, Mg, and Al. From Si, the dispersion becomes smaller \citep{frebel15, bonifacio15}. 

In this work we investigate how the yields of $20$~$M_{\odot}$ massive stars models (also called the source stars) with different initial velocities and experiencing explosions with strong fallback
compare with the abundance patterns (considering C, N, O, Na, Mg, and Al) of 272 EMP stars in the range $-4<$ [Fe/H] $<-3$. 
The abundance fitting of each of these EMP stars allows us to determine the characteristics of the best source stars, particularly their initial velocity distribution. This distribution is compared to distributions based on the observation of nearby OB stars. The mixing processes (dredge-up, thermohaline mixing) that may have altered the initial surface chemical composition of EMP stars are taken into account. 

The paper is organized as follows. In Sect.~\ref{models} we discuss the models, especially the interplay between rotation and nucleosynthesis. In Sect.~\ref{empcomp} we introduce the EMP sample and fitting method. We present the main results   and discuss them in Sect.~\ref{res} and Sect.~\ref{disc}, respectively.

\section{Source star models}\label{models}

\subsection{Physical ingredients}\label{physing}

The source star models considered here were computed with the Geneva stellar evolution code \citep[e.g.][]{eggenberger08}. 
We computed 20~$M_{\odot}$ models with $\upsilon_{\rm ini}/ \upsilon_{\rm crit} =$ 0, 0.1, 0.2, 0.3, 0.4, 0.5, 0.6, and 0.7\footnote{The critical velocity $\upsilon_{\rm crit}$ is reached when the gravitational acceleration is counterbalanced by the centrifugal force. It is expressed as $\upsilon_{\rm crit} = \sqrt{\frac{2}{3}\frac{GM}{R_{\rm p,c}}}$ with $R_{\rm p,c}$ the polar radius at the critical limit.}. 
An initial mass of 20~$M_{\odot}$ may be considered   a representative mass of a standard massive star population, and might have been a typical initial mass during the era of the first stars \citep[e.g.][]{susa14, ishigaki18}.
The initial source star metallicity was set to $Z=10^{-5}$ and the initial metal mixture is $\alpha$-enhanced \citep[details can be found in Sect.~2.1 of][]{frischknecht16}. The initial abundances are summarized in Table~\ref{table:1}.
The prescription for radiative mass-loss rates is from \cite{vink01} when $\log T_{\rm eff} \geq 3.9$. The radiative mass loss scales with metallicity as $\dot{M} \propto Z^{0.85}$. When $\log T_{\rm eff} < 3.9$, the mass-loss recipe from \cite{jager88} is used. $D_{\rm shear}$ is from \cite{talon97} and $D_{\rm h}$ from \cite{zahn92}. 
The efficiency of rotational mixing is calibrated such that the surface N/H value at core H depletion of a 15~$M_{\odot}$ model at solar metallicity with $\upsilon_{\rm ini} = 300$ km~s$^{-1}$ is enhanced by a factor of 3 compared to its initial surface N/H value. These surface enrichments qualitatively agree with observation of $10 - 20$~$M_{\odot}$ rotating stars \citep[e.g.][]{gies92, villamariz05, hunter09}. 

This study focuses on the light elements which are thought to be significantly affected by rotational mixing during the life of the star (C, N, O, F, Ne, Na, Mg, and Al; see Introduction). A minimal reaction network containing $n$, $^{1}$H, $^{3,4}$He, $^{12,13,14}$C, $^{14,15}$N, $^{16,17,18}$O, $^{18,19}$F, $^{20,21,22}$Ne, $^{23}$Na, $^{24,25,26}$Mg, $^{26,27}$Al, $^{28}$Si, $^{32}$S, $^{36}$Ar, $^{40}$Ca, $^{44}$Ti, $^{48}$Cr, $^{52}$Fe, and $^{56}$Ni is used. 
It allows us to follow the abundances of the elements of interest together with the reactions that contribute significantly to generate nuclear energy \citep[e.g.][]{hirschi04, ekstrom12}. 
The important nuclear reaction rates for this work are from \cite{angulo99} for the CNO cycle, except $^{14}$N($p,\gamma$)$^{15}$O, which is  from \cite{mukhamedzhanov03}. The rates related to the Ne-Na and Mg-Al cycles are from \cite{iliadis01} except $^{19}$F($p,\gamma$)$^{20}$Ne \citep{angulo99}, $^{20}$Ne($p,\gamma$)$^{21}$Na \citep{angulo99}, $^{22}$Ne($p,\gamma$)$^{23}$Na \citep{hale02}, and $^{27}$Al($p,\gamma$)$^{28}$Si \citep{cyburt10}. Other important reactions for the present work are $^{22}$Ne($\alpha,n$)$^{25}$Mg \citep{jaeger01}, $^{22}$Ne($\alpha,\gamma$)$^{26}$Mg \citep{jaeger01}, $^{17}$O($\alpha,n$)$^{20}$Ne \citep{angulo99}, and $^{17}$O($\alpha,\gamma$)$^{21}$Ne \citep{angulo99}.

The models are computed until the end of the core oxygen burning phase (when the central $^{16}$O mass fraction drops below $10^{-4}$). 
Table~\ref{table:2} shows the main characteristics of the models, which are labelled  vvX, where X refers to the initial rotation rate.

\subsection{Explosion with strong fallback}\label{secfallback}

Hydrodynamical calculations of core-collapse supernovae show that the explosion energy, gravitational potential, and asphericity control the amount of material falling back into the compact object \citep{fryer07, moriya10}. 
Large fallback can occur for low explosion energies, but also for energetic jet-like explosions where the matter around the jet axis is ejected while the matter around the equatorial plane experiences significant fallback onto the central remnant \citep{maeda03, tominaga09}.
It has been suggested that the light curve of SN2008ha could be explained by a 13~$M_{\odot}$ star that experienced a large amount of fallback \citep{moriya10}. 
Low or zero metallicity massive stars may experience a higher degree of fallback compared to solar metallicity stars because they are more compact, and   consequently might explode with more difficulty   \citep[e.g.][]{woosley02}.  Employing a piston at the edge of the iron-core with $E_{51} = 1.2$, \cite{woosley95} reported remnant masses of $\sim 2$ and $\sim 4$~$M_{\odot}$ for a solar and zero metallicity 20~$M_{\odot}$ model, respectively.
The idea of a large fallback has also been suggested  to account for the abundances of the most metal-poor stars \citep[e.g.][]{umeda02}.
In particular, \cite{iwamoto05} found that the patterns of HE1327-2326 ([Fe/H] $=-5.6$) and HE0107-5240 ([Fe/H] $=-5.2$) may be explained by Pop III 25~$M_{\odot}$ SNe with $E_{51} \sim 0.7$ and with\footnote{At the time of the explosion $M_{\rm cut}$, the mass cut is the mass coordinate delimiting the part of the star that is expelled from the part that is locked into the remnant.} $M_{\rm cut} \sim 6$~$M_{\odot}$ \citep[they also assumed mixing at the time of the explosion, following][]{umeda02}.

In this work, the source stars are supposed to experience explosions with strong fallback. 
Only the layers above the C-burning shell (typically above $4~M_{\odot}$ for a 20~$M_{\odot}$) are supposed to be expelled, without having been significantly processed by explosive nucleosynthesis.
The EMP stars subsequently formed with some of this ejecta plus some of the initial background of metals (among it the elements from $\sim$~Si to $\sim$~Fe) left by one or a few previous source(s), possibly Pop III massive stars.

Although explosions with strong fallback may be a common phenomena in the early Universe, more standard SNe should also be expected. 
Considering solely explosions with strong fallback is likely an important limitation of this study.
Considering various kinds of explosions should provide additional solutions (unseen in this study) for reproducing the abundance patterns of EMP stars.
The assumption regarding the explosion made here allows us to test to what extent rotating models experiencing explosions with strong fallback can reproduce the abundance patterns of EMP stars.

   \begin{figure*}
   \centering
   \begin{minipage}[c]{.33\linewidth}
       \includegraphics[scale=0.3]{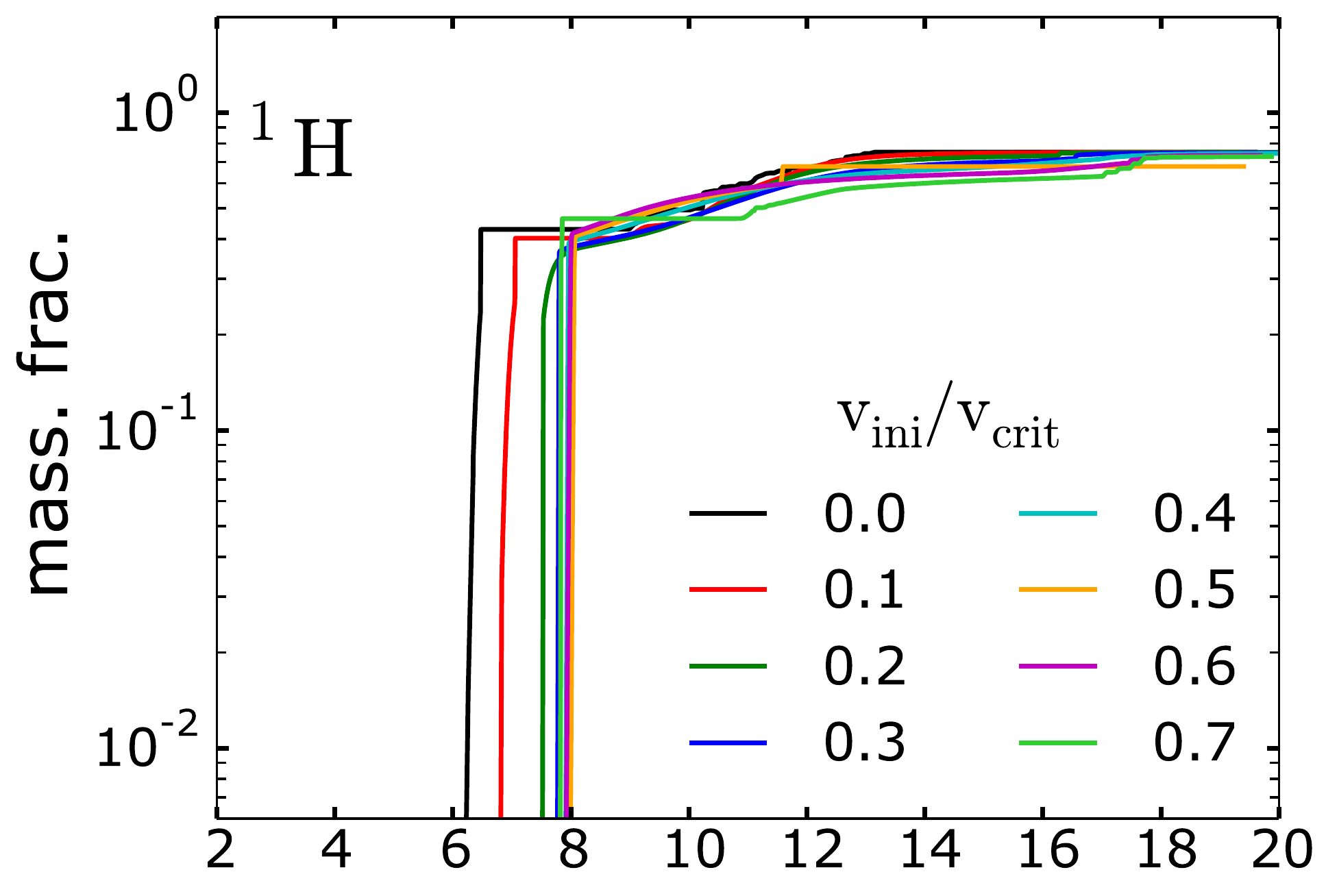}
   \end{minipage}
   \begin{minipage}[c]{.33\linewidth}
       \includegraphics[scale=0.3]{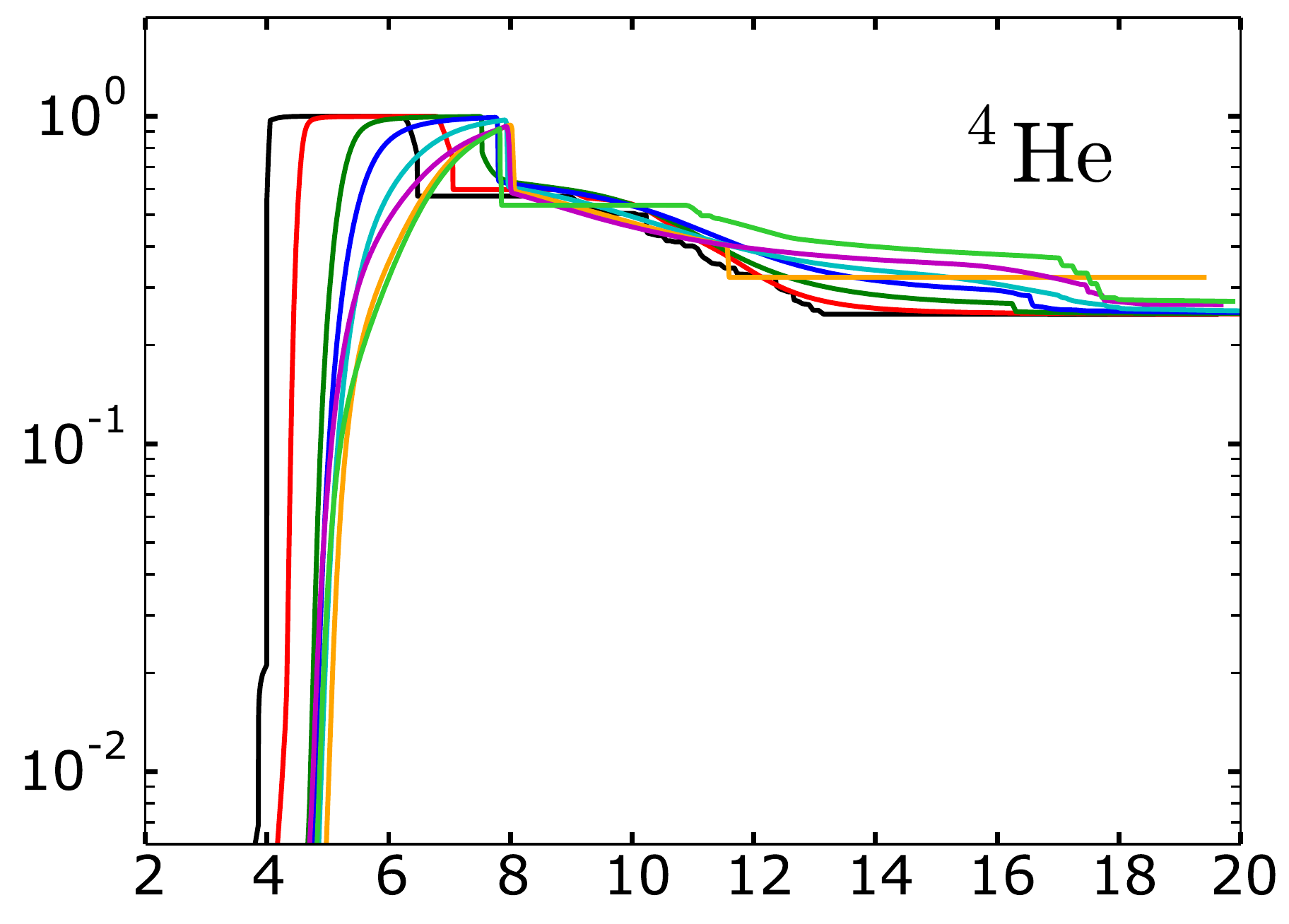}
   \end{minipage}
   \begin{minipage}[c]{.33\linewidth}
       \includegraphics[scale=0.3]{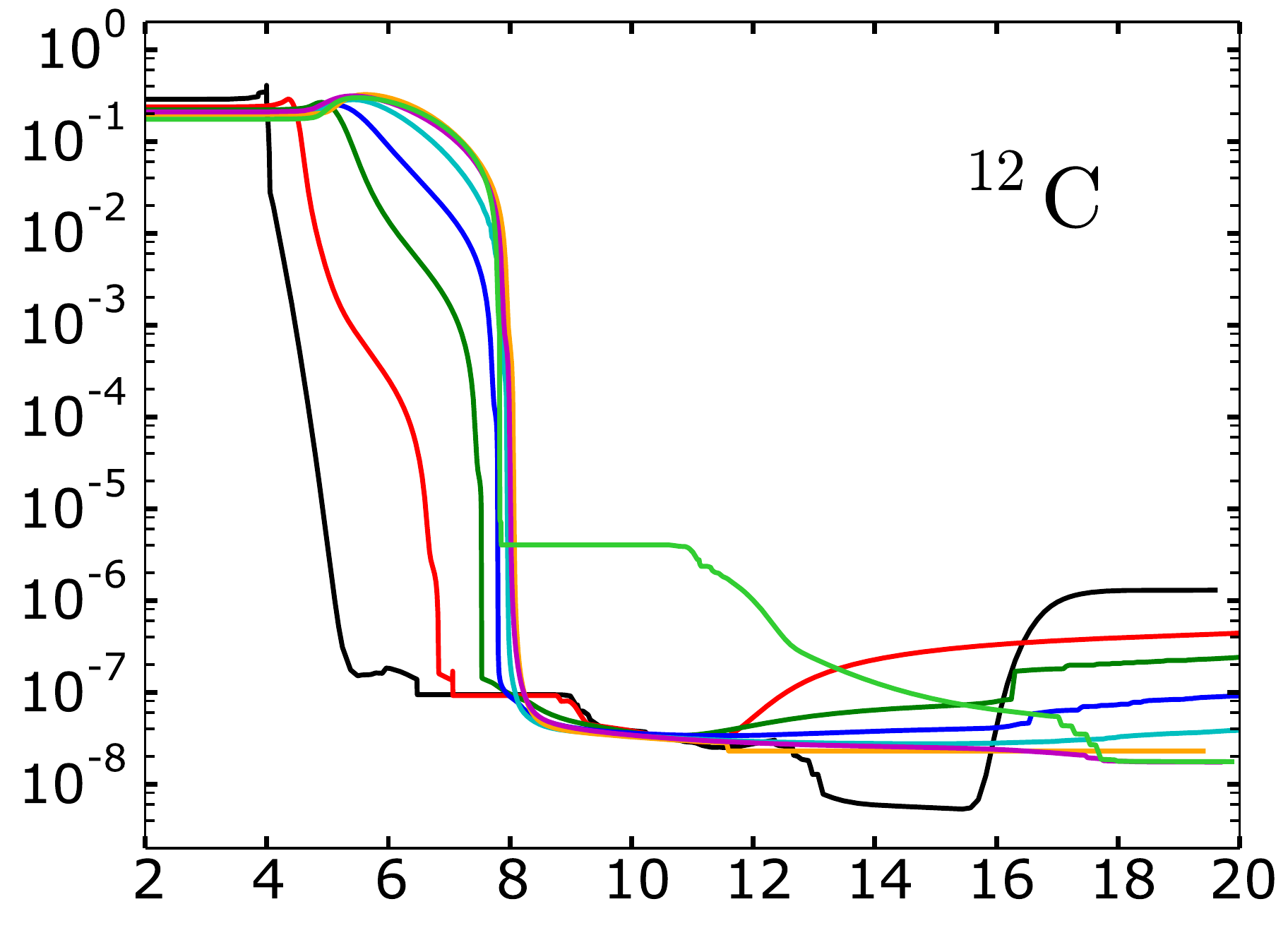}
   \end{minipage}
   \begin{minipage}[c]{.33\linewidth}
       \includegraphics[scale=0.3]{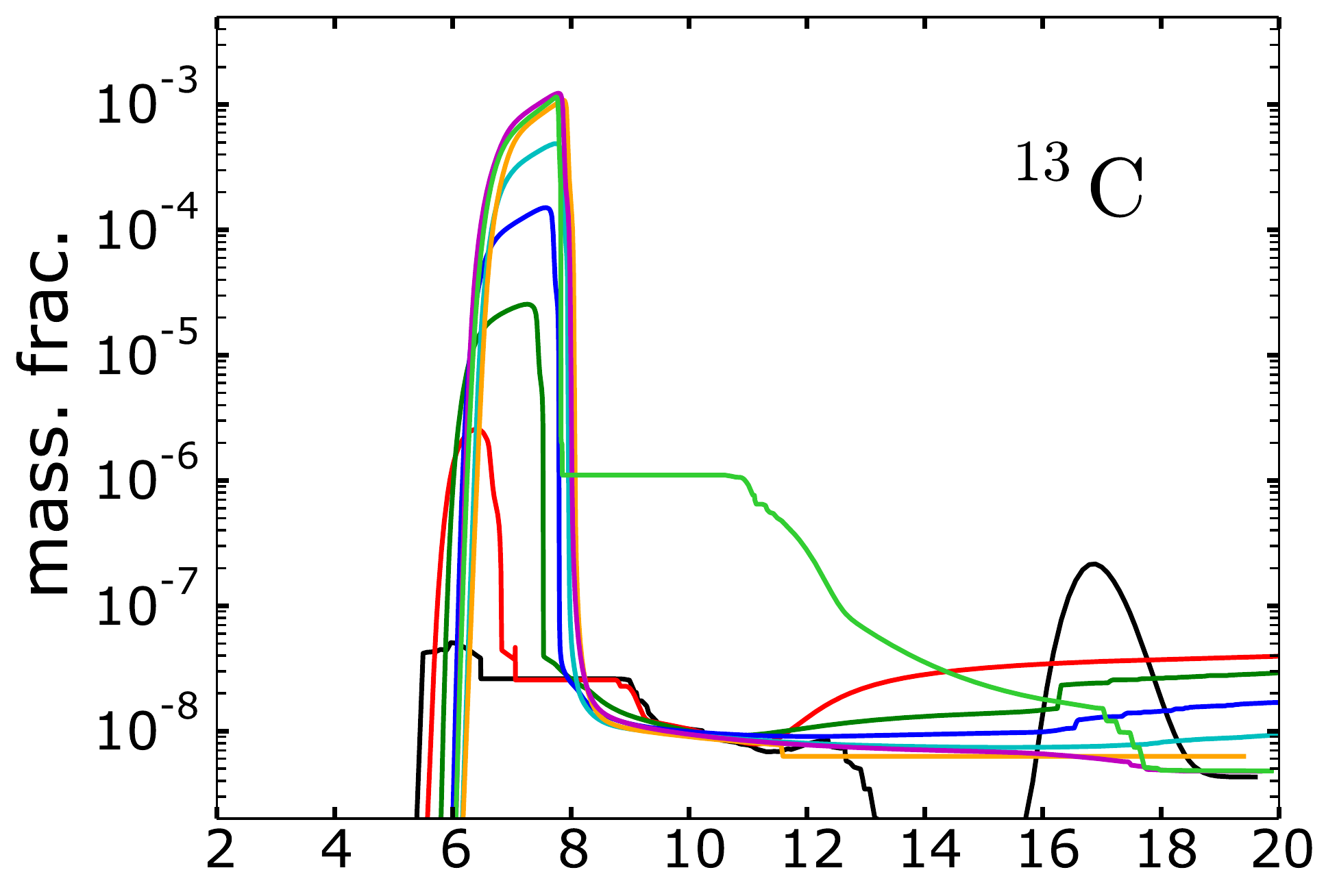}
   \end{minipage}
   \begin{minipage}[c]{.33\linewidth}
       \includegraphics[scale=0.3]{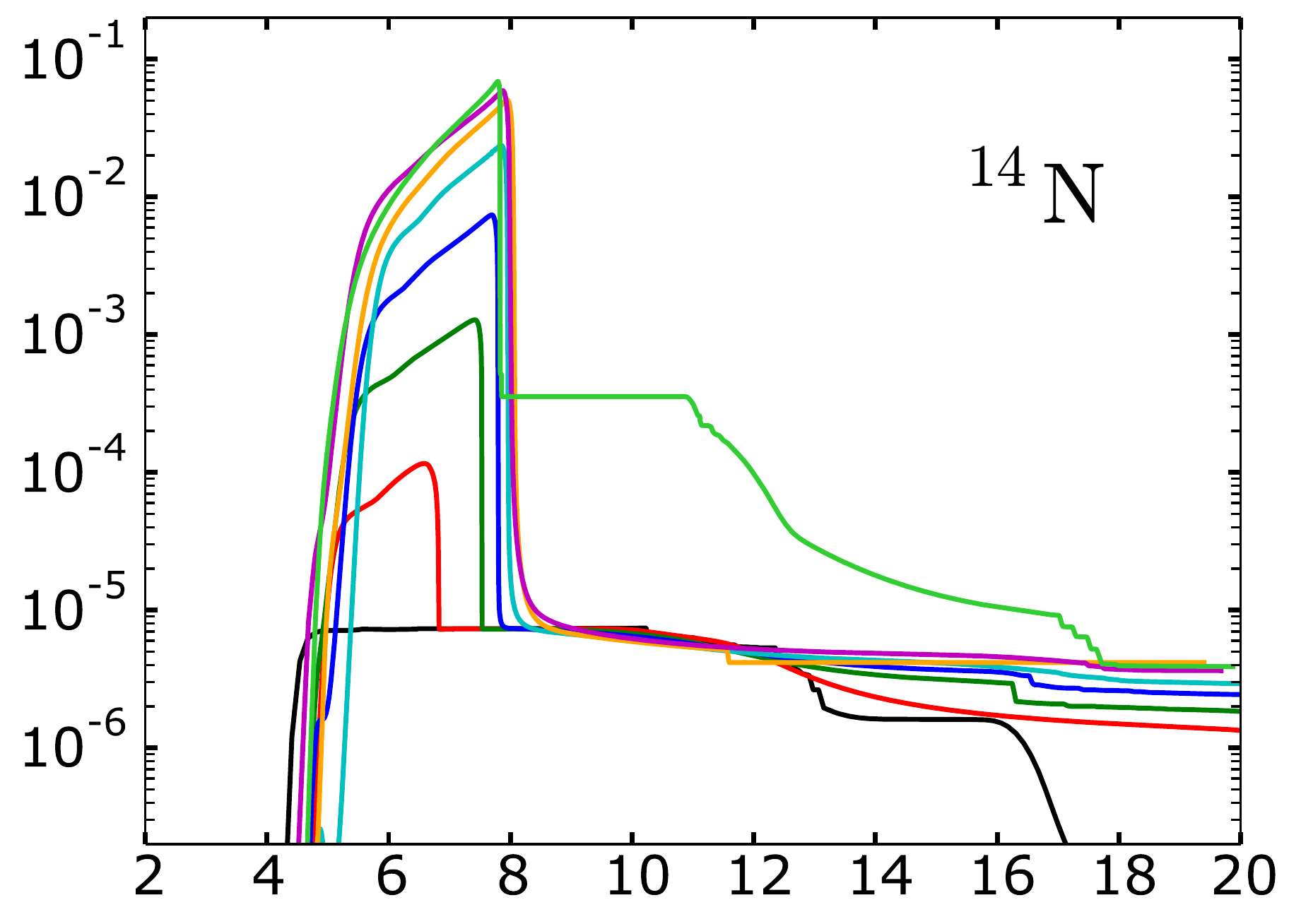}
   \end{minipage}
   \begin{minipage}[c]{.33\linewidth}
       \includegraphics[scale=0.3]{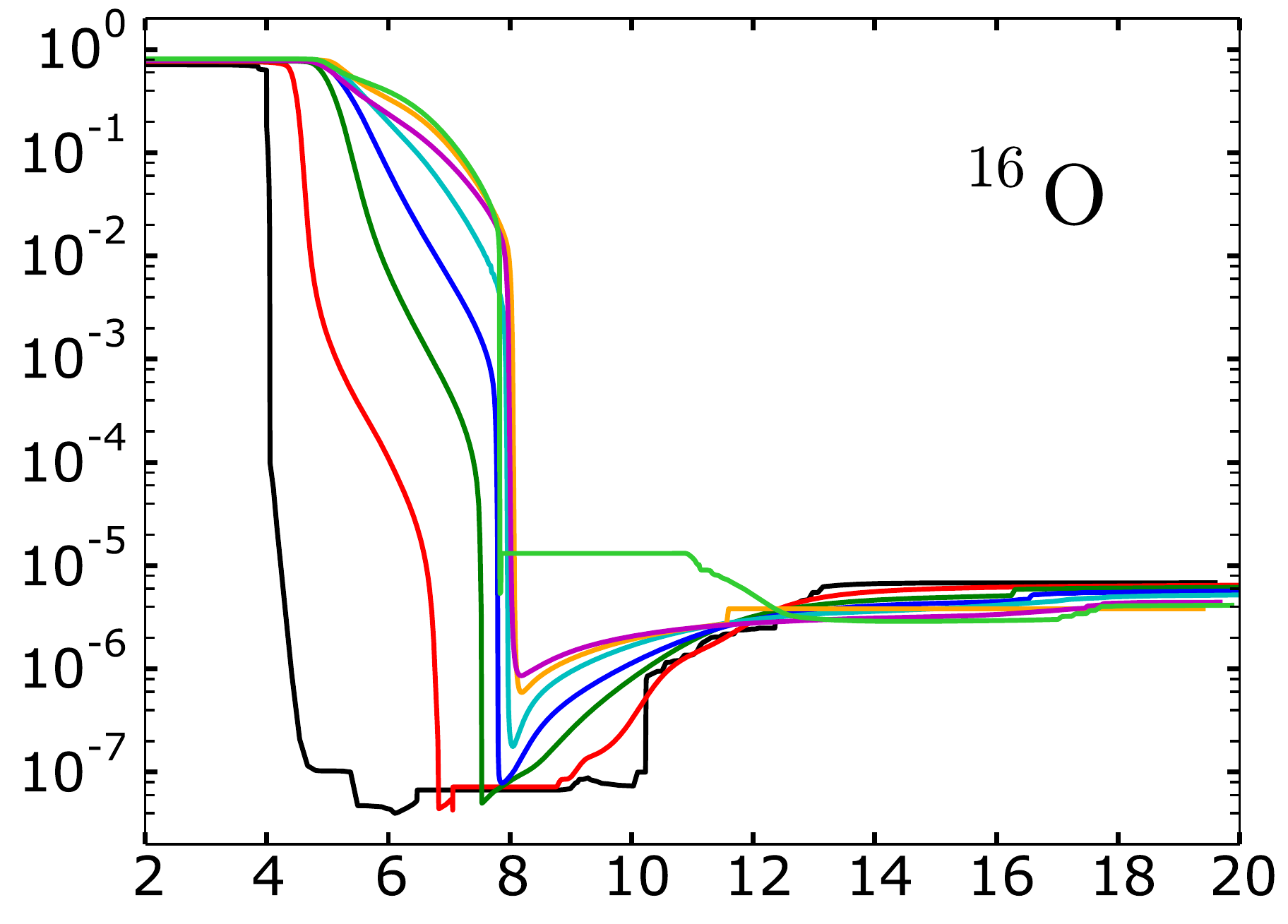}
   \end{minipage}
   \begin{minipage}[c]{.33\linewidth}
       \includegraphics[scale=0.3]{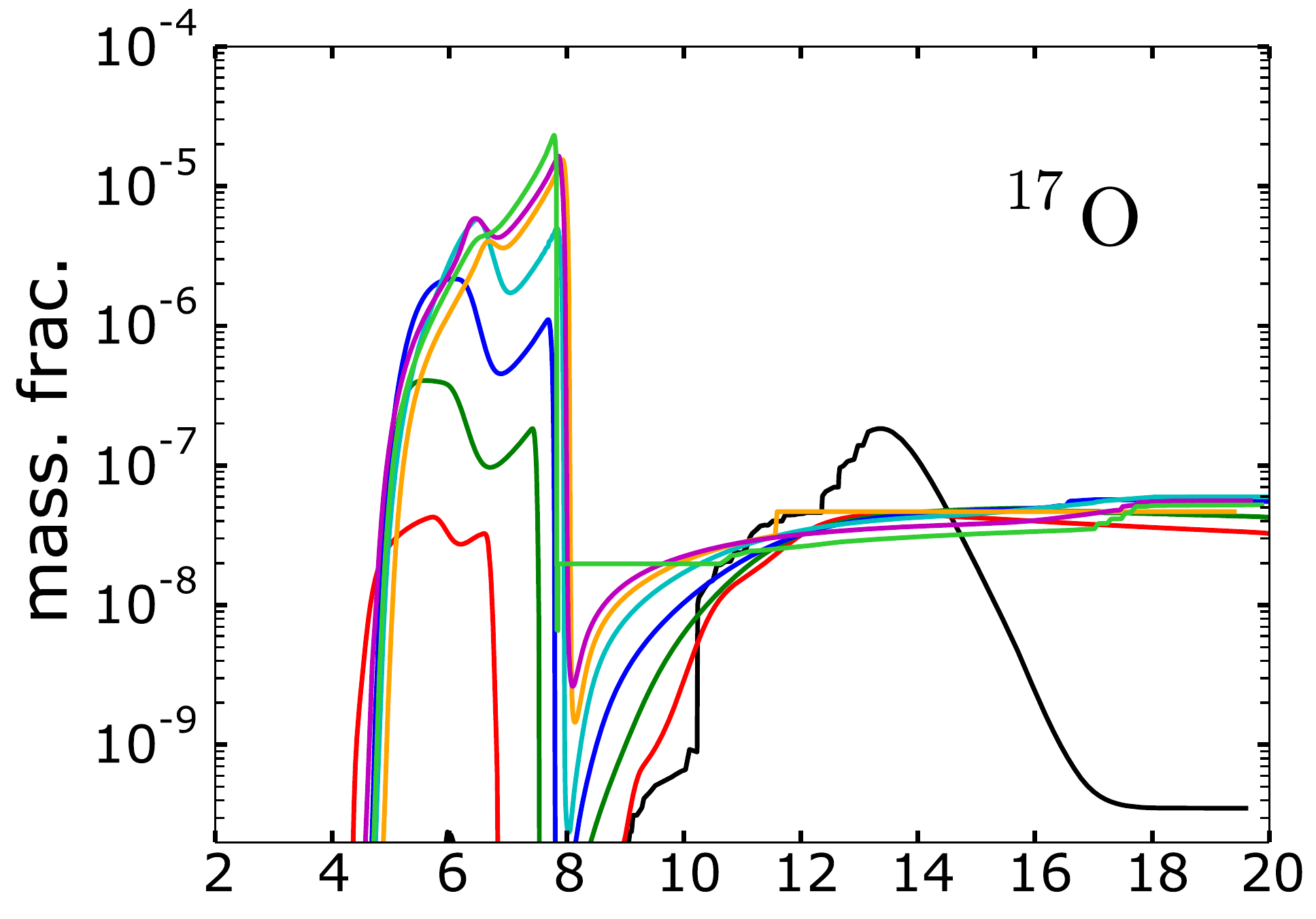}
   \end{minipage}
      \begin{minipage}[c]{.33\linewidth}
       \includegraphics[scale=0.3]{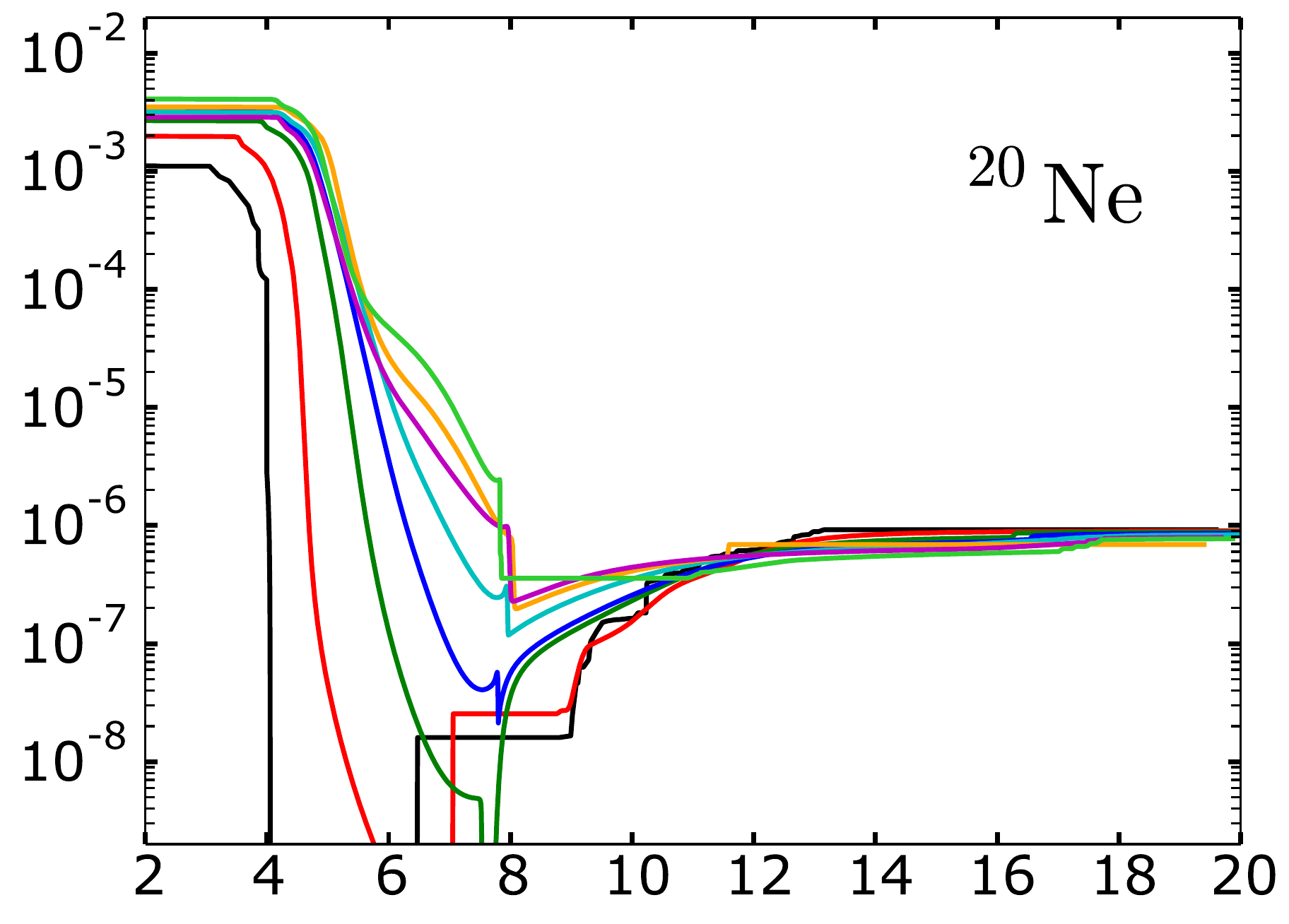}
   \end{minipage}
   \begin{minipage}[c]{.33\linewidth}
       \includegraphics[scale=0.3]{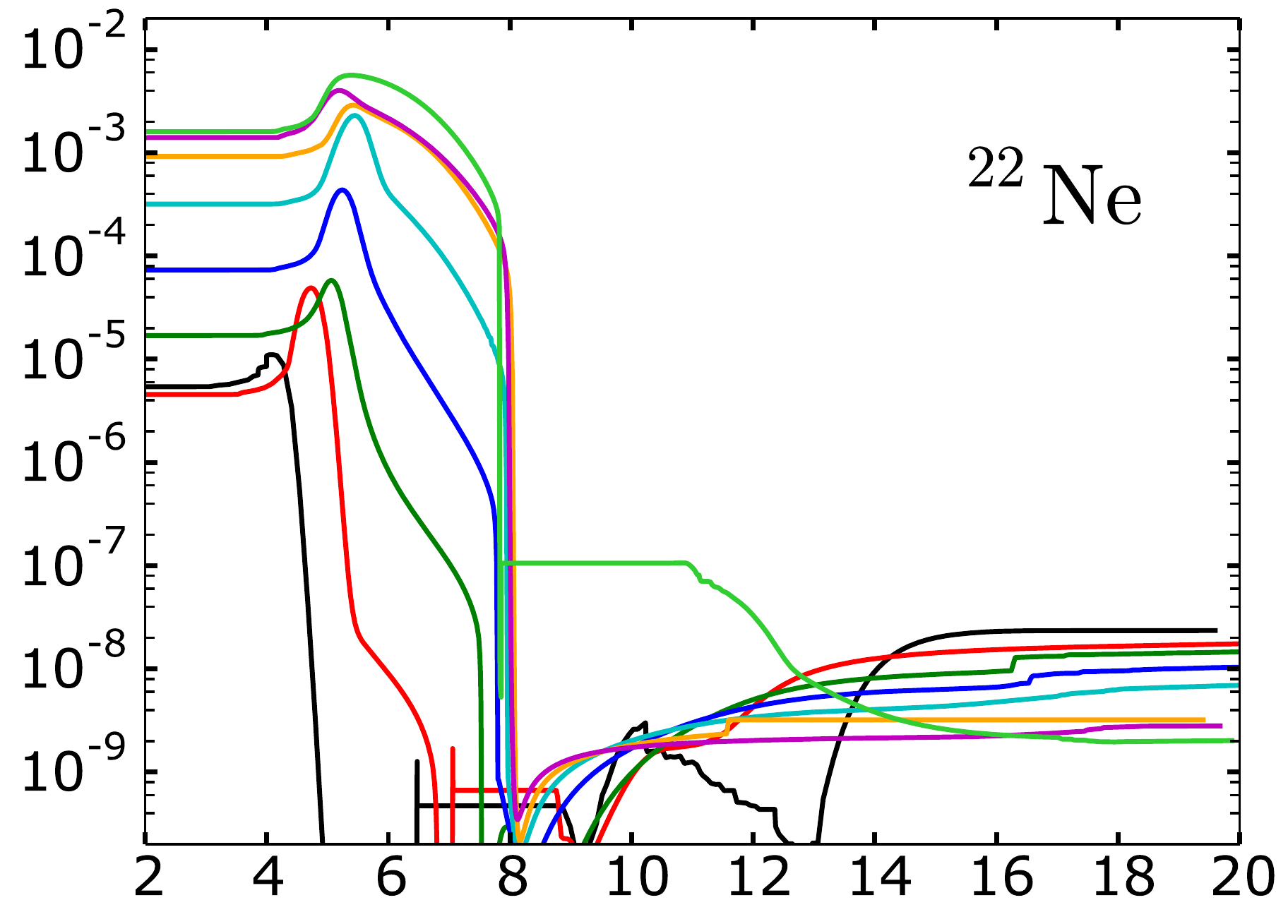}
   \end{minipage}
   \begin{minipage}[c]{.33\linewidth}
       \includegraphics[scale=0.3]{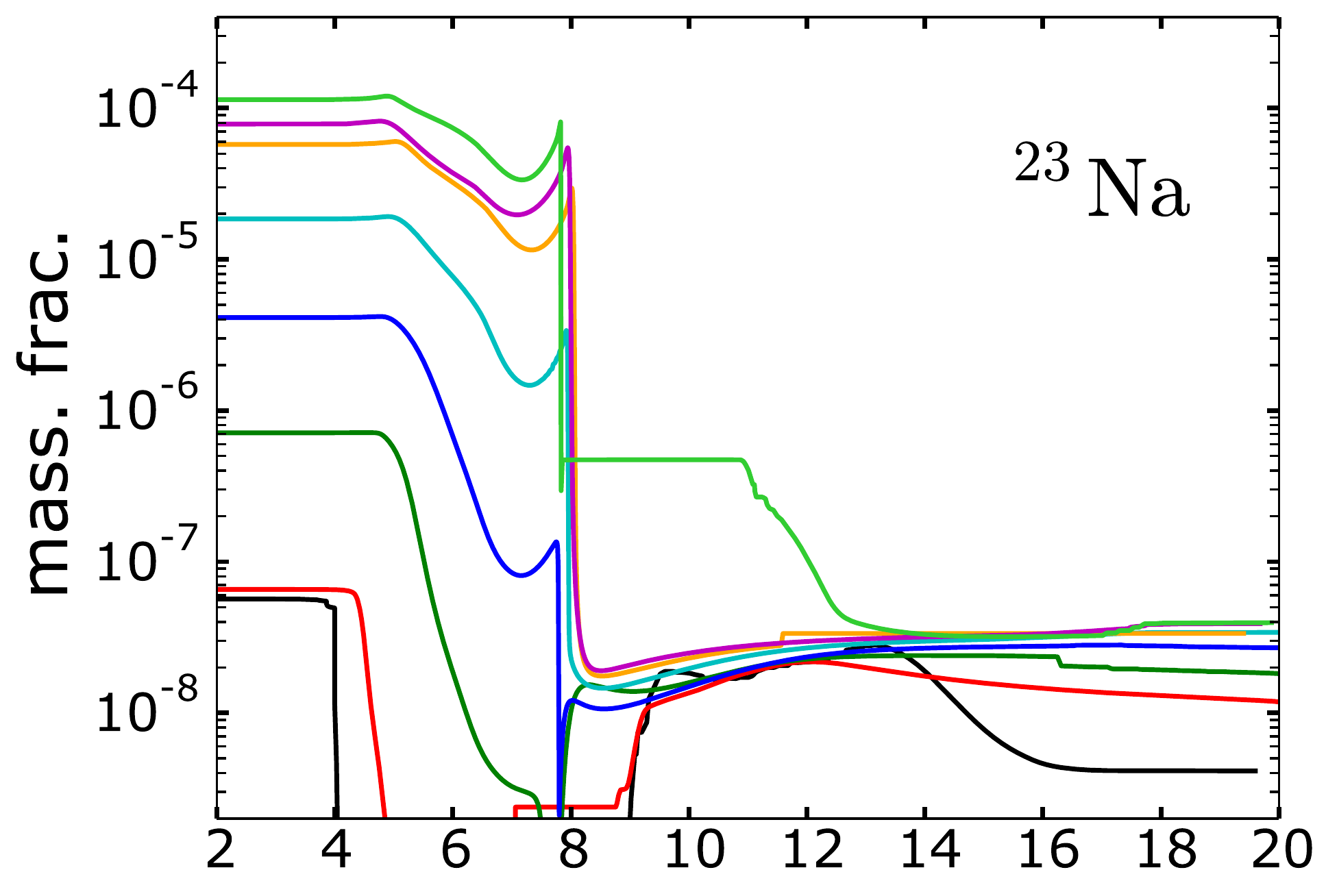}
   \end{minipage}
   \begin{minipage}[c]{.33\linewidth}
       \includegraphics[scale=0.3]{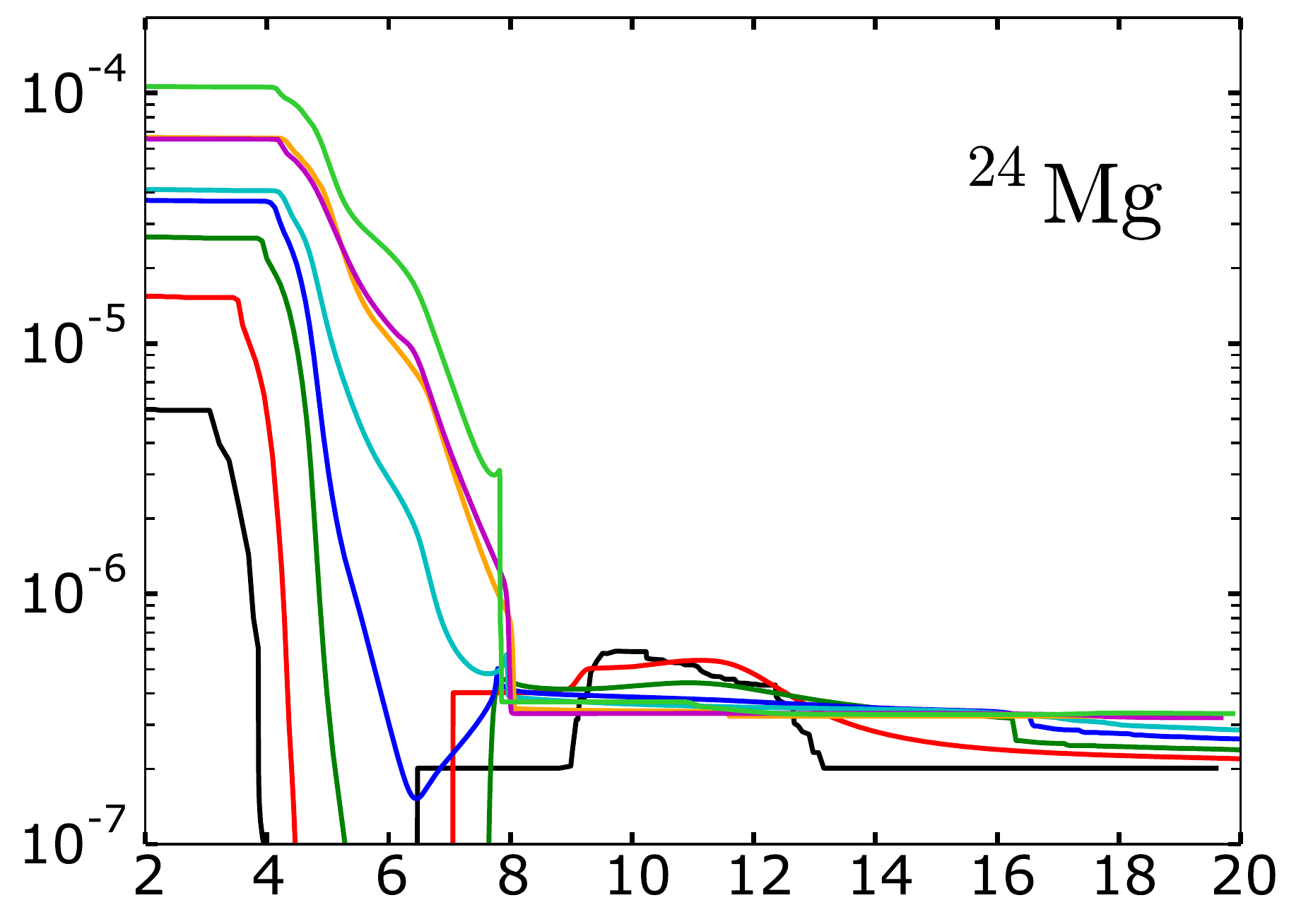}
   \end{minipage}
   \begin{minipage}[c]{.33\linewidth}
       \includegraphics[scale=0.3]{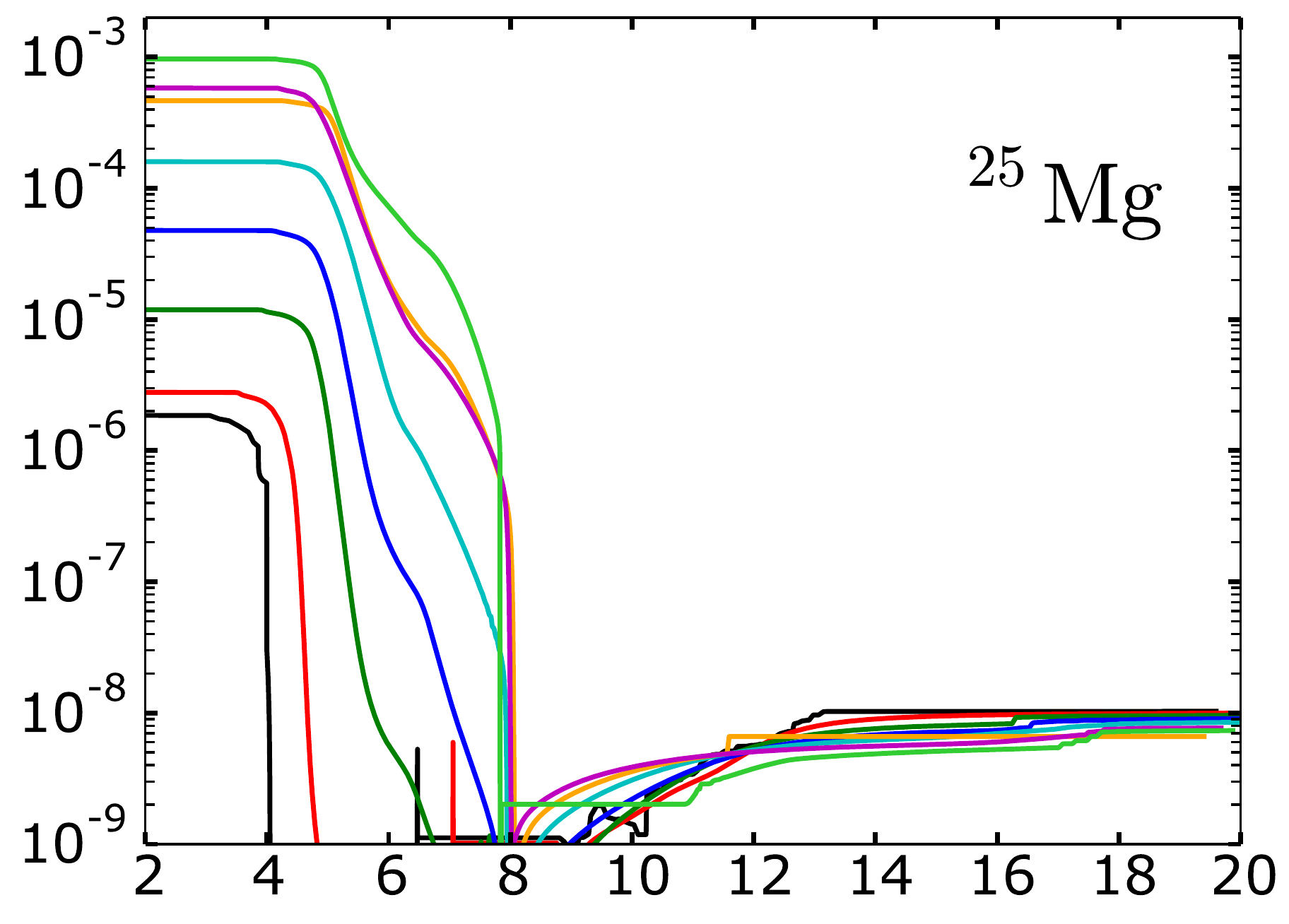}
   \end{minipage}
   \begin{minipage}[c]{.33\linewidth}
       \includegraphics[scale=0.3]{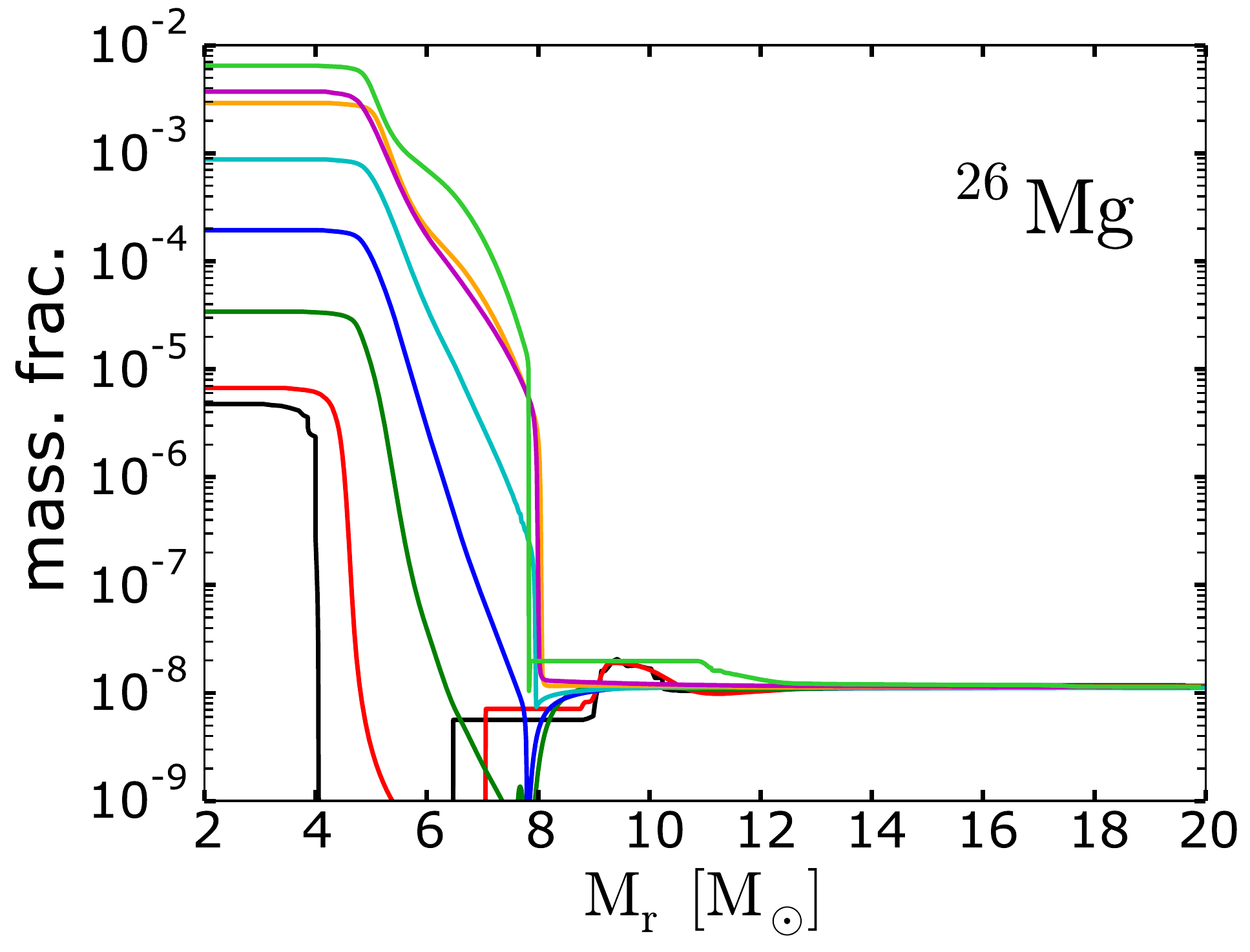}
   \end{minipage}
   \begin{minipage}[c]{.33\linewidth}
       \includegraphics[scale=0.3]{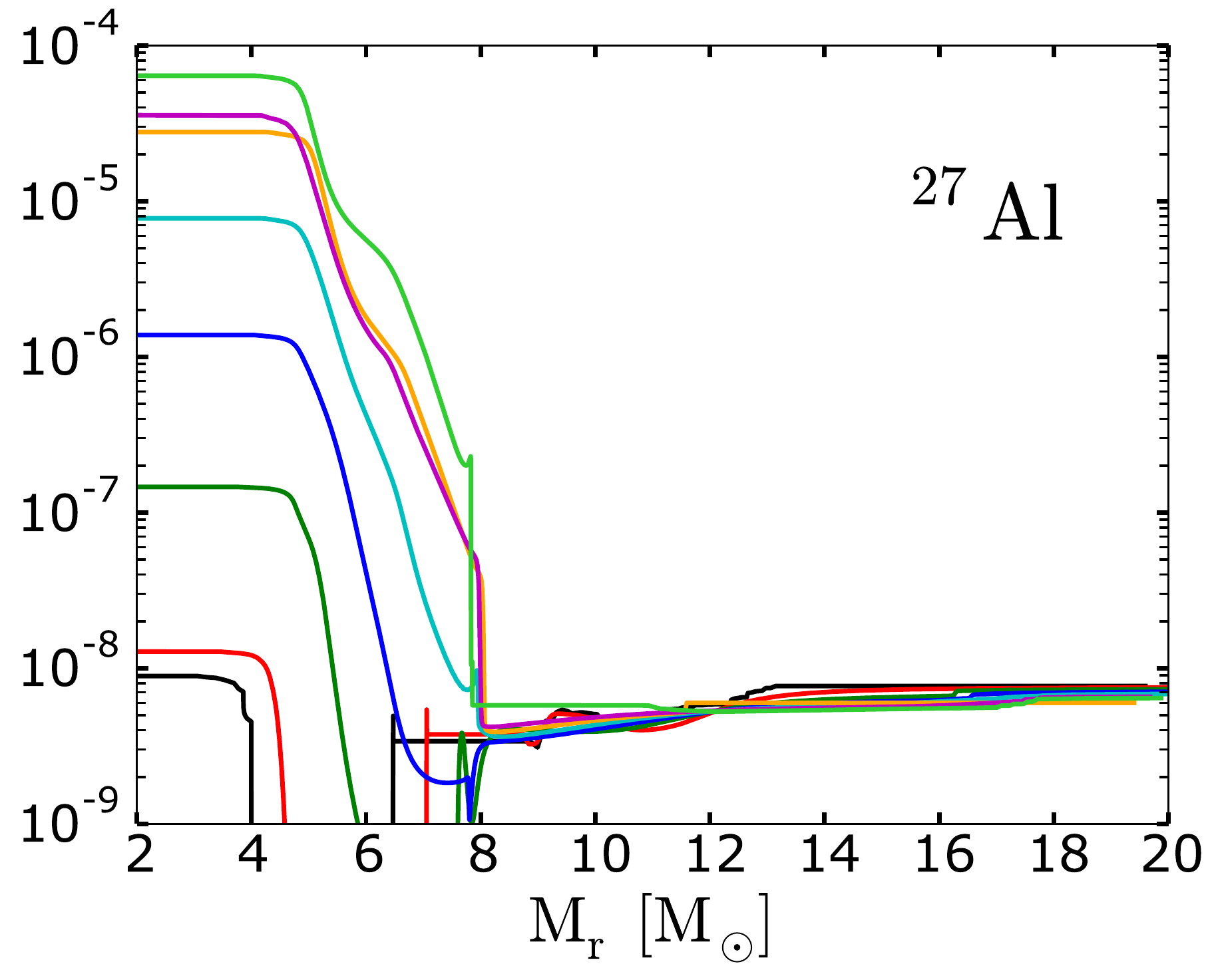}
   \end{minipage}
   \begin{minipage}[c]{.33\linewidth}
       \includegraphics[scale=0.3]{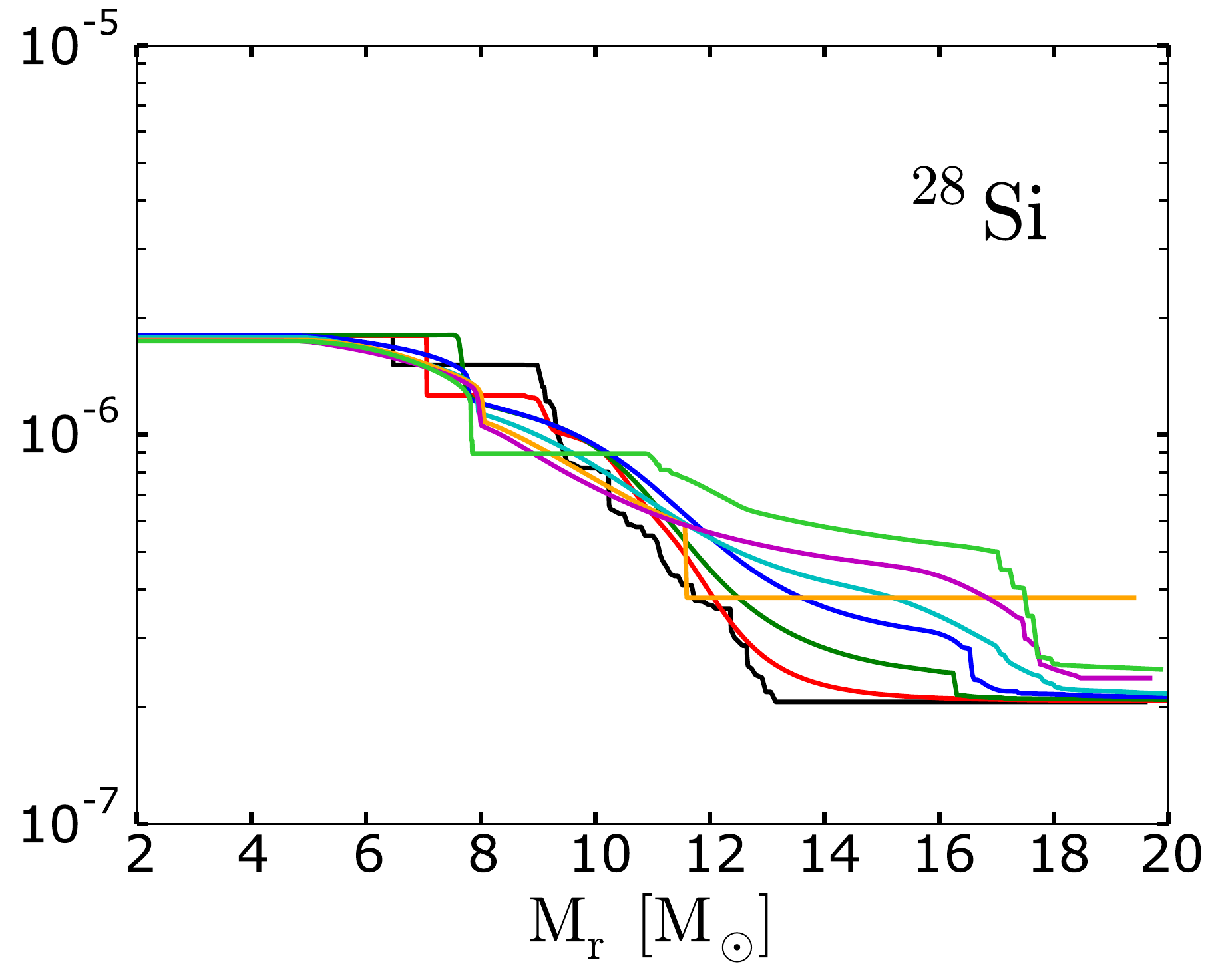}
   \end{minipage}
   \caption{
   Internal abundance profiles in mass fraction of the 20~$M_{\odot}$ models with various initial rotation rates, at the end of the core He-burning phase. 
   }
\label{abrotmod}
    \end{figure*}

  \begin{table}
\scriptsize{
\caption{Initial abundances (in mass fraction) of the stellar models. The last line gives the ratio of the initial sum of CNO elements in the models, compared to the Sun.
\label{table:1}}
\begin{center}
\resizebox{5.2cm}{!} {
\begin{tabular}{lc} 
\hline 
Isotope  & Mass fraction  \\
\hline 
$^{1}$H           &            7.516e-01             \\
$^{3}$He           &            4.123-05             \\
$^{4}$He          &            2.484e-01             \\
$^{12}$C           &          1.292e-06               \\
$^{13}$C           &          4.297e-09               \\
$^{14}$N           &           1.022e-07              \\
$^{15}$N           &           4.024e-10              \\
$^{16}$O           &           6.821e-06              \\
$^{17}$O           &            3.514e-10             \\
$^{18}$O           &           2.001e-09              \\
$^{19}$F           &            8.385e-11             \\
$^{20}$Ne           &          9.205e-07               \\
$^{21}$Ne           &         7.327e-10                \\
$^{22}$Ne           &          2.354e-08              \\
$^{23}$Na           &          4.135e-09               \\
$^{24}$Mg           &         2.012e-07                \\
$^{25}$Mg           &         1.030e-08                \\
$^{26}$Mg           &         1.179e-08                \\
$^{27}$Al           &           7.694e-09              \\
$^{28}$Si           &          2.060e-07               \\
 \hline
 (C+N+O) / (C+N+O)$_{\odot}$  &   9.444e-04   \\
 \hline
\end{tabular}
}
\end{center}
}
\end{table}

  \begin{table}
\scriptsize{
\caption{Properties of the 20 $M_{\odot}$ source star models: model label (column 1), $\upsilon_\text{ini}/\upsilon_\text{crit}$ (column 2), $\Omega_\text{ini}/\Omega_\text{crit}$ (column 3), initial equatorial velocity (column 4), mean equatorial velocity during the MS (column 5), total lifetime (column 6), mass of the model at the end of the evolution (column $7$). 
\label{table:2}}
\begin{center}
\resizebox{8.5cm}{!} {
\begin{tabular}{l|cccc|c|c} 
\hline 
Model  &  $\upsilon_\text{ini}/\upsilon_\text{crit}$  & $\Omega_\text{ini}/\Omega_\text{crit}$ & $\upsilon_\text{ini}$ & $ \langle \upsilon_\text{eq} \rangle_{\rm MS}$  & $\tau_\text{tot}$ & $M_\text{fin}$  \\ 
 &   & & [km/s] & [km/s] & [Myr] & [$M_\odot$]  \\
\hline 
vv0           &               0.0               &       0               &  0        &       0    &       8.94    &            19.998                    \\
vv1            &               0.1               &       0.15          &  88        &       70    &       9.43    &         19.998                   \\
vv2            &               0.2               &       0.32          &  188        &       146    &       9.81     &       19.960              \\
vv3            &               0.3               &       0.46         &  276        &       223    &       10.07     &       19.726             \\
vv4            &               0.4               &       0.69          &  364        &       302    &       10.29     &       19.625             \\
vv5            &               0.5               &       0.71           &  454        &       384    &       10.49    &       19.189             \\
vv6            &               0.6               &       0.81          &  547        &       470    &       10.65    &       19.222            \\
vv7            &               0.7               &       0.90       &  644        &       543    &       11.04    &          19.764              \\
 \hline
\end{tabular}
}
\end{center}
}
\end{table}

\subsection{Mixing and nucleosynthesis}\label{mixnuc}

During the core H-burning and He-burning phase, the mixing induced by rotation changes the distribution of the chemical elements inside the star. In advanced stages (core C-burning and after), the burning timescale becomes small compared to the rotational mixing timescale so that rotation barely affects the distribution of chemical elements. 
During the core He-burning phase, the rotational mixing triggers exchanges of material between the convective He-burning core and the convective H-burning shell \citep{maeder15b, frischknecht16, choplin16}. In addition, the growing convective He-burning core progressively engulfs the products of H-burning.
The main steps of this mixing process are summarized below: 

\begin{enumerate}
\item In  the He-burning core, $^{12}$C and $^{16}$O are synthesized via the 3$\alpha$ process and $^{12}$C($\alpha,\gamma$)$^{16}$O.
\item The abundant $^{12}$C and $^{16}$O in the He-burning core are progressively mixed into the H-burning shell. It boosts the CNO cycle and creates primary CNO elements, especially $^{14}$N and $^{13}$C.
\item The products of the H-burning shell (among them primary $^{13}$C and $^{14}$N) are mixed back into the He-core. From the primary $^{14}$N, the reaction chain $^{14}$N($\alpha,\gamma$)$^{18}$F($e^+ \nu_e$)$^{18}$O($\alpha,\gamma$)$^{22}$Ne allows the synthesis of primary $^{22}$Ne. The reactions $^{22}$Ne($\alpha,n$) and $^{22}$Ne($\alpha,\gamma$) make $^{25}$Mg and $^{26}$Mg, respectively. 
The neutrons released by the $^{22}$Ne($\alpha,n$) reaction produce $^{19}$F, $^{23}$Na, $^{24}$Mg, and $^{27}$Al by $^{14}$N($n,\gamma$)$^{15}$N($\alpha,\gamma$)$^{19}$F, $^{22}$Ne($n,\gamma$)$^{23}$Ne($e^- \bar{\nu}_e$)$^{23}$Na, $^{23}$Na($n,\gamma$)$^{24}$Na($e^- \bar{\nu}_e$)$^{24}$Mg, and $^{26}$Mg($n,\gamma$)$^{27}$Mg($e^- \bar{\nu}_e$)$^{27}$Al, respectively.
Free neutrons can also trigger the s-process, provided enough seeds, like $^{56}$Fe, are present \citep{pignatari08,frischknecht16, choplin18}. 
\item The newly formed elements in the He-burning core can be mixed again into the H-burning shell. It can boost the Ne-Na and Mg-Al cycles: additional Na and Al can be produced.
\end{enumerate}

Figure~\ref{abrotmod} shows the results of this mixing process in the 20~$M_{\odot}$ models with various initial rotation rates. The chemical profiles are shown at the end of the core He-burning phase.
As the initial rotation increases,  the chemicals transit more efficiently from one burning region to another because of stronger rotational mixing and  the convective He-burning core tends to grow more, which also facilitates the exchanges of chemicals between the two burning regions.
For high initial rotation rates, however ($\upsilon_{\rm ini}/\upsilon_{\rm crit} \gtrsim 0.5$), the growing of the He-core is limited by the very active H-burning shell (the high activity is due to the copious amounts of $^{12}$C and $^{16}$O entering  the shell and boosting the CNO cycle). 
This effect limits the efficiency of the mixing process for high rotation and makes the production of chemicals starting to saturate for $\upsilon_{\rm ini}/\upsilon_{\rm crit} \gtrsim 0.5$ (e.g. the $^{14}$N or $^{22}$Ne profiles in Fig.~\ref{abrotmod}).

   \begin{figure*}
   \centering
       \includegraphics[scale=0.47]{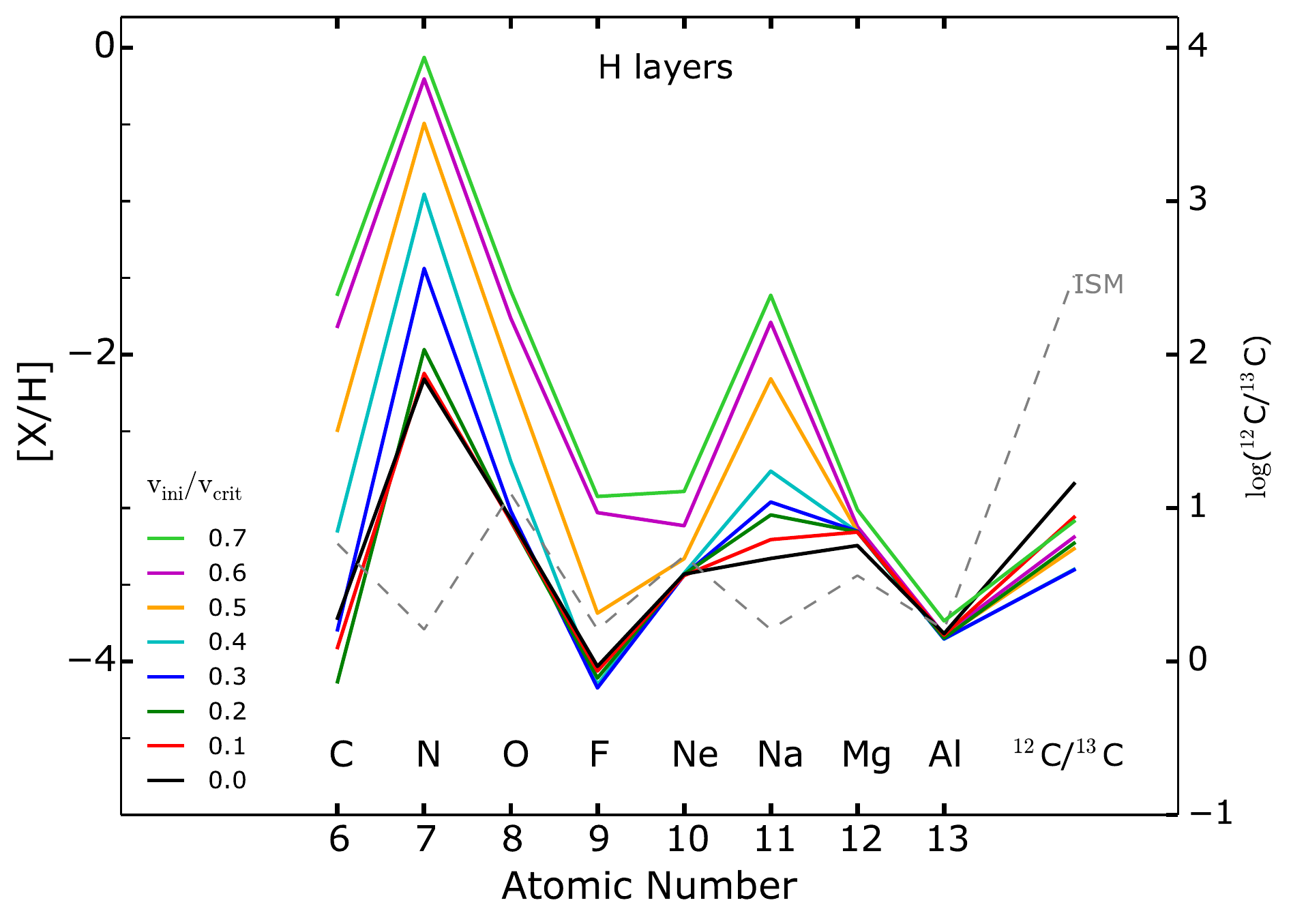}
       \includegraphics[scale=0.47]{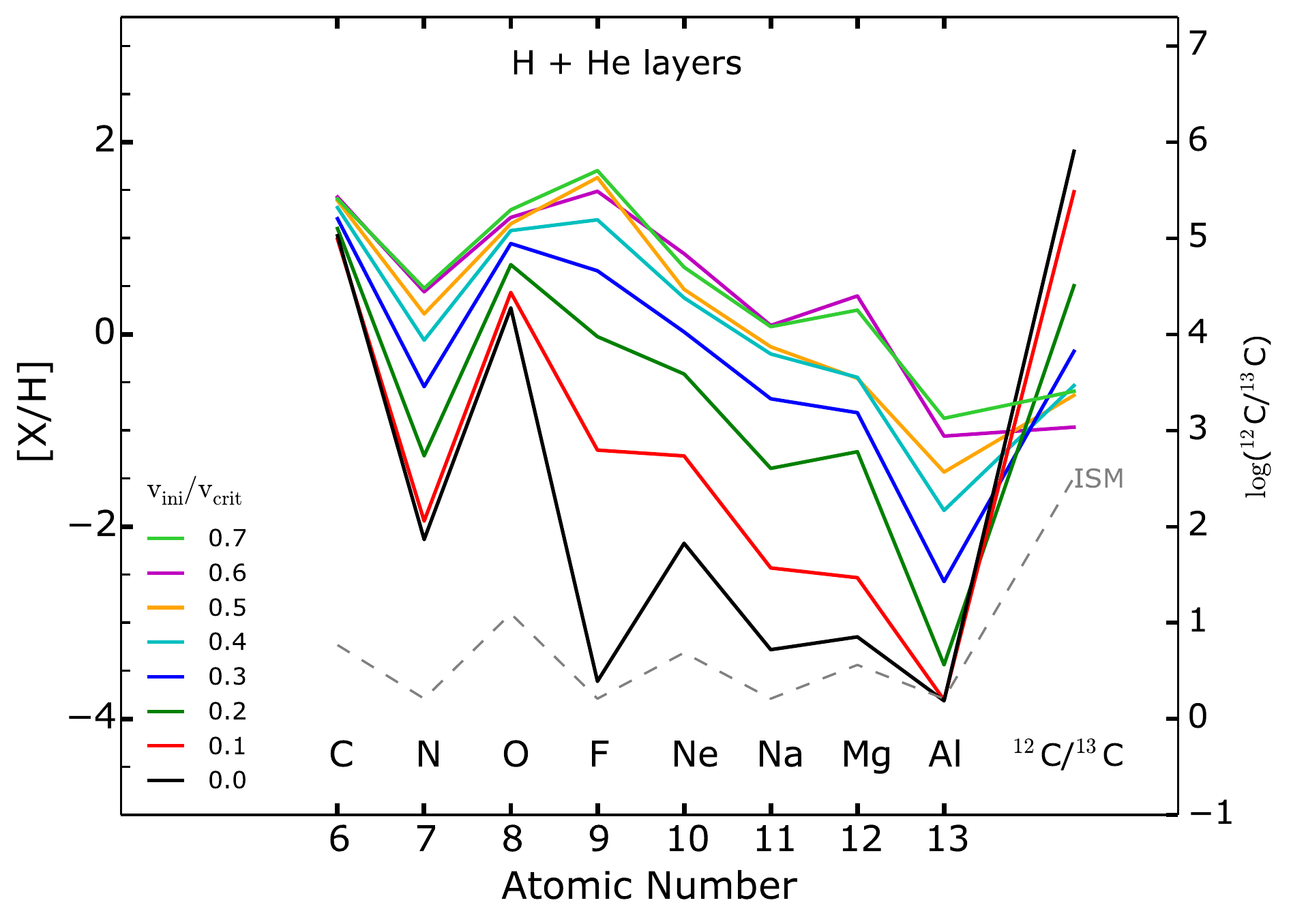}
  
   \caption{Integrated [X/H] and $\log$($^{12}$C/$^{13}$C) ratios in the H-rich layers (left panel) and H + He-rich layers (right panel) at the pre-SN stage for the 20~$M_{\odot}$ models with different initial rotation rates. The material ejected through winds during the evolution is included. The dashed line shows the initial source star composition.}
\label{xhrot}
    \end{figure*}

While the Ne-Na cycle is boosted in the H-burning shell of rotating models (see the peak of $^{23}$Na at $\sim 8$~$M_{\odot}$), neither the Mg-Al cycle nor the $^{27}$Al($p,\gamma$)$^{28}$Si reaction is significantly boosted (no similar peak in the H-burning shell)  because the temperature in the H-burning shell ($T \lesssim 45$ MK) is too low to efficiently activate these reactions. 
Moreover, the synthesis of extra Al in the H-burning shell needs extra Mg, which is only built in the He-core when $T \gtrsim 220$ MK (through an $\alpha$-capture on $^{22}$Ne). In a 20~$M_{\odot}$ model, this temperature corresponds to the end of the core He-burning phase. The extra Mg created in the core has then little time to be transported to the H-burning shell and boosts the Mg-Al cycle.

\subsection{Composition of H- and He-rich layers at the pre-SN stage}

\subsubsection*{H-rich layers}

During the evolution of stars, some material is ejected through stellar winds. Rotation is expected to affect the mass loss of massive stars \citep{maeder99b,maeder00}. In these 20 $M_{\odot}$ models, however, losses through winds stay small ($<1$ $M_{\odot}$, see Table~\ref{table:2}). The radiative mass loss--metallicity relation ($\dot{M} \propto Z^{0.85}$) plays  a major role here and prevents significant radiative mass loss episodes.
The left panel of Fig.~\ref{xhrot} shows the chemical composition of the H-rich layers at the end of the evolution. 
It comprises all the material above the bottom of the H-shell (defined where the mass fraction of X($^{1}$H) drops below 0.01). It includes the wind material.
Depending on the model, the bottom of the H-shell is located at a mass coordinate in between
6.3 and 8~$M_{\odot}$.
The typical CNO pattern appears for all the models (more N, less C and O), but the sum of CNO elements increases with rotation, as a result of $^{12}$C and $^{16}$O having diffused to the H-burning shell. Thanks to the extra Ne entering in the H-shell and boosting the Ne-Na cycle (see Sect.~\ref{mixnuc}), [Na/H] spans $\sim 2$ dex from the non-rotating to the fast rotating model. The [Mg/H] and [Al/H] ratios do not vary more than 0.5 dex. 
The $\log$($^{12}$C/$^{13}$C) ratio is close to the CNO equilibrium value of $\sim$~$0.6$.

\subsubsection*{H-rich and He-rich layers}

The right panel of Fig.~\ref{xhrot} shows the chemical composition of the H-  and He-rich layers. All the material above the bottom of the He-shell (where the mass fraction of X($^{4}$He) drops below 0.01) is considered. 
In this case, the mass cuts are  between  3.9 and 5.2 $M_{\odot}$, depending on the model. 
Compared to the previous case, the additional $\sim 2$~$M_{\odot}$ are H-free, so that it   raises the [X/H] values slightly (by < 0.5 dex). The CNO pattern is flipped compared to the ejecta of the H-rich layers only because $^{14}$N is depleted, while $^{12}$C and $^{16}$O are abundant in the region processed by He-burning (Fig.~\ref{abrotmod}). 
The He-burning products (particularly Ne, Mg, and Al) are boosted compared to the previous case. These products also increase with initial rotation as a result of the back-and-forth mixing process (see Sect.~\ref{mixnuc}). In the region processed by He-burning, $^{13}$C is depleted by ($\alpha,n$) and ($\alpha,\gamma$) reactions so that the $^{12}$C/$^{13}$C ratio is largely enhanced compared to the case where only the H-rich ejecta is considered. Rotation nevertheless decreases this ratio because of the additional $^{13}$C that was synthesized in the layers processed by H-burning.

  \subsection{Effect of the mass cut}\label{mcuteff}

Figure~\ref{mcutfig} shows the effect of varying the mass cut $M_{\rm cut}$ from $\sim 5$ to $\sim 9$~$M_{\odot}$ in the vv4 model. The highest $M_{\rm cut}$ is located in the H-envelope, the lowest value in the He-shell. The numbers in parentheses indicate the mass fraction of $^{1}$H at the mass coordinate equal to the indicated $M_{\rm cut}$. Varying the mass cut from the H- to He-shell flips the CNO pattern. It goes from a $\wedge$-shape to a $\vee$-shape pattern:  in the H-shell, the effect of CNO cycle is dominant (high N/C and N/O ratios), while in the He-shell, the C and O produced by He-burning dominate so that the C/N and C/O ratios are dramatically increased. While the $\log$($^{12}$C/$^{13}$C) is close to the CNO equilibrium value (about $0.6$) for shallow $M_{\rm cut}$, it increases with deeper mass cuts as a results of $^{13}$C-depletion and $^{12}$C-richness of the layers processed by He-burning. Deeper mass cuts also increase [Ne/H] or [Na/H] ratios, for instance, as a result of the addition of He-burning material in the ejecta.

\section{Linking EMP stars with their source stars}\label{empcomp}

If they were present, the stars discussed in the previous section should have left their chemical imprints on the next stellar generations.
We therefore investigate whether the material processed by rotation may be found in the chemical composition of observed EMP stars. 
For the comparison, the elements C, N, O, Na, Mg, and Al and the $^{12}$C/$^{13}$C ratio are considered because they  are the elements mostly affected by rotation. The possible origin of the heavier elements is discussed in Sect.~\ref{compodisc}.

   \begin{table}
\scriptsize{
\caption{Number of stars in the considered sample \citep[from the SAGA database,][]{suda08, suda17} having a given abundance available (first row) or having only a limit (second row). \label{table:3}}
\begin{center}
\resizebox{8.1cm}{!} {
\begin{tabular}{l|ccccccc} 
\hline 
& C & N & O & Na & Mg & Al & $^{12}$C/$^{13}$C \\
\hline 
Abundance                  & 272 & 75  & 35 &  166 &  271 & 248 & 30 \\
Limit                            & 0     & 54  & 56 &   2    &   0    &  1    &  9  \\
 \hline
\end{tabular}
}
\end{center}
}
\end{table}

\subsection{Metal-poor star sample}\label{sample}

The stars with $\leq -4$ [Fe/H] $\leq -3$ and with at least three abundance values determined among C, N, O, Na, Mg, Al, and $^{12}$C/$^{13}$C are selected. 
This sample comprises 272 stars. The abundance data of the metal-poor stars considered in this work mostly come from the SAGA database \citep{suda08, suda17}. The recently observed stars fulfilling the criteria mentioned above were added: G64-12 \citep{placco16a}, LAMOSTJ2217+2104 \citep{aoki18}, SDSSJ0140+2344, SDSSJ1349+1407 \citep{bonifacio18}, SDSSJ0826+6125, SDSSJ1341+4741 \citep{bandyopadhyay18}.
The individual references are given in   Appendix~A. Table~\ref{table:3} shows the number of available abundance and limits for the EMP stars considered.

\subsection{Internal mixing processes in EMP stars}\label{intsample}

The link between the EMP star and its source(s) is made more difficult by the fact that EMP stars themselves may have undergone internal mixing processes from their birth   \citep[e.g. dredge-up, thermohaline mixing, rotation, or atomic diffusion,][]{richard02, charbonnel07, stancliffe09}. These processes may have caused surface abundance modifications, so that the abundance of an element derived from observations may be different to the abundance at the birth of the EMP star.

The first dredge-up  and the thermohaline mixing, may be the most important processes. They happen in evolved EMP stars and can decrease the C surface abundance and $^{12}$C/$^{13}$C ratio and increase the N surface abundance \citep[e.g.][]{charbonnel94, stancliffe07, eggleton08, lagarde19}.
In this work, the evolutionary effects on the surface carbon abundance were corrected following \cite{placco14c}, who proposed a correction based on stellar models to apply to the carbon abundance of stars with [Fe/H] $<-2$. This correction likely allows   the initial surface carbon abundance of metal-poor stars to be recovered.
By comparing the \cite{placco14c} sample with our sample, we corrected the C abundance of 
173 EMP stars. The correction $\Delta$[C/H] applied to the [C/H] ratio varies between 0 and 0.77. After correction, the sample comprises 87 CEMP stars (EMP stars are considered   CEMP if [C/Fe] $\geq 0.7$).
For N and $^{12}$C/$^{13}$C, no similar correction is available yet. 
Here we set the [N/H] and $^{12}$C/$^{13}$C ratios as upper and lower limits, respectively, for the stars on the upper giant branch ($\log g < 2$). These stars may indeed have experienced important modification of their surface N abundance and $^{12}$C/$^{13}$C ratio\footnote{In our sample, 144 (128) EMP stars have $\log g < 2$ ($ \geq 2$). Among the stars with $\log g \geq 2$, 24 have a determined N abundance, 7 have a determined $^{12}$C/$^{13}$C ratio and 6 have both a determined N abundance and $^{12}$C/$^{13}$C ratio.
}. This assumption, and more
generally internal mixing processes in EMP stars, is discussed further in Sect.~\ref{mixemp}.

We note that most of the CEMP stars considered here belong to group II \citep{yoon16} because the metallicity range considered ($-4<$ [Fe/H] $<-3$) contains mainly group II stars. Group III stars are found at [Fe/H] $\lesssim-4,$ and group I mostly at [Fe/H] $\gtrsim -3$ (their figure 1). In our sample, $\sim 8 \%$ of the considered CEMP stars have $A(C) > 7$, where group I stars are found.

   \begin{figure}
   \centering
       \includegraphics[scale=0.47]{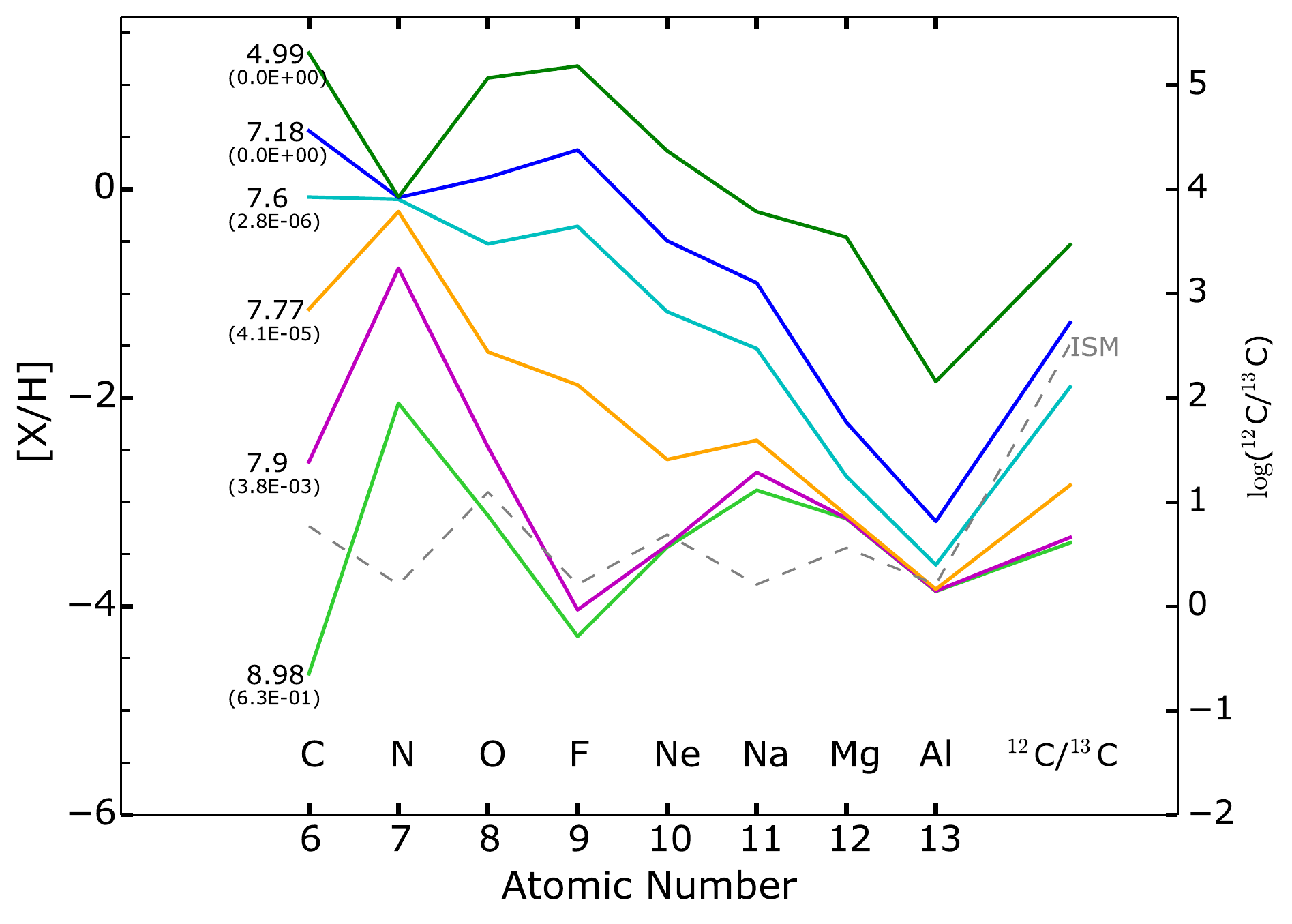}  
   \caption{Effect of the mass cut for the vv4 model. The mass cut (in $M_{\odot}$) of the corresponding abundance pattern is indicated on the left. The numbers in parenthesis indicate the $^{1}$H mass fraction in the star at the mass coordinate equal to the mass cut.}
\label{mcutfig}
    \end{figure}

\subsection{Fitting procedure}\label{fitproc}

To fit the abundances of EMP stars with massive star models, we followed the same procedure described in \cite{heger10} and \cite{ishigaki18}, especially for calculating the $\chi^2$. Considering N data points, U upper limits, and L lower limits, the $\chi^2$ can be computed as
\begin{equation}
\begin{split}
\chi^{2} =& \sum^{N}_{i=1} \frac{(D_i - M_i)^2}{\sigma^2_{o,i} + \sigma^2_{t,i}}\\
                &+\sum^{N+U}_{i=N+1} \frac{(D_i - M_i)^2}{\sigma^2_{o,i} + \sigma^2_{t,i}} \Theta(M_i - D_i)\\
                &+\sum^{N+U+L}_{i=N+U+1} \frac{(D_i - M_i)^2}{\sigma^2_{o,i} + \sigma^2_{t,i}} \Theta(D_i - M_i)
\end{split}
\label{eqab}
,\end{equation}
where $D_{i}$ and $M_{i}$ are the [X/H] and $\log$($^{12}$C/$^{13}$C) ratios derived from observations and predicted by source star models, respectively; $\Theta(x)$ is the Heaviside function;  and
$\sigma_{o,i}$ and $\sigma_{t,i}$ are the observational and theoretical uncertainties. When not available, $\sigma_{o,i}$ is set to $\bar{\sigma}_{o,i}$, which is the mean uncertainty of the sample for the abundance $i$. 
Mean uncertainties are 0.18, 0.26, 0.17, 0.12, 0.14, and 0.18 dex for [C/H], [N/H], [O/H], [Na/H], [Mg/H], and [Al/H], respectively.
The theoretical uncertainties $\sigma_{t,i}$ are set to 0.15 for [C/H], [N/H], and [O/H] and 0.3 for [Na/H], [Mg/H], and [Al/H]. 
Larger uncertainties for Na, Mg, and Al are considered because of the larger uncertainties on the yields of these elements. The nuclear reaction rates important for the Ne-Na and Mg-Al cycles still suffer significant uncertainties that can affect stellar yields \citep{decressin07, choplin16}. The uncertainties associated with ($\alpha$,$\gamma$) or ($\alpha$,$n$) reactions operating in He-burning zones, such as $^{22}$Ne($\alpha$,$\gamma$) or $^{22}$Ne($\alpha$,$n$), are also non-negligible \citep[e.g.][]{frischknecht12}. For $\log$($^{12}$C/$^{13}$C), we set a general sigma value of 0.3 including uncertainties of  models and of observations. Theoretical uncertainties are further discussed in Sect.~\ref{thuncer}. 

For each EMP star, the minimum $\chi^2$ is searched among the eight source star models. For each model the mass cut $M_{\rm cut}$ and mass of the added interstellar medium (ISM) $M_{\rm ISM}$ are left  as free parameters. The $M_{\rm cut}$ parameter is varied between the bottom of the He-shell (located at $\sim 4-5$ $M_{\odot}$ depending on the model) and the stellar surface, and  $M_{\rm ISM}$ is varied between $10^2$ and $10^6$ $M_{\odot}$. 
By adding ISM material, we assume some dilution of the source star ejecta with the surrounding ISM. It should occur at the time of the source star explosion. The added ISM material has the same composition as the initial source star composition (dashed line in Figs.~\ref{xhrot} and \ref{mcutfig}). The initial ISM composition and its possible impact on our results are further discussed in Sect.~\ref{mismdis}.

\subsection{Weighting the source star models}\label{secwei}

As we  show in the next section, it happens that for a given EMP star, more than one of the eight 20 $M_{\odot}$ source star models can give a reasonable solution. 
To account for this when deriving, for instance, the velocity distribution of the best source star models, we associate a weight with each fit, scaling with the goodness of   fit. The goodness of   fit is determined thanks to the p-value $p$, 
which is directly equal to the weight and can be written as
\begin{equation}
p(\chi^2, N-m) = 1 - F(\chi^2, N-m) = 1 - \frac{\gamma (\frac{N-m}{2}, \frac{\chi^2}{2})}{\Gamma (\frac{N-m}{2})}
,\end{equation}
with $F$ the $\chi^2$ cumulative distribution function, $\Gamma$ the gamma function, and $\gamma$ the lower incomplete gamma function. Here 
$N$ is the number of measured abundances for the considered EMP star 
and $m=2$ is the number of free parameters for a given 20~$M_{\odot}$ model ($M_{\rm cut}$ and $M_{\rm ISM}$). 
From $\chi^2$, $N$, and $m$, the reduced $\chi^2$ can be calculated as $\chi_{\nu}^2 = \chi^2 / (N-m)$. 

Since high $\chi^2$ values give negligible weights, the models that give bad fits are automatically discarded and will not contribute to determining the overall characteristics of the best progenitors (e.g. velocity distribution). Similarly, if no good fit can be found for a given EMP star, this EMP star is not considered because of the negligible weights of all the source star models. 

We note that the weights for  given EMP stars are not normalized to one. If so, the EMP stars that cannot be fitted correctly (hence having low weights for all source star models) will contribute similarly to the well fitted EMP stars. 
However, it is  possible to normalize the weights to one if  a threshold $\chi_{\nu, \text{th}}^2$ is first set (e.g. $\chi_{\nu, \text{th}}^2 = 3$) so as to discard the EMP stars where no $\chi_{\nu}^2 < \chi_{\nu, \text{th}}^2$ can be found. 
We checked that this alternative method gives  results that are similar to those of the adopted method; the results are presented in the next section.

\section{Results}\label{res}

\begin{figure}
   \centering
       \includegraphics[scale=0.46]{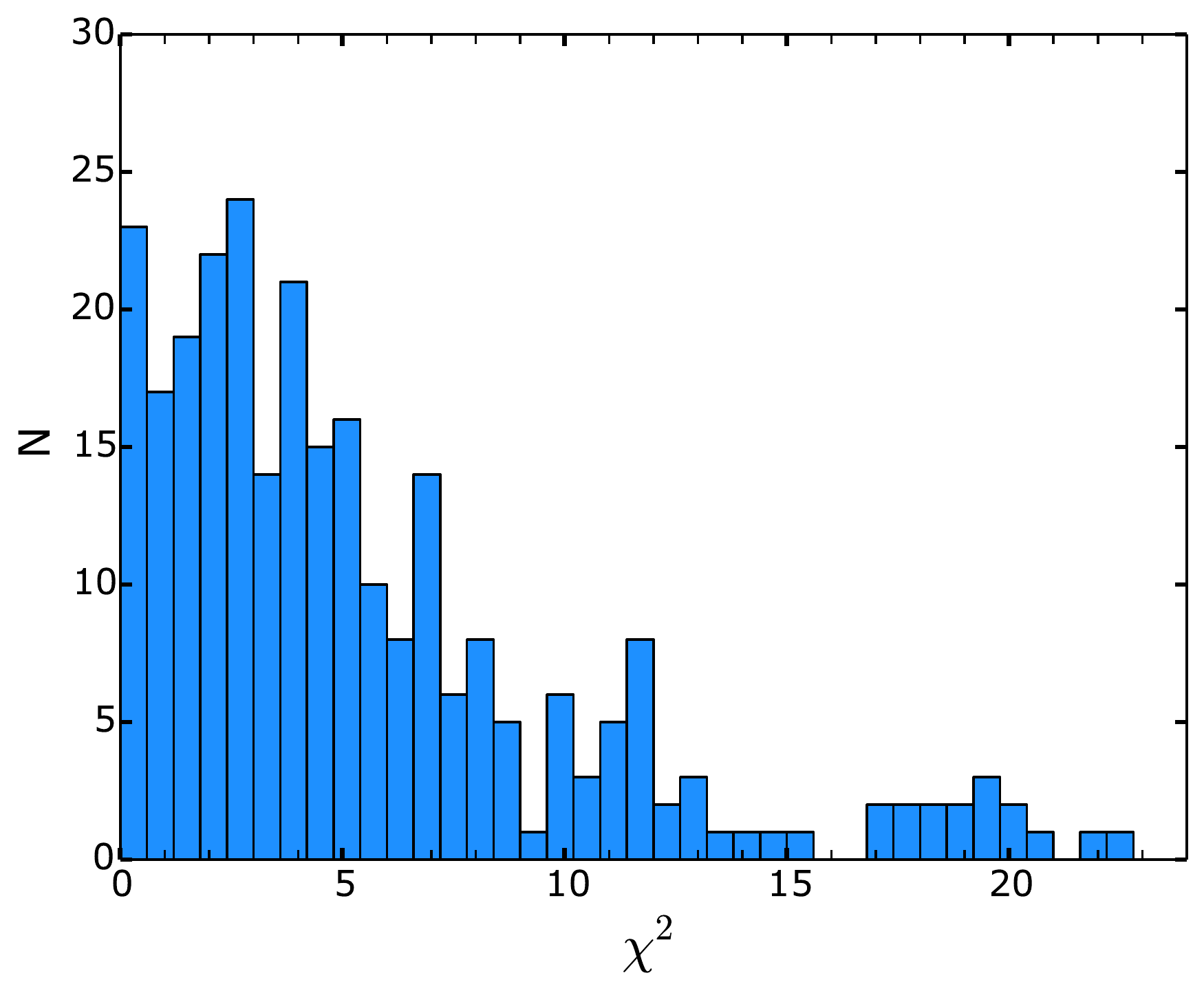}
   \caption{Distribution of the $\chi^{2}$ values for the 272 EMP stars fitted.}
\label{chi2}
    \end{figure}

Figure~\ref{chi2} shows the distribution of $\chi^2$ of the best model for the 272 star fitted. 
Two stars with very high $\chi^2$ (BS16929-005 and SDSSJ0826+6125 with $\chi^2 = 31$ and 58, respectively) are not shown in Fig.~\ref{chi2}. 
The stars that cannot be fitted correctly are discussed further in Sect.~\ref{badfit}. 
Tables~\ref{table:4} and \ref{table:5} give the fraction of fits having a $\chi^2$ and $\chi_{\nu}^2$ value below a given threshold, respectively. From Table~\ref{table:4} we see that 60~\% (39~\%) of the stars have $\chi ^2 < 5$ ($\chi^2 < 3$). The CEMP stars show overall better fits than C-normal EMP stars. We also note that the 8~\% of CEMP stars with A(C) > 7 \citep[or group I CEMP stars,][]{yoon16} follow a   $\chi^2$ distribution that is similar  to that of the entire CEMP sample. In particular, $61$~\% of them have $\chi^2 <3$.

\begin{figure*}
   \centering
   \begin{minipage}[c]{.33\linewidth}
       \includegraphics[scale=0.31]{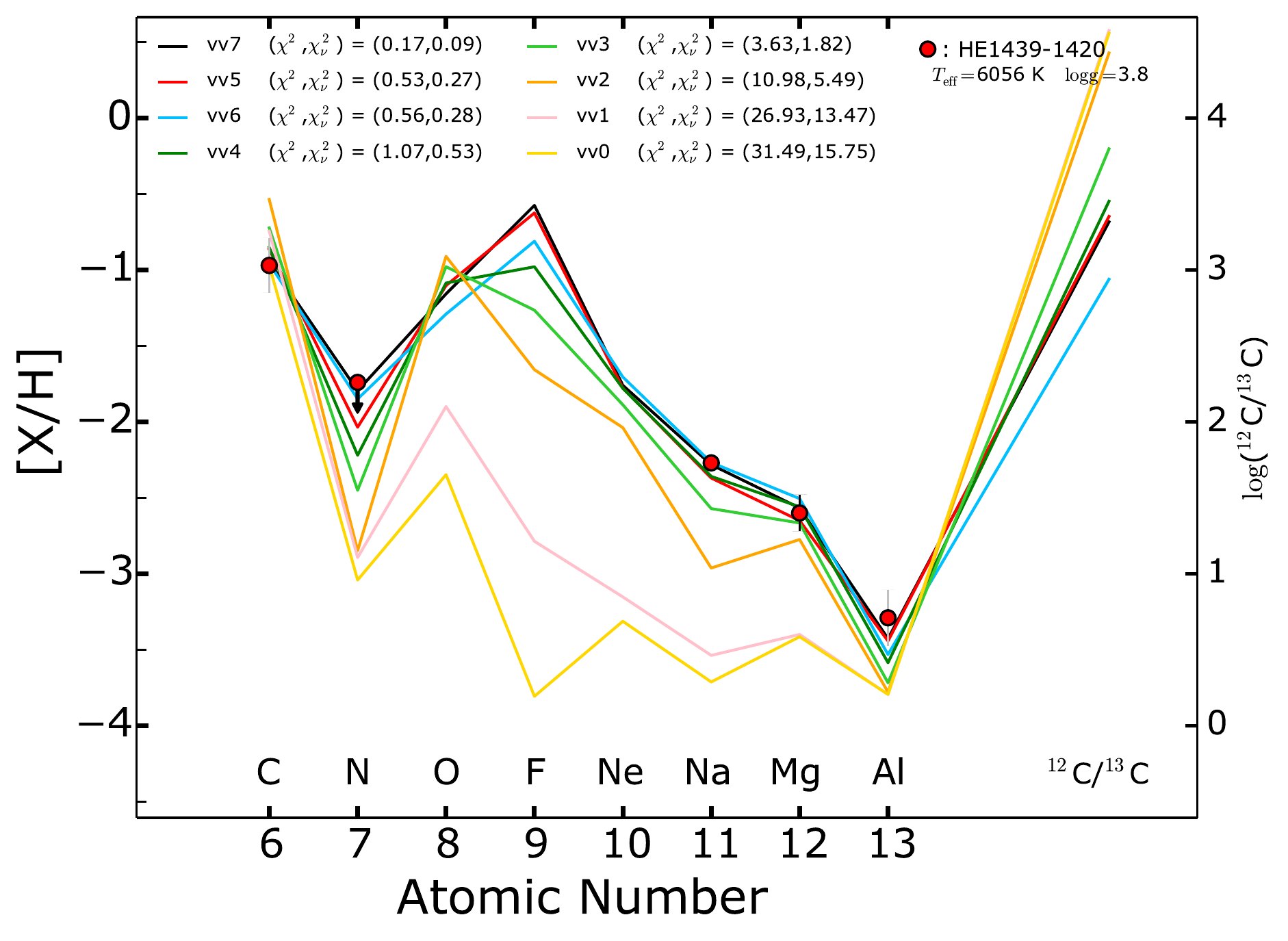}
  \end{minipage}
   \begin{minipage}[c]{.33\linewidth}
       \includegraphics[scale=0.31]{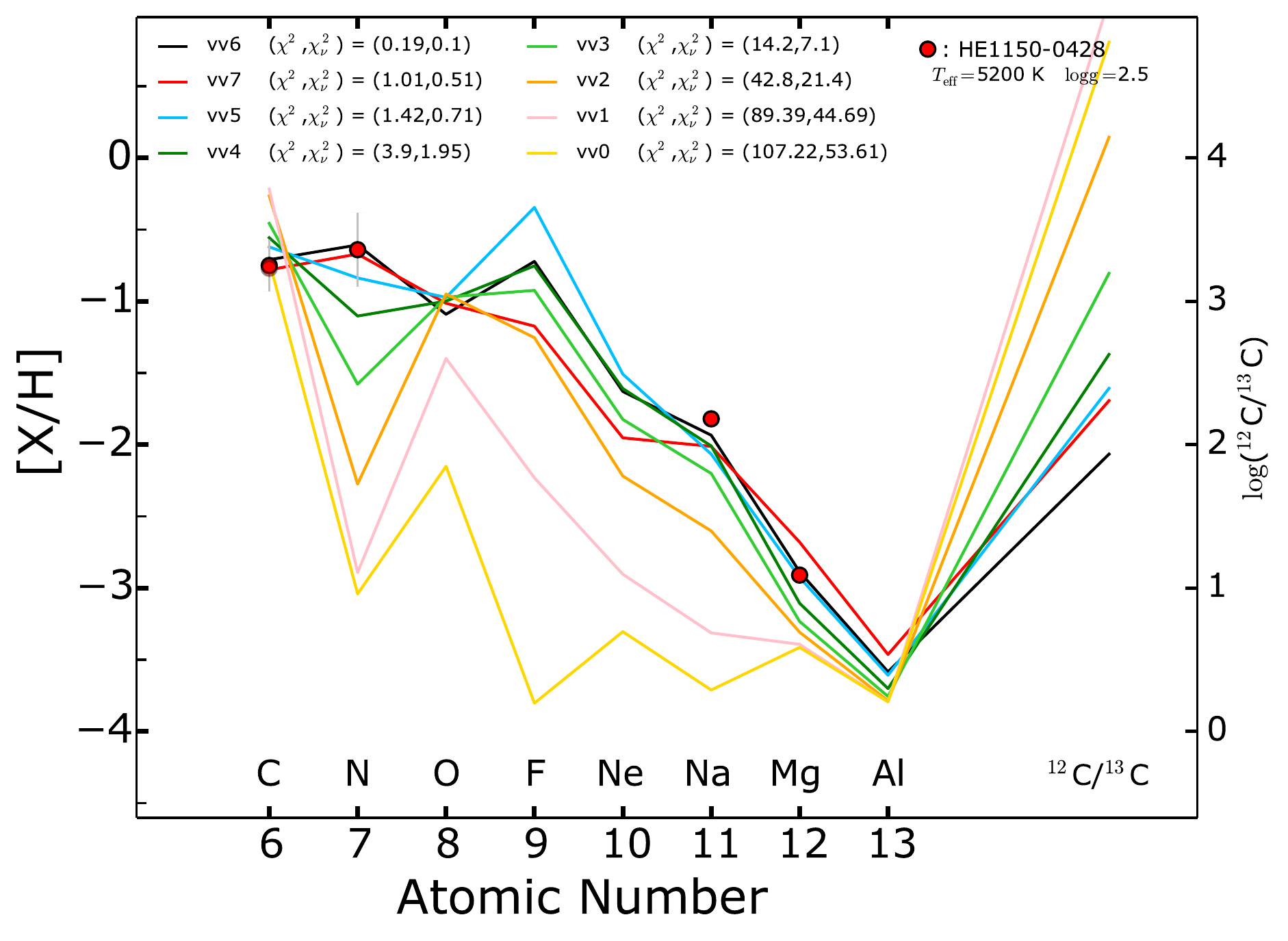}
  \end{minipage}
   \begin{minipage}[c]{.33\linewidth}
       \includegraphics[scale=0.31]{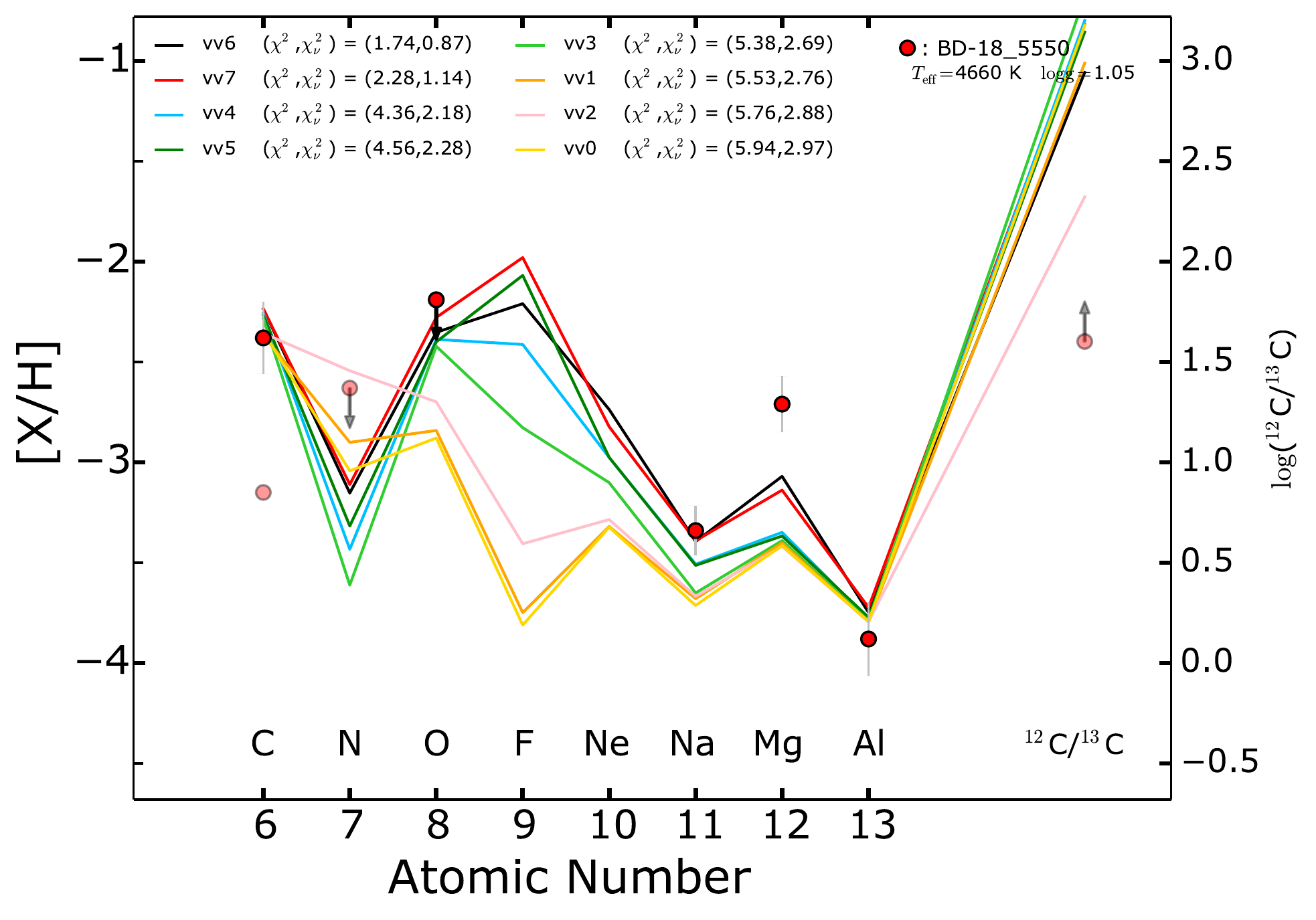}
  \end{minipage}
   \begin{minipage}[c]{.33\linewidth}
       \includegraphics[scale=0.31]{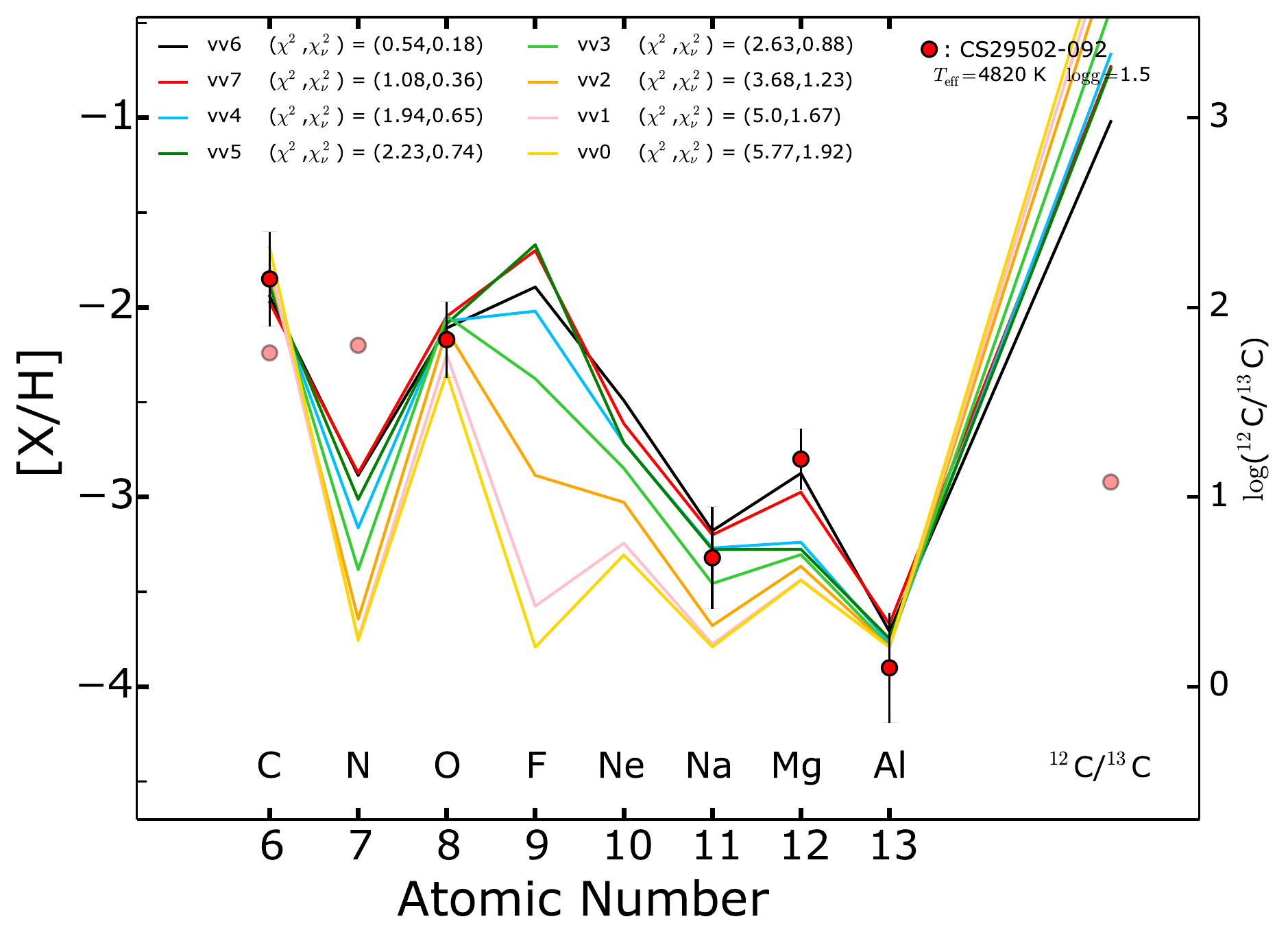}
  \end{minipage}
   \begin{minipage}[c]{.33\linewidth}
       \includegraphics[scale=0.31]{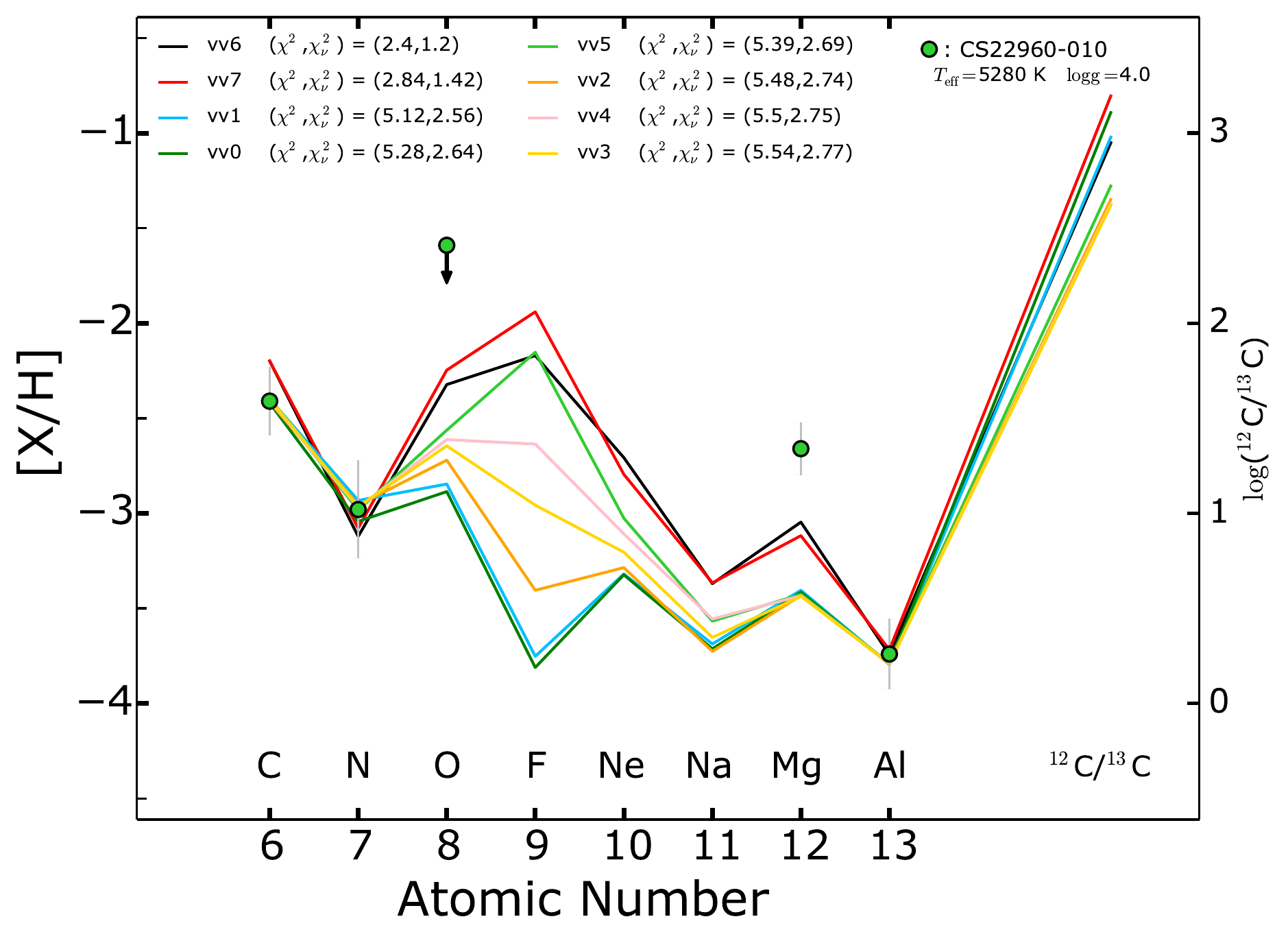}
  \end{minipage}
   \begin{minipage}[c]{.33\linewidth}
       \includegraphics[scale=0.31]{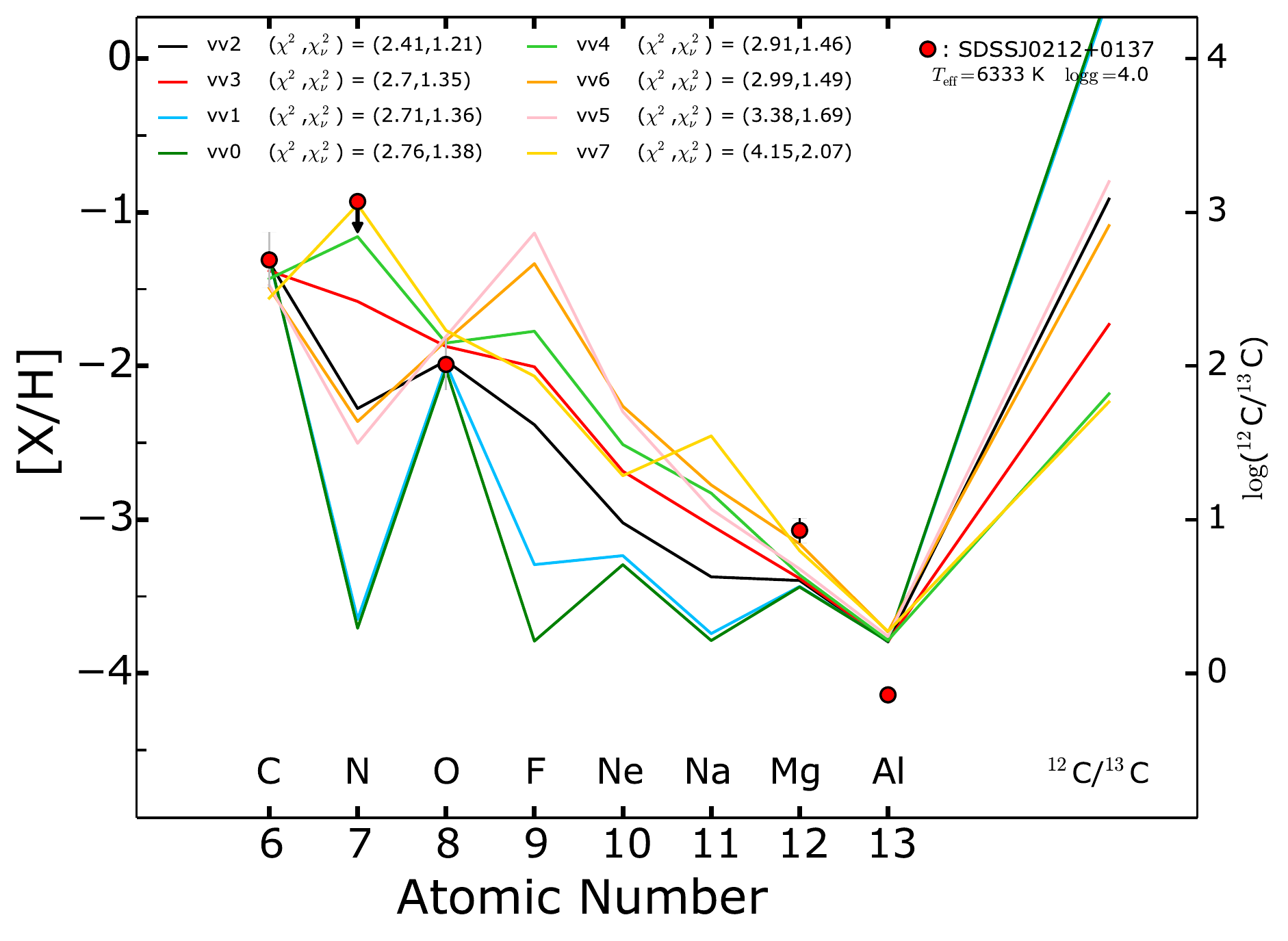}
  \end{minipage}
   \begin{minipage}[c]{.33\linewidth}
       \includegraphics[scale=0.31]{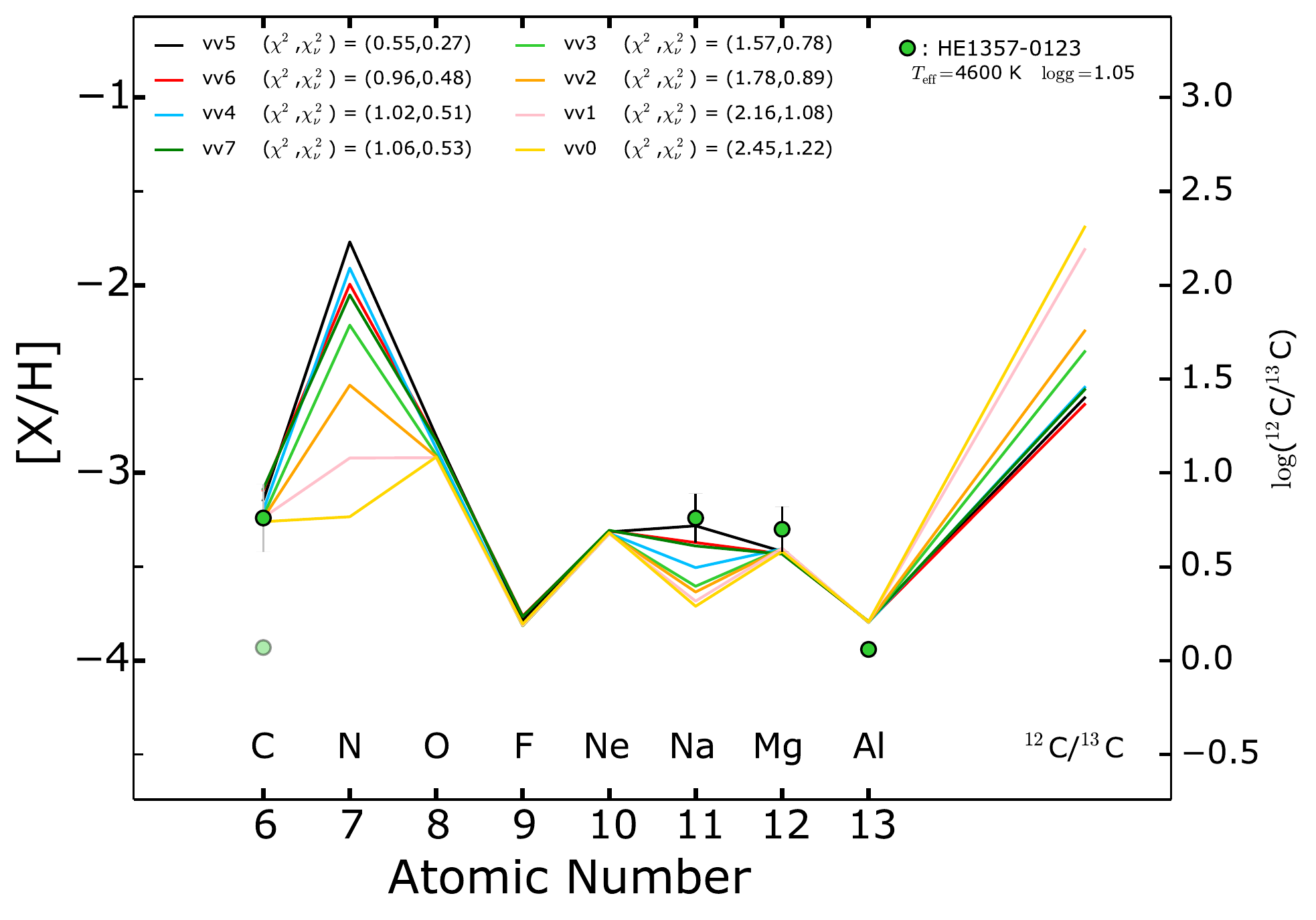}
  \end{minipage}
   \begin{minipage}[c]{.33\linewidth}
       \includegraphics[scale=0.31]{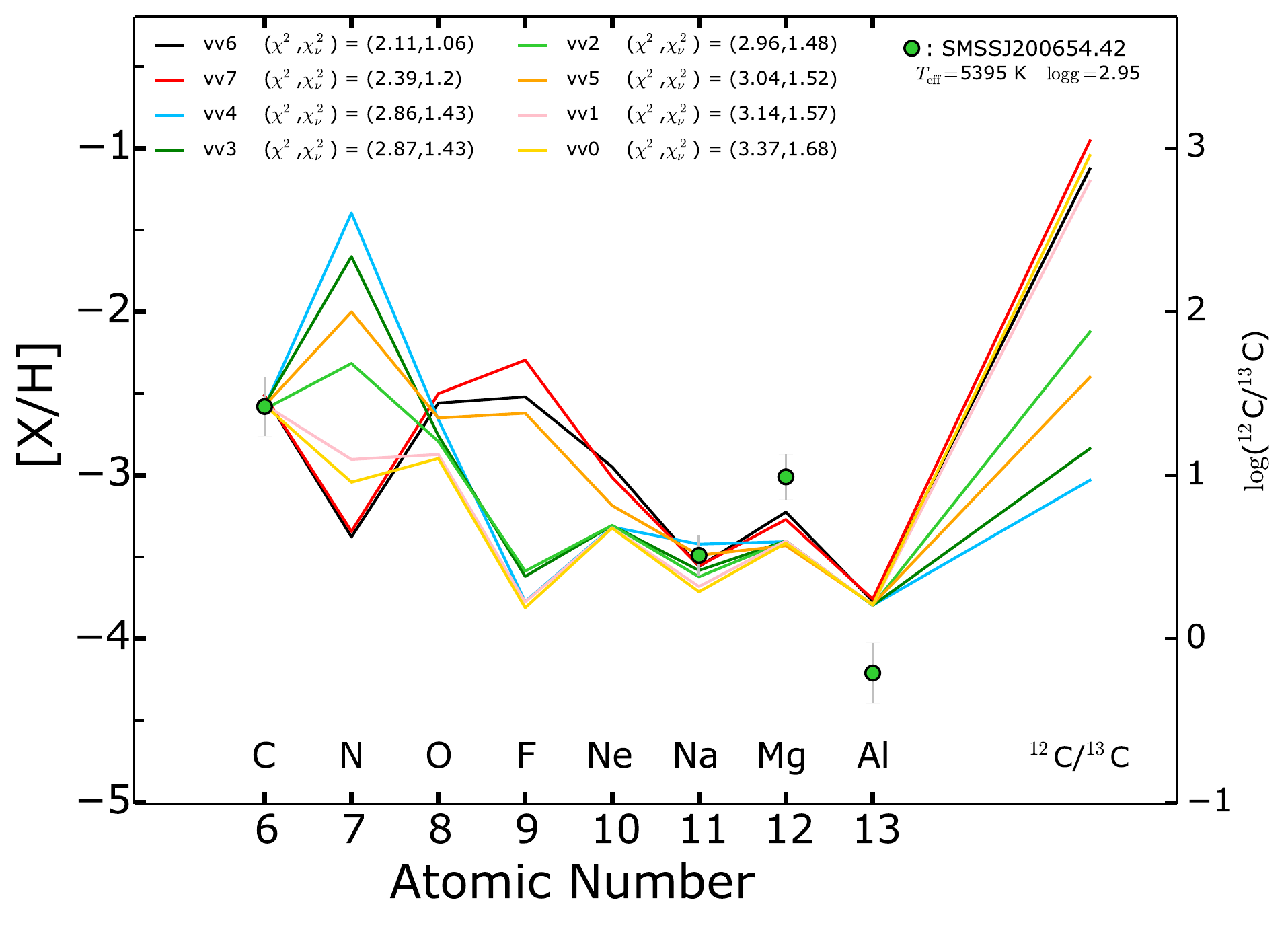}
  \end{minipage}
   \begin{minipage}[c]{.33\linewidth}
        \includegraphics[scale=0.31]{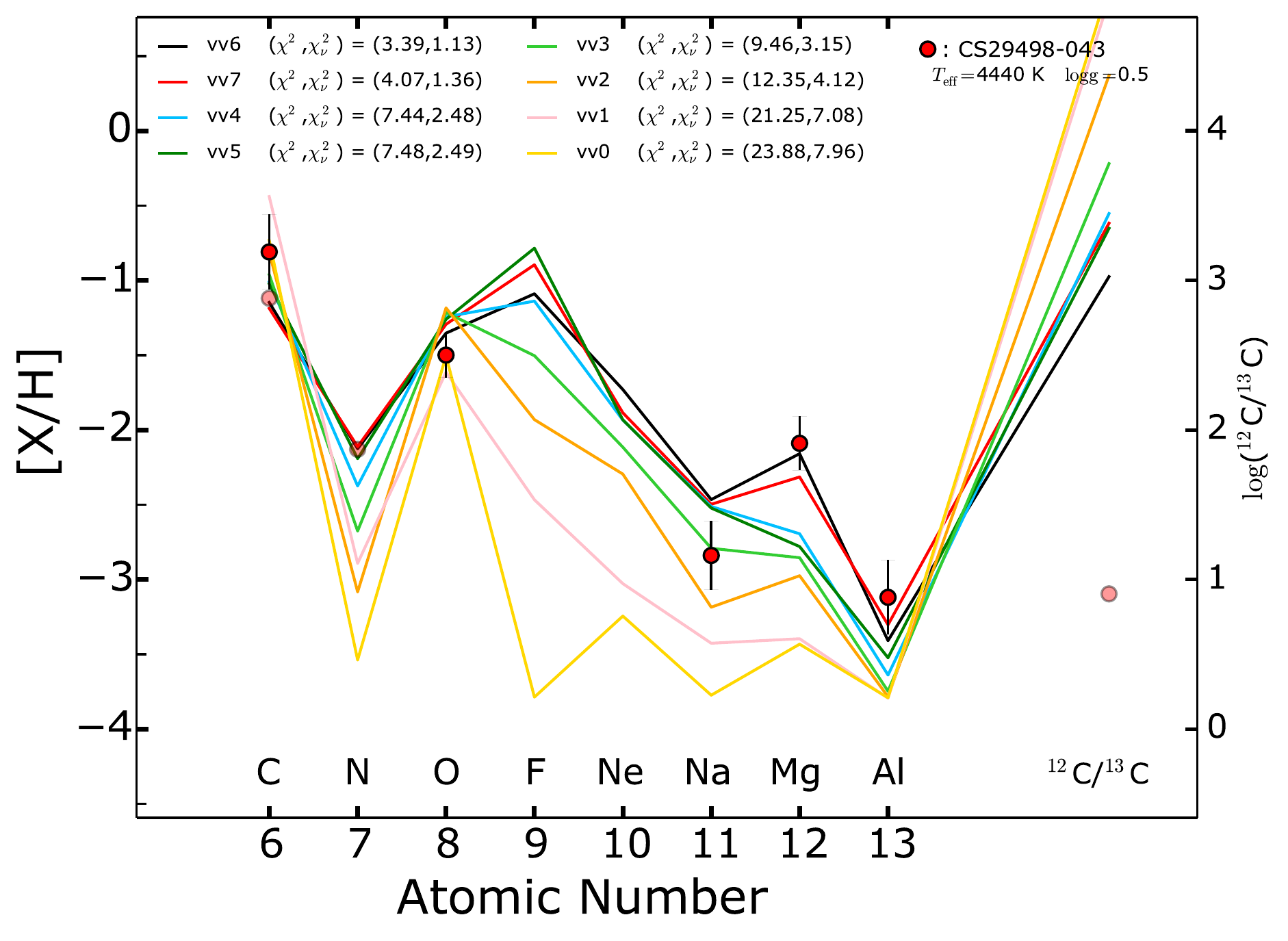}
  \end{minipage}
   \caption{Examples of EMP stars having at least one source star model with $\chi^2_{\nu} < 2$ ($\chi^2 \lesssim 3$). 
  Red and green circles denote CEMP ([C/Fe] $>0.7$) and C-normal EMP stars, respectively.
   For each EMP star, the best fit for each of the 20 $M_{\odot}$ models is shown, ranked by increasing $\chi^2$.    For evolved EMP stars ($\log g < 2$), [N/H] and $\log$($^{12}$C$^{13}$C) are shown as light red and green symbols and are  considered  upper and lower limits, respectively.
   When available, the correction on the C abundance from \cite{placco14c} is taken into account. In this case, the [C/H] ratio before (after) correction is shown by a light (normal) red or green symbol. 
     When available, the observational uncertainty is shown by a black bar. If it is not available, the mean observational uncertainty of the EMP sample (Sect.~\ref{fitproc}) is shown as a grey bar.
  }
\label{star1}
    \end{figure*}

Figure~\ref{mapchi1} in Appendix~B shows a summary plot of the abundance fitting of each of the 272 EMP stars considering each of the eight source star models.
Figure~\ref{allfit1} in Appendix~B shows the abundance fitting of all the EMP stars having at least one source star model with $\chi_{\nu}^2<2$. 
Figure~\ref{star1} shows the abundance fitting of several EMP stars with $\chi^2_{\nu} < 2$. We discuss below some of these fits individually.

\subsubsection*{HE1439-1420}
\begin{figure}
   \centering
      \includegraphics[scale=0.45]{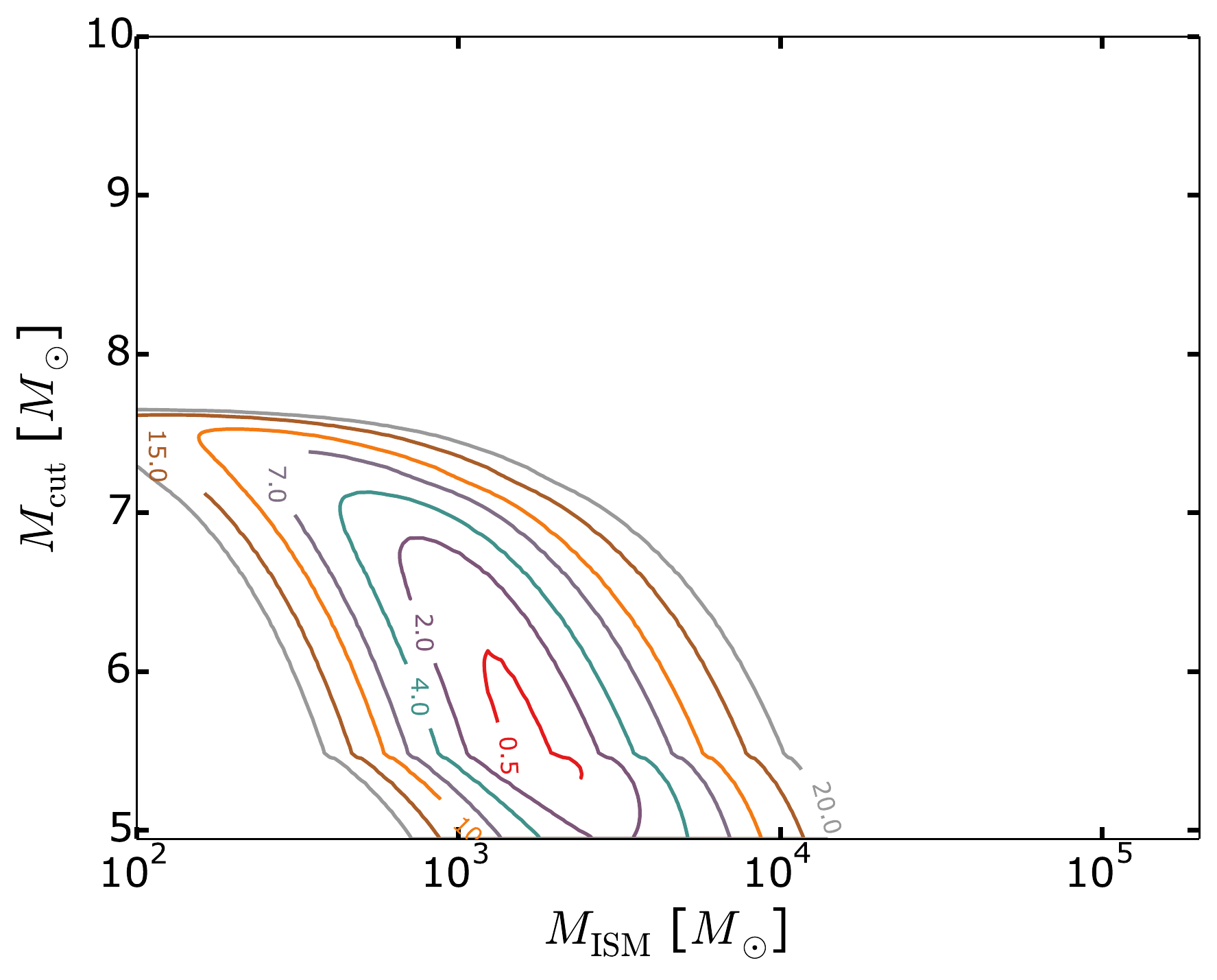} 
   \caption{$\chi^2$ contour map for the vv7 model, considering the fit of HE1439-1420 (the axis shows the two free parameters $M_{\rm cut}$ and $M_{\rm ISM}$). The numbers associated with the contours give the $\chi^2$ values. The minimum $\chi^2$ is found for $M_{\rm cut} = 5.45$ $M_{\odot}$ and $M_{\rm ISM} = 1789$ $M_{\odot}$. These values produce the black pattern shown in the top left panel of Fig.~\ref{star1}.
   }
\label{chimap1}
    \end{figure}

This star is shown in the top left panel. 
The C and N abundances of HE1439-1420 are compatible with a $\vee$-shape CNO pattern, but not with a $\wedge$-shape one. 
As shown in Figs.~\ref{xhrot} and \ref{mcutfig}, $\vee$-shape CNO patterns are characteristics of mass cuts located in the He-shell region. 
In addition, the high Na and Mg enhancement of this star are best explained by fast rotating models that  have experienced significant mixing between the H- and He-burning regions. 
The best fit is given by the vv7 model, with $M_{\rm cut} = 5.45$ $M_{\odot}$ and $M_{\rm ISM} = 1789$ $M_{\odot}$. 
The $\chi^2$ map for the vv7 model is shown in Fig.~\ref{chimap1}. 
The only minimum $\chi^2$ is located at the values mentioned previously. 
In general, it has been checked for the other EMP stars that only one minimum can be found in the $M_{\rm cut}$ $-$ $M_{\rm ISM}$ plane.

   \begin{table}
\scriptsize{
\caption{Fraction of stars with a $\chi^2$ value below a given threshold. Results are given for the entire sample (272 stars), for CEMP stars (87 stars) and for C-normal EMP stars (185 stars). \label{table:4}}
\begin{center}
\resizebox{8.1cm}{!} {
\begin{tabular}{l|ccccc} 
\hline 
& $\chi^{2}<10$  & $\chi^{2}<5$ & $\chi^{2}<3$ & $\chi^{2}<2$ & $\chi^{2}<1$ \\
\hline 
All stars           &               84 \%              &       60 \%              &  39 \%   &       25 \%  &       11 \%  \\
CEMP             &               89 \%              &       77  \%             &  59 \%   &       48 \% &       30 \%  \\
C-normal EMP     &               82 \%              &       52  \%       &  29 \%  &       14 \% &       3  \% \\
\hline 
\end{tabular}
}
\end{center}
}
\end{table}

   \begin{table}
\scriptsize{
\caption{Fraction of fits having a $\chi_{\nu}^2$ value (reduced $\chi^2$, see text) below a given threshold. Results are given for the entire sample (272 stars), for CEMP stars (87 stars) and for C-normal EMP stars (185 stars). \label{table:5}}
\begin{center}
\resizebox{8.1cm}{!} {
\begin{tabular}{l|ccccc} 
\hline 
& $\chi_{\nu}^{2}<10$  & $\chi_{\nu}^{2}<5$ & $\chi_{\nu}^{2}<3$ & $\chi_{\nu}^{2}<2$ & $\chi_{\nu}^{2}<1$ \\
\hline 
All stars           &               97 \%              &       81 \%              &  58 \%   &       39 \%  &       17 \%  \\
CEMP             &               99 \%              &       91  \%             &  77 \%   &       64 \% &       40 \%  \\
C-normal EMP     &               96 \%              &       76  \%       &  49 \%  &       26 \% &       5  \% \\
\hline 
\end{tabular}
}
\end{center}
}
\end{table}

\subsubsection*{HE1150-0428}
This star is shown in the top middle panel. The best fit is given by the vv6 model, with $M_{\rm cut} = 7.46$ $M_{\odot}$ and $M_{\rm ISM} = 100$ $M_{\odot}$. This $M_{\rm cut}$ is $\sim 0.4$ $M_{\odot}$ below the bottom of the hydrogen envelope. A deeper $M_{\rm cut}$ will give a more $\vee-$shape pattern for CNO, while a shallower $M_{\rm cut}$ will give a more $\wedge$-shape pattern. HE1150-0428, having [C/H] $\sim$ [N/H] therefore requires a $M_{\rm cut}$ around the bottom of the H-envelope, which is the region where the CNO pattern is flipping from a $\wedge$-shape to a $\vee$-shape.

\subsubsection*{BD-18\_5550} 
The high [Mg/H] of this star (top right panel) requires both fast rotation and a deep mass cut. The best model is   vv6, with $M_{\rm cut} = 4.74$ and $M_{\rm ISM} = 50941$~$M_{\odot}$.

\subsubsection*{CS29502-092}
This star (shown in the middle left panel) is evolved, with $\log g<2$. Consequently, the [N/H] and $^{12}$C/$^{13}$C are set as upper and lower limits respectively (see Sect.~\ref{intsample}). In this case, the good source star models have similar characteristics to those in the previous case (BD-18\_5550). 
The hypothesis of non-efficient internal mixing processes in EMP stars, which means that the [N/H] and $^{12}$C/$^{13}$C ratios should not be set as limits in evolved stars, is investigated in Sect.~\ref{mixemp}.\\

\subsection{Characteristics of the best source star models}\label{caracbest}

We now derive the $M_{\rm cut}$, $M_{\rm ISM}$, and velocity distribution of the best source star models. 
The following analysis determines the ability of the eight models   to fit the entire EMP star sample. Our approach may be seen as statistical: for a given EMP star and source star, the weight attributed to the fit (Sect.~\ref{secwei}) can be seen as the likelihood of this source star being the true source of this specific EMP star. 

In the case of HE1439-1420, the weights associated with  the $\chi^2$ and $\chi_{\nu}^2$ values shown in the top left panel of Fig.~\ref{star1} are as follows:
\begin{itemize}
\item 0.92 for the vv7 model, ($\chi^2$, $\chi_{\nu}^2$) $= (0.17,0.09)$;
\item 0.77 for the vv5 model, ($\chi^2$, $\chi_{\nu}^2$) $= (0.53,0.27)$;
\item 0.76 for the vv6 model, ($\chi^2$, $\chi_{\nu}^2$) $= (0.56,0.28)$;
\item 0.59 for the vv4 model, ($\chi^2$, $\chi_{\nu}^2$) $= (1.07,0.53)$;
\item 0.16 for the vv3 model, ($\chi^2$, $\chi_{\nu}^2$) $= (3.63,1.82)$;
\item $4.2 \times 10^{-3}$ for the vv2 model, ($\chi^2$, $\chi_{\nu}^2$) $= (10.98,5.49)$;
\item $1.4 \times 10^{-6}$ for the vv1 model, ($\chi^2$, $\chi_{\nu}^2$) $= (26.93, 13.47)$;
\item $1.4 \times 10^{-7}$ for the vv0 model, ($\chi^2$, $\chi_{\nu}^2$) $= (31.49,15.75)$.
\end{itemize}

\subsubsection{Mass cut distribution}\label{mcutdis}

Figure~\ref{histmc} (top panel) shows the weighted $M_{\rm cut}$ distribution of the source star models. The distribution is normalized to 1. 
The two peaks in the mass cut distribution correspond to the bottom of the H-shell ($\sim 8$~$M_{\odot}$) and the bottom of the He-burning shell ($\sim 5$~$M_{\odot}$).
Around these two mass coordinates, the stellar chemical composition is diverse and therefore a wide variety of chemical patterns, needed to account for the wide variety of EMP star chemical pattern, can be produced. 
We note that the best solutions for the CEMP stars (red distribution) are preferentially found for deeper $M_{\rm cut}$ in the layers processed by He-burning where C, Na, Mg, and Al are abundant (cf. Figs.~\ref{xhrot} and \ref{mcutfig}).

Finding most of the mass cuts at the bottom of the H-shell and He-shell may be physically motivated by the fact that the energy required to unbind the material above a given mass cut shows jumps at these locations (Fig.~\ref{histmc}, middle and bottom panels). By peaking at $\sim 5$ and $\sim 8$ $M_{\odot}$, the derivative of $E_{\rm UB}$ reveals these jumps. 
The jump at $\sim 8$ $M_{\odot}$ is smaller for slow rotating models (especially vv0 and vv1) because these models do not fully enter the red super giant (RSG) phase (Fig.~\ref{hrdfig}), where the stellar radius expands significantly. On the contrary, since the faster rotating models enter the RSG phase, their envelopes expand dramatically and become loosely bound.
Overall, we may expect a clustering of the mass cuts around $\sim 5$ and $\sim 8$ $M_{\odot}$ because the binding energy of most models significantly increases below these mass coordinates, making these deeper layers more difficult to expel.

\subsubsection{$M_{\rm ISM}$ distribution}\label{mismdis}

Figure~\ref{histdil} shows the weighted distribution of $M_{\rm ISM}$, the mass of added ISM. The distribution is normalized to 1.
The overall blue distribution peaks around low $M_{\rm ISM}$ values. 
Among the best source star models (with $\chi^2 < 3$), 59~\% have $10^2< M_{\rm ISM}< 10^3$~$M_{\odot}$ and 73~\% have $10^2 < M_{\rm ISM}< 10^4$~$M_{\odot}$. 
For such $M_{\rm ISM}$ values, the contribution of ISM is generally small compared to the source star contribution.
For CEMP stars, it is more equally distributed. It can be understood together with the mass cut distribution: the good solutions for CEMP stars are generally found at deep mass cuts; it therefore requires significant dilution with ISM to not overestimate the CEMP star abundances. 

\begin{figure}
   \centering
       \includegraphics[scale=0.46]{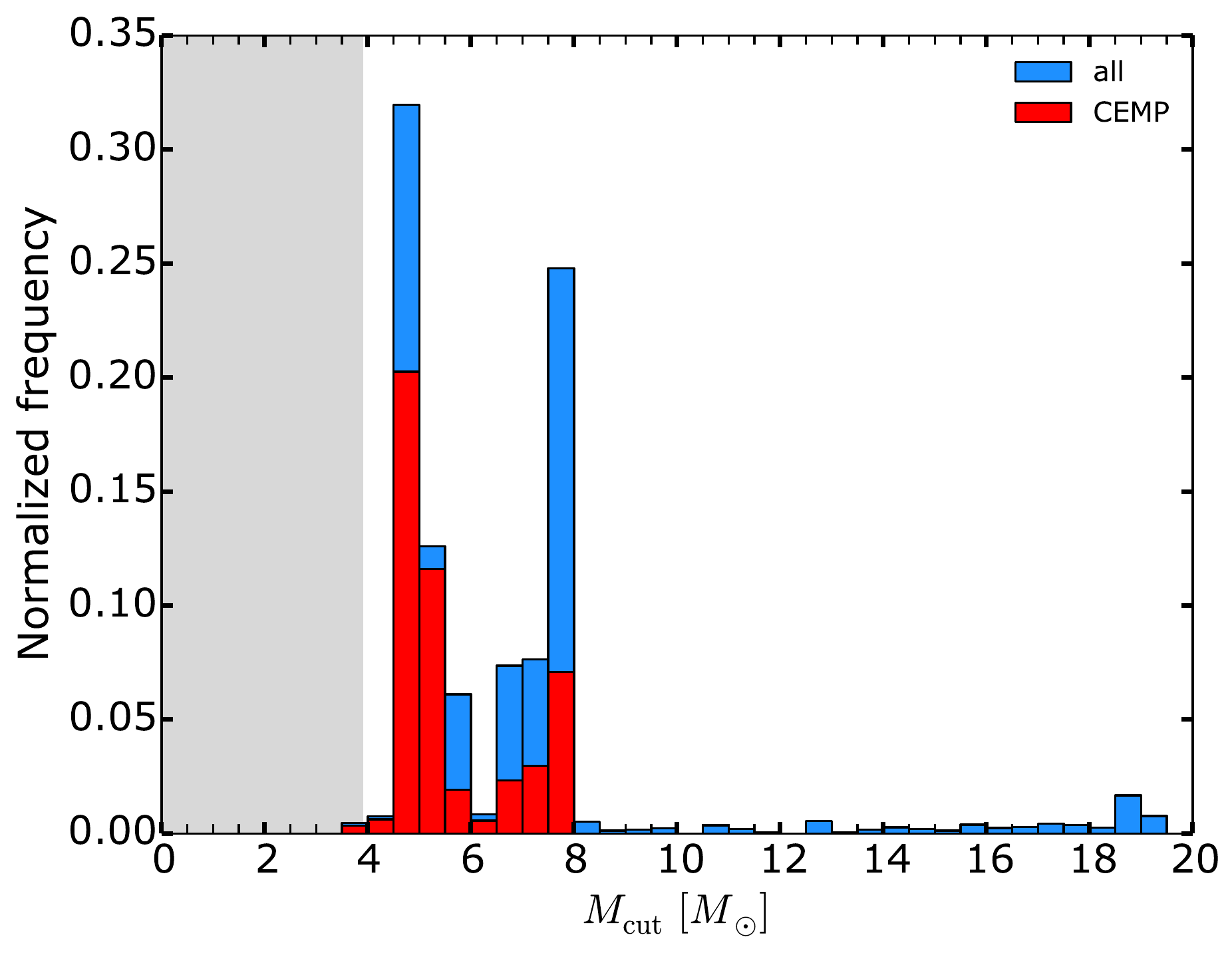}
       \includegraphics[scale=0.46]{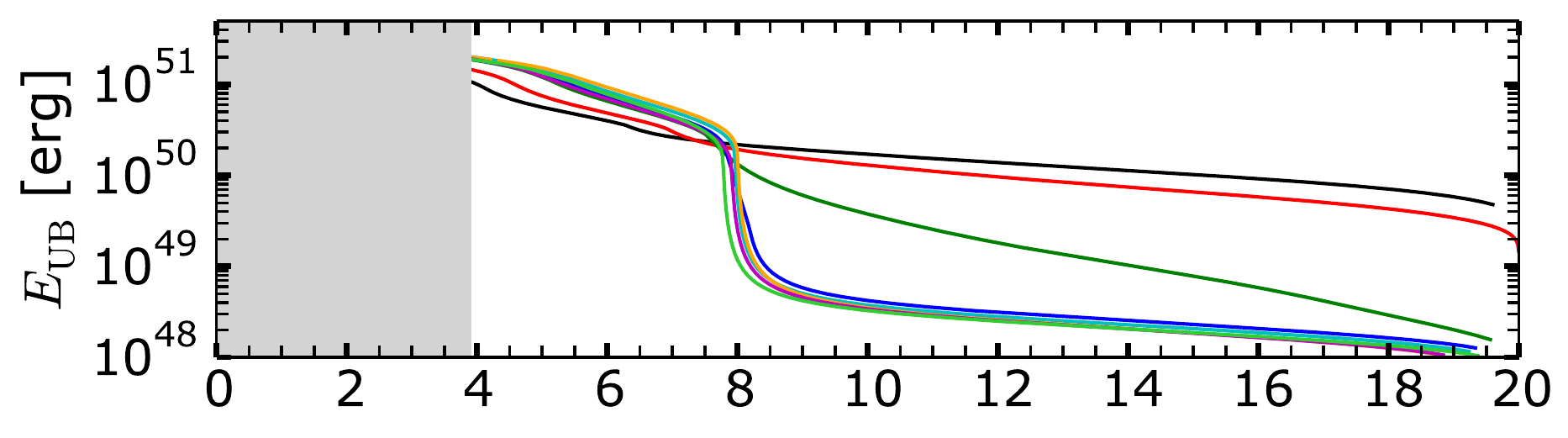}
       \includegraphics[scale=0.46]{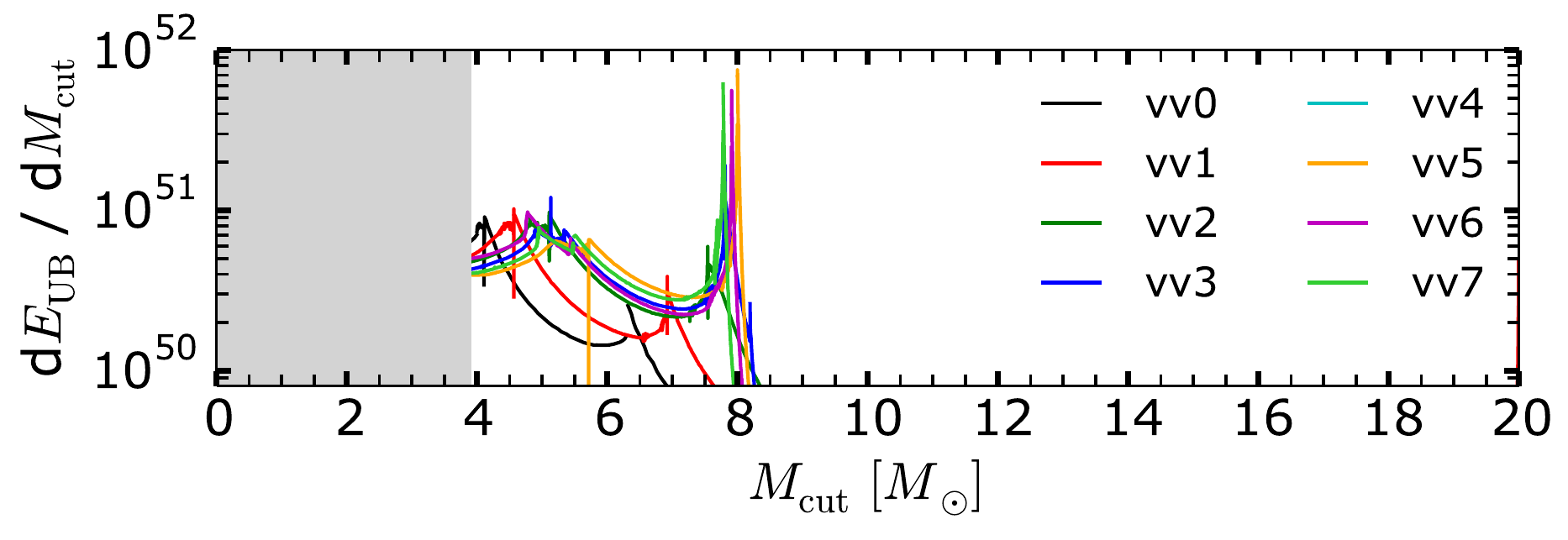}
   \caption{\textit{Top panel}: Distribution of the source stars $M_{\rm cut}$ values derived from the CEMP star sample (red) and the entire sample (blue). 
   \textit{Middle panel} : Energy required to unbind the source star material above the considered $M_{\rm cut}$. It is computed as $E_{\rm UB} (M_{\rm cut}) = \int_{M_{\rm cut}}^{M_{\rm fin}} GM_{\rm r}/r~\text{d}M_{\rm r}$ with $M_{\rm fin}$ the final source star mass (see Table~\ref{table:2}). 
   \textit{Bottom panel} : Derivative of $E_{\rm UB}$ as a function of $M_{\rm cut}$. 
   The shaded regions show the approximate unexplored $M_{\rm cut}$ region (below the bottom of the He-shell).
   }
\label{histmc}
    \end{figure}

\begin{figure}
   \centering
       \includegraphics[scale=0.46]{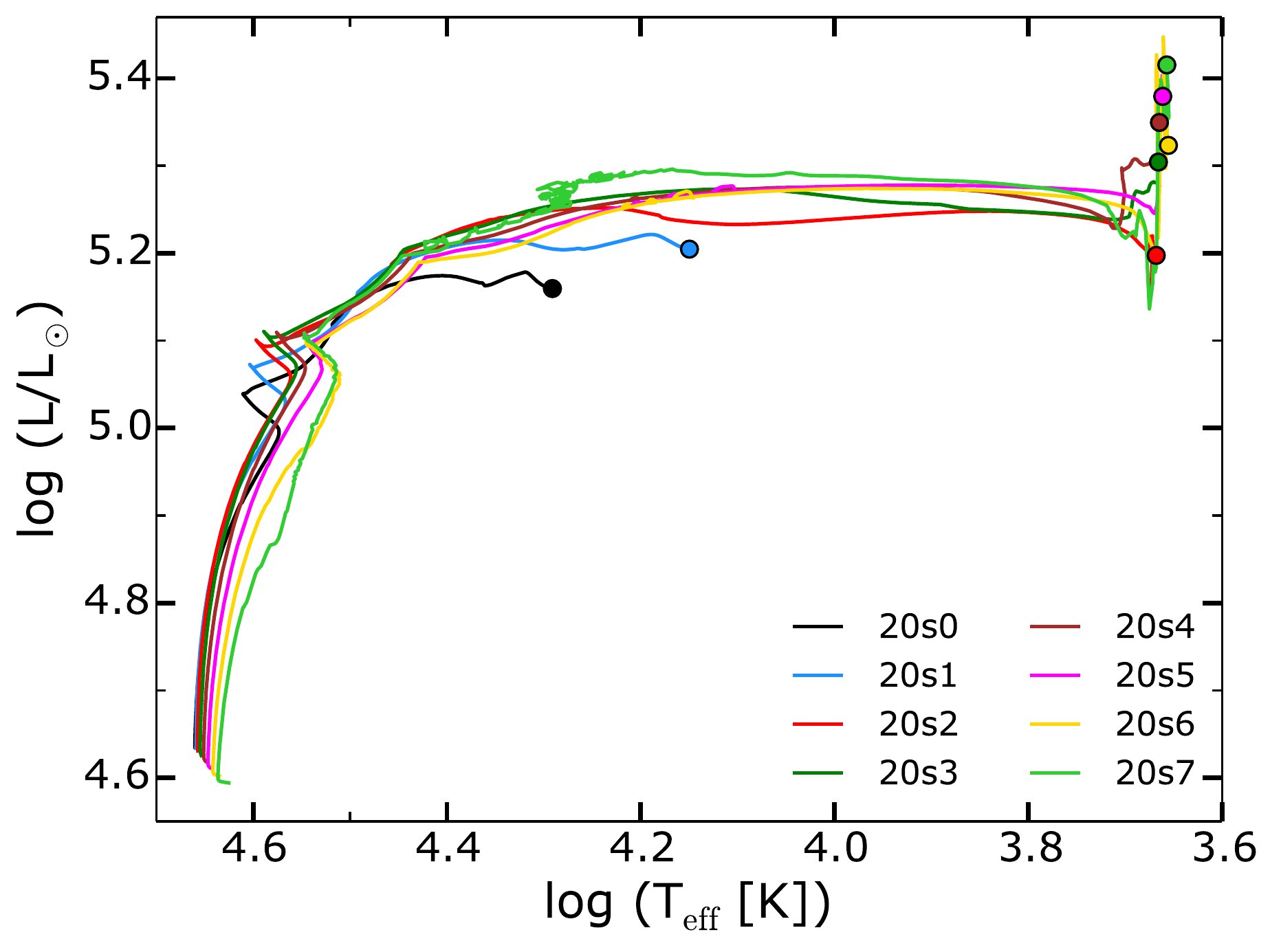}
   \caption{Tracks of the models in the Hertzsprung--Russell diagram. The circles denote the endpoints of the evolution. 
   }
\label{hrdfig}
    \end{figure}

\begin{figure}
   \centering
       \includegraphics[scale=0.46]{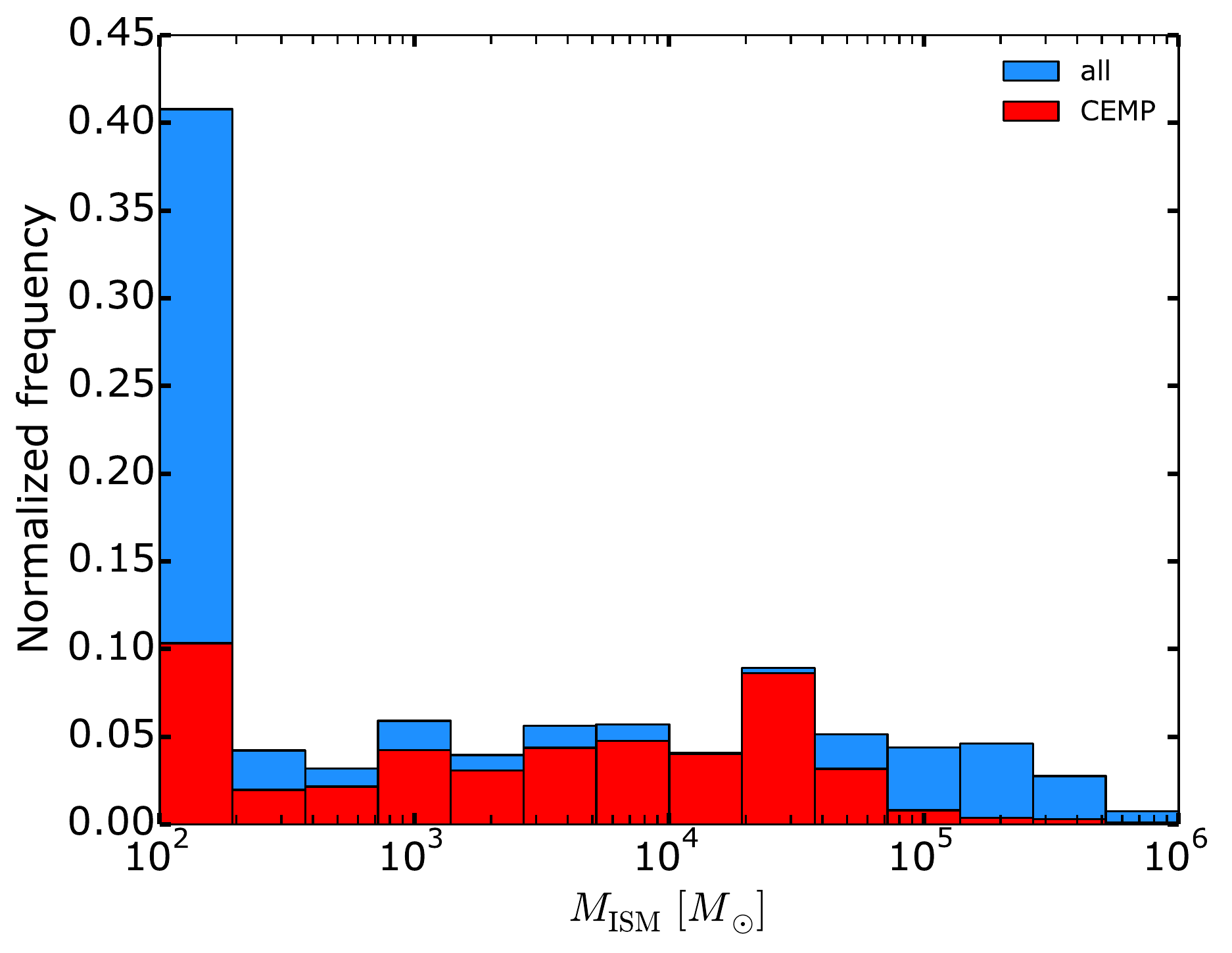}
   \caption{Distribution of the $M_{\rm ISM}$ values derived from the CEMP sample (red) and the entire sample (blue). 
   }
\label{histdil}
    \end{figure}

In most cases, even with large $M_{\rm ISM}$, the source star material still dominates. For instance, BD-18\_5550 (Fig.~\ref{star1}, top right panel) is best fitted by the vv6 model with ($\chi^2, \chi^2_{\nu}$) = (1.74, 0.87), $M_{\rm cut} = 4.74$~$M_{\odot}$, and $M_{\rm ISM} = 5.1 \times 10^{4}$ $M_{\odot}$. This $M_{\rm ISM}$ value corresponds to $\sim 0.07$~$M_{\odot}$ of carbon. In the source star ejecta there is $\sim 0.59$~$M_{\odot}$ of carbon\footnote{Even if the mass of carbon is high, the resulting [C/H] ratio may be low, as a result  of the large amount of added hydrogen.}. Hence, the resulting carbon content reflects more  source star material than  ISM material.
When $M_{\rm cut} \sim 8$~$M_{\odot}$ (i.e. around the second peak, Fig.~\ref{histmc}), less carbon is ejected from the source star. The star BS16920-017 is best fit  by the vv5 model with ($\chi^2, \chi^2_{\nu}$) = (1.69, 0.84), $M_{\rm cut}=7.99$~$M_{\odot}$ and $M_{\rm ISM} = 10^{2}$~$M_{\odot}$. In this case the carbon mass from the stellar ejecta and the ISM are similar, about $10^{-4}$~$M_{\odot}$. However, the total amount of metals is higher in the stellar ejecta ($\sim 3.2 \times 10^{-3}$~$M_{\odot}$) compared to the added ISM material ($\sim 9.6 \times 10^{-4}$~$M_{\odot}$). This is mainly because there is more nitrogen in the stellar ejecta ($\sim 2.5 \times 10^{-3}$~$M_{\odot}$) than in the added ISM ($\sim 10^{-5}$~$M_{\odot}$). The stellar ejecta is enriched in nitrogen because of the operation of the CNO cycle, which was boosted by the progressive arrival of $^{12}$C and $^{16}$O from the He-burning core.

Generally, apart from the most outer source star layers, the model loses memory of its initial chemical composition as evolution proceeds. The overproduction of C, N, O, Na, Mg, and Al in the source star is mainly due to the primary channel, which means that they were synthesized from the initial H and He content, not the initial metal content. 
Consequently, and also because the source star material generally dominates over the ISM material, the results should not depend critically on the initial metal mixture considered.

\subsubsection{Velocity distribution}\label{veldist}

Figure~\ref{histrot} shows the weighted $\upsilon_{\rm ini}/\upsilon_{\rm crit}$ distribution of the source star models. The distribution is normalized to 1.
There is a clear difference between the distribution derived from the C-normal   and CEMP   stars. 
For C-normal stars the amount of good fits is comparable, whatever the initial rotation rate. It gives a rather flat $\upsilon_{\rm ini}/\upsilon_{\rm crit}$ distribution; instead, for CEMP stars the ability to find good fits globally increases from non-rotation to fast rotation by a factor of $\sim 6-7$. 
This difference is due to the fact that C-normal stars generally have low [X/H] ratios, which can often be reproduced by any source star models. It eventually provides similar weights to all the source star models.
The CEMP stars are enriched in carbon and also very often in N, O, Na, Mg, and  Al, whose production increases with the initial source star rotation. While a large C or O abundance can be achieved in every source star model (Fig.~\ref{xhrot}, right panel), a large N, Na, Mg or Al abundance is preferentially obtained in fast rotating models. It gives higher weights to rotating models, hence the difference between the red and green distribution in Fig.~\ref{histrot}.

The overall $\upsilon_{\rm ini}/\upsilon_{\rm crit}$ distribution follows a trend in between the C-normal EMP and CEMP distributions. The number of good fits increases by a factor of $\sim 3$ from no rotation to fast rotation.
For indicative purposes, three fits of the $\upsilon_{\rm ini}/\upsilon_{\rm crit}$ distribution are shown in Fig.~\ref{histrot}. The grey fit is of the form $a x + b$, with $(a,b) = (0.192, 0.058)$. The blue fit is a fourth-degree polynomial: $\sum_{n=0}^{4} a_{n} x^{n}$ with $(a_0, a_1, a_2, a_3, a_4) = (0.064, 0.31, -1.71, 4.91, -3.90)$. The red fit is the sum of a normal distribution and a skew normal distribution. It is of the form
\begin{equation}
\alpha~g(x; a_{0},b_{0}) + \beta~f(x; a_{1},b_{1}, c_{1}) 
\label{fitt}
,\end{equation}
where
\begin{equation}
g(x;a,b) = e^{-(x-a)^2 / b^2}
\end{equation}
and
\begin{equation}
f(x;a,b,c) = g(x;a,b) \left[ 1+erf\left(c \frac{x-a}{b}\right)  \right].
\end{equation}
The coefficients in Eq.~\ref{fitt} are $(\alpha, \beta, a_0, b_0, a_1, b_1, c_1) = (0.14, 3.50 \times 10^{-2}, 5.77 \times 10^{-1}, 6.41 \times 10^{-1}, 5.42 \times 10^{-1}, 2.44 \times 10^{-1}, 6.0)$.
Although the red fit has the smallest residuals, we cannot exclude the other fits, which also provide reasonable agreements to the distribution.

\subsection{Impact on the individual elements on the fitting}\label{impact}

In order to check which chemical element(s)  can have the largest impact in determining the best EMP source stars, we first inspected two individual stars (CS29498-043 and CS29502-092, shown in Fig.~\ref{star1}) which have all seven  abundances determined\footnote{However, these stars are  evolved with $\log g<2$ so that [N/H] and $^{12}$C/$^{13}$C are set as upper and lower limits, respectively.}. We fitted these stars while removing each time one of the seven elements from the fit. Overall, the fits are similar and the ranking of the source star models remains the same, except when removing Mg from the fit. Without Mg, the best fits for these two stars tend to be slower rotators.
This shows that Mg may be an important element for constraining the rotation of the EMP source stars.

To check  further, we performed the fitting analysis again while removing each time one of the seven elements from the analysis. When removed, an element was not considered to fit any of the EMP stars. 
Removing either N, O, or $^{12}$C/$^{13}$C did not affect significantly the distributions shown in  Figs.~\ref{histmc}, \ref{histdil}, and \ref{histrot}.
On the contrary, C and Mg (Na and Al to a smaller extent) mostly determine the shape of the distributions. This is mainly due to the fact that the number of stars with a measured N, O, or $^{12}$C/$^{13}$C ratio is low compared to the number of stars having a C, Na, Mg, or Al abundance (see Table~\ref{table:3}). 
In addition, some stars with a measured N or $^{12}$C/$^{13}$C ratio have $\log g < 2$ so that only limits were considered (see Sect.~\ref{intsample}). 
Also, the observational uncertainties $\sigma_{\rm o}$ for N are on average higher than for C, Na, Mg, and Al (see Sect.~\ref{fitproc}). Overall, it  gives less weight to the N abundance compared to other abundances. 
In the end, C and Mg (Na and Al to a smaller extent) have the highest impact on the derived distributions.

\subsection{Robustness of the fitting analysis}\label{robust}

\subsubsection{Size of the abundance data}

\begin{figure}
   \centering
       \includegraphics[scale=0.46]{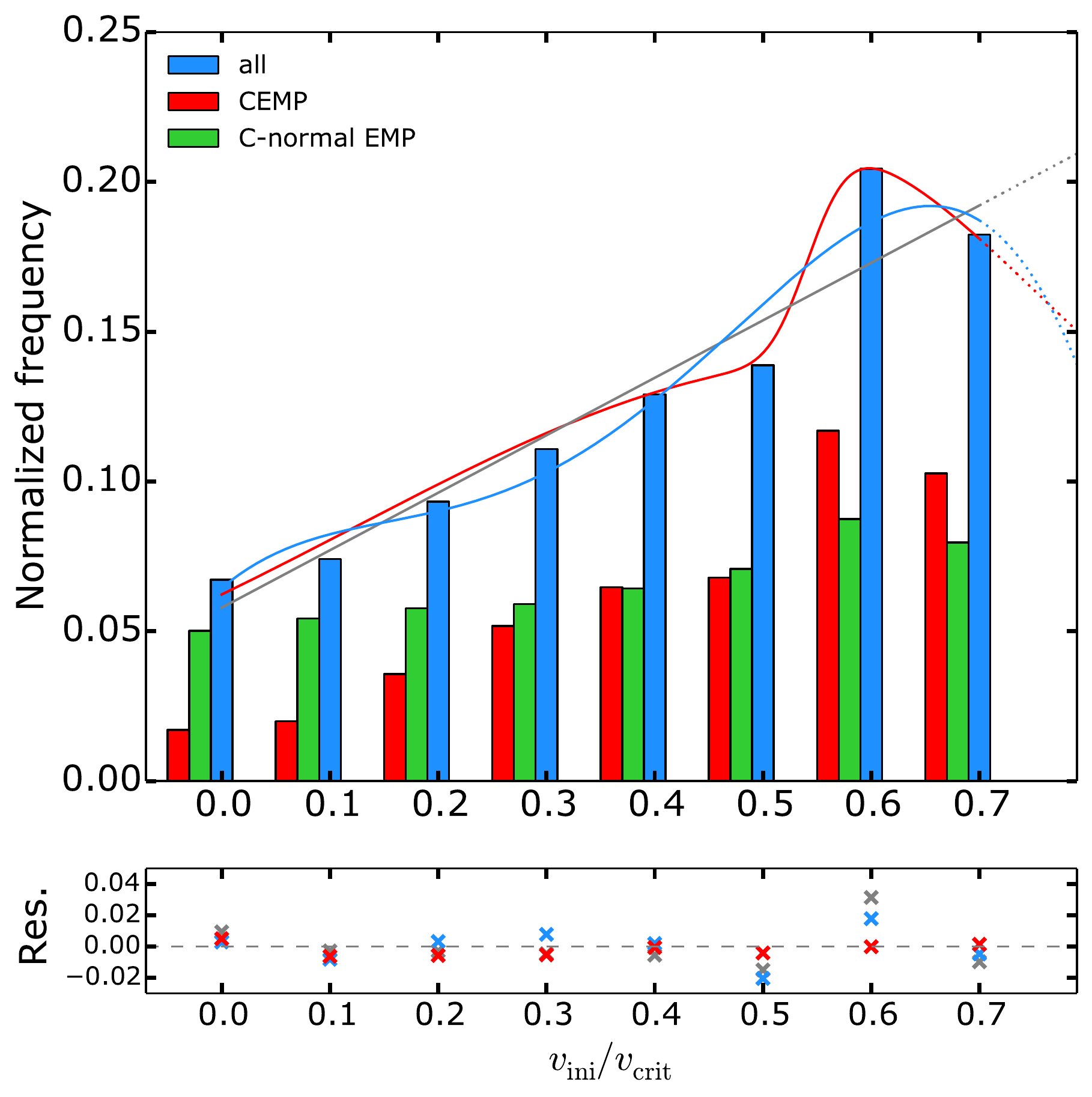}
   \caption{Distribution of the source stars $\upsilon_{\rm ini}/\upsilon_{\rm crit}$ values derived from the CEMP sample (red), the C-normal EMP sample (green), and the entire sample (blue).
   Three fits of the distribution are shown (red, blue, and grey curves; see text for details) together with their associated residuals (bottom panel). }
\label{histrot}
    \end{figure}

The amount of abundance data available varies from one EMP star to another. To check the robustness of our results against the number of abundance constraints, we performed   the fitting analysis again by selecting only the EMP stars having at least five measured abundances. While 119 EMP stars have at least five abundances or limits, only 38 have at least five determined abundances. Considering only these 38 stars for the analysis gives a similar increasing trend to that seen in Fig.~\ref{histrot}, and a similar double peaked mass cut distribution (Fig.~\ref{histmc}). Compared to Fig.~\ref{histdil}, the $M_{\rm ISM}$ distribution is more equally distributed between $10^2$ and $\sim 10^5$~$M_{\odot}$. These overall similar results suggest that the derived distributions are robust against the number of abundance constraints.

\subsubsection{Varying the theoretical uncertainties}\label{thuncer}

The chemical yields of stellar models depend on processes such as convection, nuclear reaction rates, or the physics of rotation for instance. These processes contain uncertainties that cannot be estimated easily (see also discussion in Sect.~\ref{secphys}).
In this work, the model uncertainties $\sigma_{t,i}$ were treated approximately and set to fixed values (see Sect.~\ref{fitproc}). 

As a simple (and limited) approach to investigate the impact of uncertainties on the results, we performed the previous analysis while changing the $\sigma_{t,i}$.
We found that increasing the $\sigma_{t,i}$ gives a similar increasing trend for the $\upsilon_{\rm ini}/ \upsilon_{\rm crit}$ distribution (Fig.~\ref{histrot}), but progressively flattens the distribution. 
This occurs because with larger uncertainties, the source stars models that had high $\chi^2$ values (mostly non-rotating or slow rotating models) now have   lower $\chi^2$ values and therefore greater weights. It increases the contribution of these models.
For instance, setting $\sigma_{t,i} = 0.2$ for [C/H], [N/H], and [O/H] (instead of 0.15) and $\sigma_{t,i} = 0.4$ for [Na/H], [Mg/H], and [Al/H] (instead of 0.3) gives a contrast (ratio of the highest $\upsilon_{\rm ini}/ \upsilon_{\rm crit}$ frequency to the lowest $\upsilon_{\rm ini}/ \upsilon_{\rm crit}$ frequency) of 2.3, against 3.0 in the standard case. 
Increasing $\sigma_{t,i}$ again to 0.25 for C, N, and O and 0.5 for Na, Mg, and Al gives a contrast of 1.9.
Overall, the specific slope (or shape) of the $\upsilon_{\rm ini}/ \upsilon_{\rm crit}$ depends on the adopted uncertainties, but the dependance likely stays modest.

\subsubsection{Internal mixing in EMP stars}\label{mixemp}

   \begin{table}
\scriptsize{
\caption{Same as Table~\ref{table:4}, but  considering very inefficient internal mixing processes in EMP stars (see text). \label{table:6}}
\begin{center}
\resizebox{8.1cm}{!} {
\begin{tabular}{l|ccccc} 
\hline 
& $\chi_{\nu}^{2}<10$  & $\chi_{\nu}^{2}<5$ & $\chi_{\nu}^{2}<3$ & $\chi_{\nu}^{2}<2$ & $\chi_{\nu}^{2}<1$ \\
\hline 
All stars           &               95 \%              &       77 \%              &  53 \%   &       33 \%  &       14 \%  \\
CEMP             &               98 \%              &       84  \%             &  67 \%   &       57 \% &       33 \%  \\
C-normal EMP     &               94 \%              &       74  \%       &  46 \%  &       21 \% &       5  \% \\
\hline 
\end{tabular}
}
\end{center}
}
\end{table}

The efficiency of the internal mixing processes in low mass stars is still discussed, particularly thermohaline mixing 
\citep[e.g.][]{denissenkov11, traxler11, wachlin14, sengupta18}.
If these processes were very efficient in EMP stars, we should see a clear separation between the C, N, and $^{12}$C/$^{13}$C abundances of evolved and unevolved EMP stars, but this is not the case \citep{choplin17a}. As a test, we did the analysis again, assuming inefficient internal mixing processes in EMP stars (i.e. their C, N, and $^{12}$C/$^{13}$C abundances did not change since their birth). In this case the carbon correction from \cite{placco14c} was not considered and the N abundance and $^{12}$C/$^{13}$C ratio were not considered as limits if the EMP star was evolved ($\log g < 2$).
This new analysis slightly decreases the number of good fits (Table~\ref{table:6}) because additional constraints were added (in the evolved EMP stars, N and $^{12}$C/$^{13}$C are now data points and not limits). 
The $\upsilon_{\rm ini}/\upsilon_{\rm crit}$ distribution is barely affected. The maximum change in the distribution shown in Fig.~\ref{histrot} is about 4~\%. 
This shows that, based on the current understanding on internal mixing processes (mainly dredge-up and thermohaline mixing), the efficiency of these processes, for the current sample of EMP stars, may not impact much the results presented here.

We also carried out the analysis again while considering only the not-so-evolved EMP stars because the abundances of evolved EMP stars suffer additional uncertainties, and thus their inclusion in the analysis may reduce its robustness. Among the 272 EMP stars, we selected the 128 EMP stars having $\log g \geq 2$. The $\upsilon_{\rm ini}/\upsilon_{\rm crit}$, $M_{\rm cut}$ and $M_{\rm ISM}$ distributions were found to be comparable to the distributions derived from the entire sample. 
In particular, the overall $\upsilon_{\rm ini}/\upsilon_{\rm crit}$ distribution is very similar to that of Fig.~\ref{histrot}.

\subsection{EMP stars with high $\chi^2$}\label{badfit}

\subsubsection*{Stars with [Na, Mg, Al / C] $\gtrsim 0$}
\begin{figure*}
   \centering
   \begin{minipage}[c]{.33\linewidth}
       \includegraphics[scale=0.3]{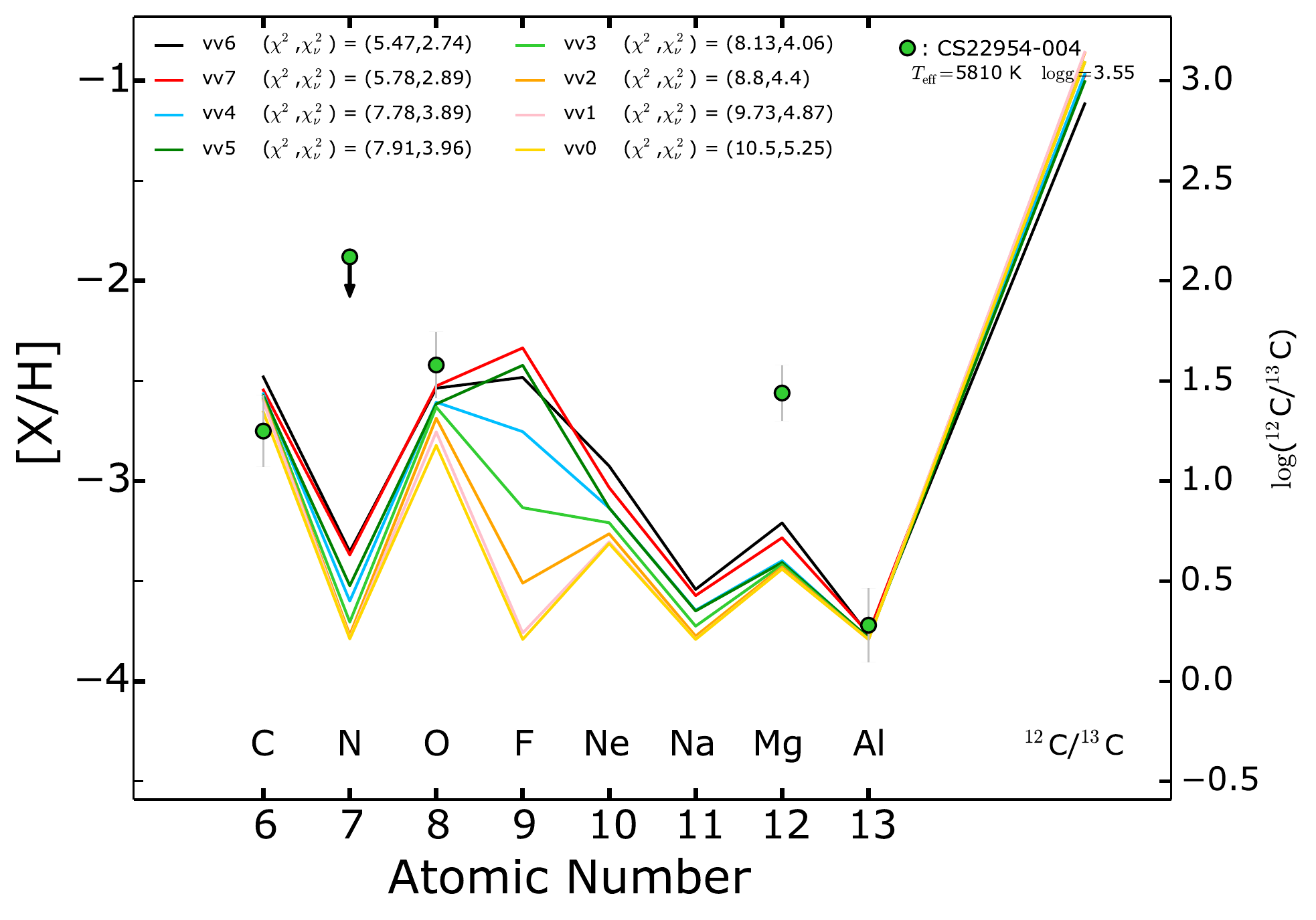}
  \end{minipage}
   \begin{minipage}[c]{.33\linewidth}
       \includegraphics[scale=0.3]{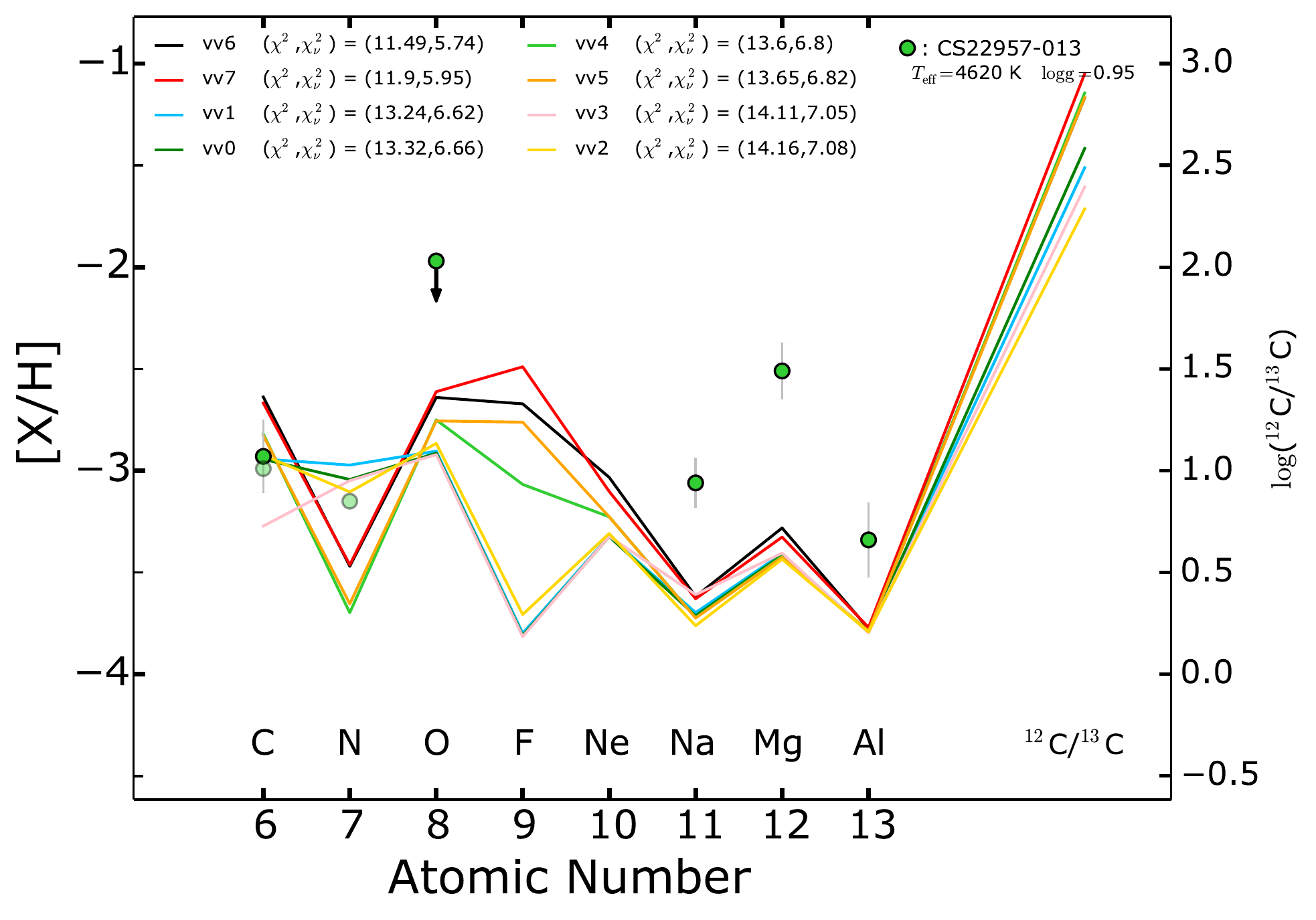}
  \end{minipage}
   \begin{minipage}[c]{.33\linewidth}
       \includegraphics[scale=0.3]{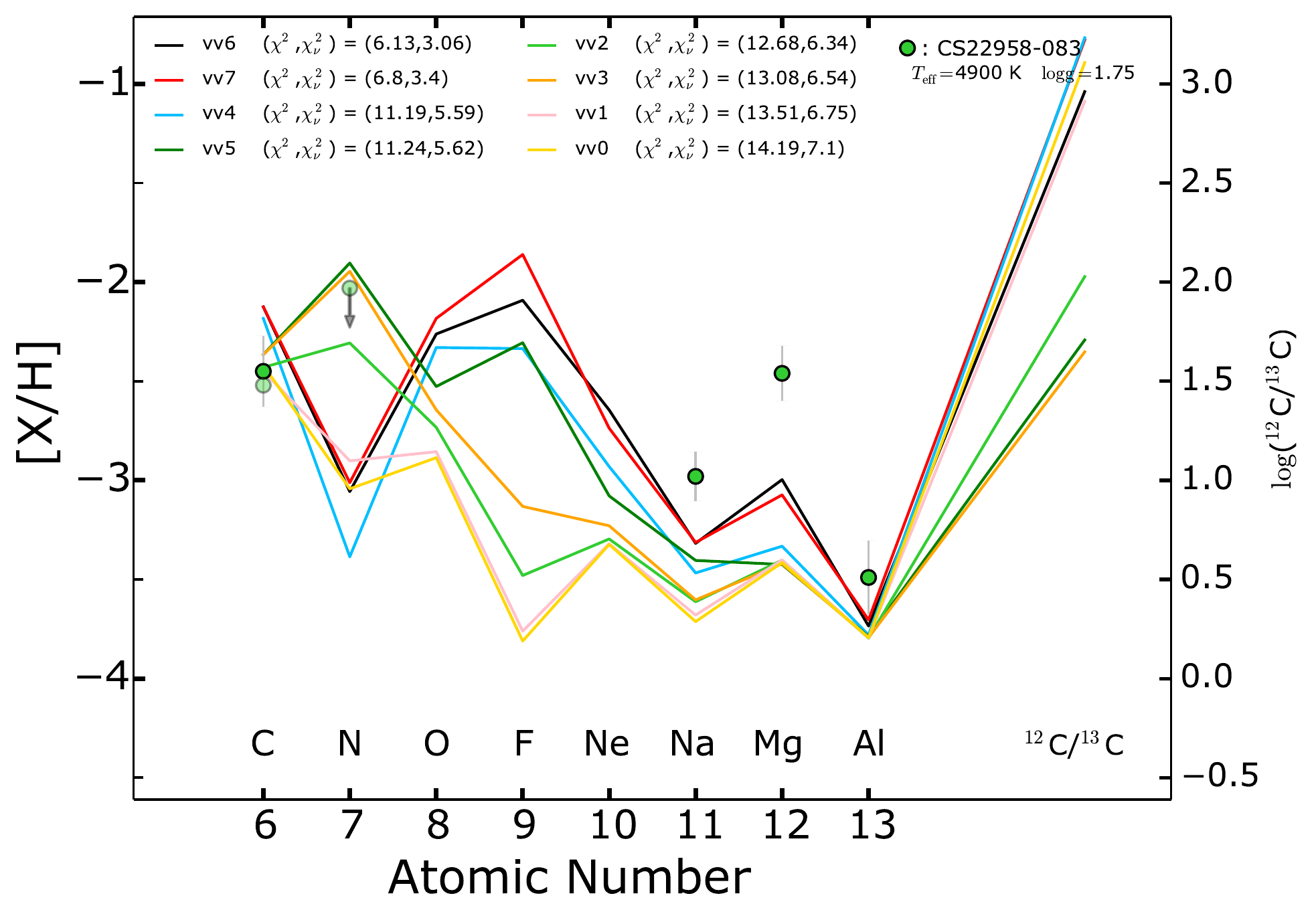}
  \end{minipage}
   \begin{minipage}[c]{.33\linewidth}
       \includegraphics[scale=0.3]{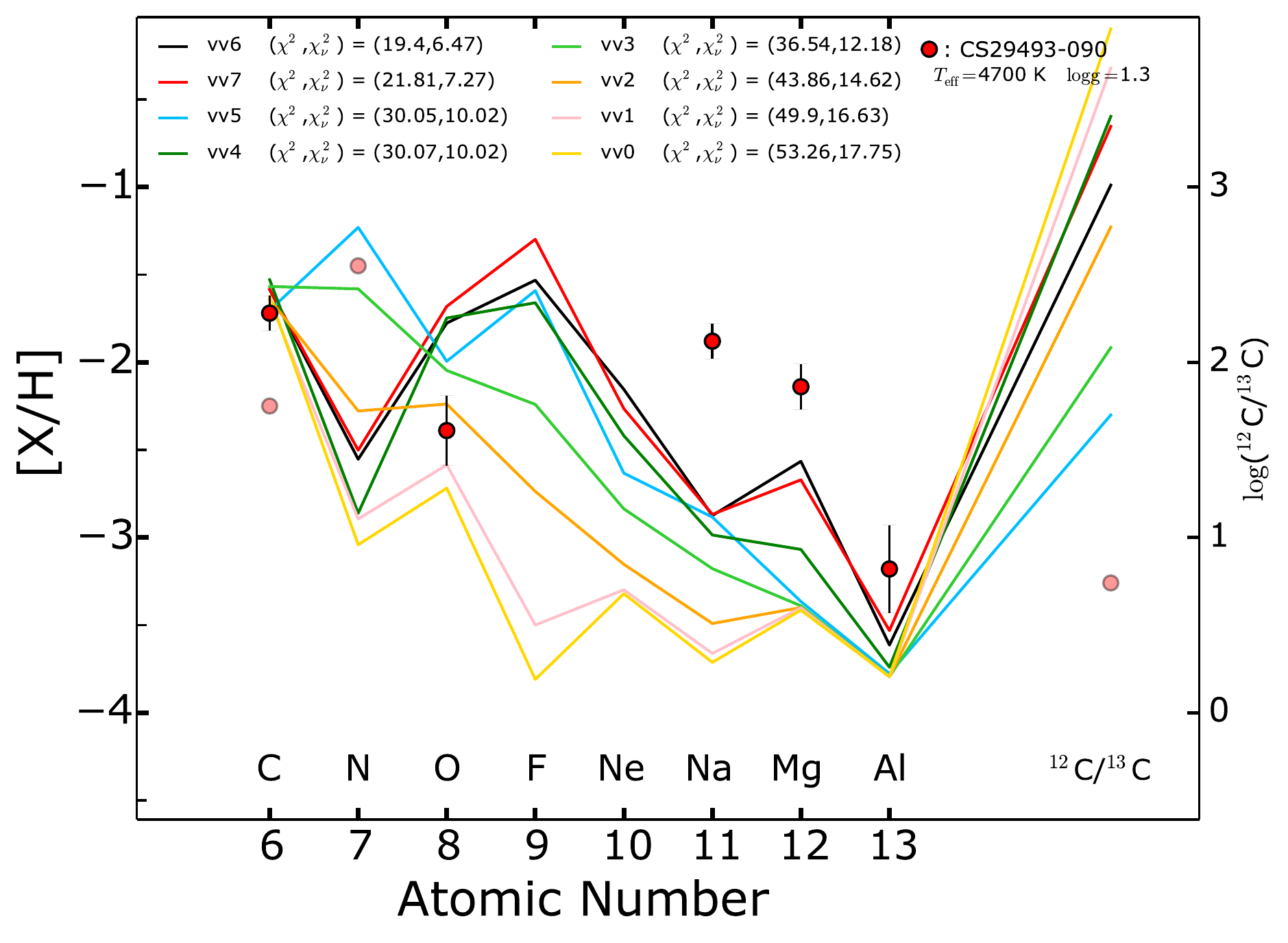}
  \end{minipage}
   \begin{minipage}[c]{.33\linewidth}
       \includegraphics[scale=0.3]{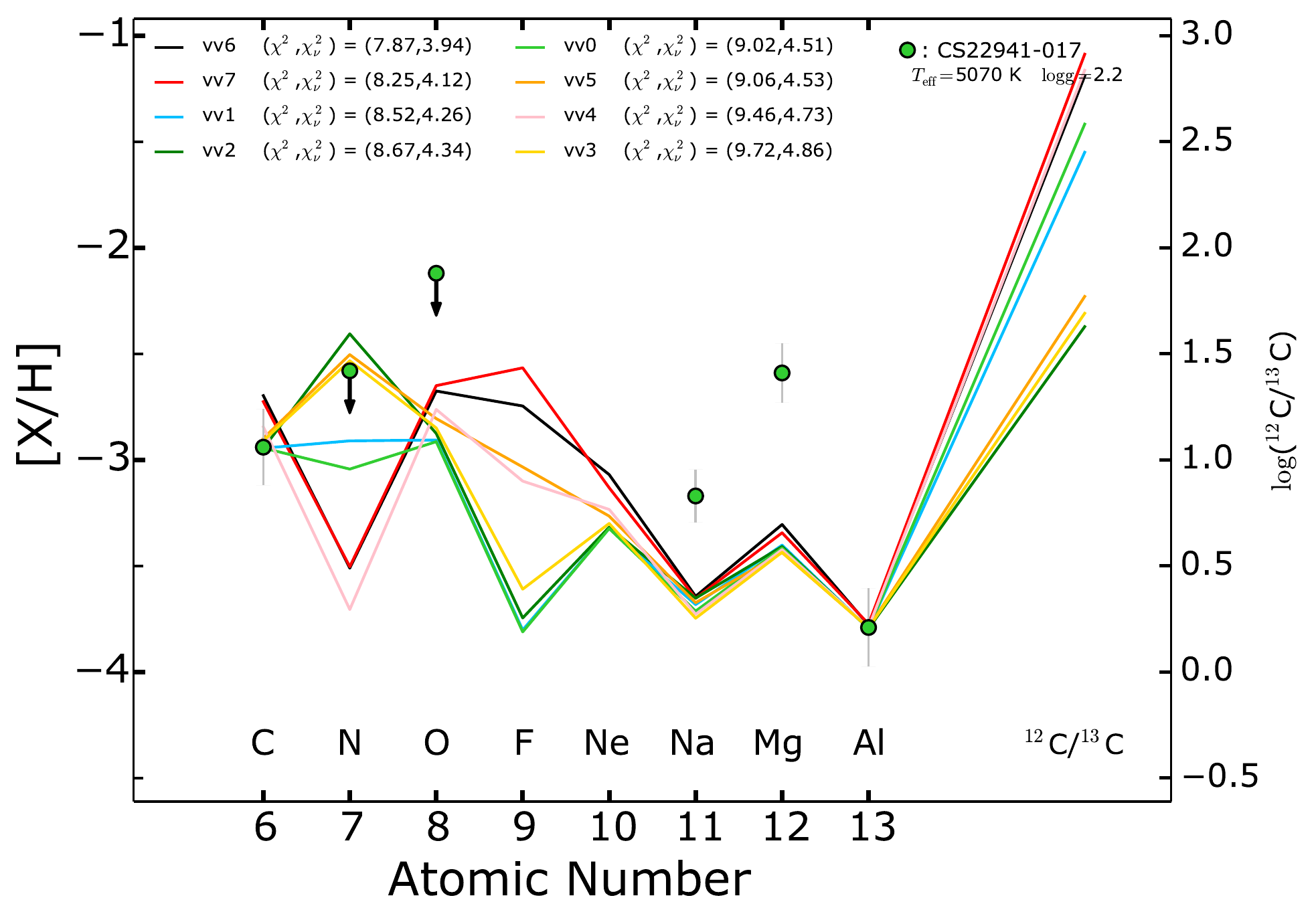}
  \end{minipage}
   \begin{minipage}[c]{.33\linewidth}
       \includegraphics[scale=0.3]{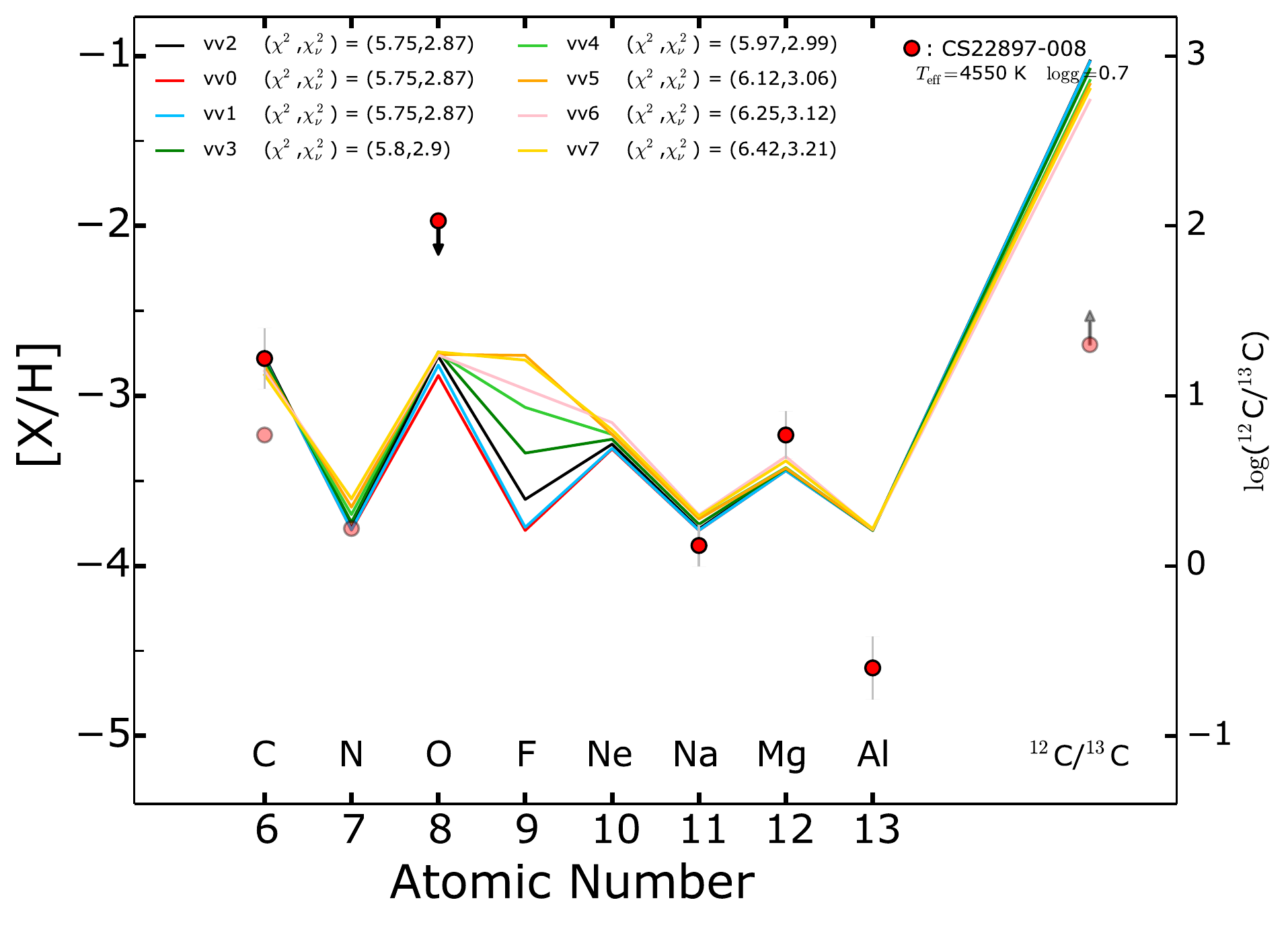}
  \end{minipage}
   \begin{minipage}[c]{.33\linewidth}
       \includegraphics[scale=0.3]{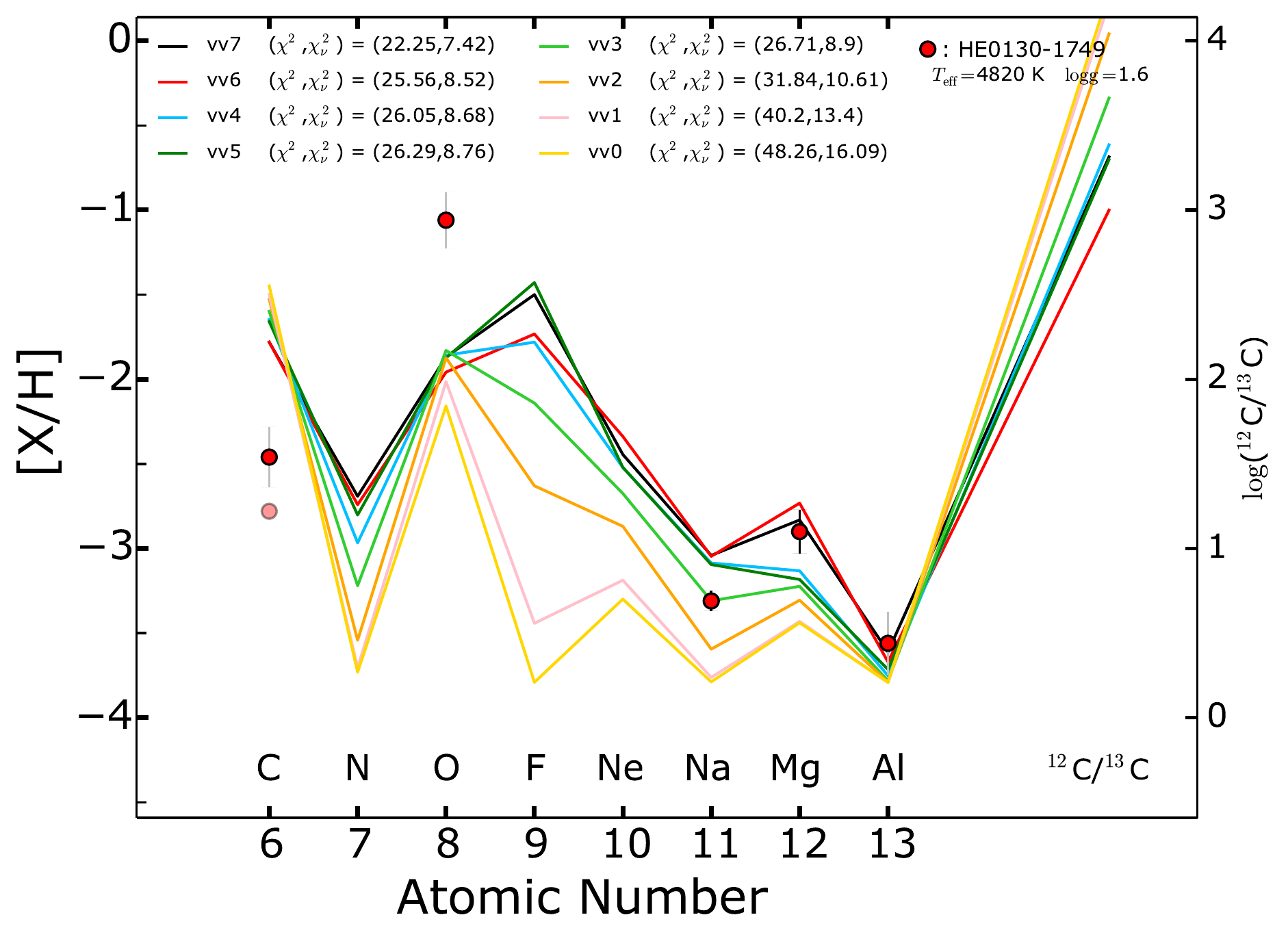}
  \end{minipage}
   \begin{minipage}[c]{.33\linewidth}
       \includegraphics[scale=0.3]{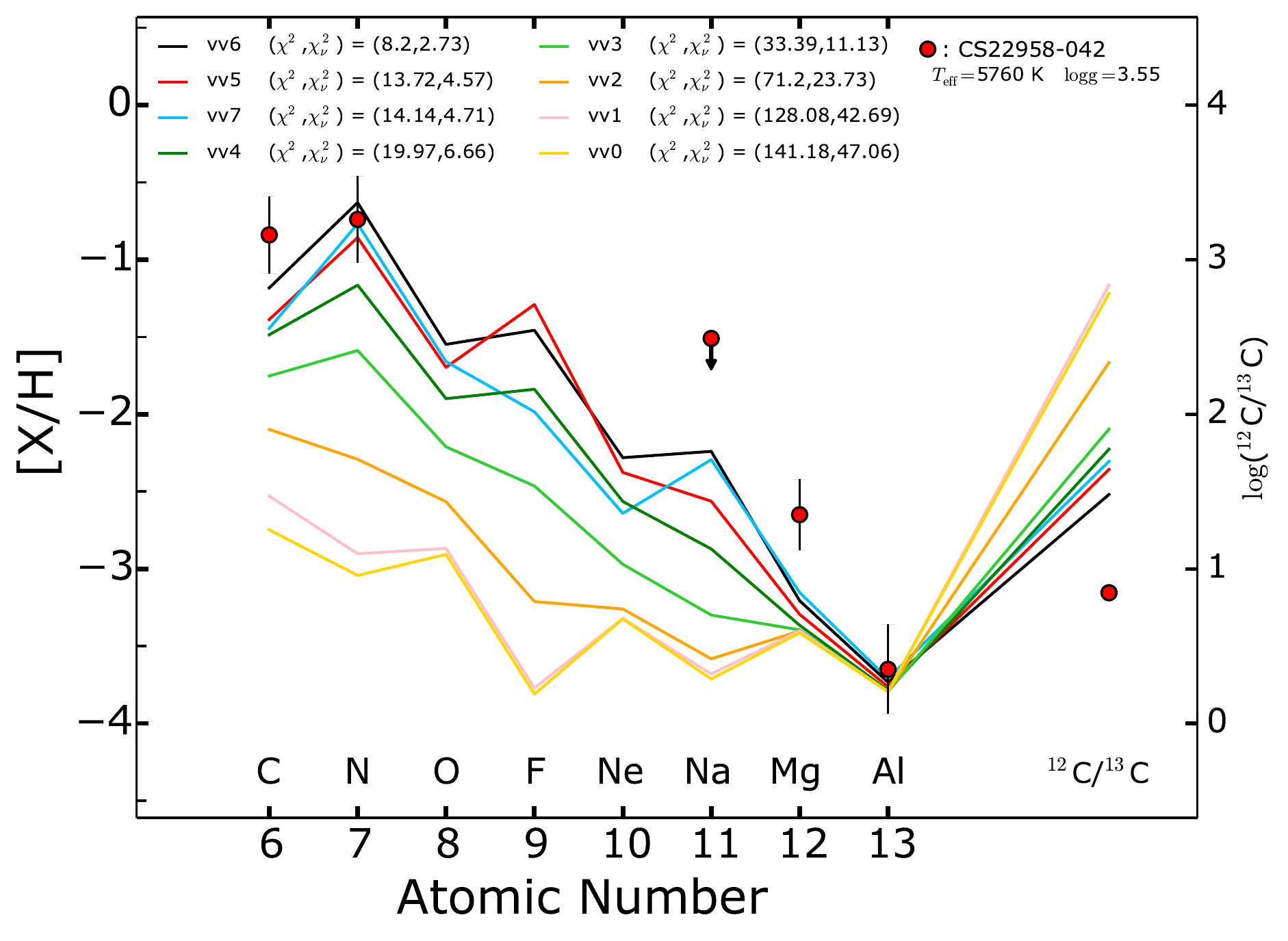}
  \end{minipage}
   \begin{minipage}[c]{.33\linewidth}
       \includegraphics[scale=0.3]{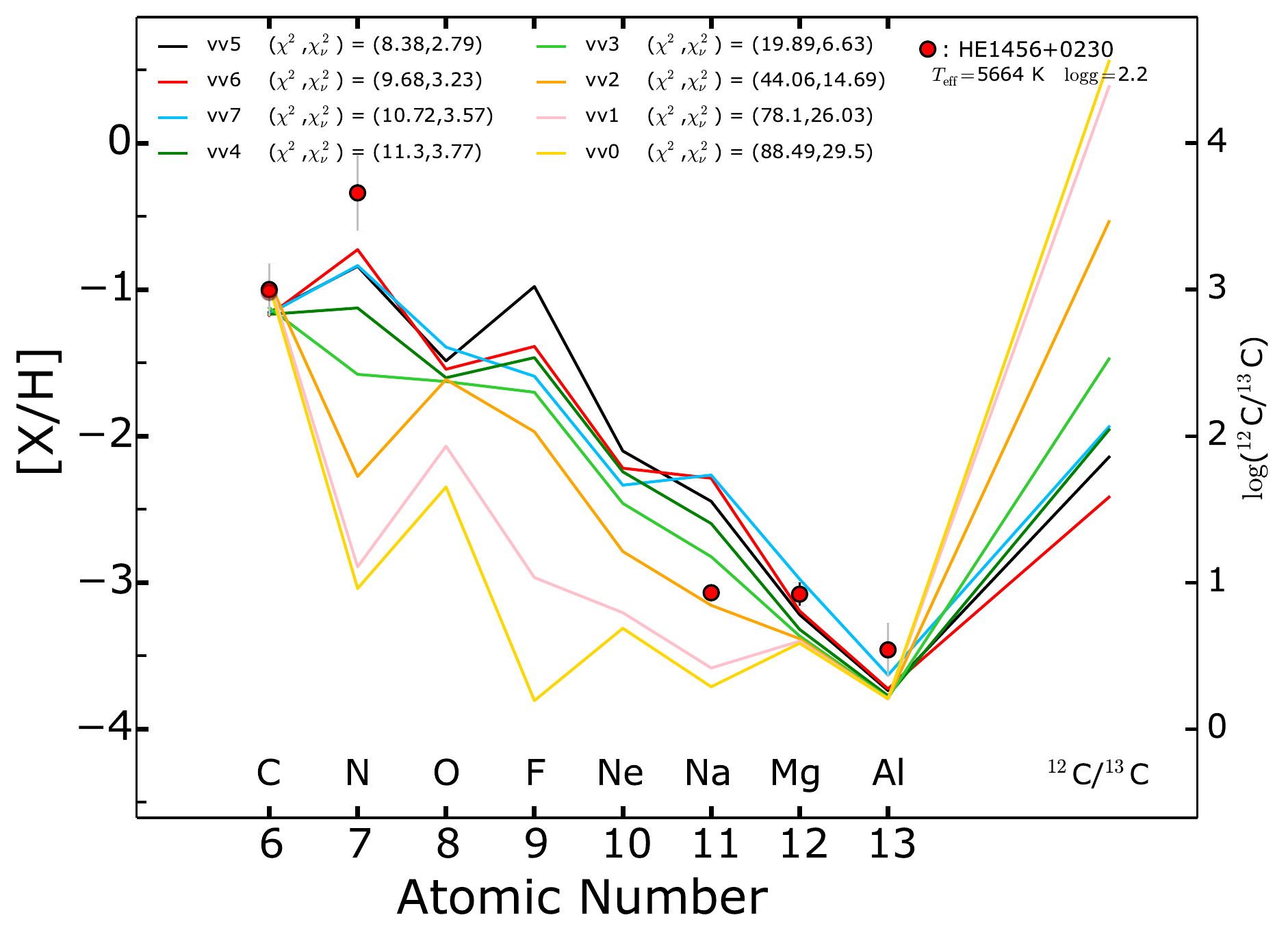}
  \end{minipage}
   \caption{Same as Fig.~\ref{star1}, but for EMP stars with $\chi_{\nu}^2 > 2$ ($\chi^2 > 5$).
   }
\label{bad1}
    \end{figure*}

The shared feature of about $70~\%$ 
of the EMP stars having a high $\chi^2$ is a relatively high [Mg/C] ratio (a high [Na/C] or [Al/C] ratio to a lesser extent), which cannot be achieved with the considered 
assumptions (see the five first panels of Fig.~\ref{bad1}). 
In the source star models considered here, [Mg/C] $<0$ (Fig.~\ref{xhrot}), except in the H-rich layers of non-rotating or slow rotating models (e.g. the black pattern in the left panel of Fig.~\ref{xhrot}), but in this case, the  low [C/H] and [Mg/H] ratios generally cannot account for the EMP star abundances. 
During the advanced stages of evolution and explosion, the most inner stellar layers are enriched in C, O, Na, Mg, Al, and heavier elements until the Fe-group \citep[e.g.][]{thielemann96, woosley02, nomoto06}.
Higher [Mg/C] ratios could be obtained in deeper source star layers. 
This EMP star group with high $\chi^2$ and high [Mg/C] ratios may indicate the need for a different source star material, likely originating from deeper layers and possibly processed by explosive nucleosynthesis. 

\subsubsection*{Stars with very low [X/H] ratios}
Some other EMP stars ($\sim 20$) 
have at least one very low [X/H] ratio, below the ISM values considered here 
(e.g. CS22897-008 with a low [Al/H] ratio; see Fig.~\ref{bad1}, middle right panel).
No solution can give such low values. Zero-metallicity source star models may provide better solutions. 
A more complete study including source stars spanning a range of different initial metallicities will be addressed in a future work.
 
 \subsubsection*{Stars with [O/C] $>0$}
Several EMP stars ($\sim 5-10$) have very high [O/C] ratios (e.g. HE0130-1749, Fig.~\ref{bad1},   bottom left panel). In most of the cases, the source star models predict [O/C] $\leq 0$. Since HE0130-1749 is evolved ($\log g = 1.6$), the initial [C/H] was likely higher than the observed value. However, the [O/C] ratio after the \cite{placco14a} correction is  still well above 0. If the internal mixing processes were much more efficient in HE0130-1749, we might expect a higher [C/H] ratio, which would improve the fit.

 \subsubsection*{Other stars with high $\chi^2$}
The other problematic EMP stars are generally specific cases. 
The unevolved star CS22958-042 has a low $^{12}$C/$^{13}$C (Fig.~\ref{bad1}, bottom centre  panel) that can only be reproduced in the H-rich layers of source stars (Fig.~\ref{xhrot}). 
However, these H-rich layers have [N/C] $\sim 1-2$, while CS22958-042 has [N/C] $\sim 0$. Producing both a low [N/C] with a low $^{12}$C/$^{13}$C is challenging for these models. A numerical experiment was carried out in \cite{choplin17a} in order to try improving the fit for such stars. This peculiar abundance trend can be reproduced if  a late mixing event is included in the source star, occurring between the hydrogen- and helium-burning shell, shortly before the end of the source star evolution. 
In the EMP star sample, only a few stars have a measured C, N, and $^{12}$C/$^{13}$C ratio together. Moreover, some of these stars are evolved and may have undergone important modifications of these abundances. 
More N abundances and $^{12}$C/$^{13}$C ratio measurements in rather unevolved stars are required to  further test the idea of a late mixing event in the source star.

Finally, HE1456+0230 (bottom right panel) has a high [C/H], [N/H] together with [N/C] $>0$. This is consistent with   material processed by H-burning of a rotating source star. However, in such   material, [Na/H] is high, which is not the case for HE1456+0230. Generally, our models predict that a high [N/H] with [N/C] $>0$ should be accompanied with some Na enhancements, together with a low $^{12}$C/$^{13}$C ratio.

\section{Discussions}\label{disc}

\subsection{CEMP and C-normal EMP stars, and source star matter ejection}

While 64~\% (40~\%) of the CEMP stars have $\chi_{\nu}^2 < 2$ ($\chi_{\nu}^2 < 1$), only 26~\% (5~\%) of C-normal EMP stars have $\chi_{\nu}^2 < 2$ ($\chi_{\nu}^2 < 1$, Table~\ref{table:5}). 
This important difference between CEMP and C-normal EMP stars suggests that the assumptions made in this study are adequate for most of the CEMP stars, but are not suitable for a large fraction of C-normal EMP stars. In particular, it is likely that the assumption of an explosion with strong fallback (Sect.~\ref{secfallback}) plays a major role and is more suitable for CEMP than C-normal EMP stars. Different kinds of SNe (e.g. more standard core collapse SNe), that could also have existed in the early Universe, may be predominantly responsible for the abundances of C-normal EMP stars. On the contrary, the great fraction of CEMP stars can be explained by source stars that experienced an explosion with strong fallback.
We note that the idea of strong fallback was already associated with CEMP stars based on the fact that they have high [C/Fe] ratios and that, in the massive source star, C is located in shallower layers than Fe \citep[e.g.][]{umeda03}. Similar conclusions are derived here without considering Fe abundances.

\subsection{Comparison with the velocity of nearby OB stars}\label{rotobstars}

\subsubsection{$\langle \upsilon_{\rm eq} \rangle _ {\rm MS}$ distributions}

\begin{figure}
   \centering
       \includegraphics[scale=0.46]{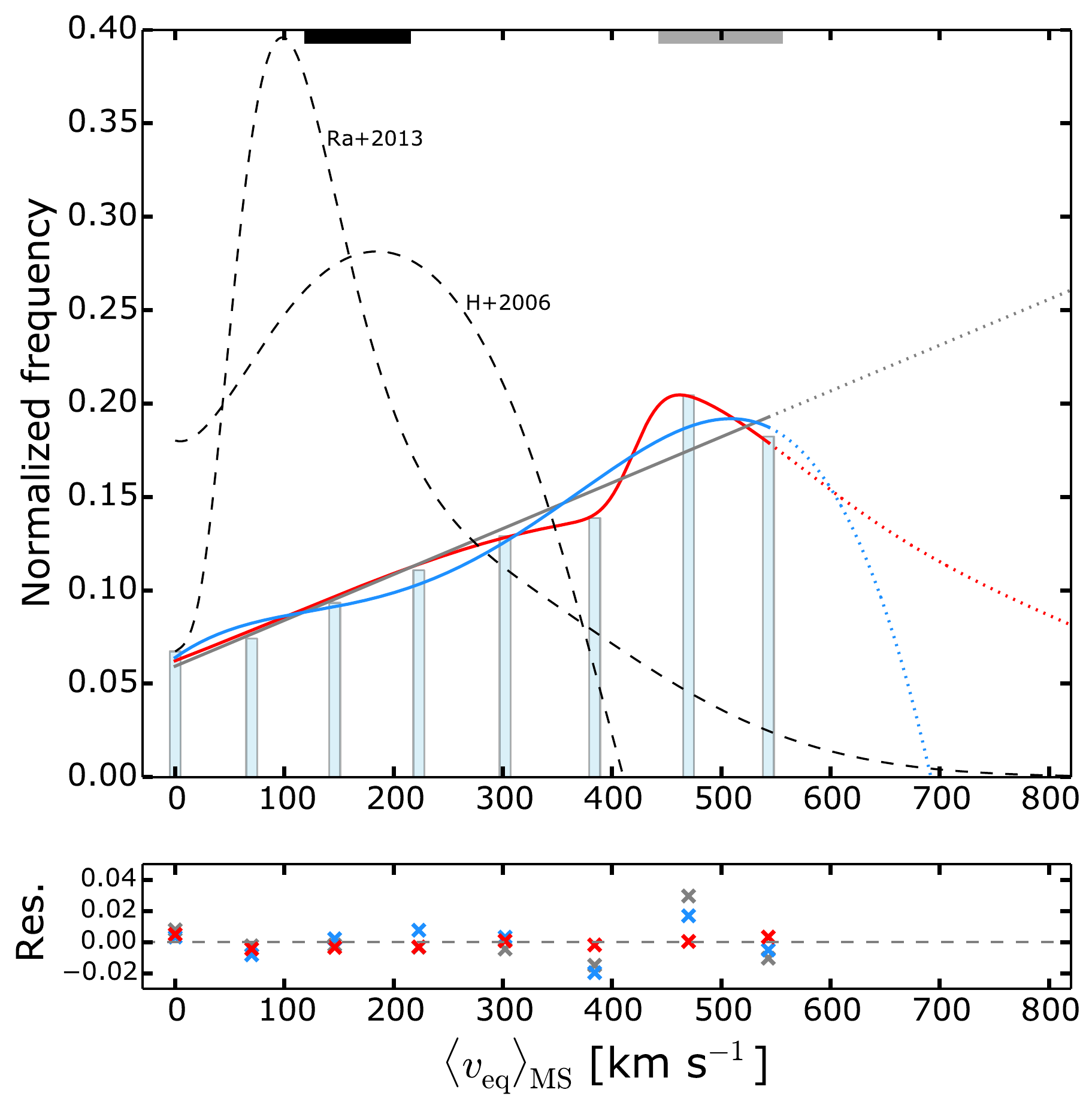}
   \caption{Distribution of the mean equatorial velocity during the MS for the EMP source star models. 
      The three fits are the same as in Fig.~\ref{histrot}, but adapted to the new x-axis (see text for details).   
   The dashed black distributions show two equatorial velocity distributions reported in \citet[][H+2006]{huang06} and \citet[][Ra+2013]{ramirez13}, based on the observation of Galactic and LMC O-type and B-type stars (see text for details). 
   The segments at the top show the $\upsilon_{\rm eq}$ range during the MS for the vv6 model (grey) and for a 20~$M_{\odot}$ model computed at solar metallicity with $\upsilon_{\rm ini}/\upsilon_{\rm crit} = 0.3$ (black).
   }
\label{histMS}
    \end{figure}

It is worth comparing the derived velocity distribution (Fig.~\ref{histrot}) to distributions based on the observation of solar or near-solar metallicity massive stars.
Figure~\ref{histMS} shows the distribution of $\langle \upsilon_{\rm eq} \rangle _ {\rm MS}$, the mean velocity during the main sequence (MS), for our source star models (the $\langle \upsilon_{\rm eq} \rangle _ {\rm MS}$ values are reported in Table~\ref{table:2}). 
These values are  representative of $\upsilon_{\rm eq}$ at any stage of the MS because of the modest variation of $\upsilon_{\rm eq}$ during the MS. This is shown by the grey segment at the top of Fig.~\ref{histMS} that represents the $\upsilon_{\rm eq}$ range during the MS for the vv6 model. 
The three fits in Fig.~\ref{histMS} are the same as in Fig.~\ref{histrot} (see Sect.~\ref{veldist} for details), but adapted to the new x-axis. The parameters are now $(a,b) = (2.46 \times 10^{-4}, 5.94 \times 10^{-2})$ for the grey fit; $(a_0, a_1, a_2, a_3, a_4) = (6.40 \times 10^{-2}, 4.14 \times 10^{-4}, -2.84 \times 10^{-6}, 1.02 \times 10^{-8}, -1.04 \times 10^{-11})$ for the blue fit; and $(\alpha, \beta, a_0, b_0, a_1, b_1, c_1) = (0.14, 3.5 \times 10^{-2}, 4.5 \times 10^{2}, 5.0 \times 10^{2}, 4.2 \times 10^{2}, 1.8 \times 10^{2}, 6.0)$ for the red fit.

The dashed distribution labelled H+2006 shows the equatorial velocity distribution reported in \cite{huang06}, based on the observation of 496 presumably single OB-type stars observed in 19 different Galactic open clusters. 
The dashed distribution labelled Ra+2013 shows the distribution reported in \cite{ramirez13}, based on the observation of 216 presumably single O-type stars observed in the 30 Doradus region of the Large Magellanic Cloud. 
The OB star samples span a range of spectral types which likely correspond to a range of evolutionary stages during the MS. 
As a first-order comparison, we consider here that the velocity distributions of these two OB star samples are representative of $\langle \upsilon_{\rm eq} \rangle_{\rm MS}$ (x-axis of Fig.~\ref{histMS}). 
This may be a reasonable assumption since these two OB star samples likely contain  objects with masses mostly below $30-40$~$M_{\odot}$ and the surface equatorial velocity of solar metallicity models with $M_{\rm ini}~\lesssim~32$~$M_{\odot}$ does not vary strongly during the MS \citep[][especially their Fig.~10]{ekstrom12}.
The black segment at the top of Fig.~\ref{histMS} illustrates this by showing the range of $\upsilon_{\rm eq}$ during the MS for a 20 $M_{\odot}$ solar metallicity model computed at $\upsilon_{\rm ini}/ \upsilon_{\rm crit} = 0.3$.

The H+2006 and Ra+2013 distributions peak at 200 and 100 km~s$^{-1}$. 
Whether the distribution derived from EMP stars peaks at $\sim 500$ km~s$^{-1}$ or would still increase toward higher velocities cannot be determined. 
In any case, a higher fraction of fast rotators and a flatter distribution can be noticed, compared to the distributions derived from observations.
We note that while the comparison is made through $\langle \upsilon_{\rm eq} \rangle_{\rm MS}$, the velocity differences are likely significant  because the variations in $\upsilon_{\rm eq}$ during the MS are modest compared to the peak difference (see the two segments at the top of Fig.~\ref{histMS}).

\subsubsection{Two different metallicity regimes}

\begin{figure}
   \centering
       \includegraphics[scale=0.48]{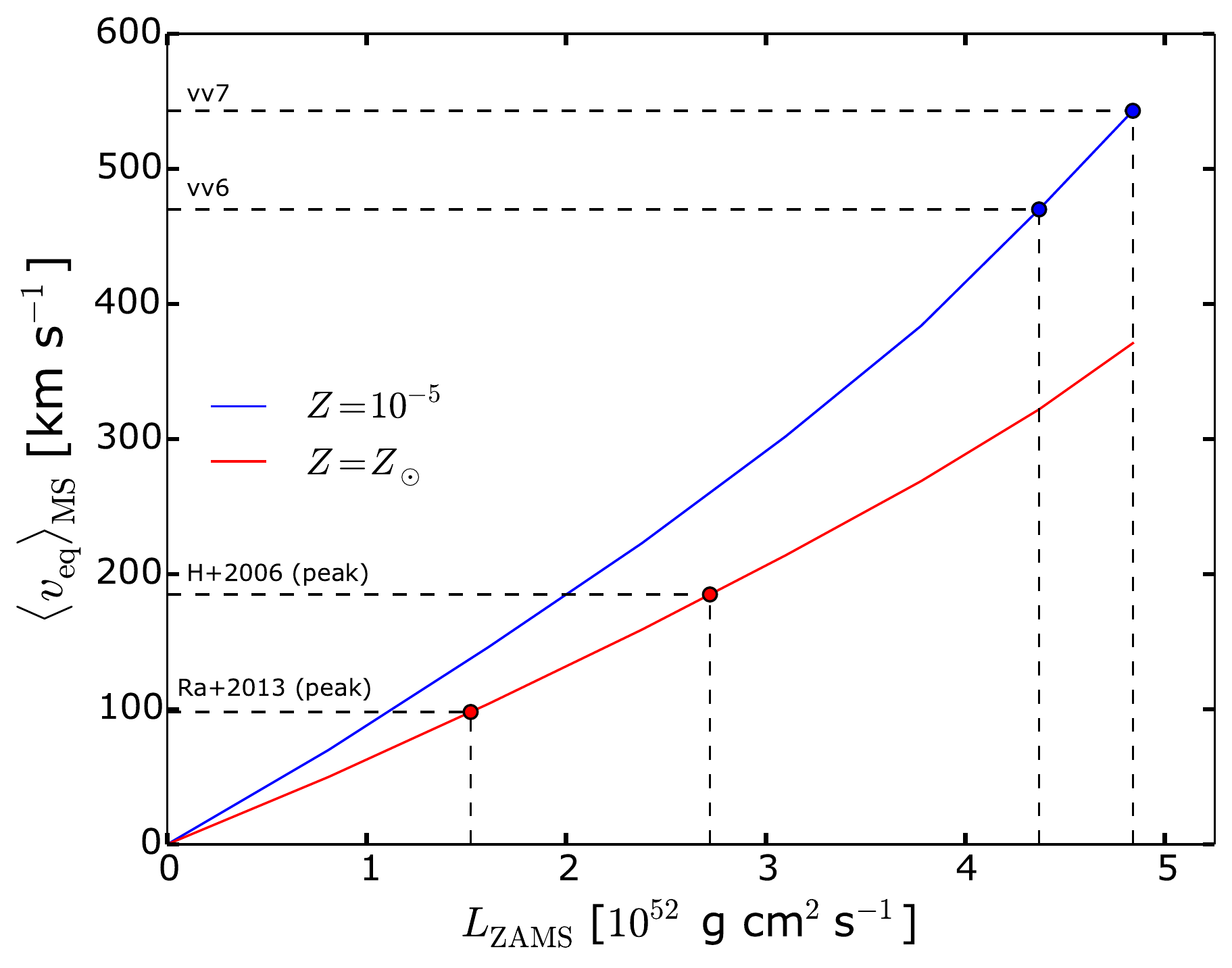}
   \caption{Mean equatorial velocity during the MS as a function of $L_{\rm ZAMS}$, the total angular momentum content at the ZAMS. The two curves show 20~$M_{\odot}$ models computed at solar metallicity (red) and at $Z=10^{-5}$ (blue, the models of Table~\ref{table:2}). The two peaks of the \citet[][H+2006]{huang06} and \citet[][Ra+2013]{ramirez13} velocity distributions are indicated, together with the vv6 and vv7 models. 
   }
\label{angmom}
    \end{figure}

The H+2006 and Ra+2013 distributions are based on OB stars, which are either in open clusters \citep[$-0.5<$~\text{[}Fe/H\text{]}~$<+0.5$ in most of the cases, e.g.][]{paunzen10, heiter14, neptopil16} or in the 30 Doradus region \citep[with metallicity $\sim 0.6~Z_{\odot}$][]{lebouteiller08}. 
This work investigates a much lower metallicity range, sub-solar by a factor of $10^3-10^4$.
Low metallicity stars are less metal-rich so that they are less opaque and more compact. 
As a consequence, for a given angular momentum content, they tend rotate faster than solar metallicity stars  \citep[e.g.][]{maeder01}. 
Figure~\ref{angmom} shows the models of Table~\ref{table:2} (blue curve) together with the same models computed at solar metallicity (red curve). For a given initial angular momentum content $L_{\rm ZAMS}$, $\langle \upsilon_{\rm eq} \rangle_{\rm MS}$ is indeed higher at low metallicity.
The increasing fraction of fast rotators at low metallicity is supported by observations of rotating massive stars in different metallicity environments (\citealt[]{maeder99, martayan07, hunter08}; however, the metallicity regime in these studies is not comparable to the very low metallicity regime investigated here).

Figure~\ref{angmom} also shows that the higher fraction of fast rotators in the distribution derived from EMP stars might not be understood solely as a result of the physics of rotating massive stars. 
According to the additional solar metallicity models we have computed, the peaks of the distributions of \cite{huang06} and \cite{ramirez13} correspond to an initial angular momentum of $\sim 1.5 - 3 \times 10^{52}$ g~cm$^2$~s$^{-1}$ (red circles in Fig.~\ref{angmom}). The distribution derived in this work peaks at $\sim 4 - 5 \times 10^{52}$ g~cm$^2$~s$^{-1}$ (blue circles). 
This might indicate that low metallicity massive stars were born with more angular momentum. It might suggest that the removal of angular momentum during massive star formation was less efficient in the early Universe. This may happen if the magnetic braking during star formation is rather inefficient \citep{hirano18}.

\subsection{Origin of trans-iron elements}\label{compodisc}

Most metal-poor stars having [Fe/H]~$<-3$ do not show strong enhancements in trans-iron elements. 
In our sample, 4 stars have [Sr/Fe] $>1$, 14 have [Ba/Fe] $>1,$ and 7 have [Eu/Fe] $>1$. 
One possible origin of  these elements (or at least some of them) could be due to the s-process operating in the massive source stars, mainly during the core He-burning stage \citep[e.g.][]{cameron60, peters68, couch74, lamb77, langer89, raiteri91a}. 
To produce trans-iron elements, this process requires some initial amount of heavy seeds (e.g. Fe) to start with because the neutron flux is too weak to reach trans-iron elements starting from light seeds. Consequently, at zero metallicity, an almost null amount of trans-iron elements is expected. The standard s-process in non-rotating massive stars is likely inefficient below $Z \sim 10^{-4}$ \citep{prantzos90}. However, it has been shown that 
rotation can enhance the neutron flux and therefore boosts the s-process \citep[e.g.][]{pignatari08, frischknecht12, choplin18, limongi18, banerjee19}. This can lead to a significant production of trans-iron elements even at low metallicity. 
In what proportions this process can produce light  (e.g. Sr) and heavy (e.g. Pb) trans-iron elements is still uncertain. 
The inclusion of the s-process in our analysis will be the object of a future work.
Other processes or sources may also have contributed to the early enrichment of trans-iron elements, such as AGB stars \citep[e.g.][]{herwig04, cristallo09}, jet-like explosions driven by magneto-rotational instabilities \citep[e.g.][]{winteler12, nishimura15}, or neutron star mergers \citep[e.g.][]{thielemann17}.

\subsection{Initial source star masses}\label{inimass}

The back and forth mixing process likely operates more efficiently in $< 60$~$M_{\odot}$ models compared to higher masses models \citep{choplin18}, one reason being that higher mass stars have shorter lifetimes hence there is less time for the back and forth mixing process to operate.
On the other hand, because higher mass stars have more massive H-shells and He-cores, their mass yields can be higher. Moreover, strong H--He shell interactions can occur in some models and can largely impact the nucleosynthesis \citep[e.g.][]{ekstrom08}.
The mass dependency of yields is then complex and it generally varies non-monotonically with initial mass \citep[e.g.][]{hirschi07, ekstrom08, yoon12, takahashi14}.
Without a detailed study, it is therefore not possible to predict how our results will depend on the initial source star mass.
It is  likely, however,  that EMP stars with high [Na, Mg, Al/H] ratios may be better reproduced by $\sim 20$~$M_{\odot}$ than $\sim 60$~$M_{\odot}$ models because an overproduction of Na, Mg, or Al requires enough time for the back and forth mixing to operate (Sect.~\ref{mixnuc}). The shorter core He-burning stage of higher mass stars\footnote{The duration of the core He-burning stage of a Pop III 20~$M_{\odot}$ is about twice as long as a Pop III 85~$M_{\odot}$ \citep{ekstrom08}.} may prevent a significant overproduction of Na, Mg, or Al through the back and forth mixing process.

\subsection{Rotational mixing uncertainties}\label{secphys}

Although we investigate the impact of uncertainties in Sect.~\ref{thuncer}, we employ a limited approach that cannot fully treat the strong stellar model uncertainties, especially those linked to the rotational mixing. 
In rotating models, the transport of chemical elements is governed by different diffusion coefficients (e.g. horizontal turbulence or shear mixing). Depending on the prescription used for these coefficients \citep[e.g.][]{zahn92, mathis04, talon97, maeder97, maeder13} the production of chemical elements (especially N) can vary significantly \citep[][]{meynet13}.

Thanks to asteroseismic observations, it has also been shown that current low mass stellar models likely miss an angular momentum transport process \citep[e.g.][]{cantiello14, eggenberger17}. This additional process may also be missing in massive stars. The nature and efficiency of this process is actively discussed \citep[e.g.][]{eggenberger19, fuller19}. If added in the present models, it may provide different chemical yields.
\cite{brott11b} also suggested that an additional mixing process should be included in massive stars  to account for the observed population of slow rotating massive stars with high surface nitrogen abundances.
The results of this work are subject to important model uncertainties that cannot be fully accounted for with simple approaches, such as the method used in Sect.~\ref{thuncer}.

\section{Summary and conclusions}\label{concl}

In this work, we have investigated whether the peculiar chemical signature of long-dead early massive stars including rotation may be found in the observed extremely metal-poor stars.
We computed 20 $M_{\odot}$ models at metallicity $Z=10^{-5}$ with eight initial rotation rates ($0 < \upsilon_{\rm ini}/\upsilon_{\rm crit} < 0.7$).

The rotational mixing operating between the He-burning core and H-burning shell affects the abundances of light elements from C to Al. At the pre-SN stage, the material above the C-burning shell ($M_{\rm r} \gtrsim 4-5$~$M_{\odot}$) shows a variable chemical composition that strongly depends on the initial rotation.
In the H-rich layers, the [C/H], [N/H], [F/H], and [Na/H] ratios are increased by $\sim 1-2$ dex from the non-rotating to the fast rotating case.
In the H + He-rich layers, the [N/H], [F/H], [Ne/H], [Na/H], [Mg/H], and [Al/H] ratios are increased by $\sim 2-4$ dex from the non-rotating to the fast rotating case.

We compared the chemical composition of this material with the abundances of EMP stars with $-4<$~[Fe/H]~$<-3$.
We assumed that the massive stars experienced an explosion with strong fallback.
Among the 272 EMP stars considered, $\sim 40-50~\%$ were found to have an abundance pattern consistent with our massive source star models. About $60 - 70~\%$ ($20-30~\%$) of the CEMP (C-normal EMP) stars can be reasonably well reproduced.
In addition, while the abundance patterns of C-normal stars are roughly equally well reproduced by non-rotating and rotating source stars, the patterns of CEMP stars are better explained by fast rotators.

The velocity distribution of the best source star models reaches a maximum at $\upsilon_{\rm ini}/\upsilon_{\rm crit}= 0.6-0.7$ (corresponding to initial equatorial velocities of $\sim 550 - 640$ km~s$^{-1}$). 
Compared to the velocity distributions derived from the observation of nearby OB stars, our distribution 
is flatter and suggests a greater amount of massive fast rotators in the early Universe. 
The stellar evolution effects (such as the higher compactness of low metallicity stars) may not be sufficient to explain the observed differences. Our results suggest that the initial angular momentum content of early massive stars might have been higher than for solar metallicity massive stars.
The possibly higher fraction of fast rotators at low metallicity may have important consequences for the SN types (and rates) from early massive stars, the integrated light of high redshift galaxies and the contribution of early massive stars to reionization.

Caution when using these results is required because of the limiting assumptions made here, especially the explosions with strong fallback (Sect.~\ref{secfallback}). Our results appear to be particularly sensitive to the Mg source star yield (Sect.~\ref{impact}), which could change significantly  if considering different kind of explosions.
As already suggested in the past \citep[e.g.][]{meynet10, maeder15a, choplin17a}, the $^{12}$C/$^{13}$C ratio and N abundance should be among the best elements to probe the mixing process(es) at work in early massive stars.
Consequently, more $^{12}$C/$^{13}$C and N measurements (preferentially in non-evolved EMP stars) are desirable to better probe these processes. 
Finally, caution is also required because of the strong stellar model uncertainties, particularly on the physics of rotational mixing (Sect.~\ref{secphys}).

\begin{acknowledgements} 
The authors thank an anonymous referee for the constructive and helpful comments.
A.C. acknowledges funding from the Swiss National Science Foundation under grant P2GEP2\_184492.
\end{acknowledgements}

\bibliographystyle{aa}
\bibliography{biblio.bib}

\section*{Appendix A}

Individual references for the considered stars include
\cite{jacobson15}, \cite{roederer14a}, \cite{andrievsky07}, \cite{hollek11}, \cite{spite06}, \cite{bonifacio07}, \cite{bonifacio09}, \cite{lai08}, \cite{spite05}, \cite{cohen13}, \cite{honda04}, \cite{lai07}, \cite{andrievsky10}, \cite{andrievsky08}, \cite{cayrel04}, \cite{honda11}, \cite{allen12}, \cite{zhang11}, \cite{barklem05}, \cite{roederer14c}, \cite{roederer09}, \cite{venn04}, \cite{aoki09}, \cite{sbordone10}, \cite{preston06}, \cite{masseron12}, \cite{spite14}, \cite{aoki05}, \cite{sivarani06}, \cite{mello14}, \cite{masseron06}, \cite{boesgaard11}, \cite{placco16a}, \cite{ren12}, \cite{placco14a}, \cite{hansen15}, \cite{hansen14}, \cite{yong13}, \cite{aoki07}, \cite{frebel07}, \cite{cohen06}, \cite{li15}, \cite{li15}, \cite{matsuno17}, \cite{aoki18}, \cite{aoki08}, \cite{aoki13}, \cite{bonifacio18}, \cite{bonifacio15}, \cite{bandyopadhyay18}, \cite{matsuno17}, \cite{behara10}, \cite{spite13}, \cite{placco15}, \cite{aguado16}, \cite{susmitha16}, \cite{aoki10}.

\section*{Appendix B}

   \begin{figure*}
   
   \centering
   \begin{minipage}[c]{.49\linewidth}
       \includegraphics[scale=0.69]{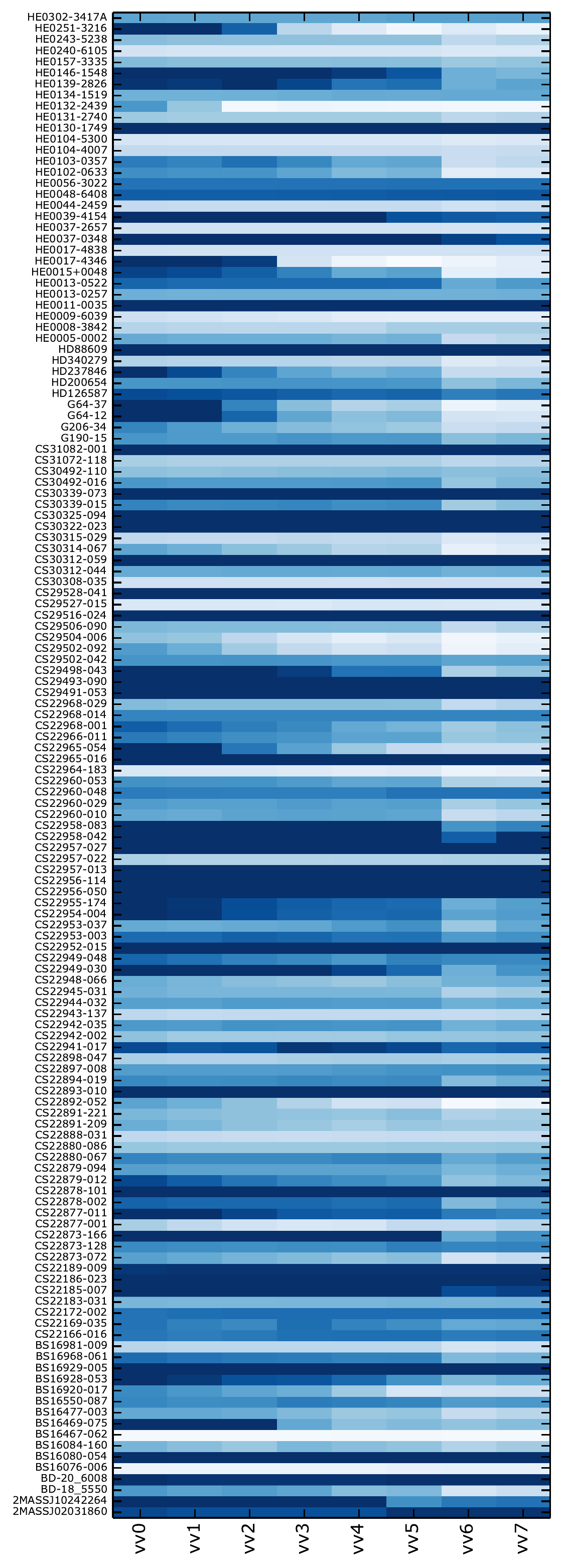}
   \end{minipage}
   \hspace{-1.2cm}
   \begin{minipage}[c]{.49\linewidth}
       \includegraphics[scale=0.69]{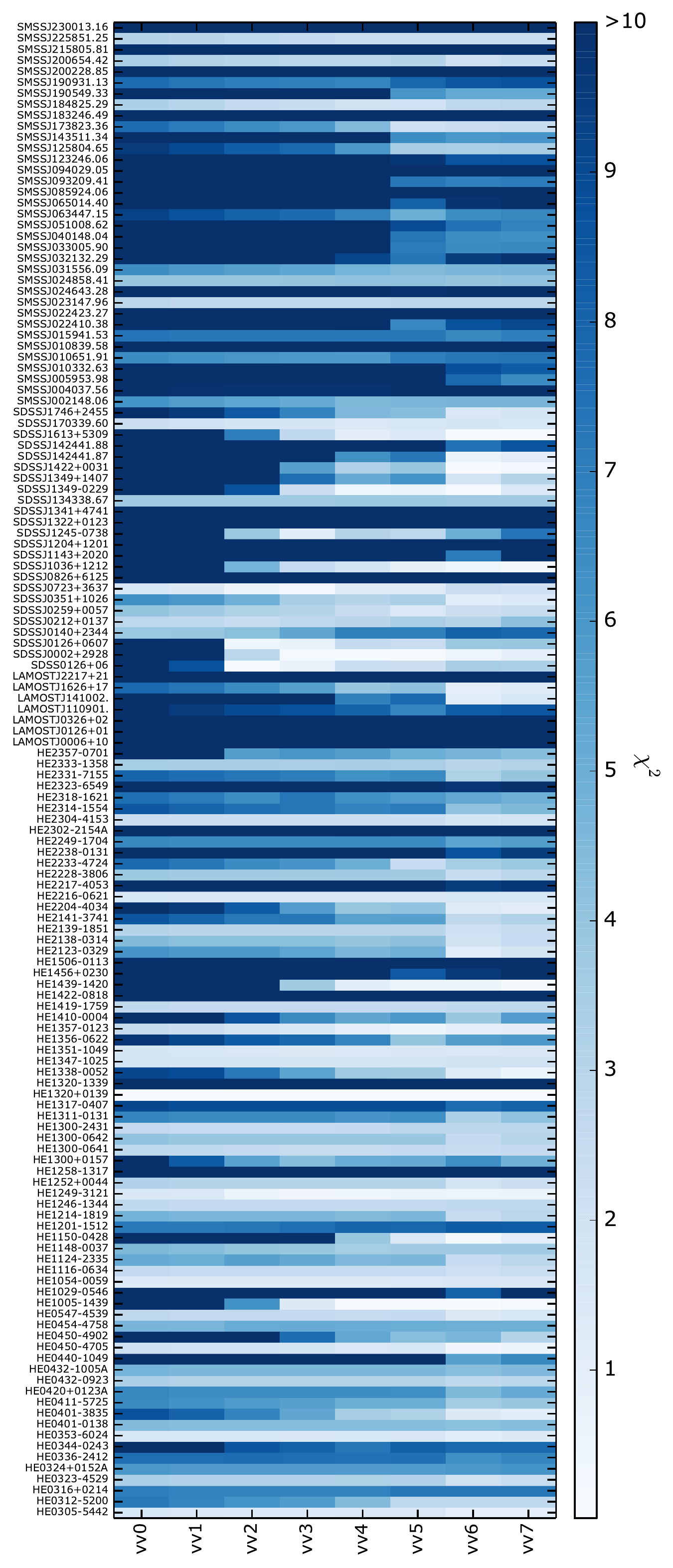}
   \end{minipage}
   \caption{
   Summary plot of the abundance fitting of the 272 EMP stars with the eight source star models. For a given EMP star (y-axis label) and a given source star model (x-axis label), the colour indicates the best $\chi^2$ found.
   } 
\label{mapchi1}
    \end{figure*}

   \begin{figure*}
   
   \centering
   \begin{minipage}[c]{.33\linewidth}
       \includegraphics[scale=0.3]{figs/XH_3best_2.pdf}
   \end{minipage}
   \begin{minipage}[c]{.33\linewidth}
       \includegraphics[scale=0.3]{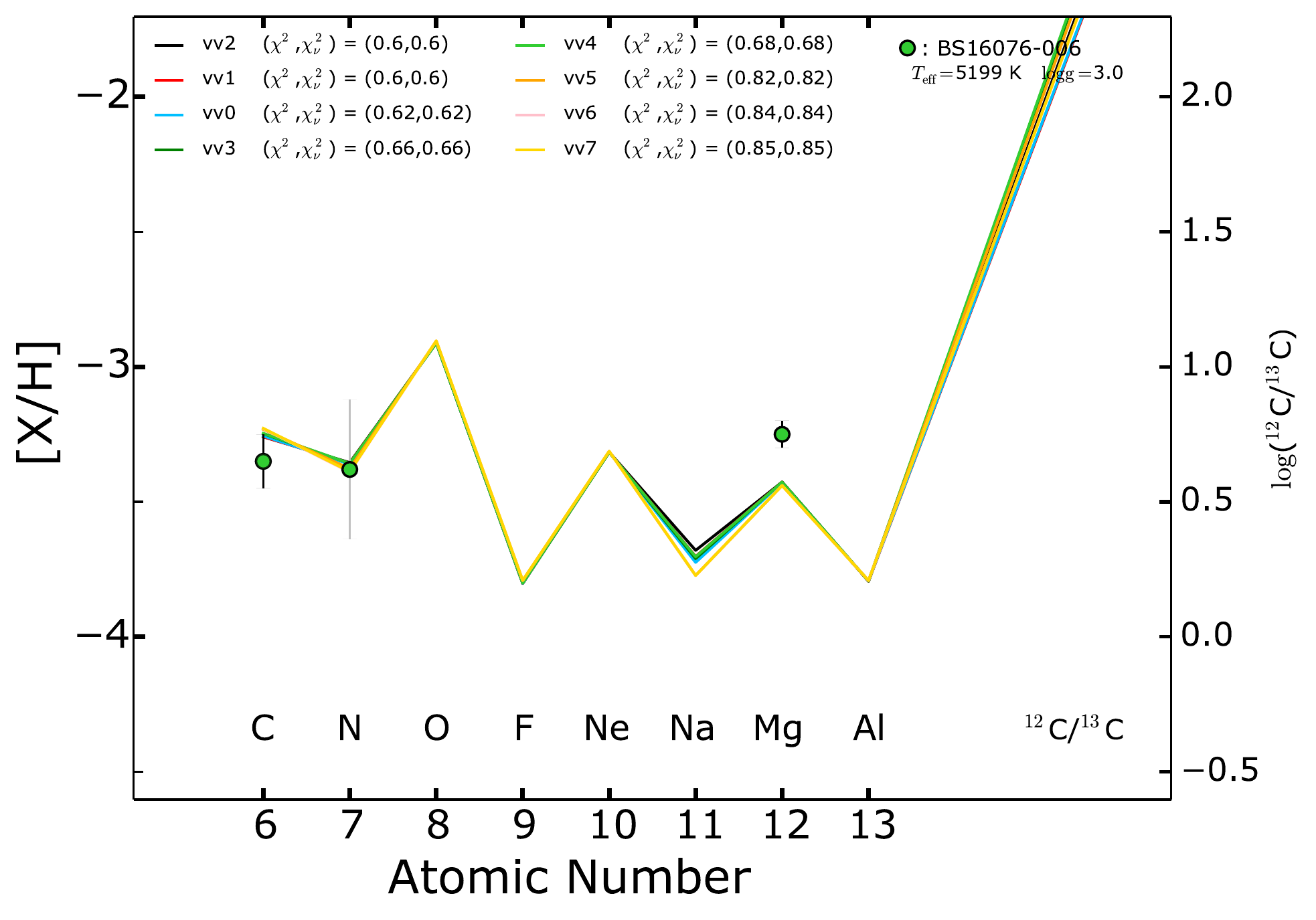}
   \end{minipage}
   \begin{minipage}[c]{.33\linewidth}
       \includegraphics[scale=0.3]{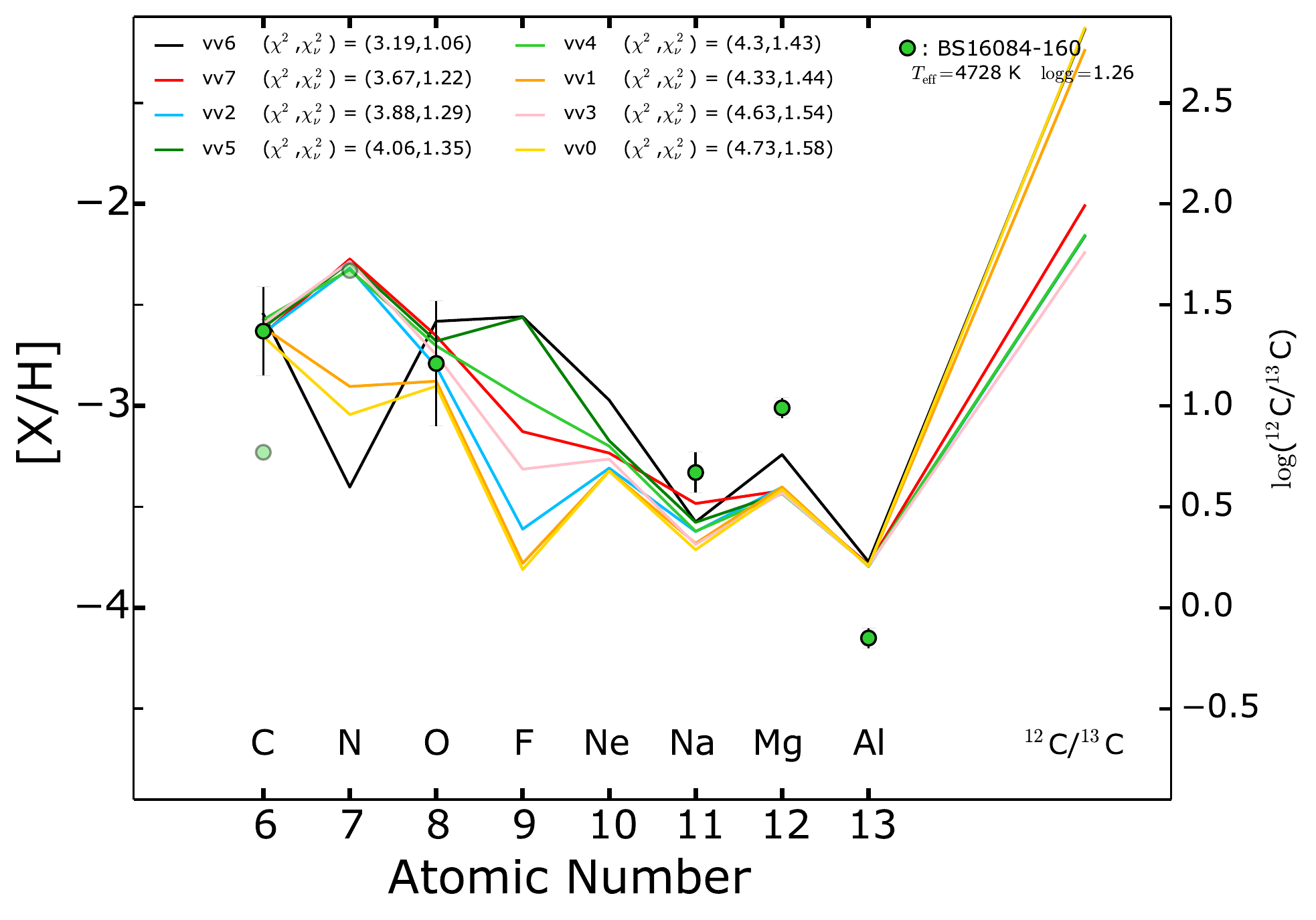}
   \end{minipage}
   \begin{minipage}[c]{.33\linewidth}
       \includegraphics[scale=0.3]{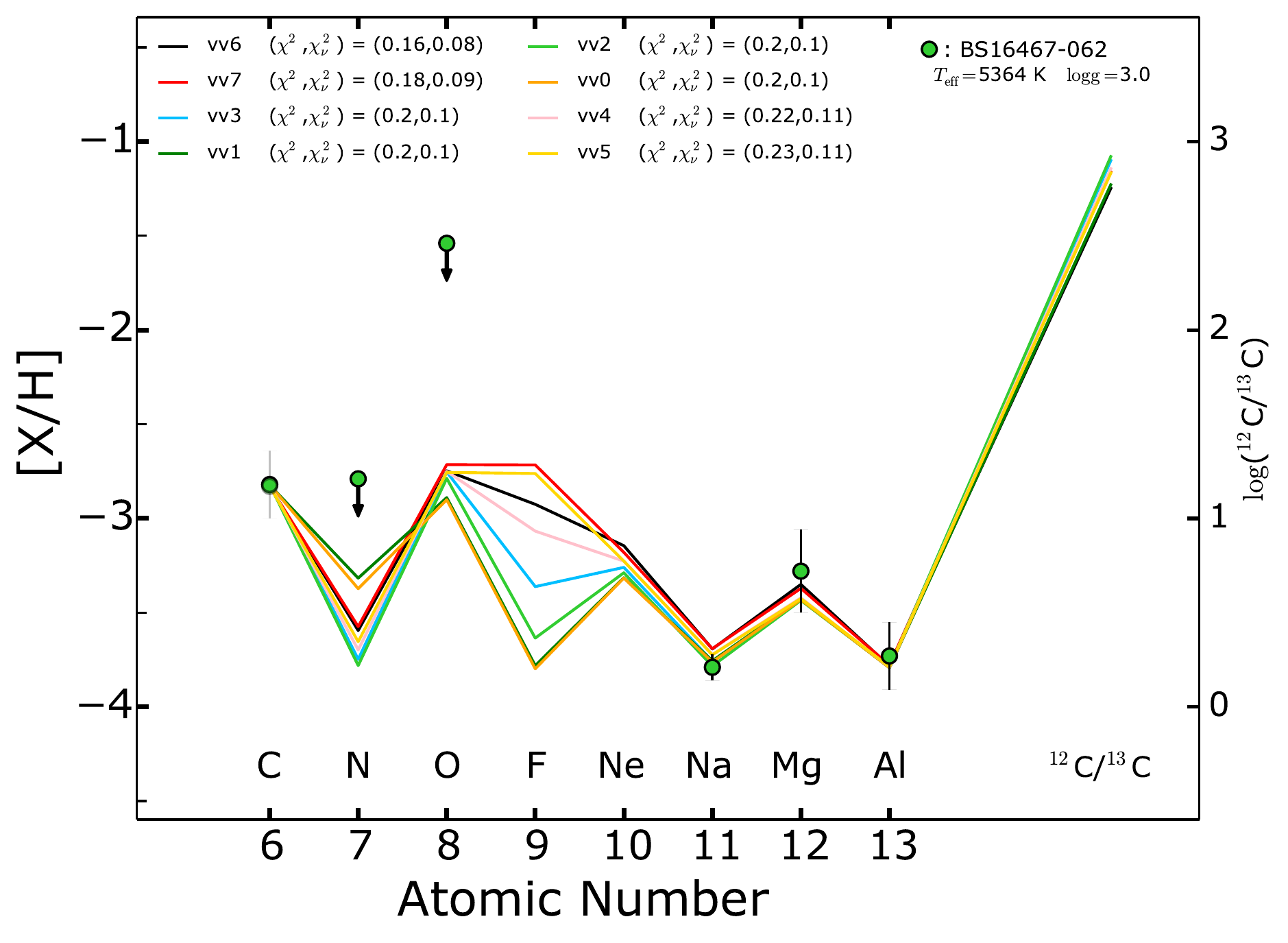}
   \end{minipage}
   \begin{minipage}[c]{.33\linewidth}
       \includegraphics[scale=0.3]{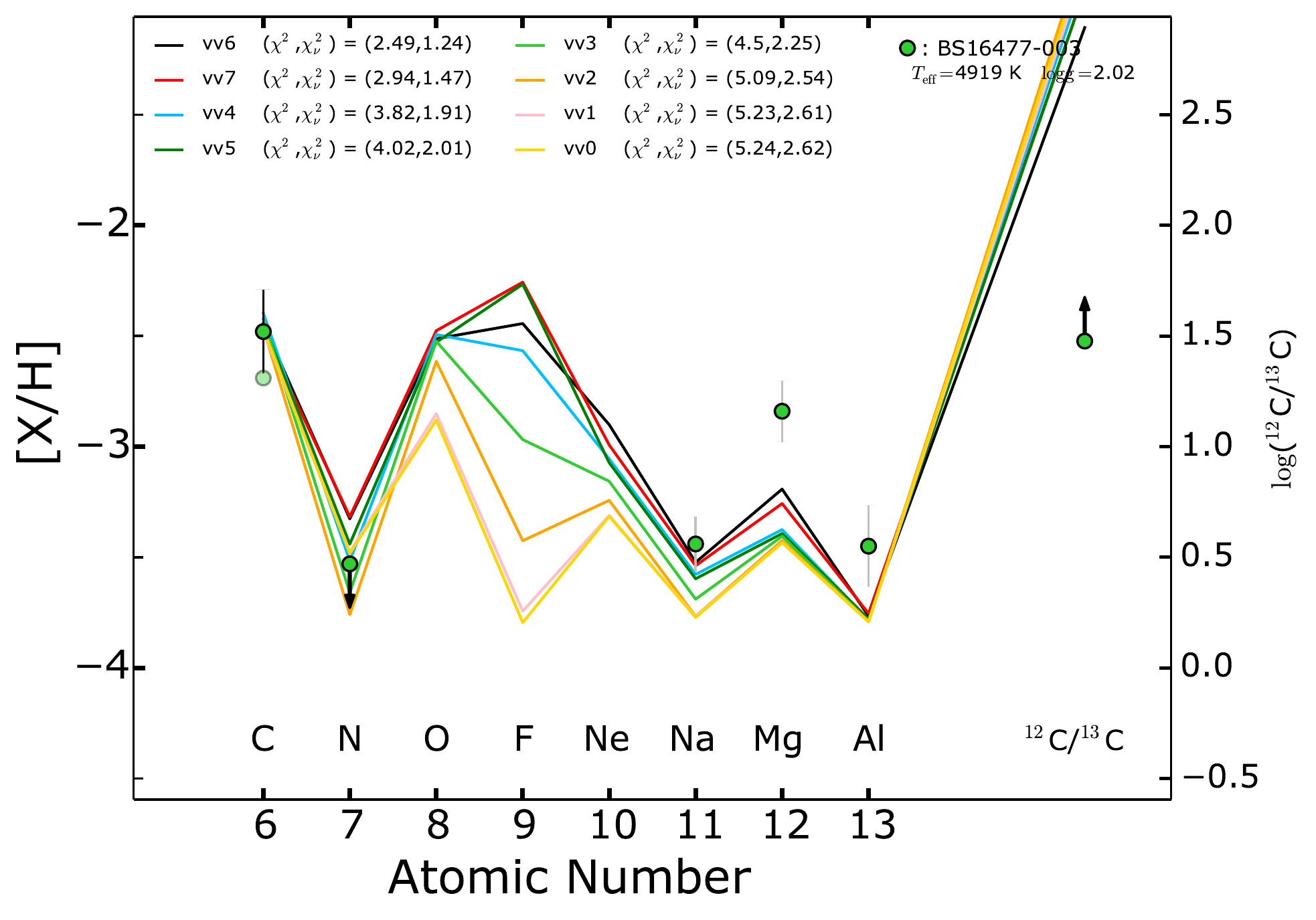}
   \end{minipage}
   \begin{minipage}[c]{.33\linewidth}
       \includegraphics[scale=0.3]{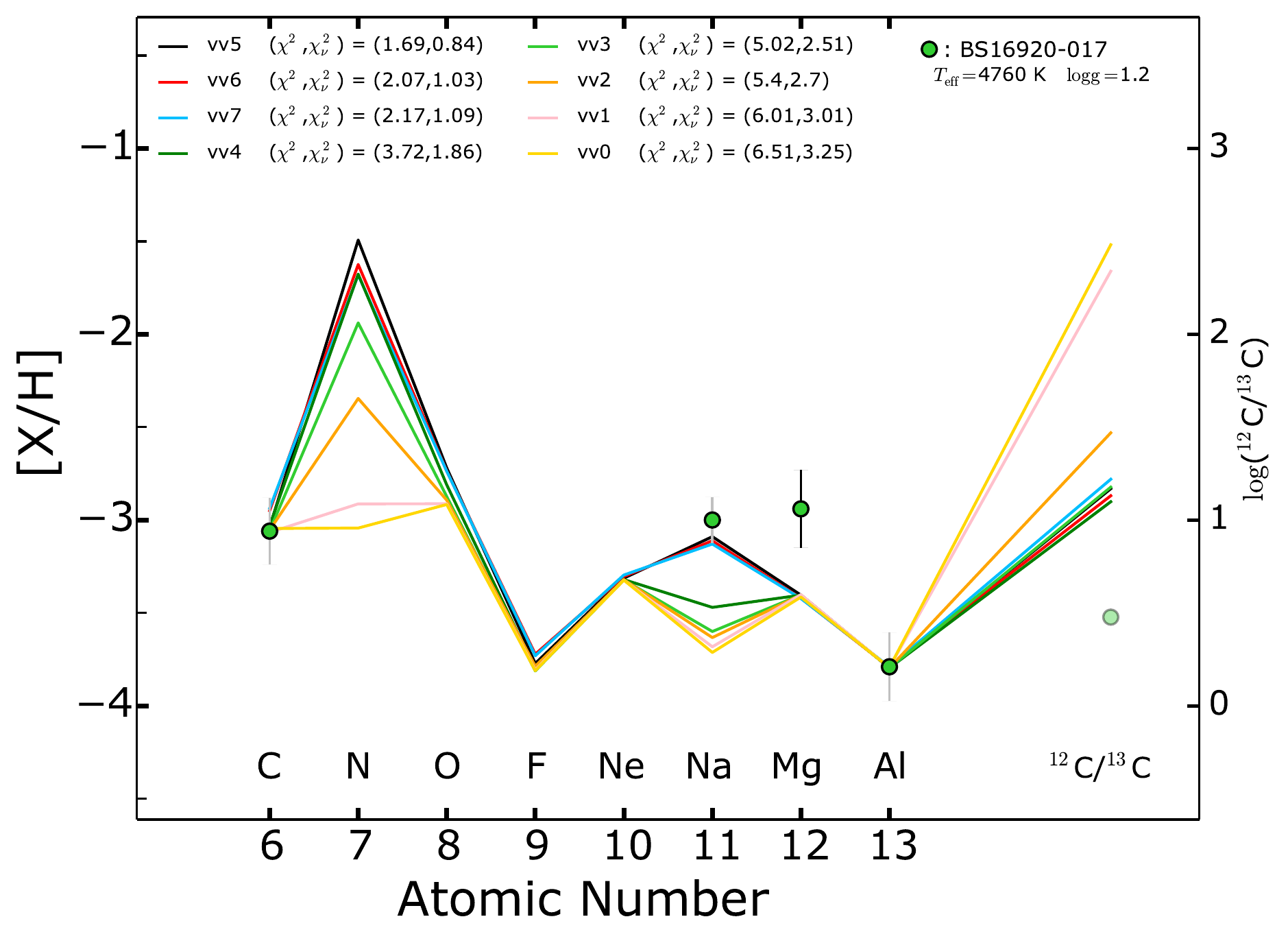}
   \end{minipage}
   \begin{minipage}[c]{.33\linewidth}
       \includegraphics[scale=0.3]{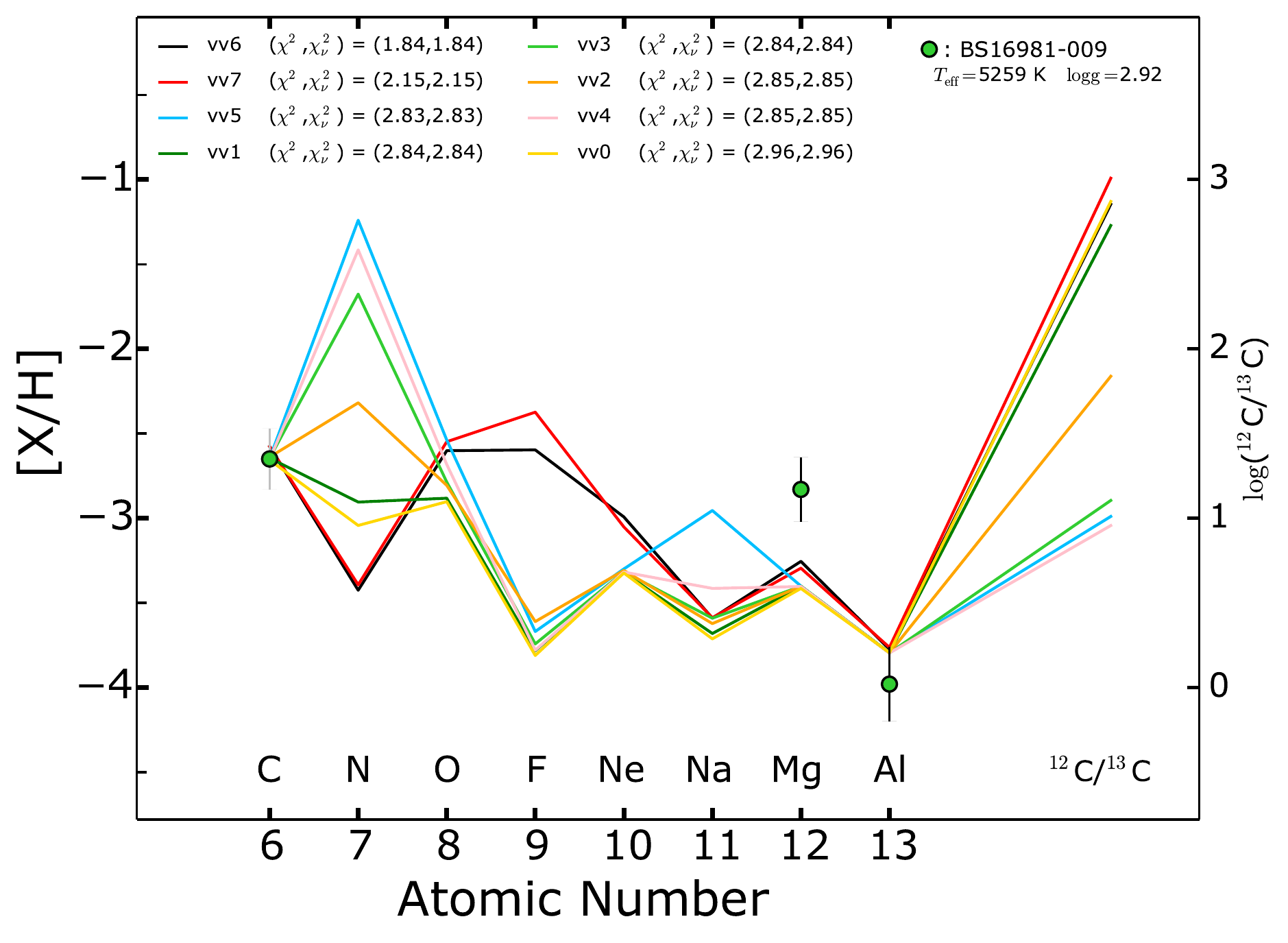}
   \end{minipage}
   \begin{minipage}[c]{.33\linewidth}
       \includegraphics[scale=0.3]{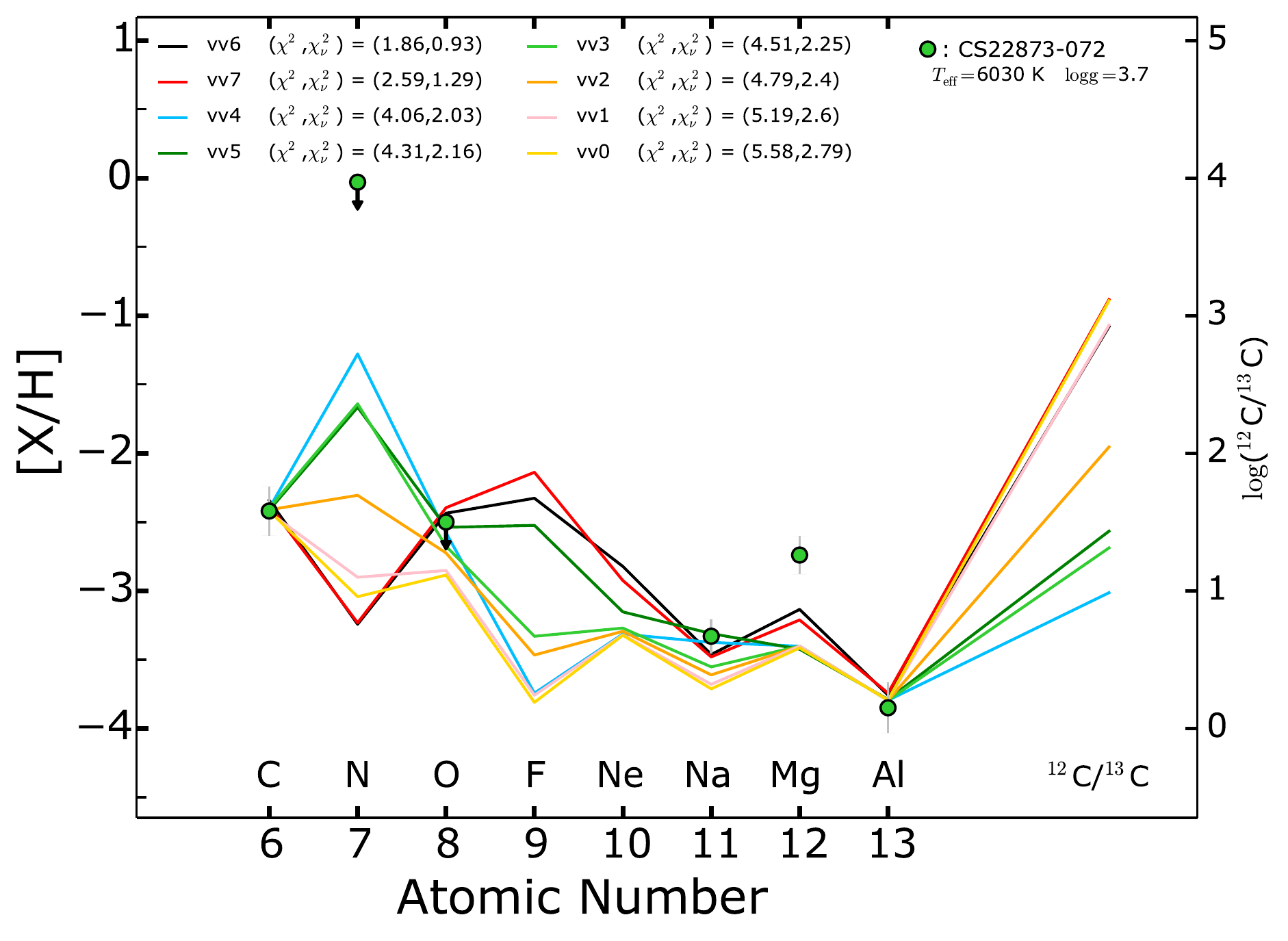}
   \end{minipage}
   \begin{minipage}[c]{.33\linewidth}
       \includegraphics[scale=0.3]{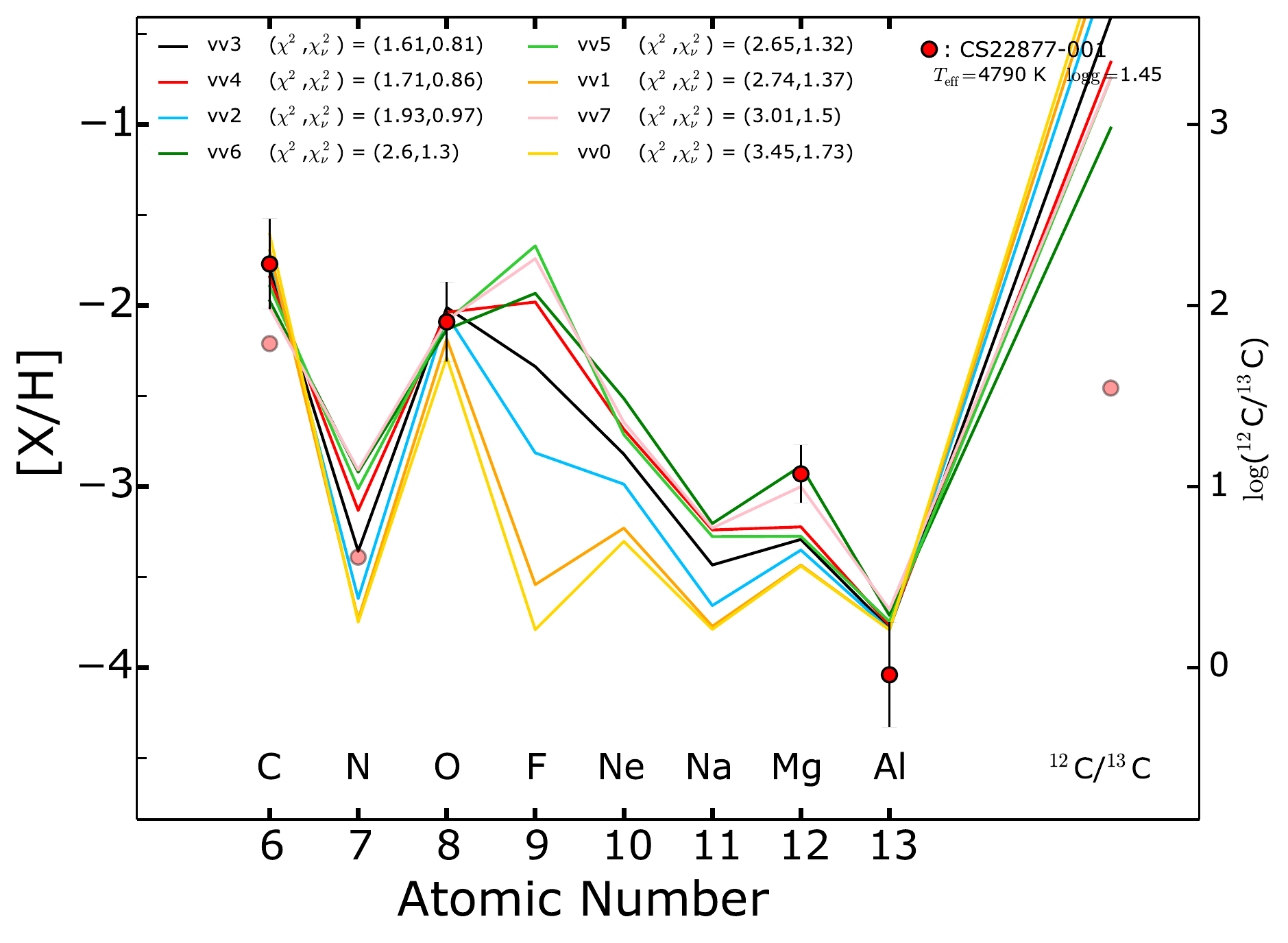}
   \end{minipage}
   \begin{minipage}[c]{.33\linewidth}
       \includegraphics[scale=0.3]{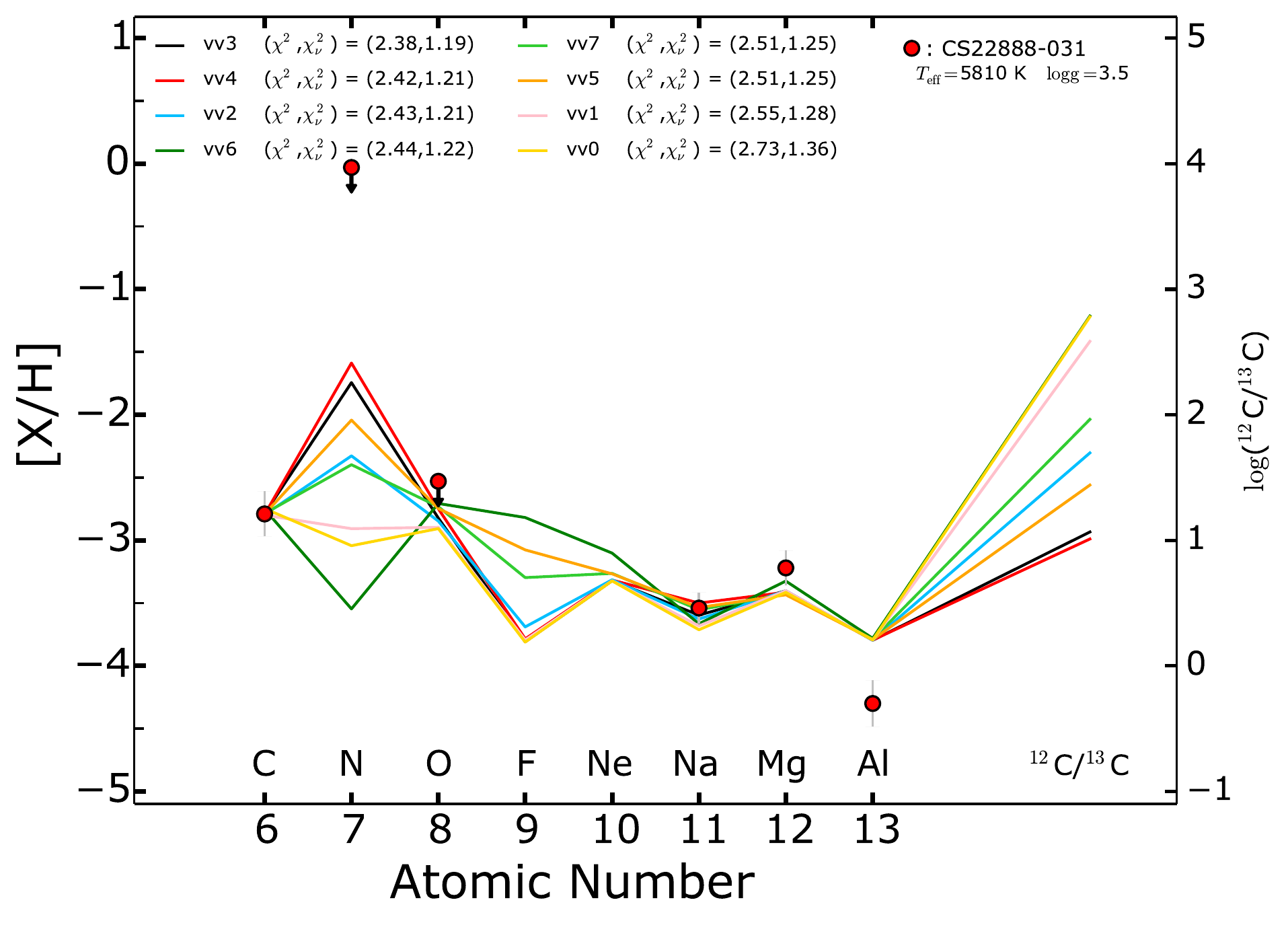}
   \end{minipage}
   \begin{minipage}[c]{.33\linewidth}
       \includegraphics[scale=0.3]{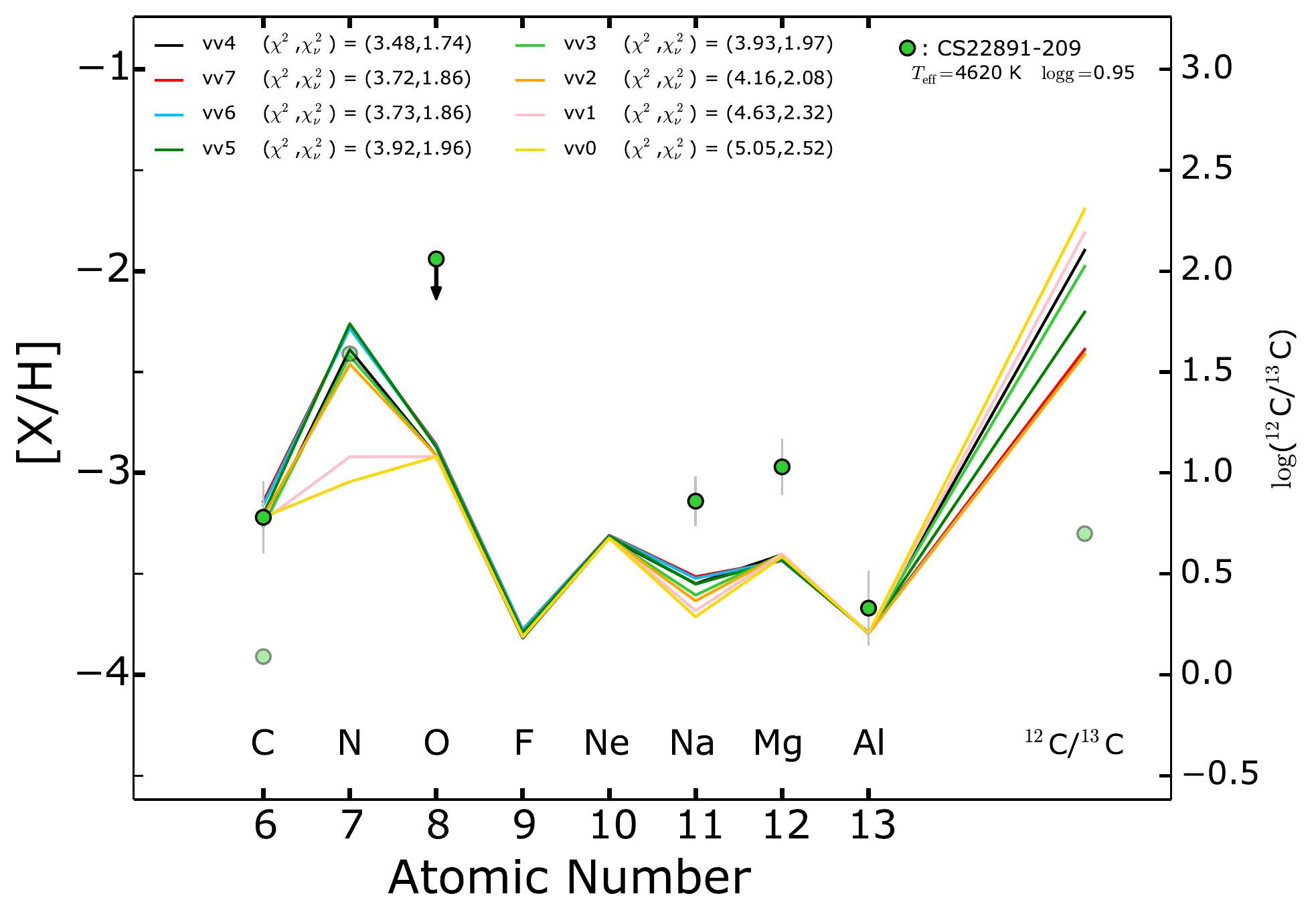}
   \end{minipage}
   \begin{minipage}[c]{.33\linewidth}
       \includegraphics[scale=0.3]{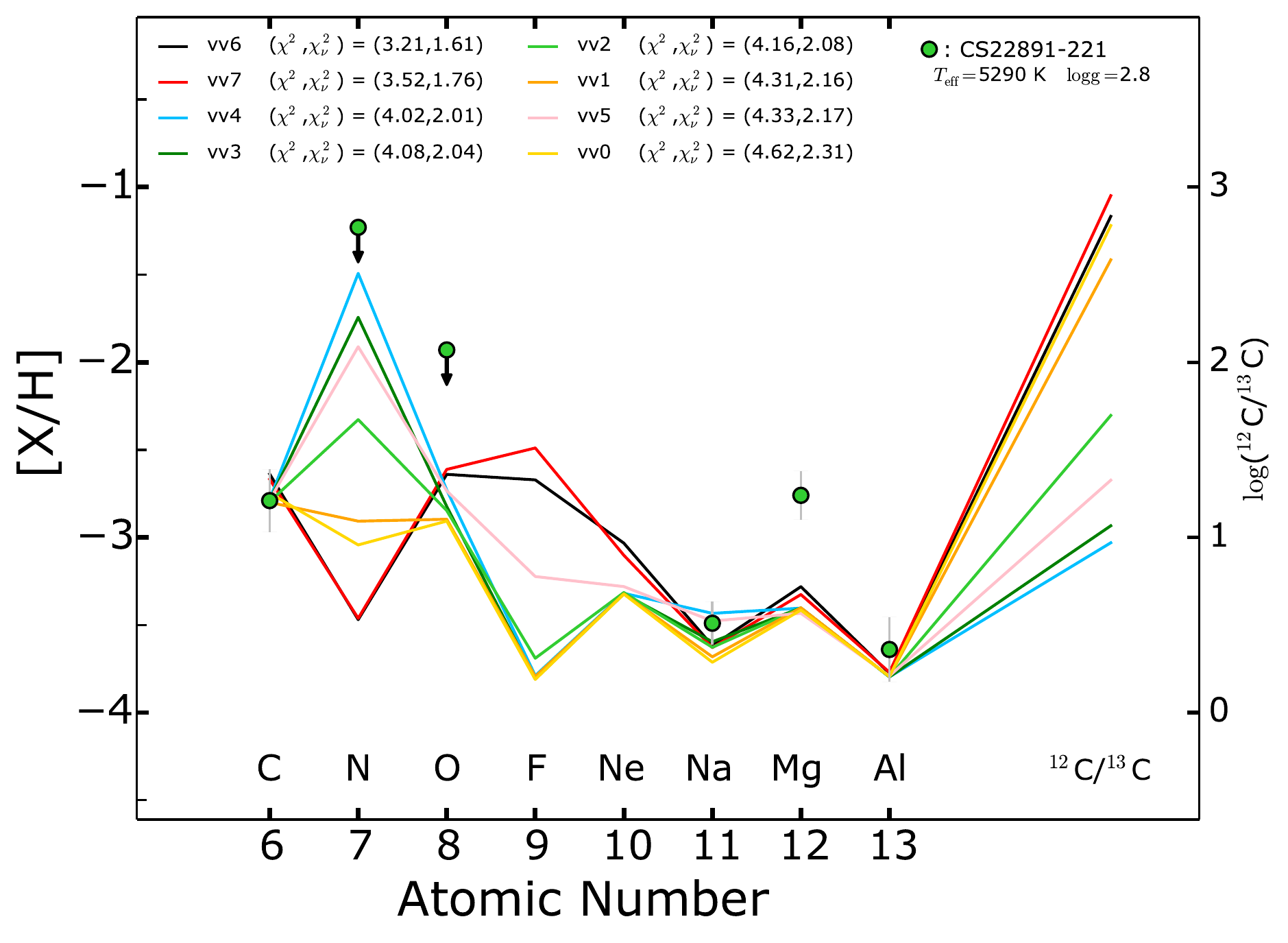}
   \end{minipage}
   \begin{minipage}[c]{.33\linewidth}
       \includegraphics[scale=0.3]{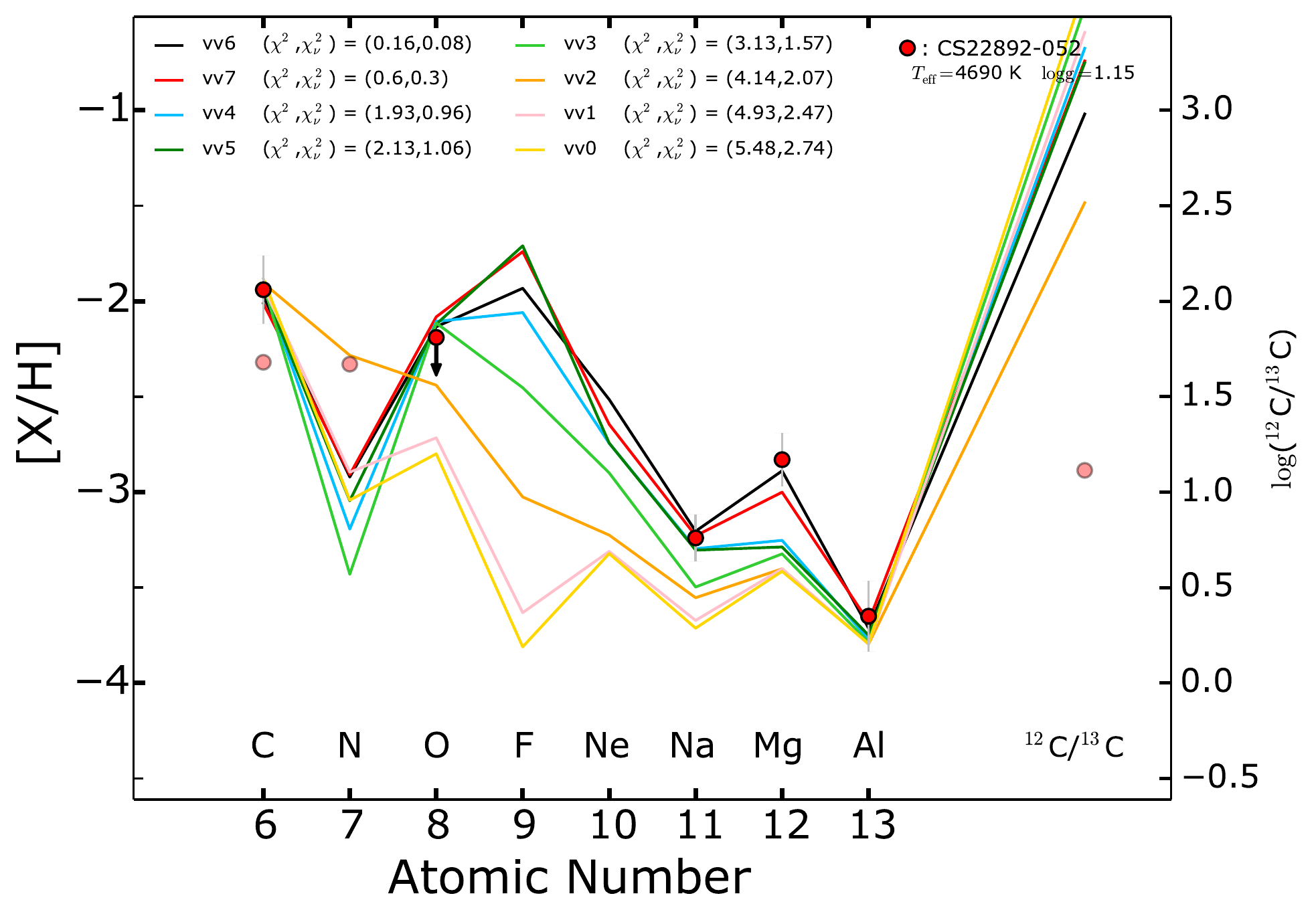}
   \end{minipage}
   \begin{minipage}[c]{.33\linewidth}
       \includegraphics[scale=0.3]{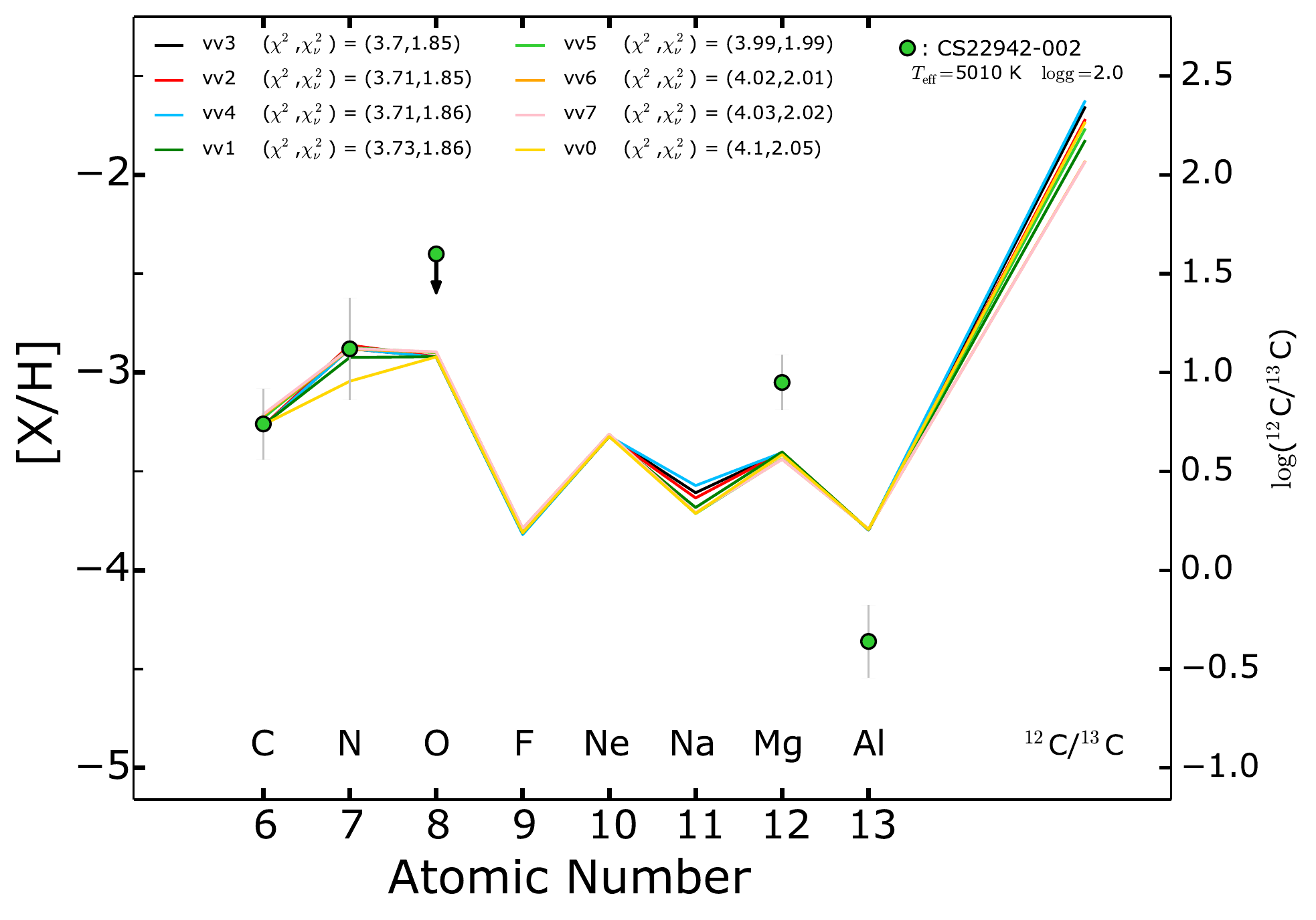}
   \end{minipage}
   \begin{minipage}[c]{.33\linewidth}
       \includegraphics[scale=0.3]{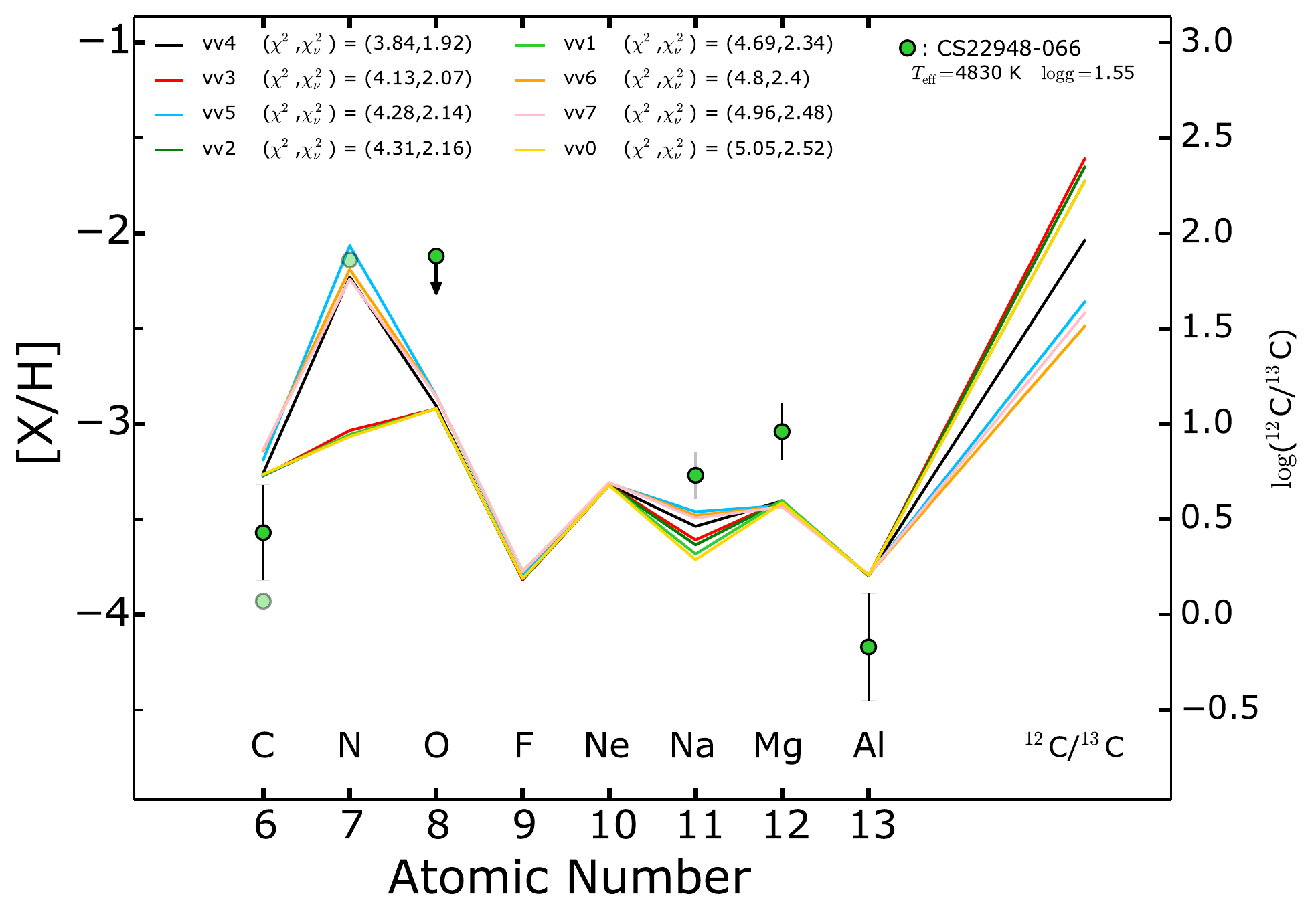}
   \end{minipage}
   \caption{EMP stars having at least one source star model with $\chi_{\nu}^2 < 2$. 
   Red and green circles denote CEMP ([C/Fe] $>0.7$) and C-normal EMP stars, respectively. 
   For evolved EMP stars ($\log g < 2$), [N/H] and $\log$($^{12}$C$^{13}$C) are shown as light red or green symbols and are considered  upper and lower limits, respectively. When available, the correction on the C abundance from \cite{placco14c} is taken into account. In this case, the [C/H] ratio before (after) correction is shown by a light (normal) red or green symbol.
   When available, the observational uncertainty is shown by a black bar. If it is not available, the mean observational uncertainty of the EMP sample (Sect.~\ref{fitproc}) is shown as a grey bar.
   } 
\label{allfit1}
    \end{figure*}

   \begin{figure*}
   \ContinuedFloat
   \centering
   \begin{minipage}[c]{.33\linewidth}
       \includegraphics[scale=0.3]{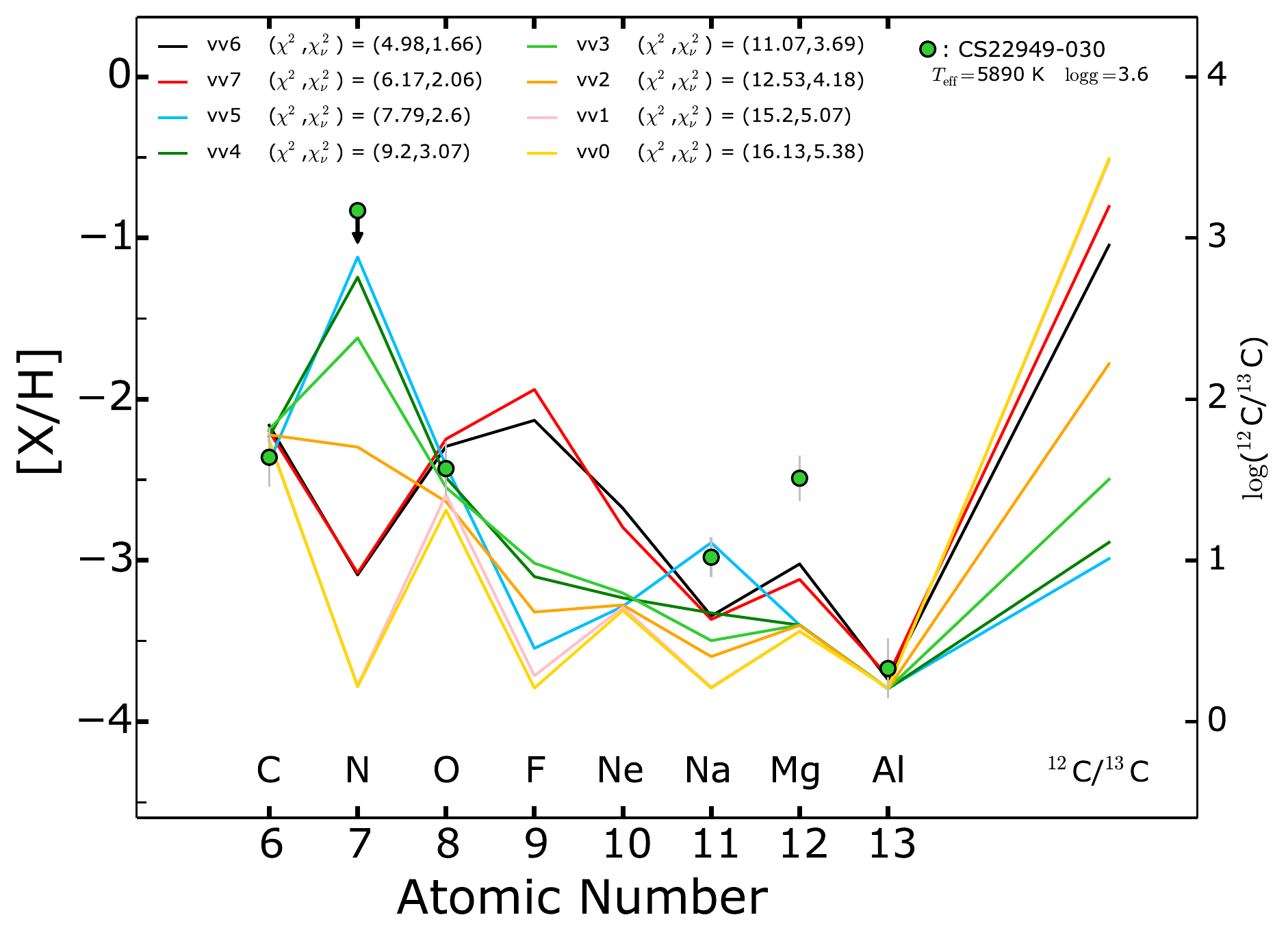}
   \end{minipage}
   \begin{minipage}[c]{.33\linewidth}
       \includegraphics[scale=0.3]{figs/XH_3best_63.pdf}
   \end{minipage}
   \begin{minipage}[c]{.33\linewidth}
       \includegraphics[scale=0.3]{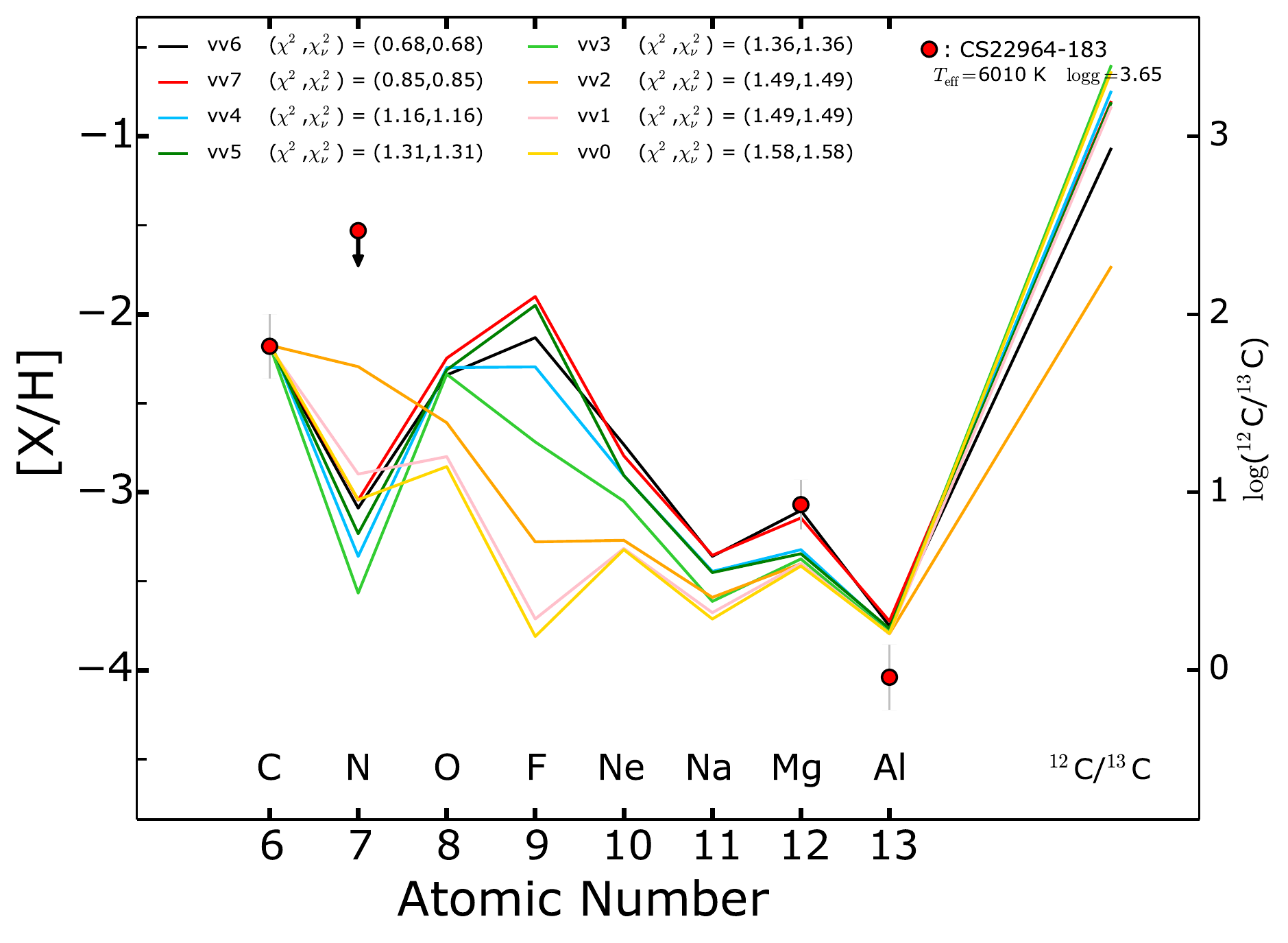}
   \end{minipage}
   \begin{minipage}[c]{.33\linewidth}
       \includegraphics[scale=0.3]{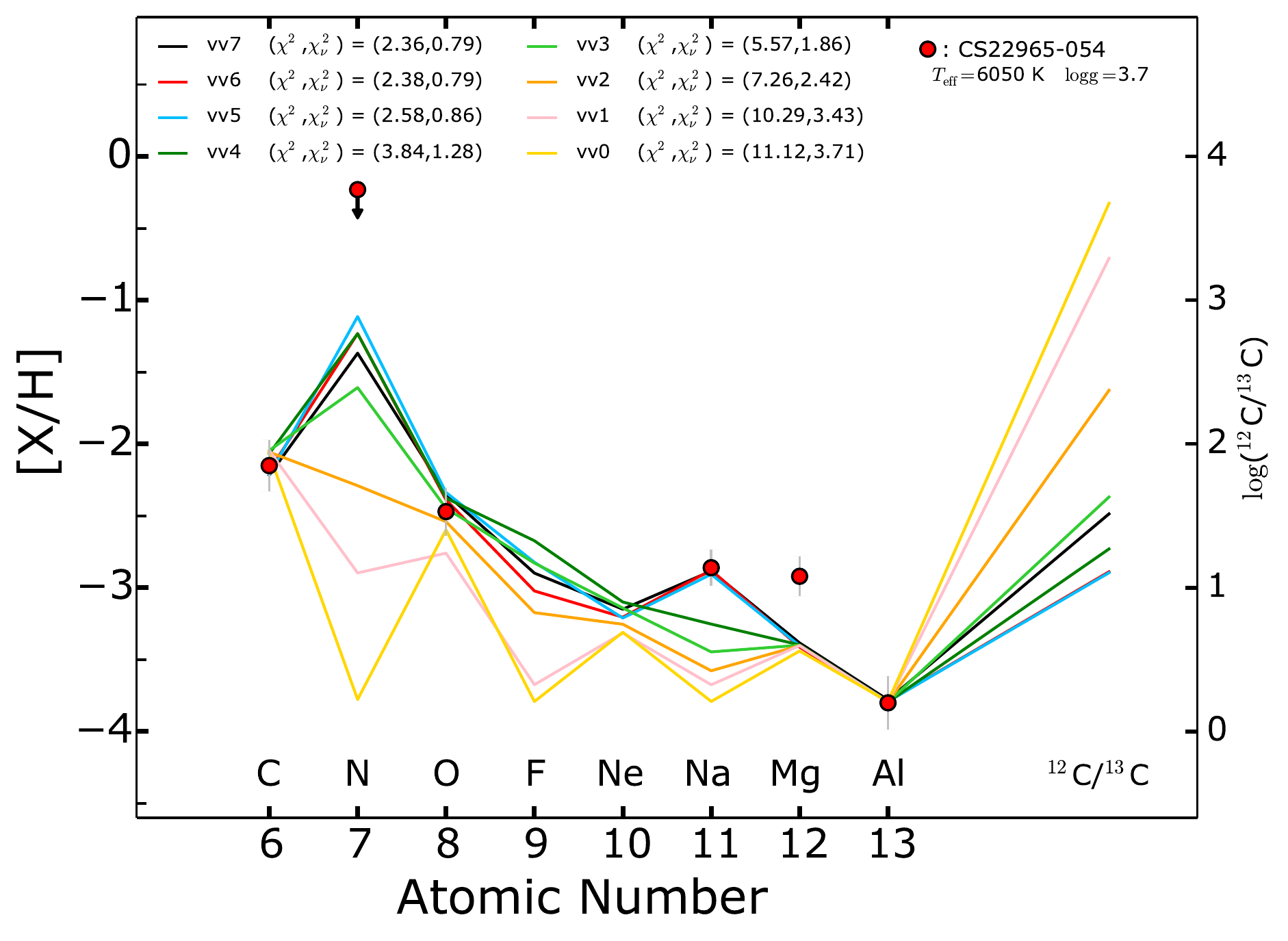}
   \end{minipage}
   \begin{minipage}[c]{.33\linewidth}
       \includegraphics[scale=0.3]{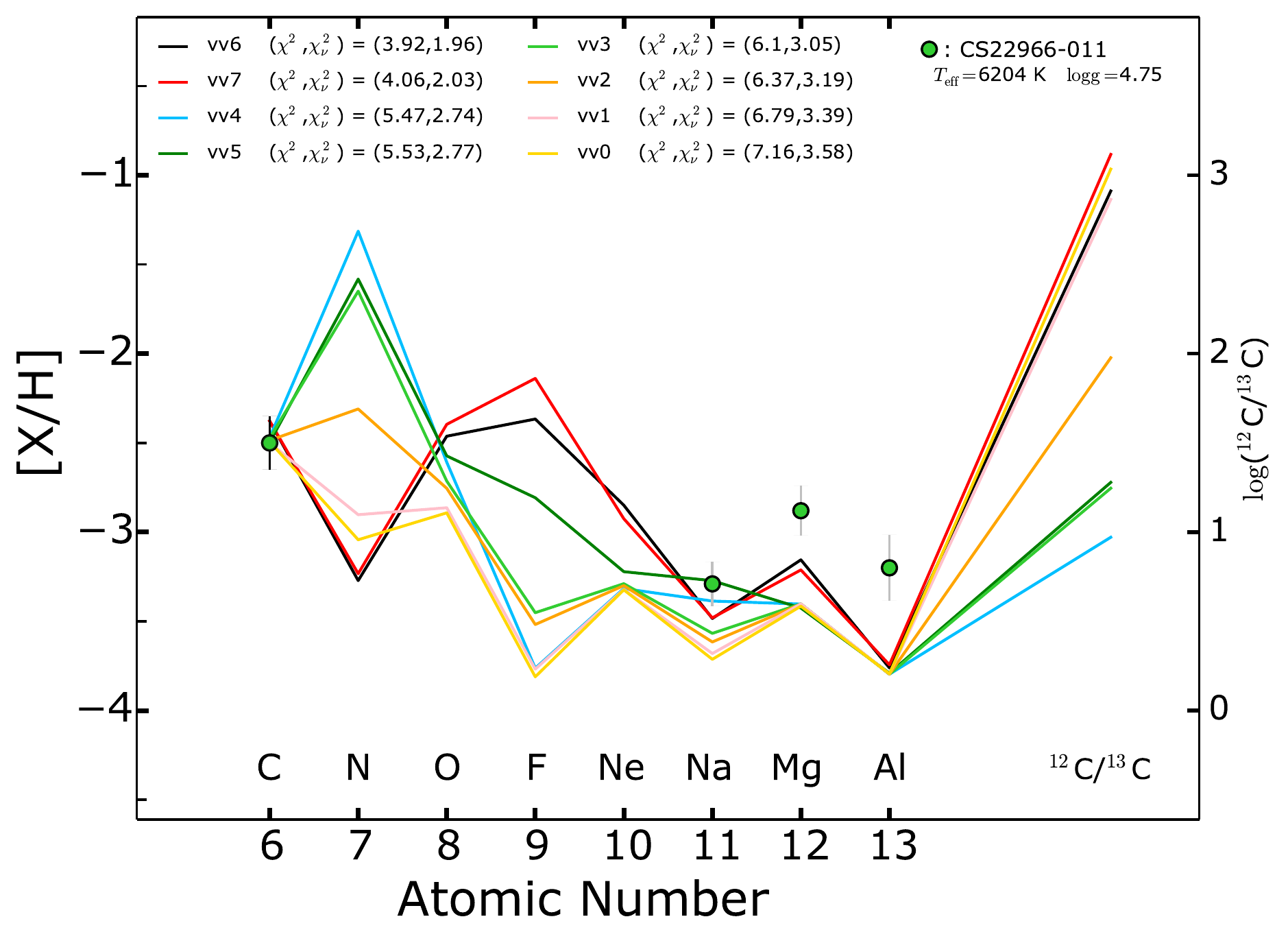}
   \end{minipage}
   \begin{minipage}[c]{.33\linewidth}
       \includegraphics[scale=0.3]{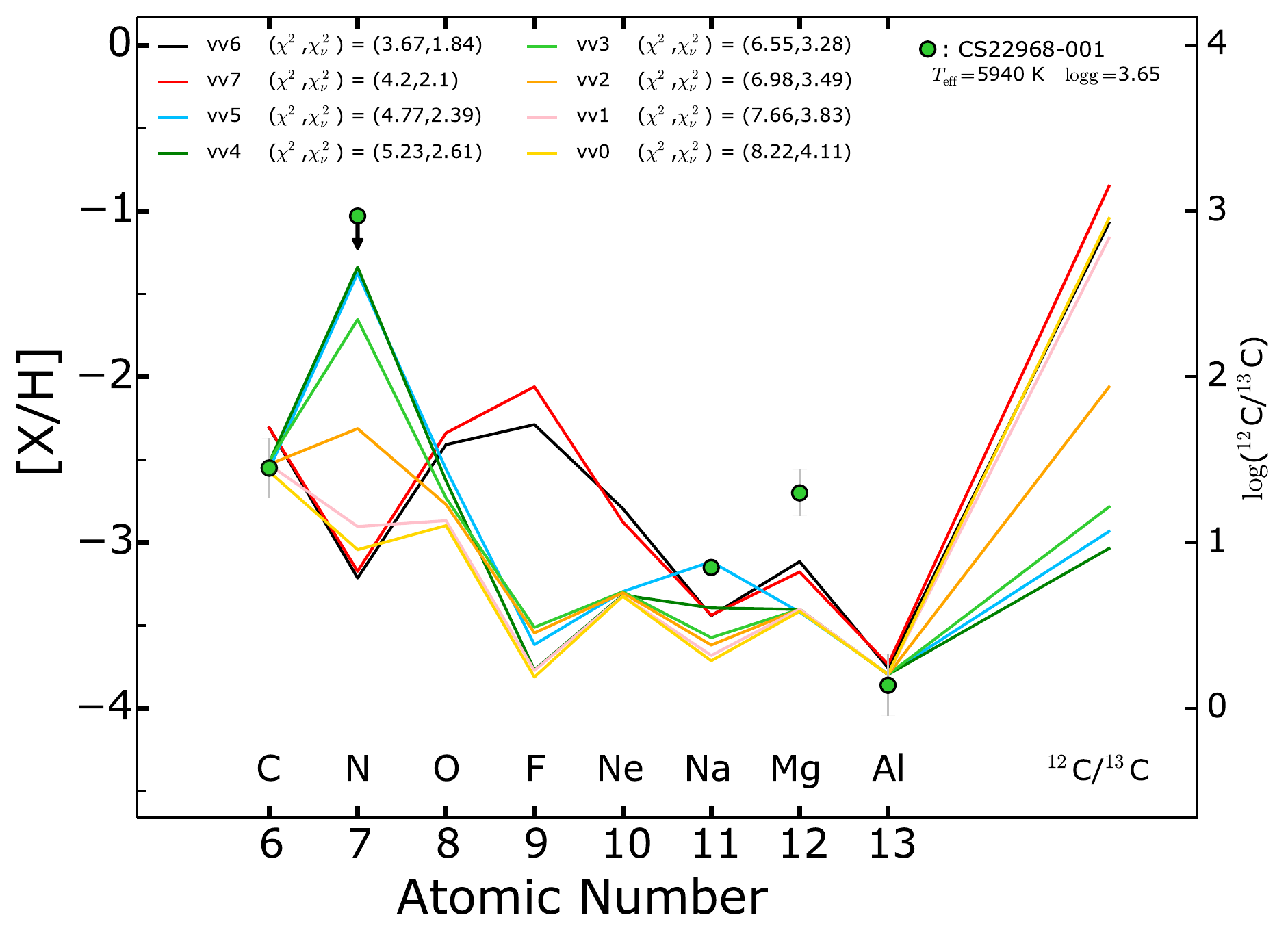}
   \end{minipage}
   \begin{minipage}[c]{.33\linewidth}
       \includegraphics[scale=0.3]{figs/XH_3best_76.pdf}
   \end{minipage}
   \begin{minipage}[c]{.33\linewidth}
       \includegraphics[scale=0.3]{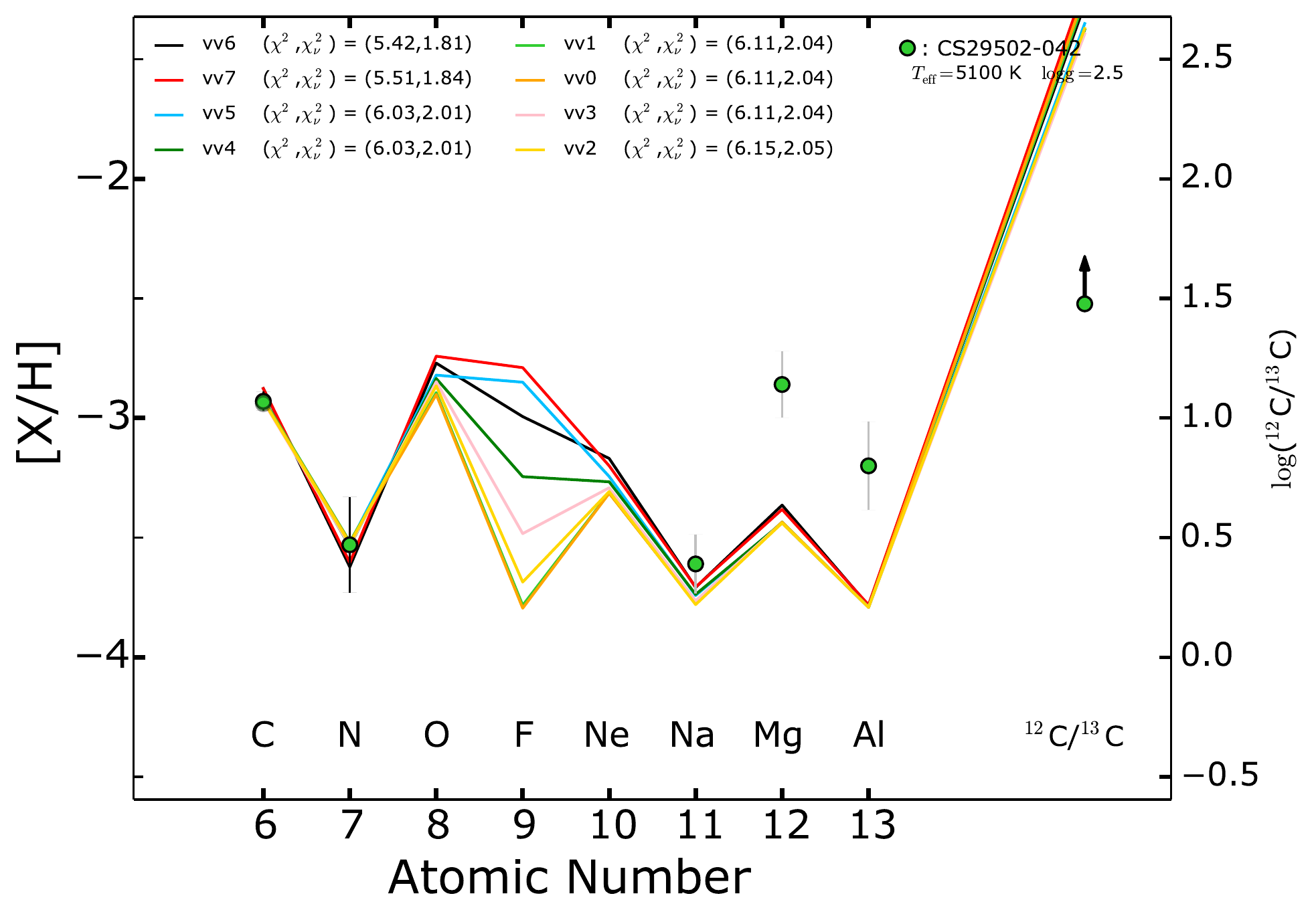}
   \end{minipage}
   \begin{minipage}[c]{.33\linewidth}
       \includegraphics[scale=0.3]{figs/XH_3best_78.pdf}
   \end{minipage}
   \begin{minipage}[c]{.33\linewidth}
       \includegraphics[scale=0.3]{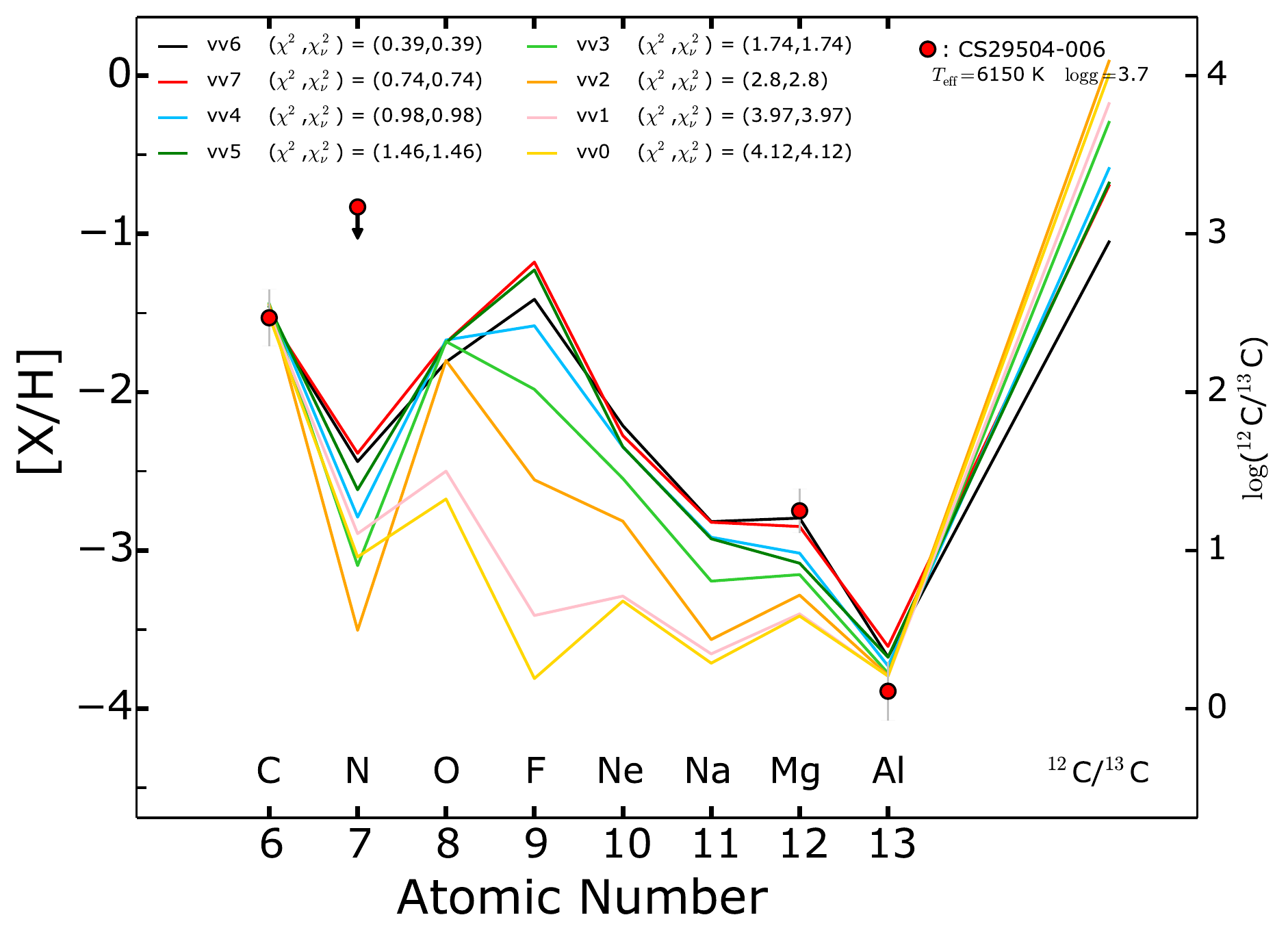}
   \end{minipage}
   \begin{minipage}[c]{.33\linewidth}
       \includegraphics[scale=0.3]{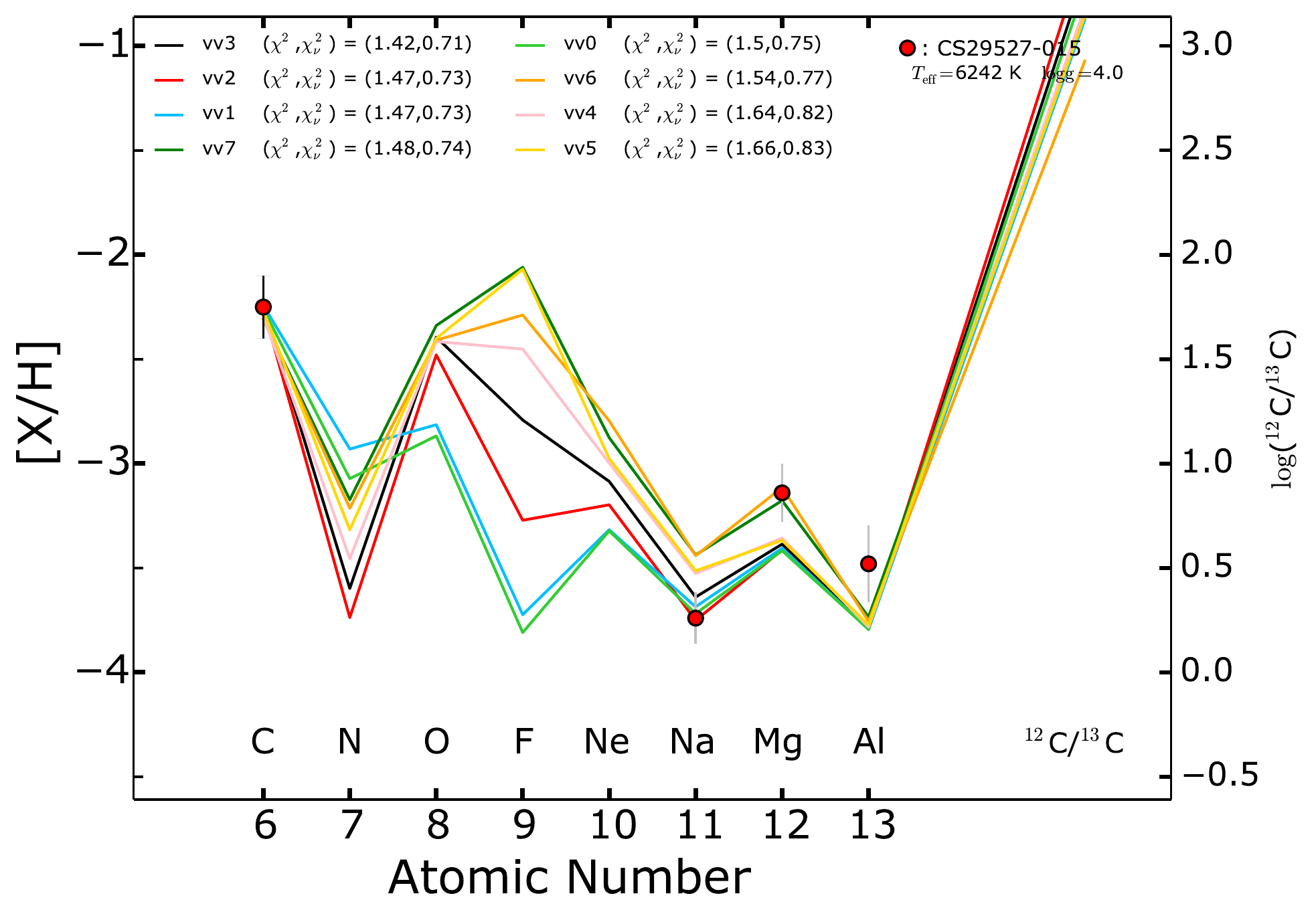}
   \end{minipage}
   \begin{minipage}[c]{.33\linewidth}
       \includegraphics[scale=0.3]{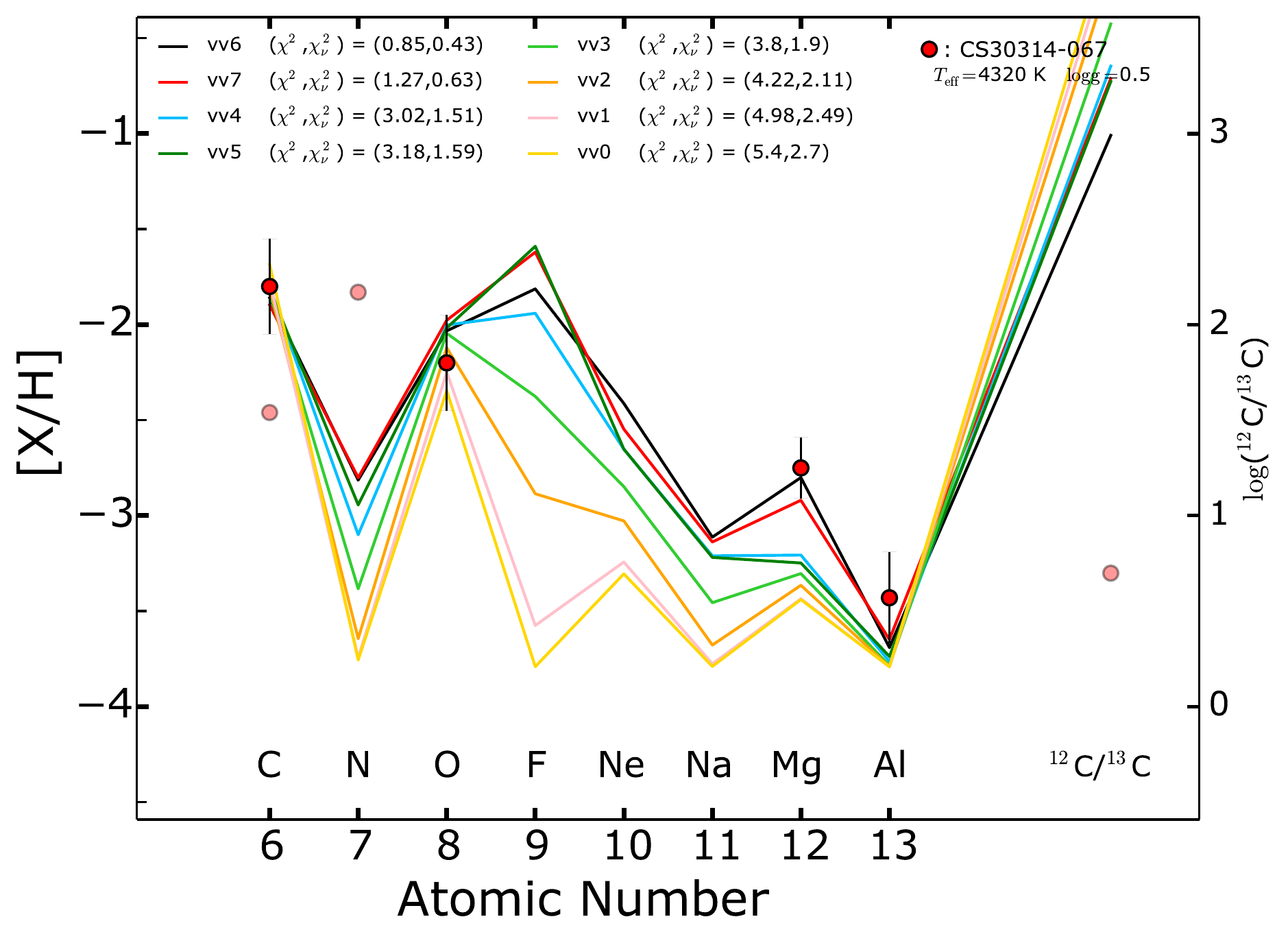}
   \end{minipage}
   \begin{minipage}[c]{.33\linewidth}
       \includegraphics[scale=0.3]{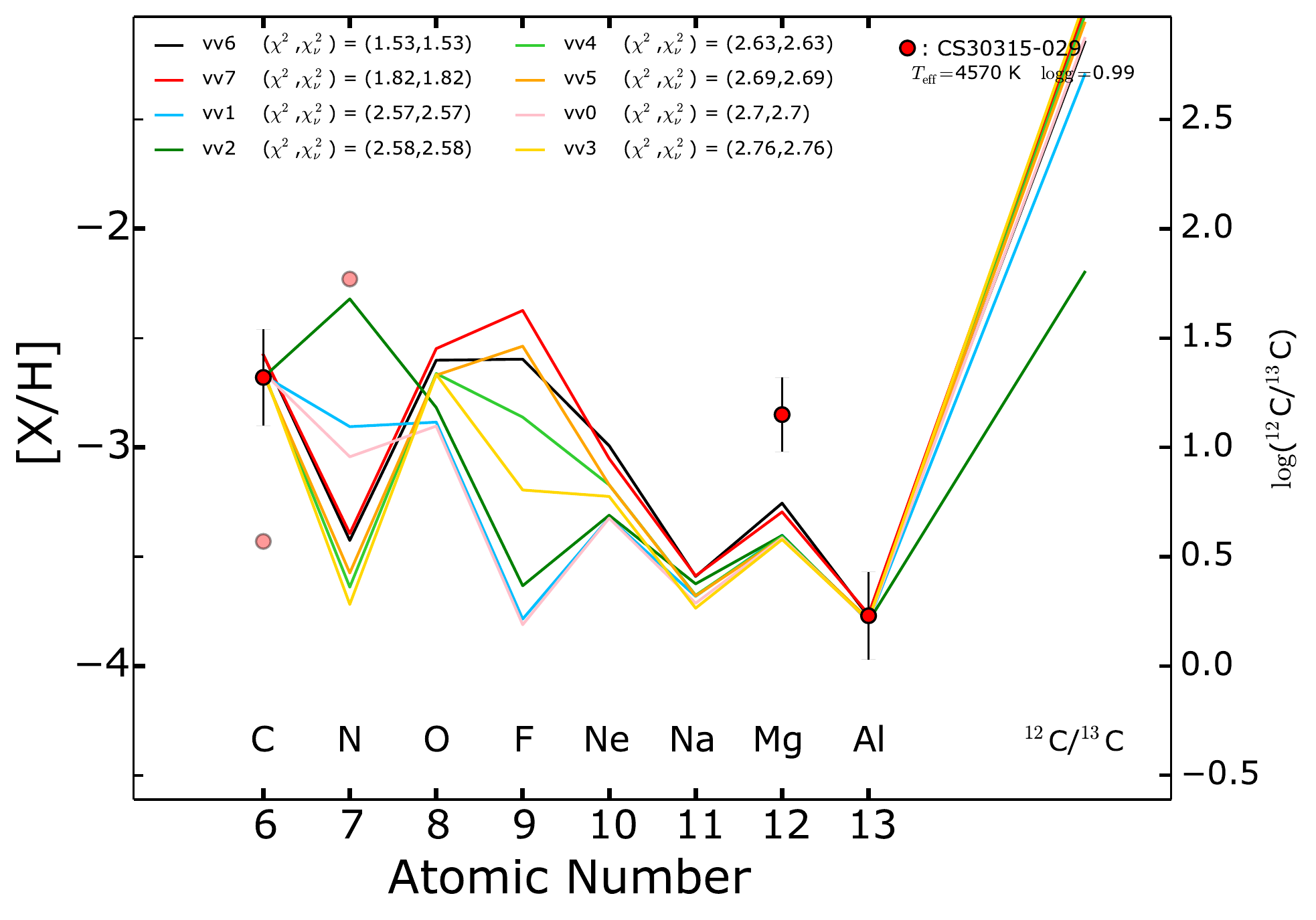}
   \end{minipage}
   \begin{minipage}[c]{.33\linewidth}
       \includegraphics[scale=0.3]{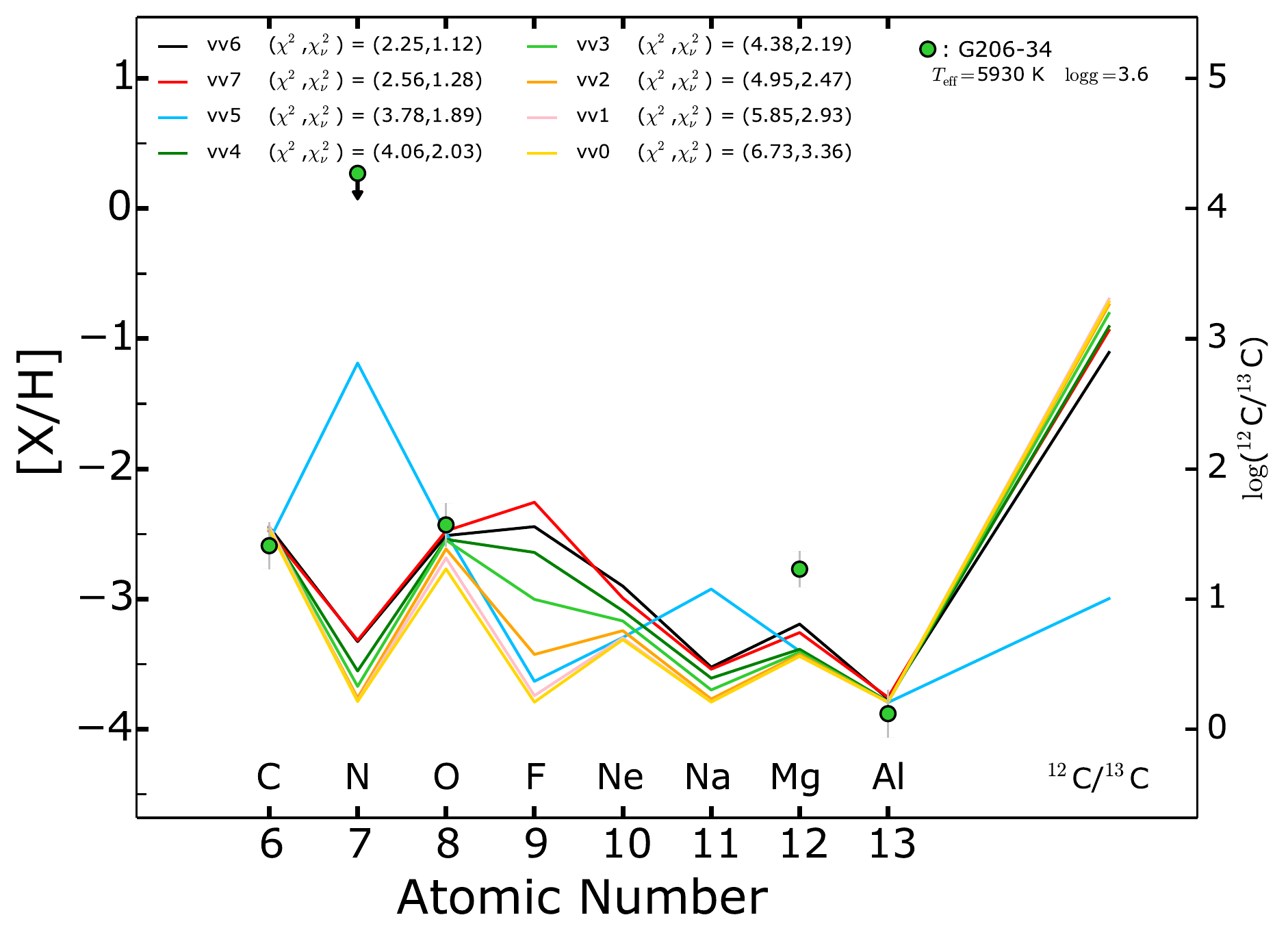}
   \end{minipage}
   \begin{minipage}[c]{.33\linewidth}
       \includegraphics[scale=0.3]{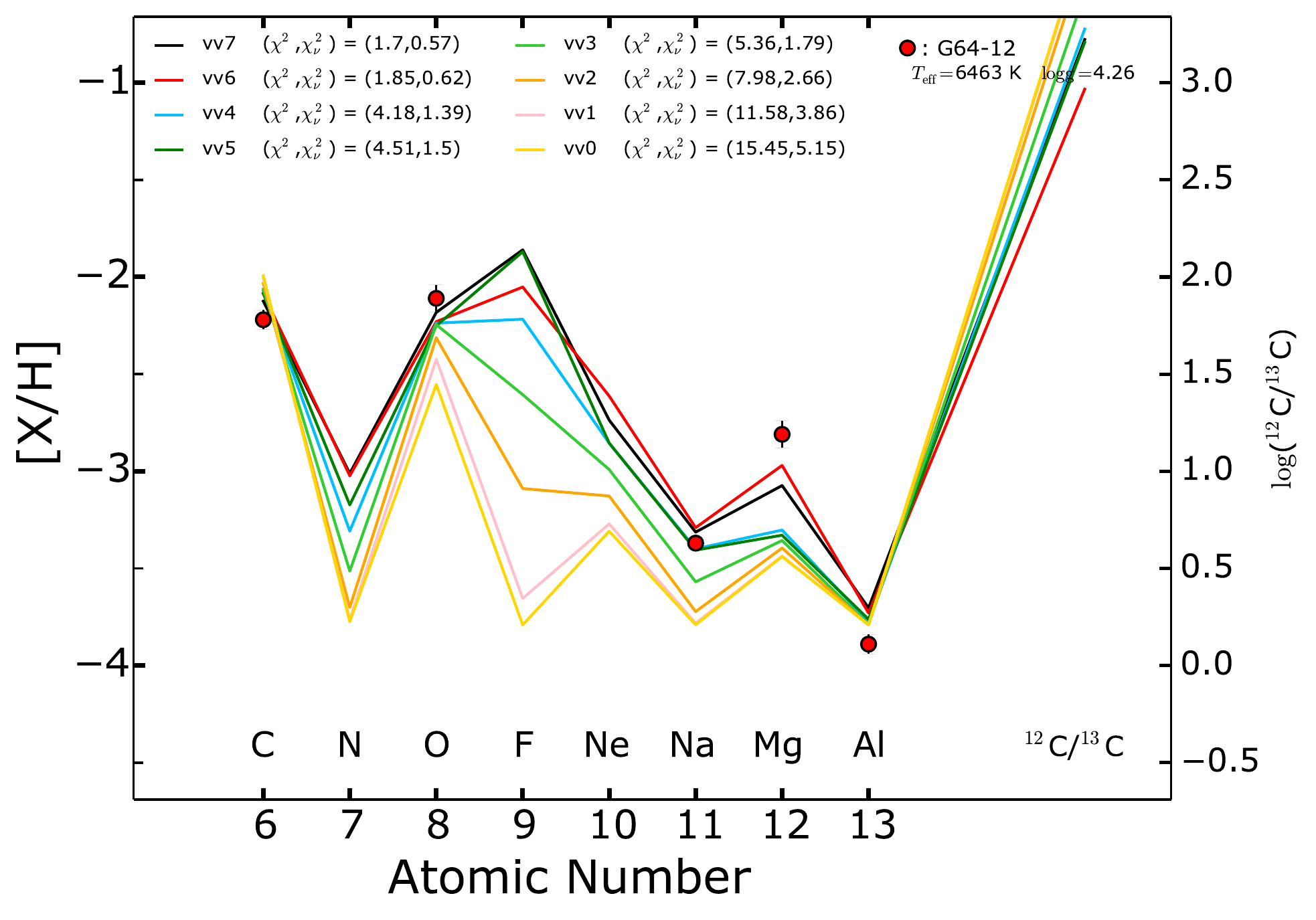}
   \end{minipage}
   \caption{Continued.}
\label{allfit1}
    \end{figure*}

   \begin{figure*}
      \ContinuedFloat
   \centering
   \begin{minipage}[c]{.33\linewidth}
       \includegraphics[scale=0.3]{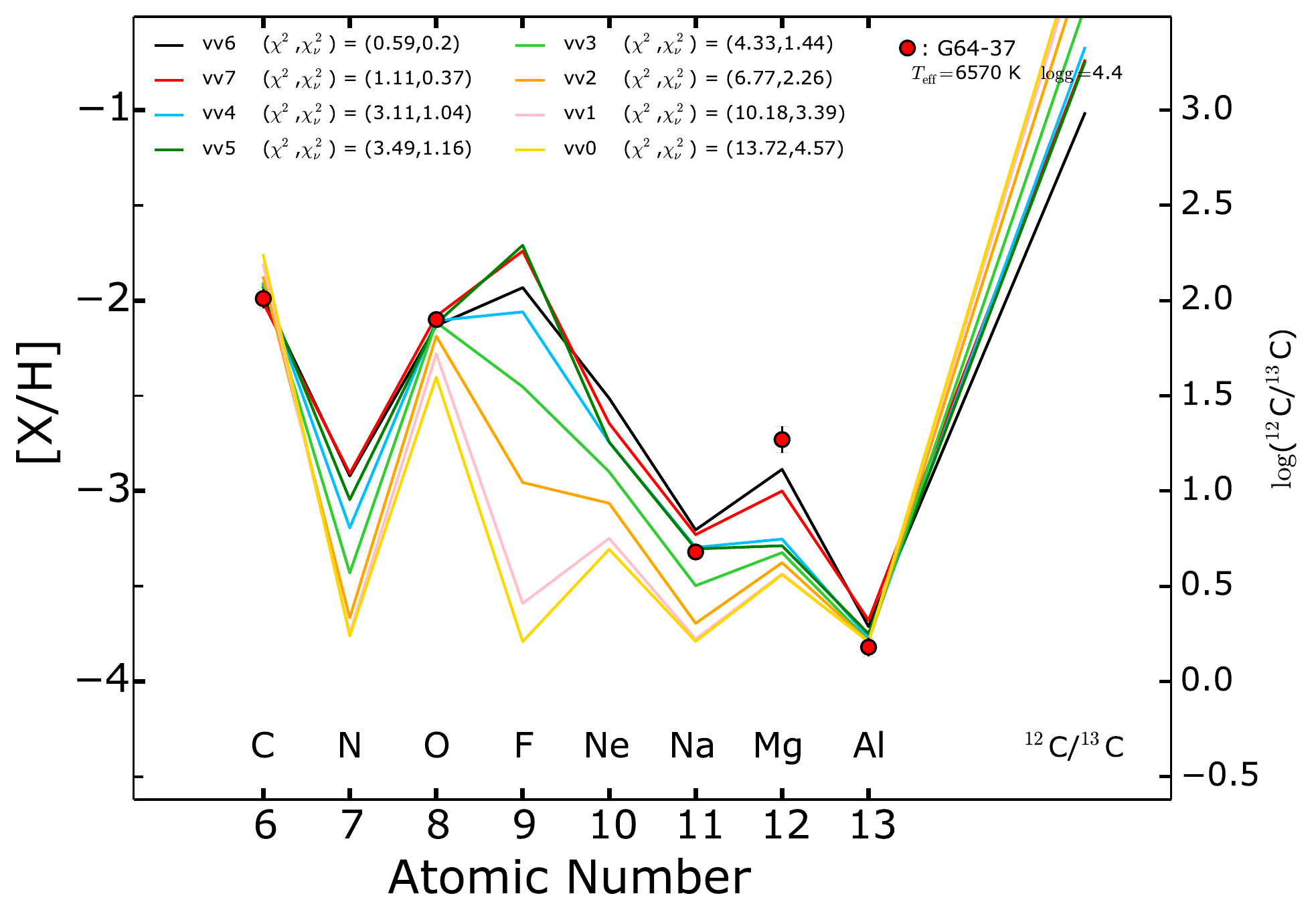}
   \end{minipage}
   \begin{minipage}[c]{.33\linewidth}
       \includegraphics[scale=0.3]{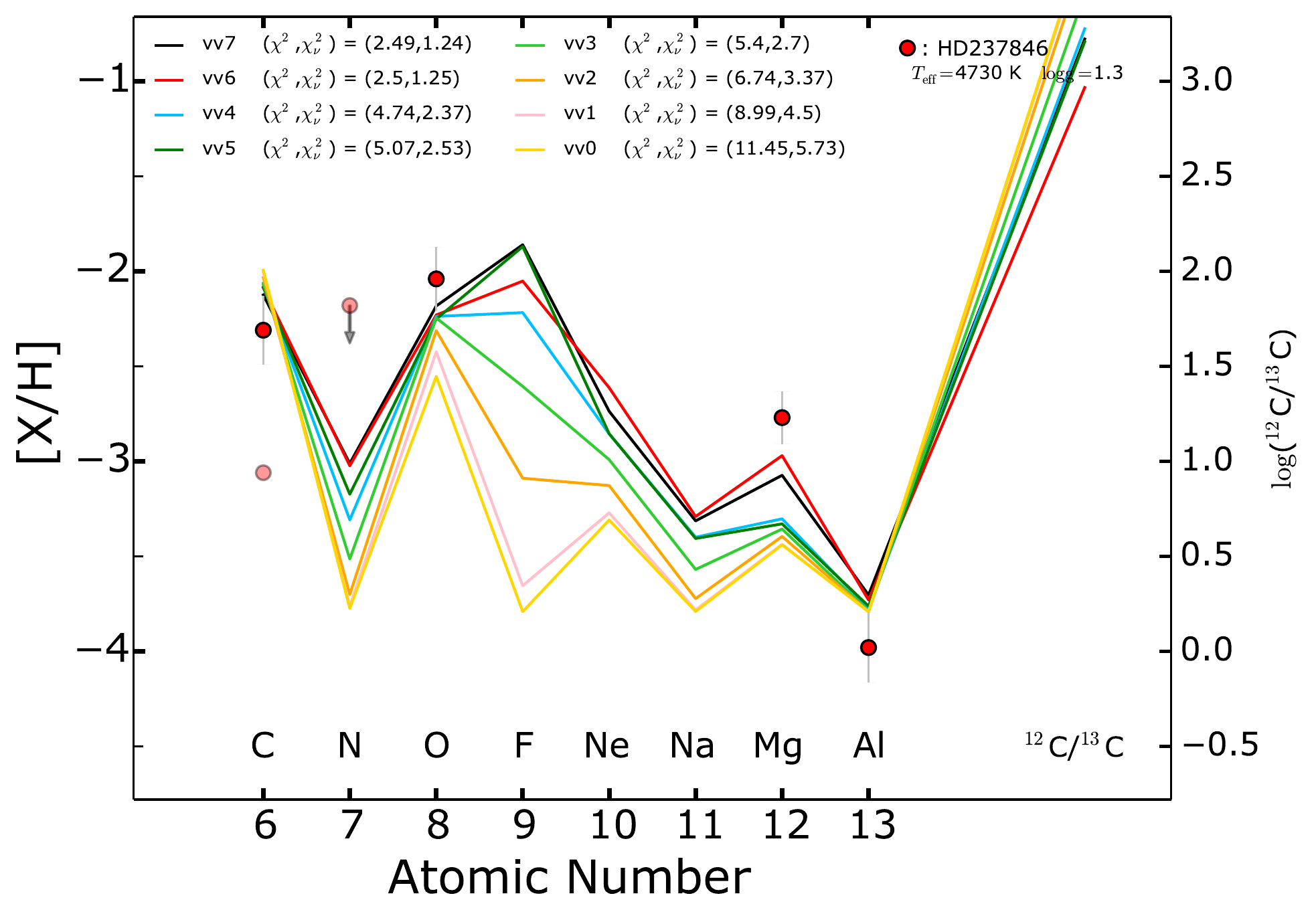}
   \end{minipage}
   \begin{minipage}[c]{.33\linewidth}
       \includegraphics[scale=0.3]{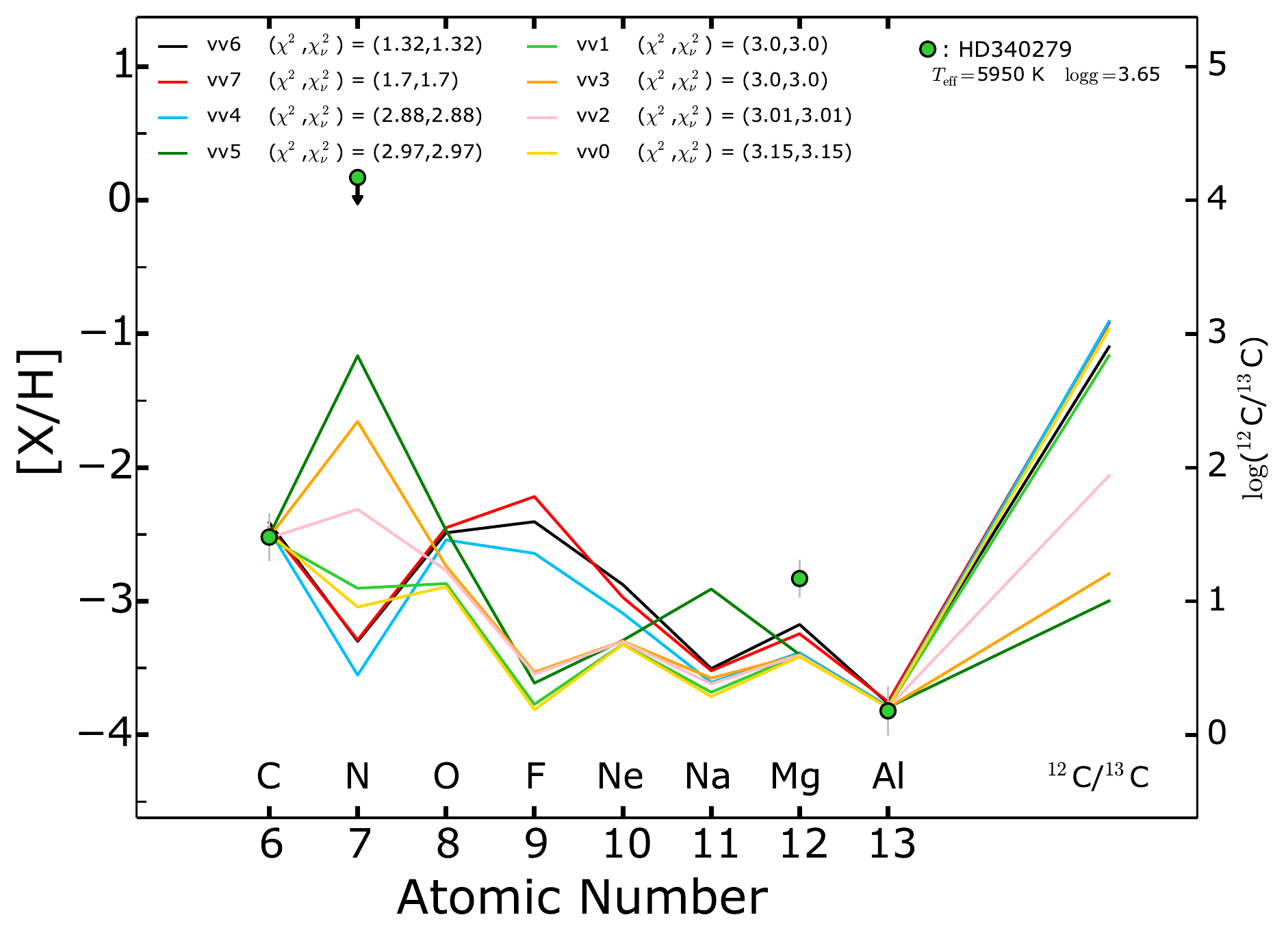}
   \end{minipage}
   \begin{minipage}[c]{.33\linewidth}
       \includegraphics[scale=0.3]{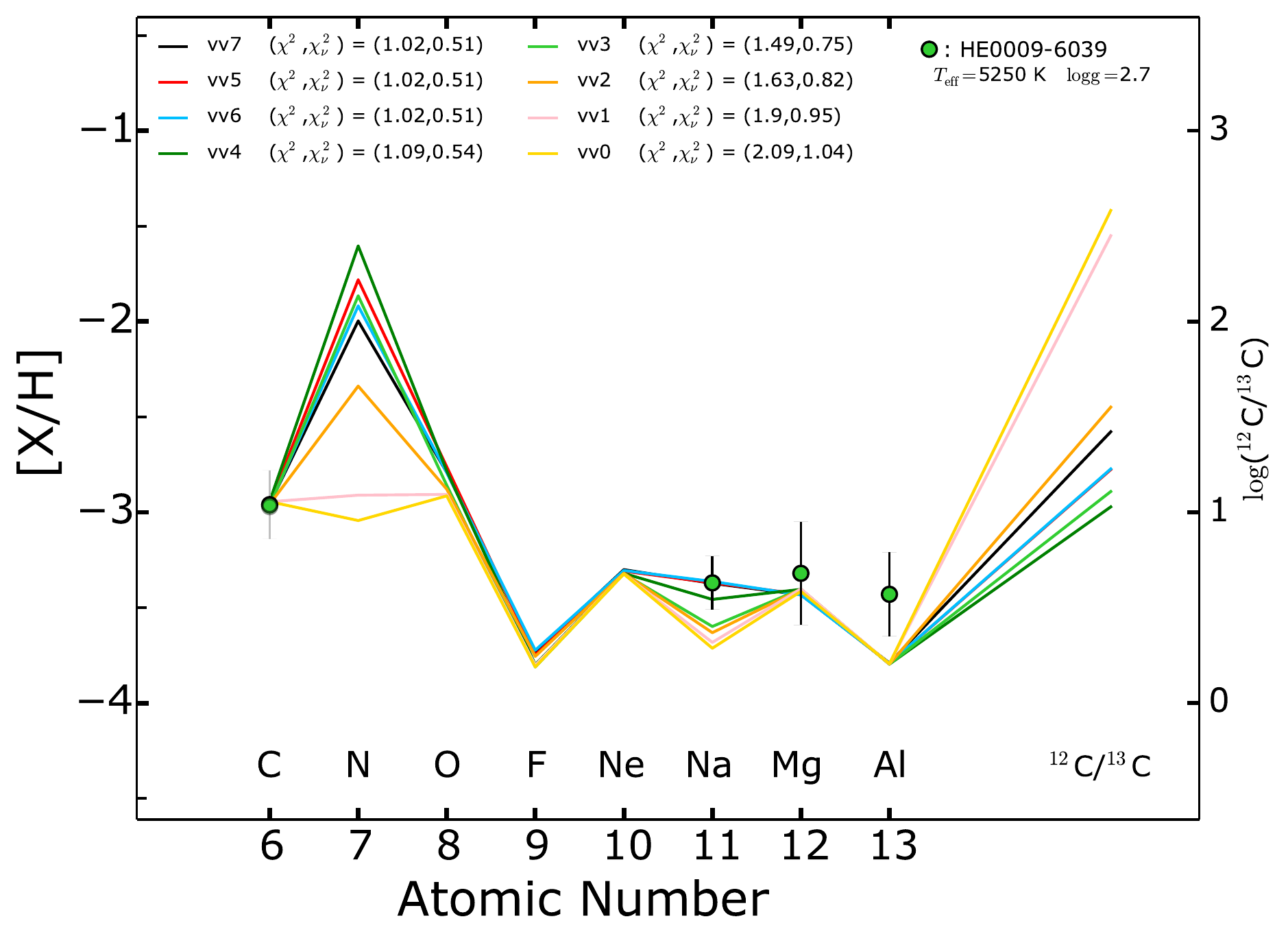}
   \end{minipage}
   \begin{minipage}[c]{.33\linewidth}
       \includegraphics[scale=0.3]{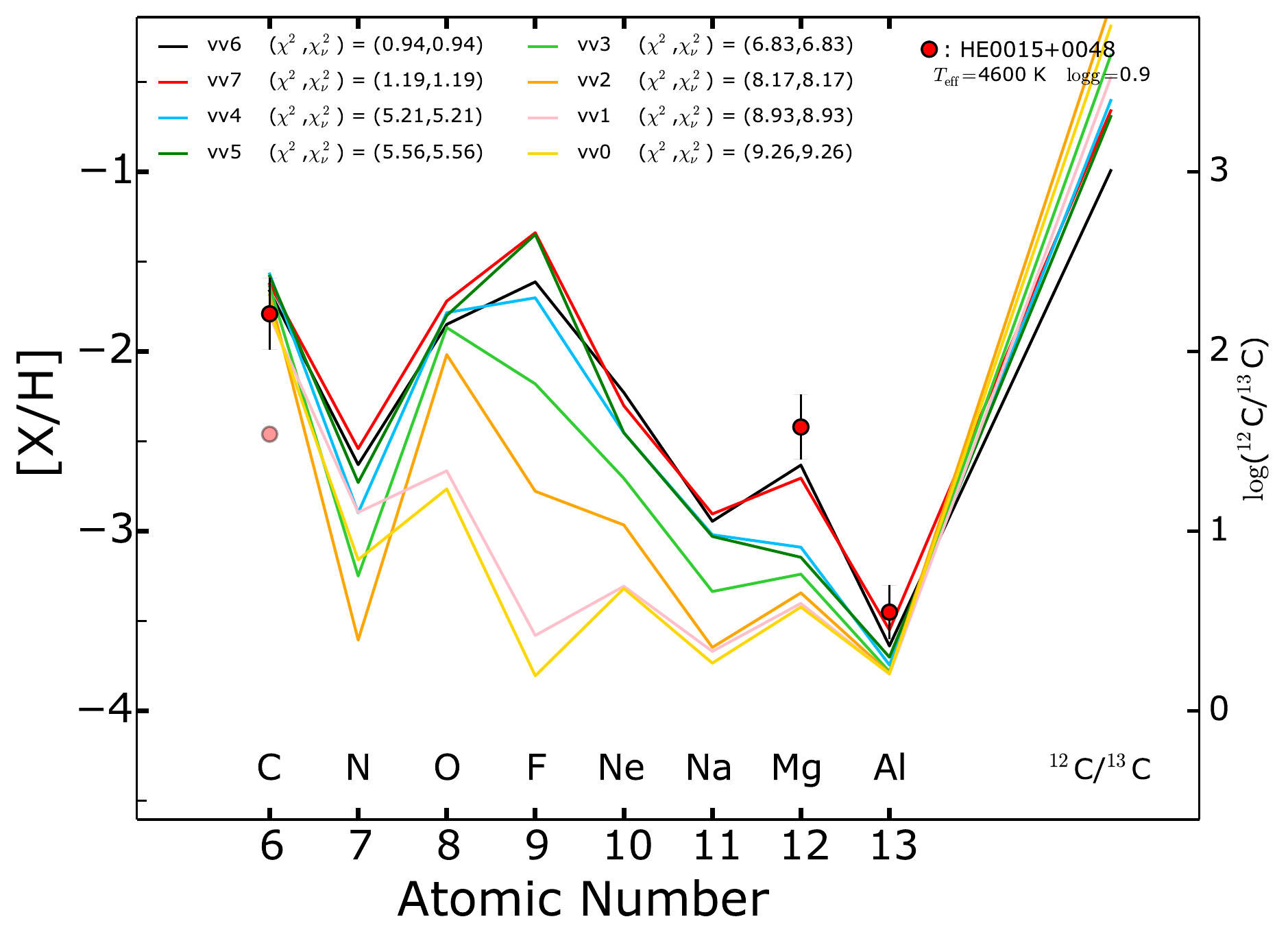}
   \end{minipage}
   \begin{minipage}[c]{.33\linewidth}
       \includegraphics[scale=0.3]{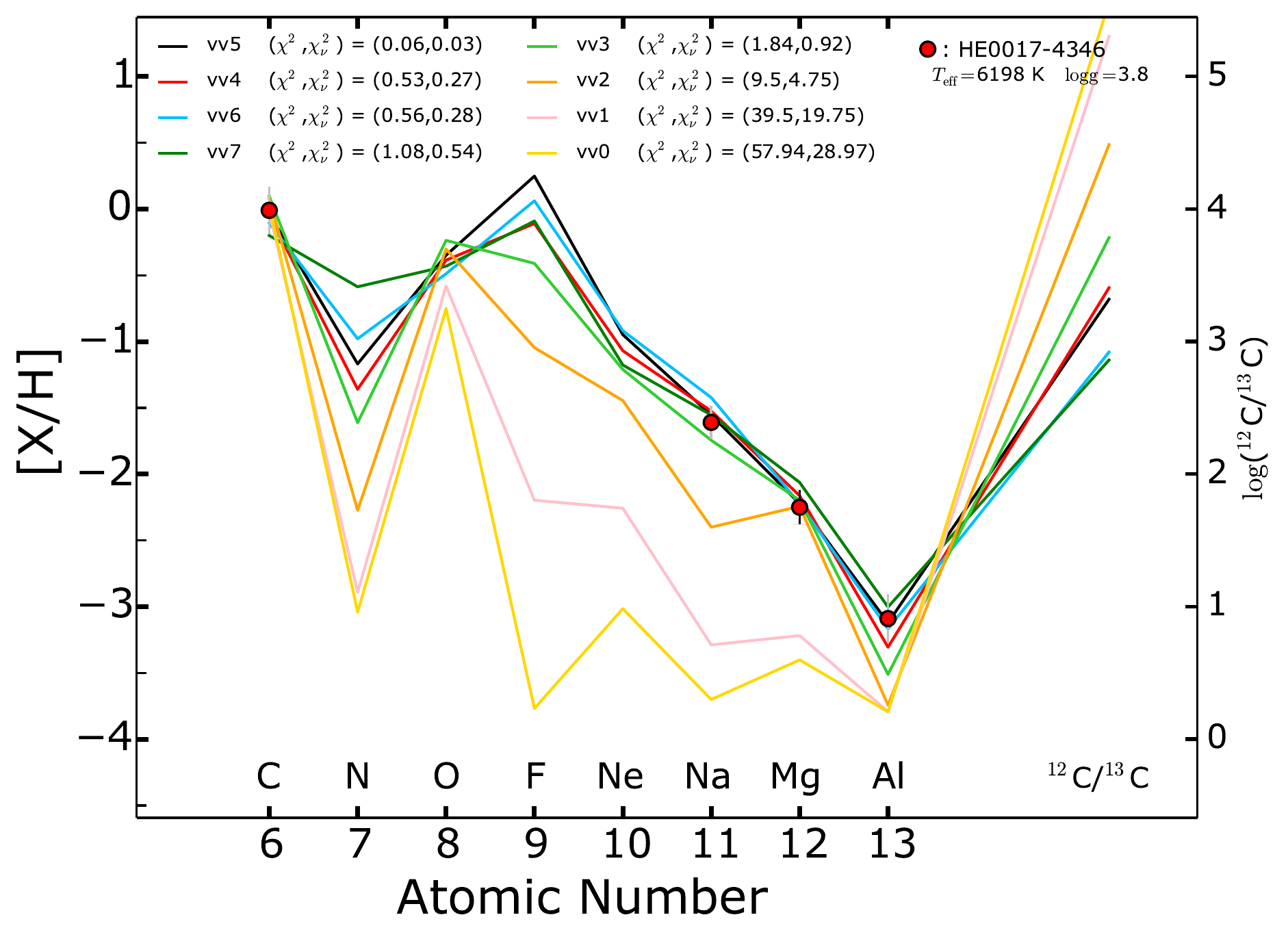}
   \end{minipage}
   \begin{minipage}[c]{.33\linewidth}
       \includegraphics[scale=0.3]{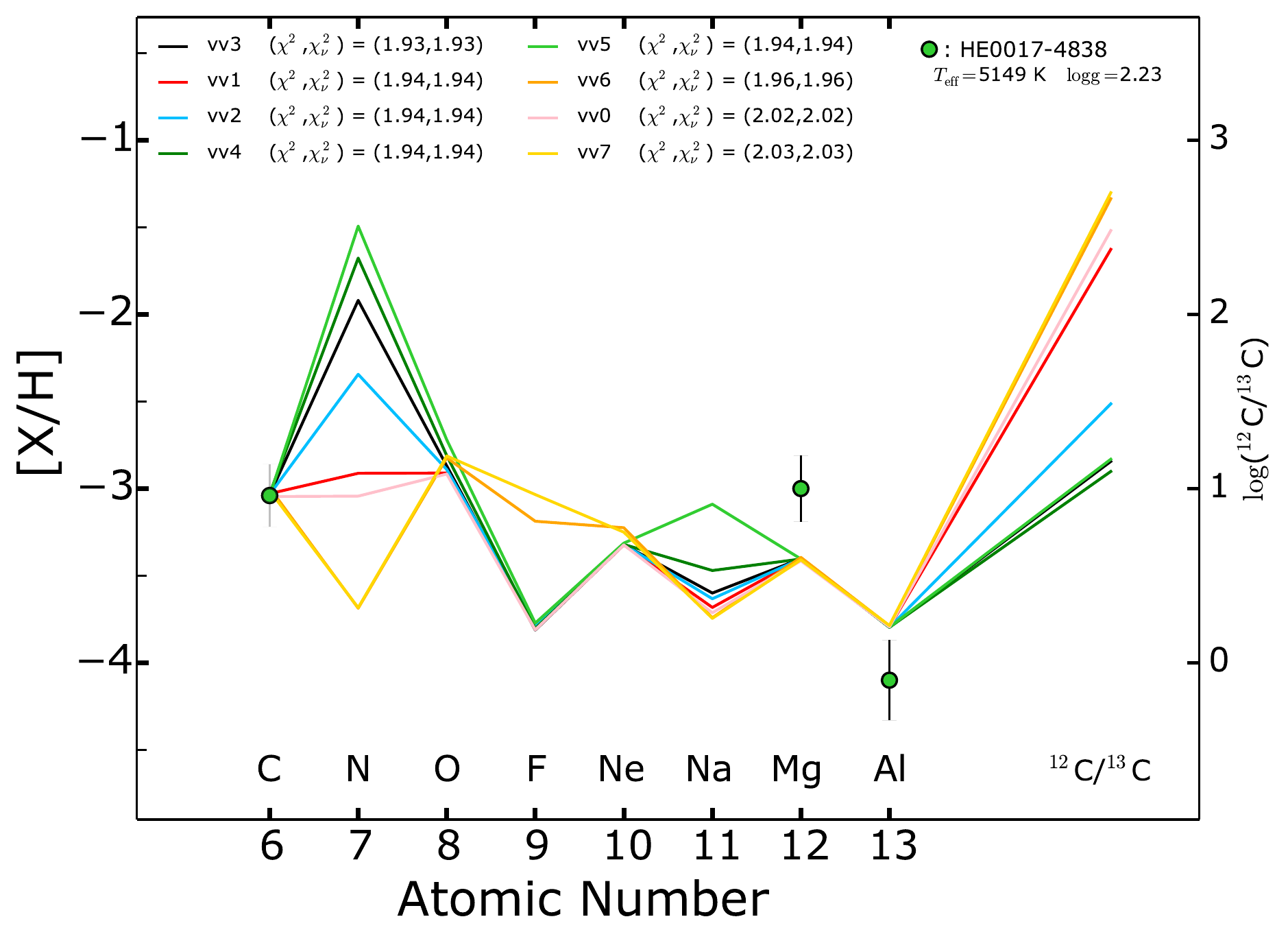}
   \end{minipage}
   \begin{minipage}[c]{.33\linewidth}
       \includegraphics[scale=0.3]{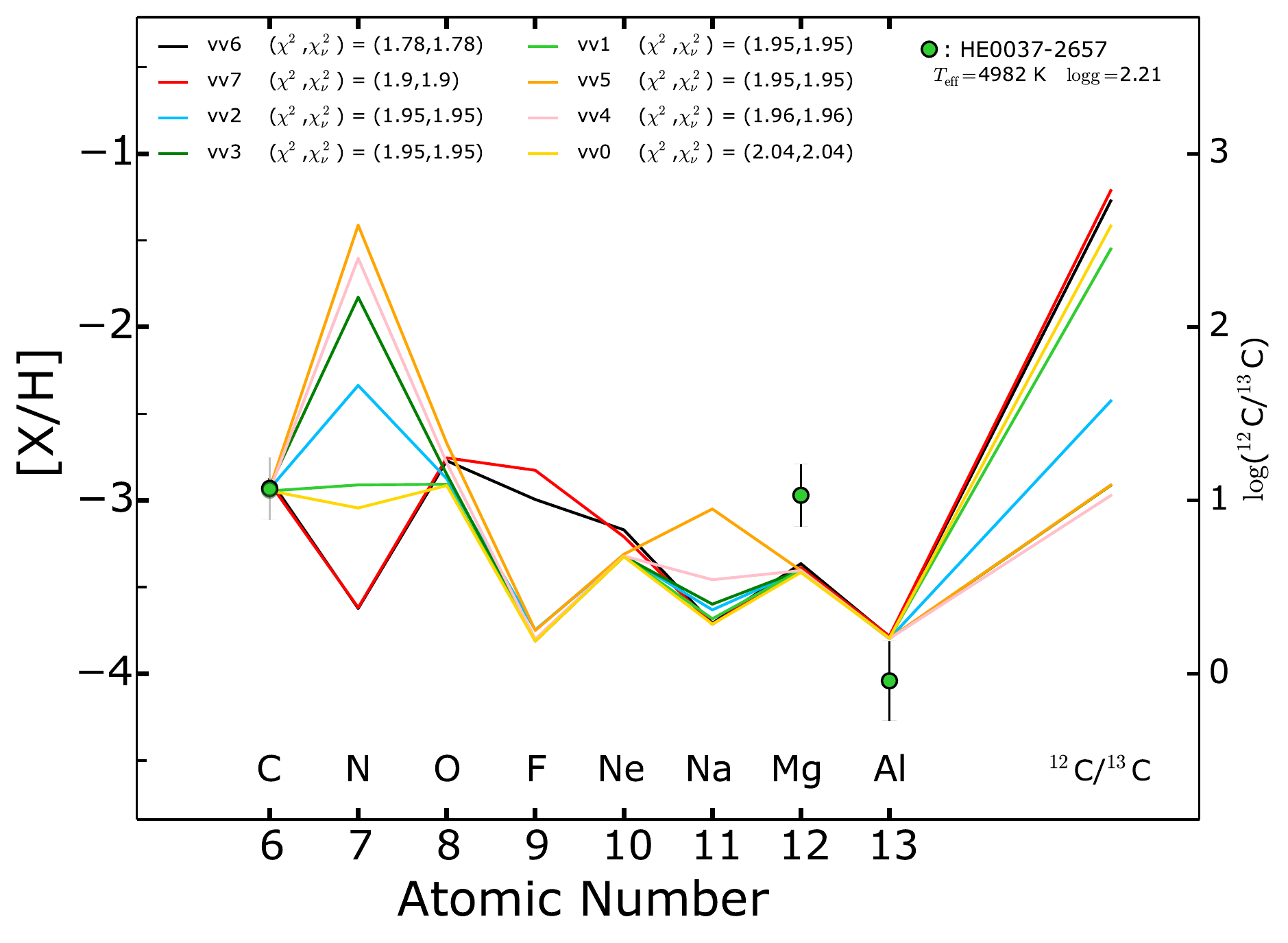}
   \end{minipage}
   \begin{minipage}[c]{.33\linewidth}
       \includegraphics[scale=0.3]{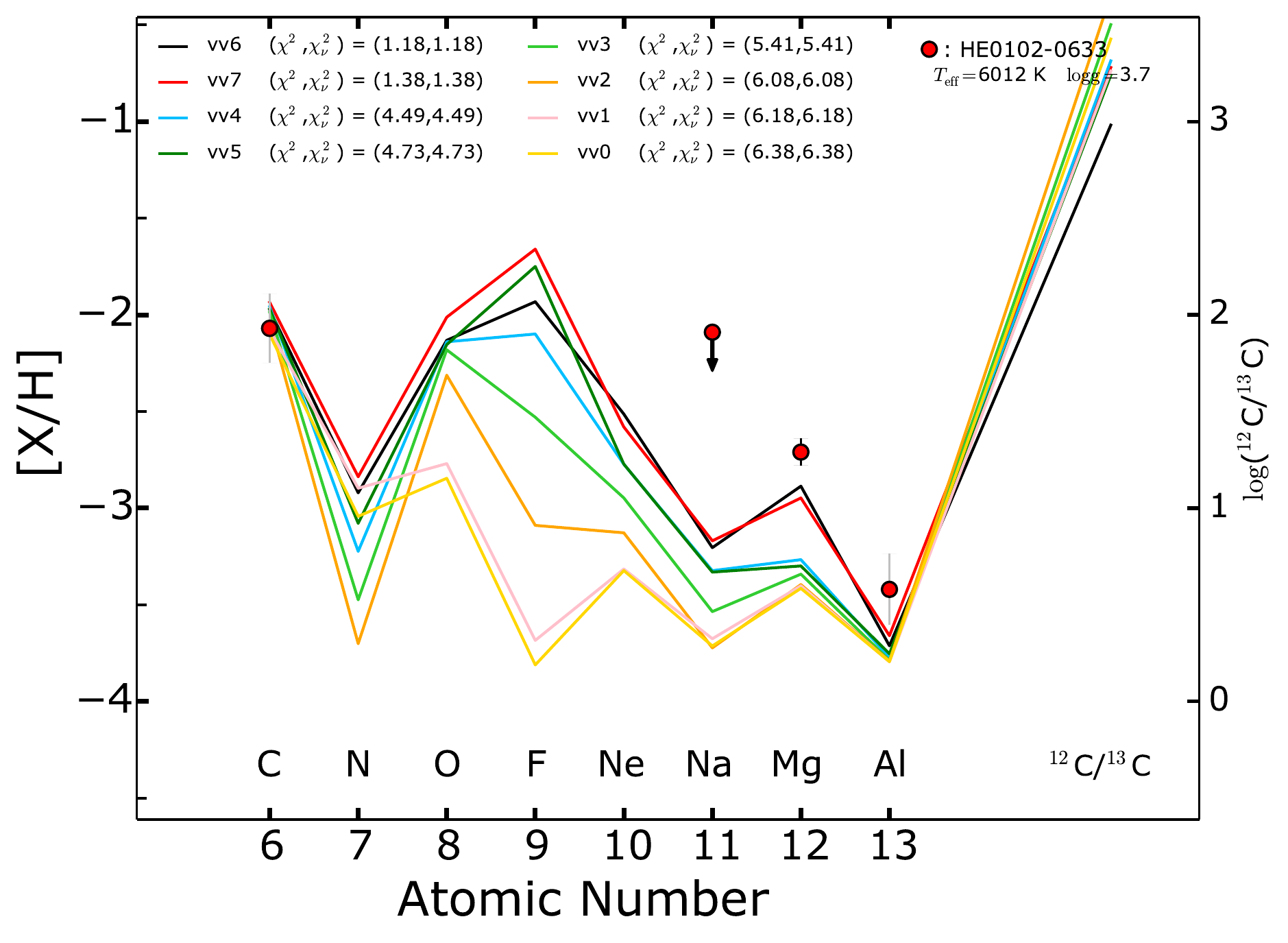}
   \end{minipage}
   \begin{minipage}[c]{.33\linewidth}
       \includegraphics[scale=0.3]{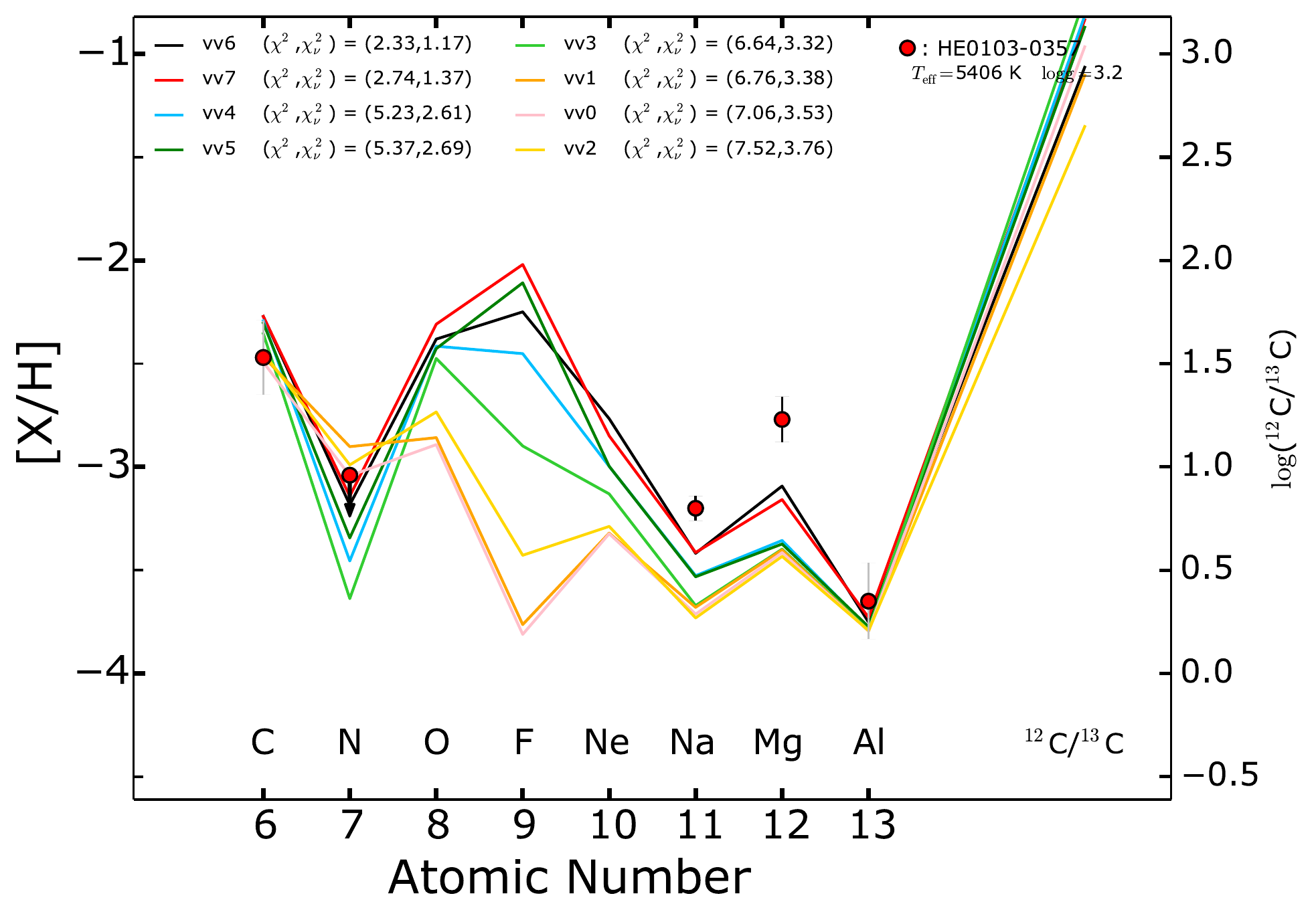}
   \end{minipage}
   \begin{minipage}[c]{.33\linewidth}
       \includegraphics[scale=0.3]{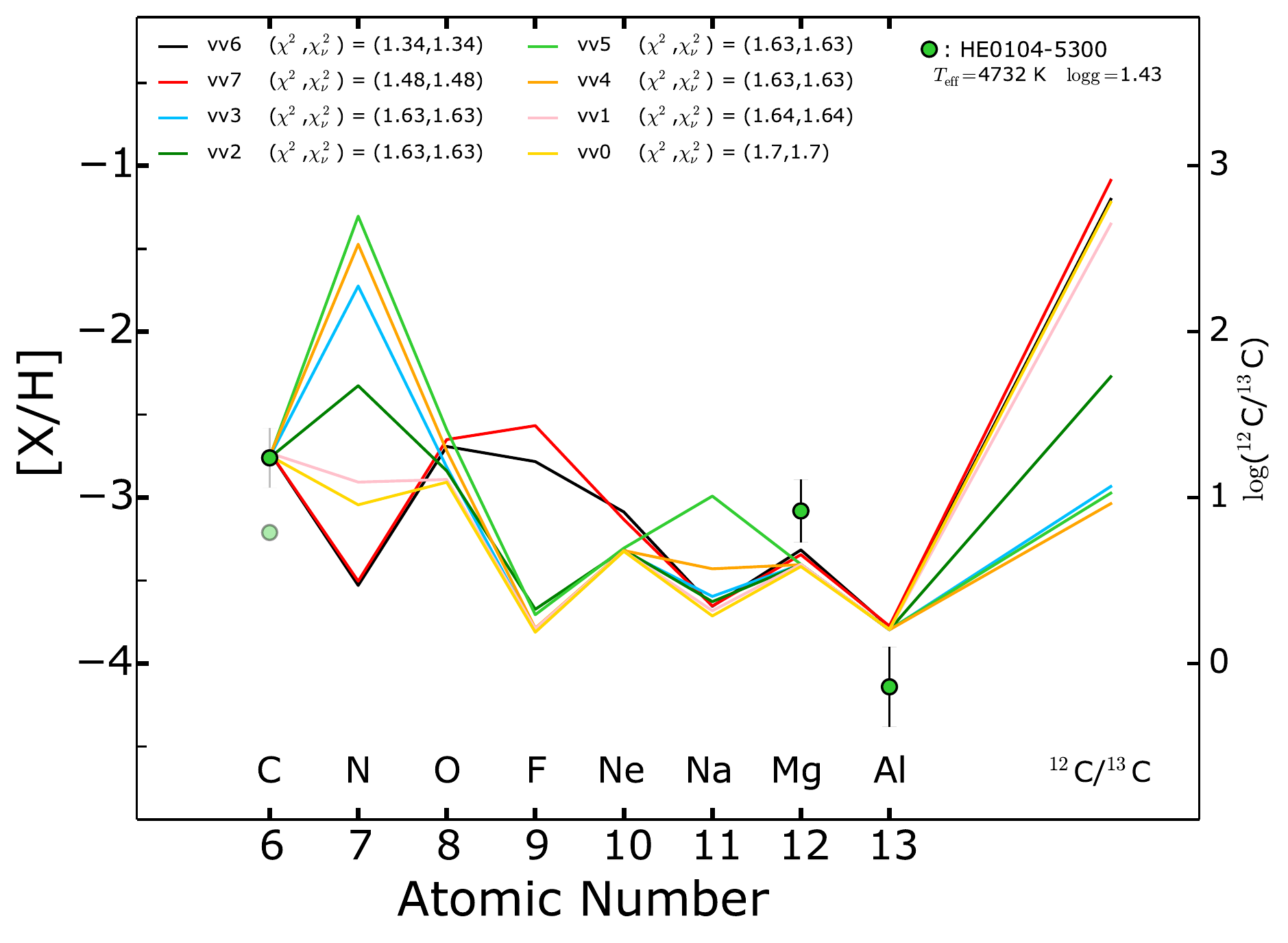}
   \end{minipage}
   \begin{minipage}[c]{.33\linewidth}
       \includegraphics[scale=0.3]{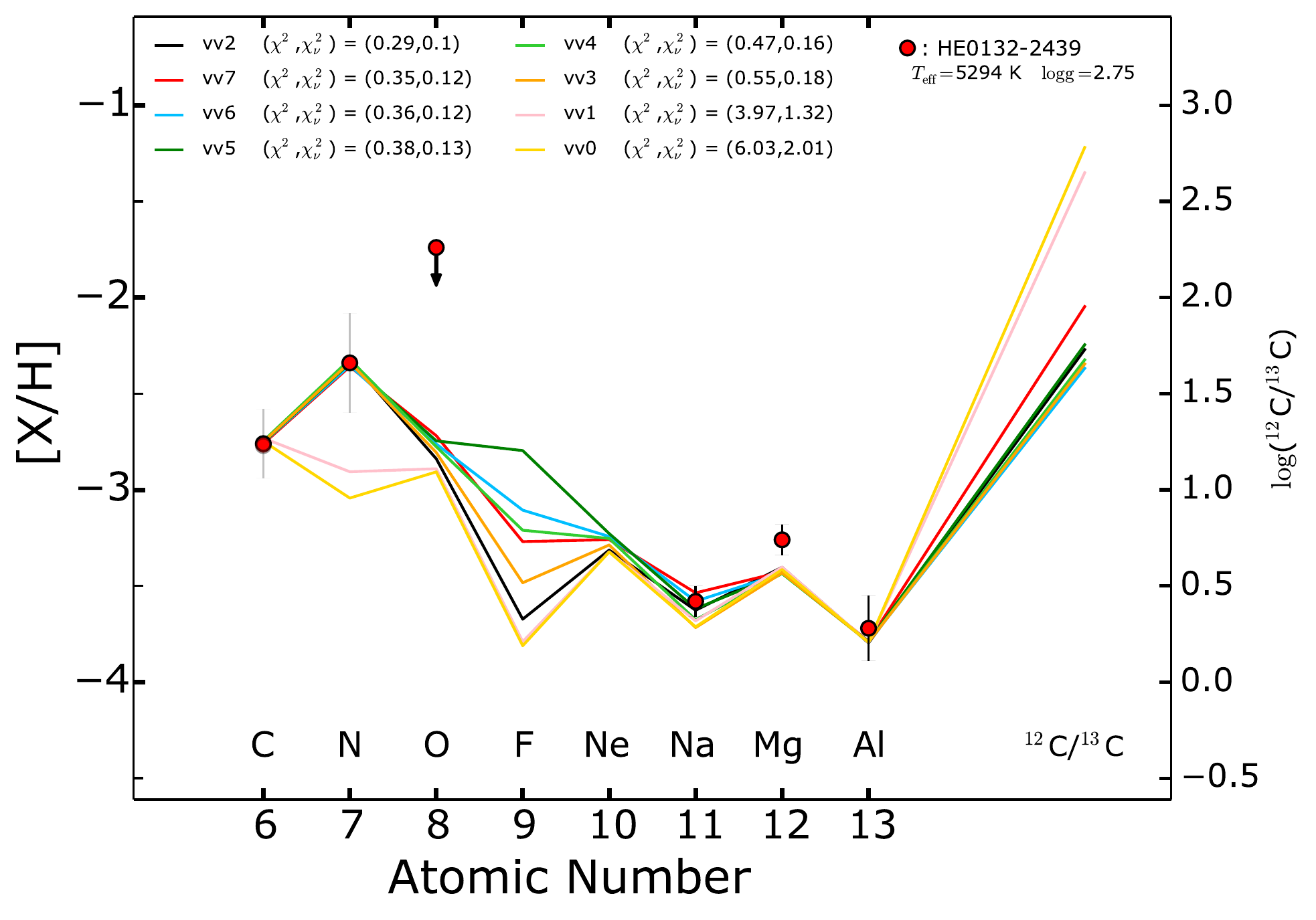}
   \end{minipage}
   \begin{minipage}[c]{.33\linewidth}
       \includegraphics[scale=0.3]{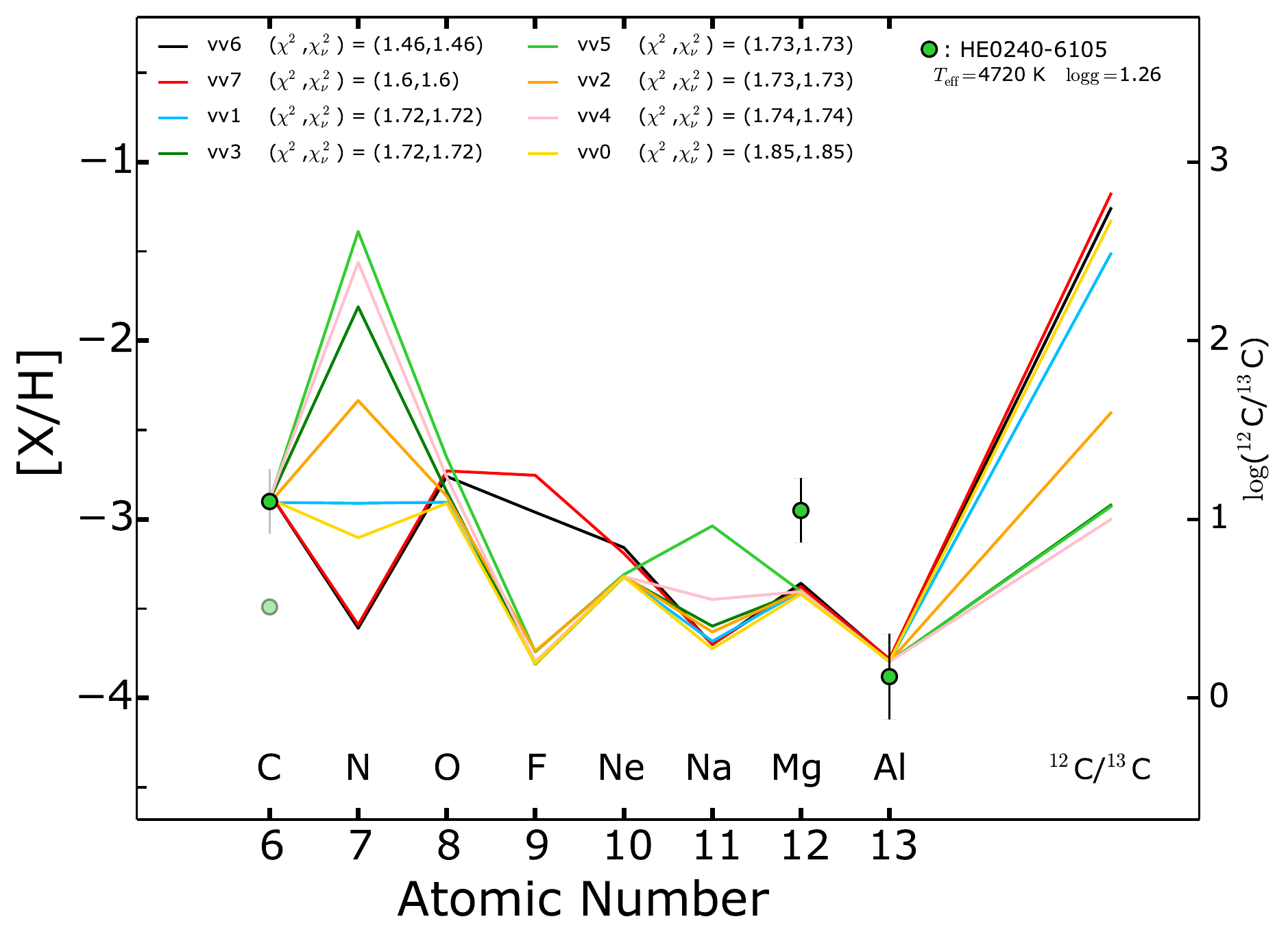}
   \end{minipage}
   \begin{minipage}[c]{.33\linewidth}
       \includegraphics[scale=0.3]{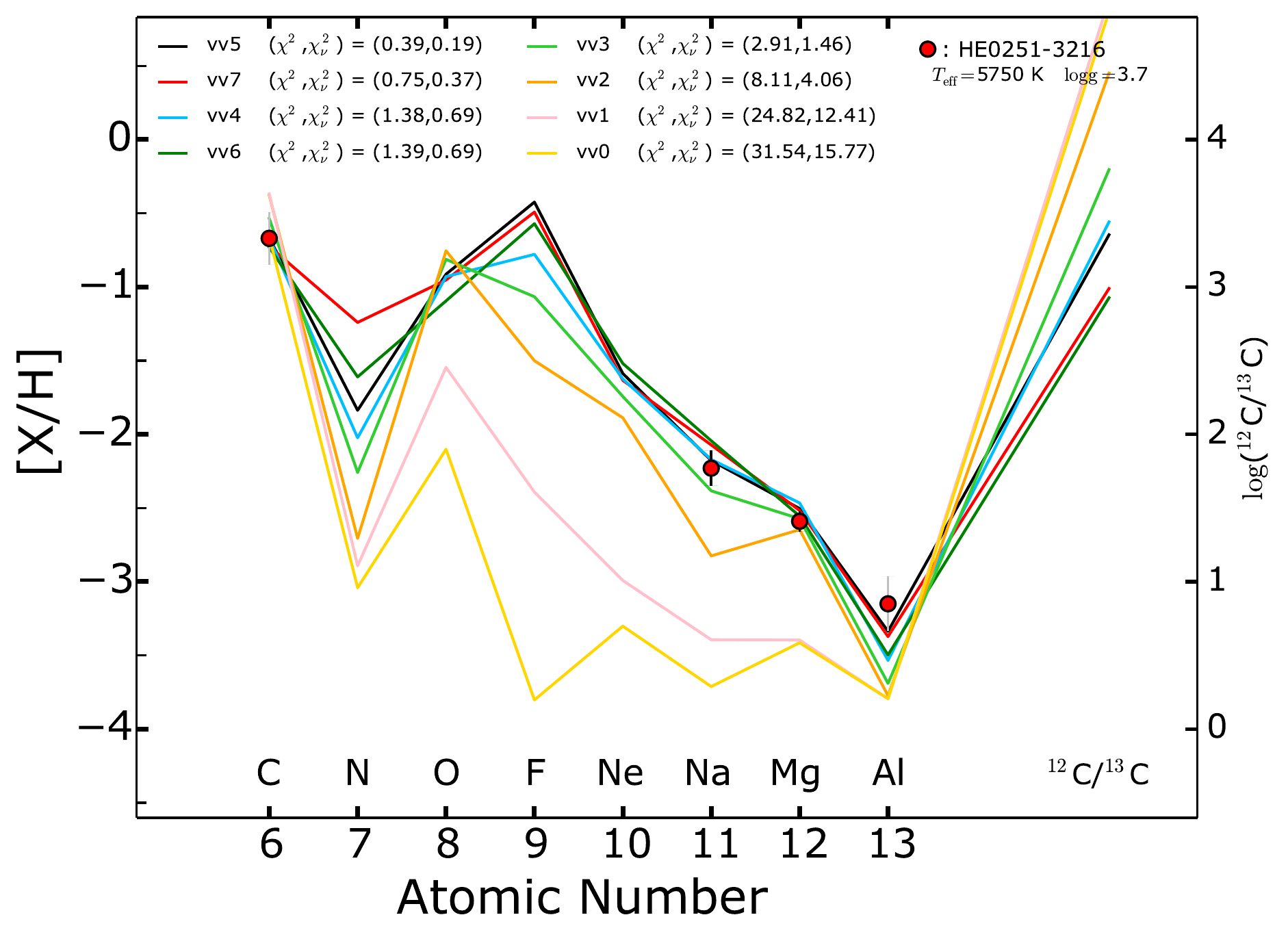}
   \end{minipage}
   \begin{minipage}[c]{.33\linewidth}
       \includegraphics[scale=0.3]{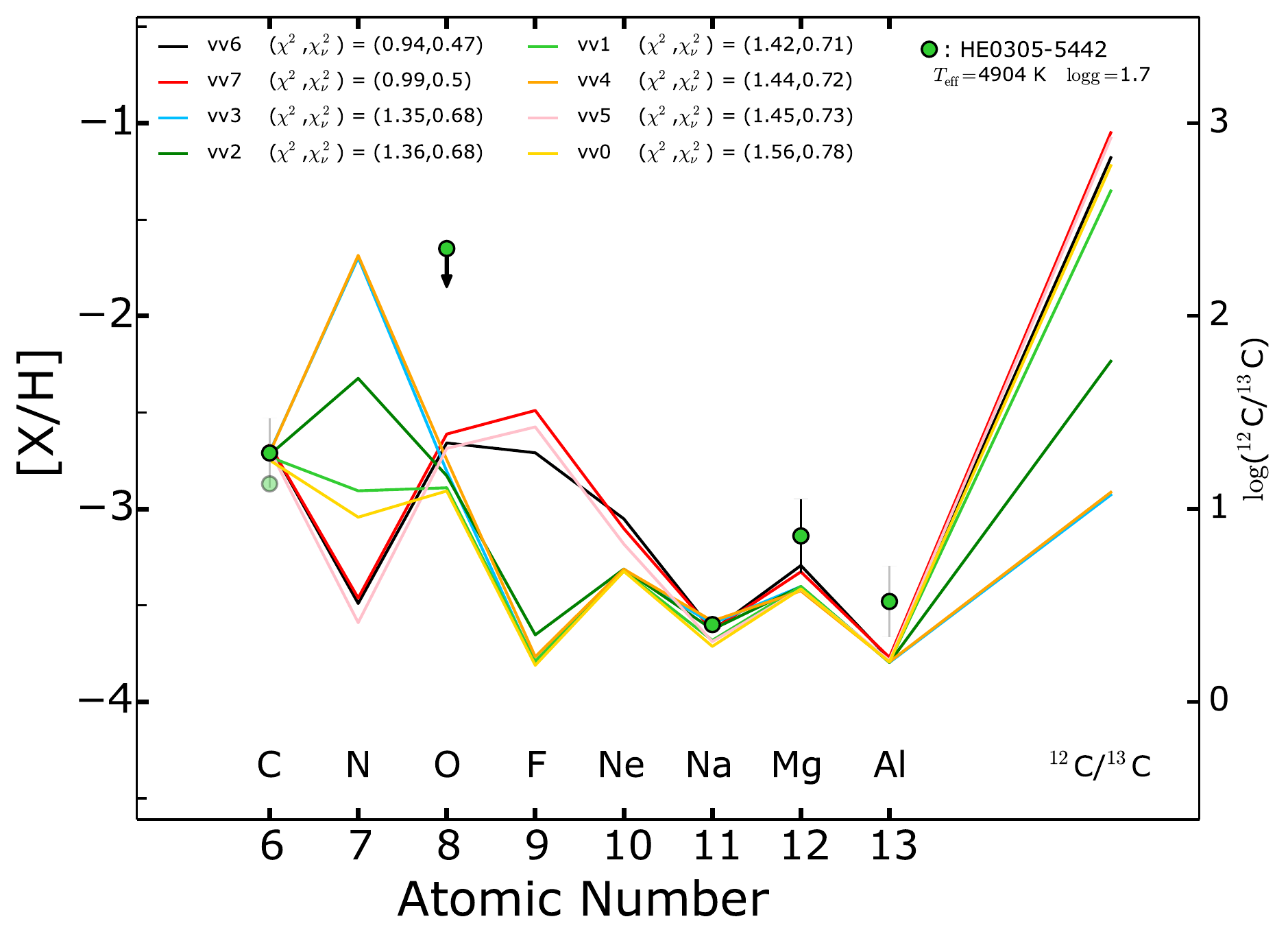}
   \end{minipage}
   \caption{Continued.}
\label{allfit1}
    \end{figure*}

   \begin{figure*}
      \ContinuedFloat
   \centering
   \begin{minipage}[c]{.33\linewidth}
       \includegraphics[scale=0.3]{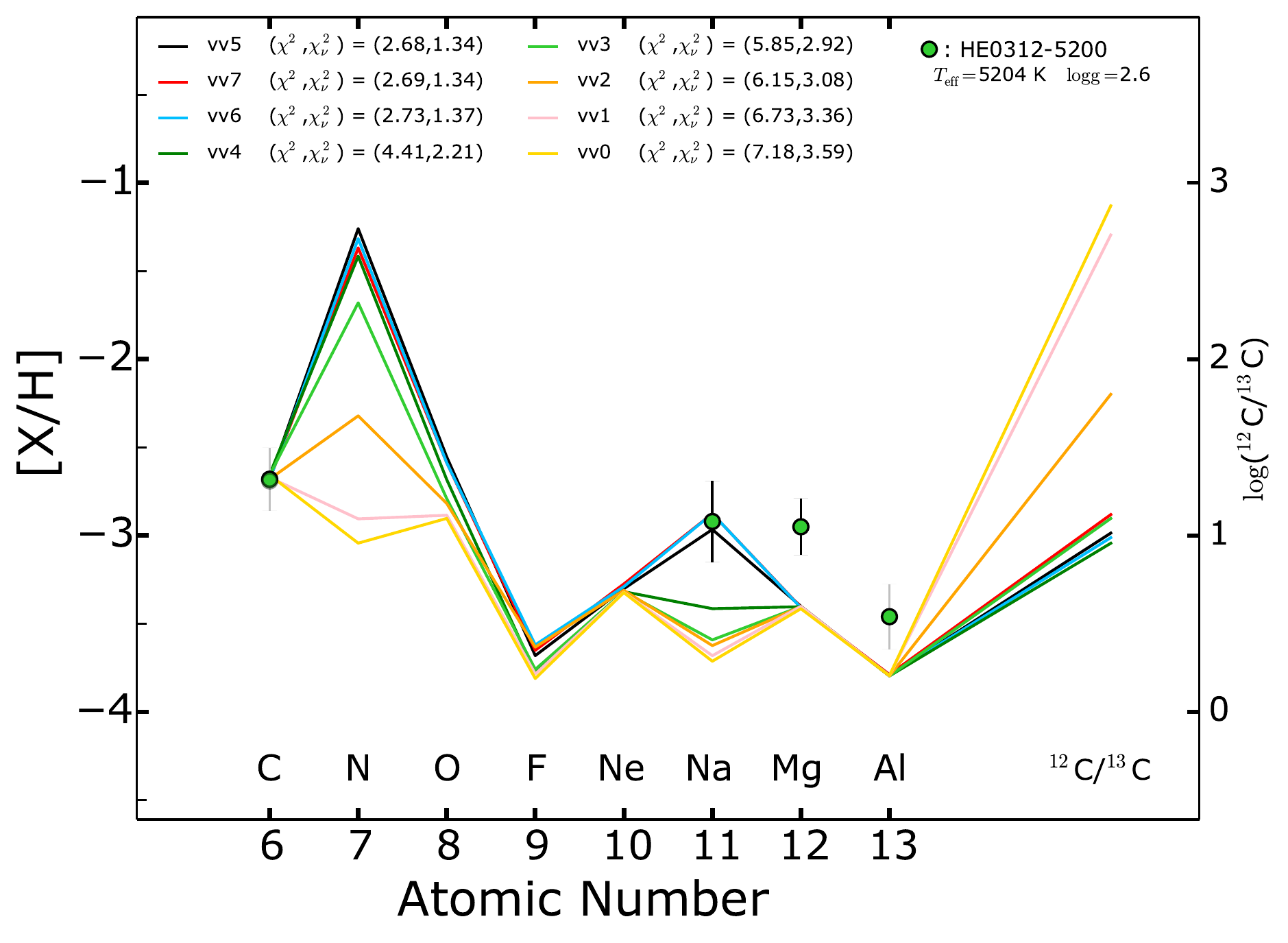}
   \end{minipage}
   \begin{minipage}[c]{.33\linewidth}
       \includegraphics[scale=0.3]{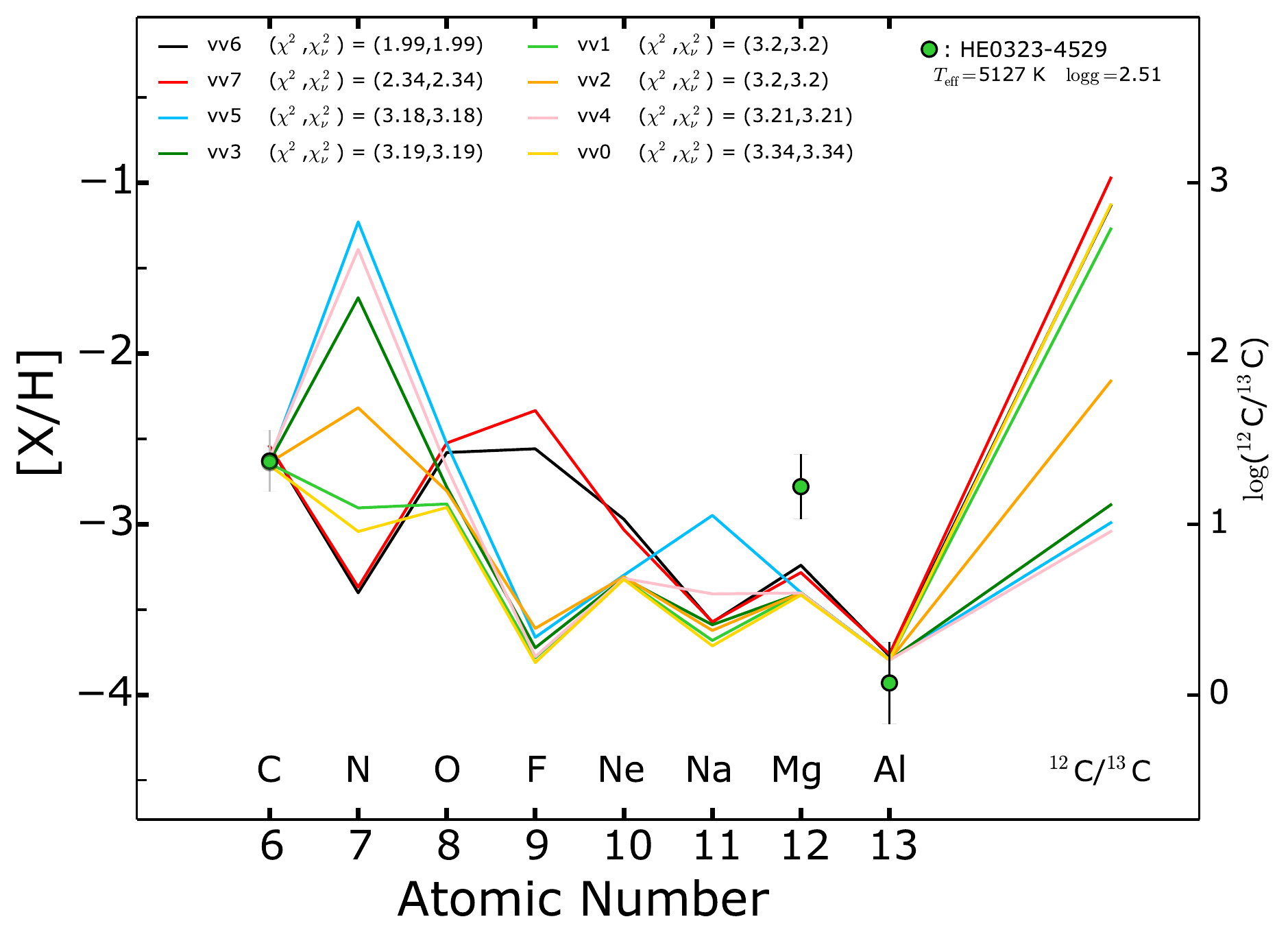}
   \end{minipage}
   \begin{minipage}[c]{.33\linewidth}
       \includegraphics[scale=0.3]{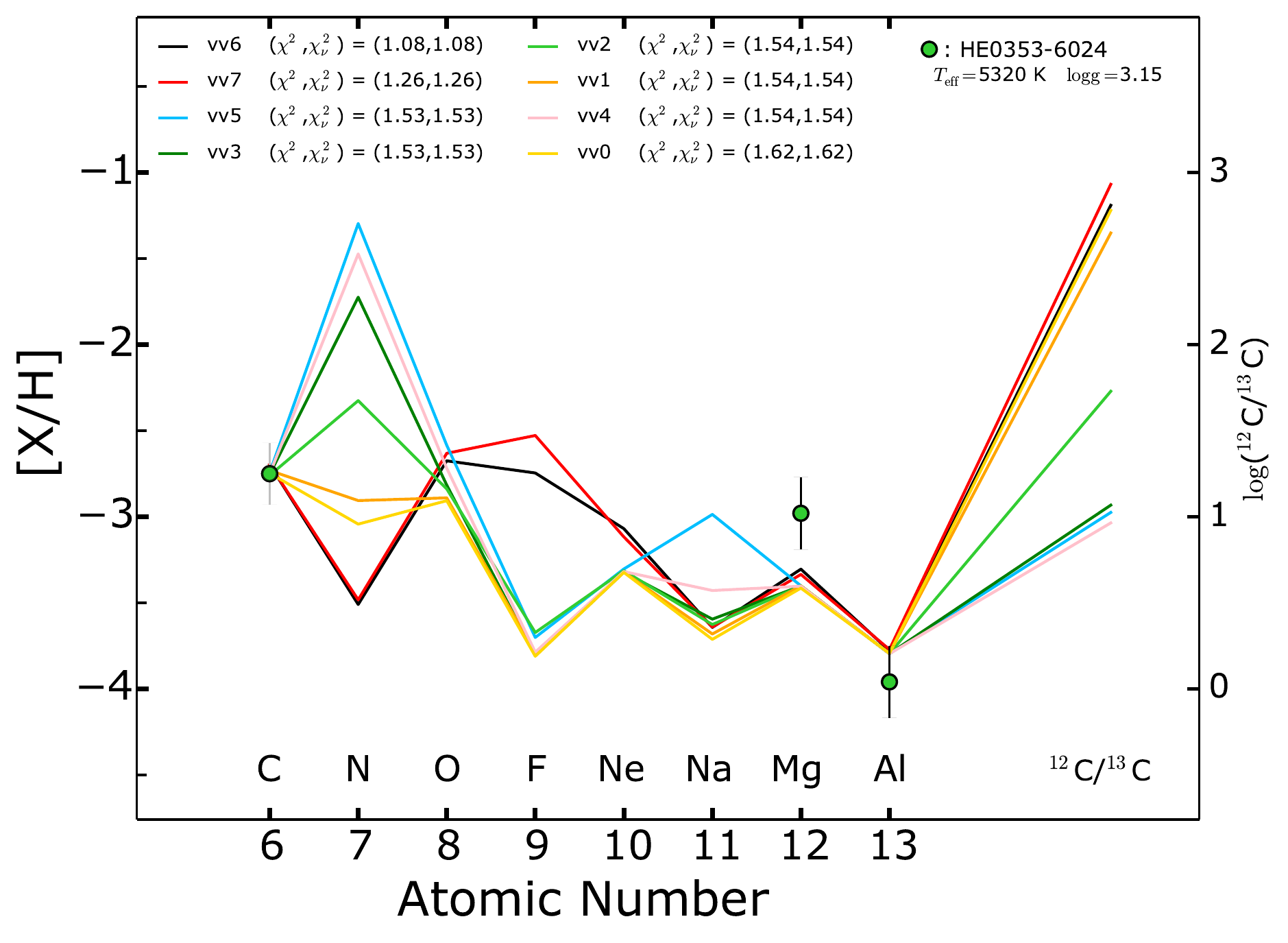}
   \end{minipage}
   \begin{minipage}[c]{.33\linewidth}
       \includegraphics[scale=0.3]{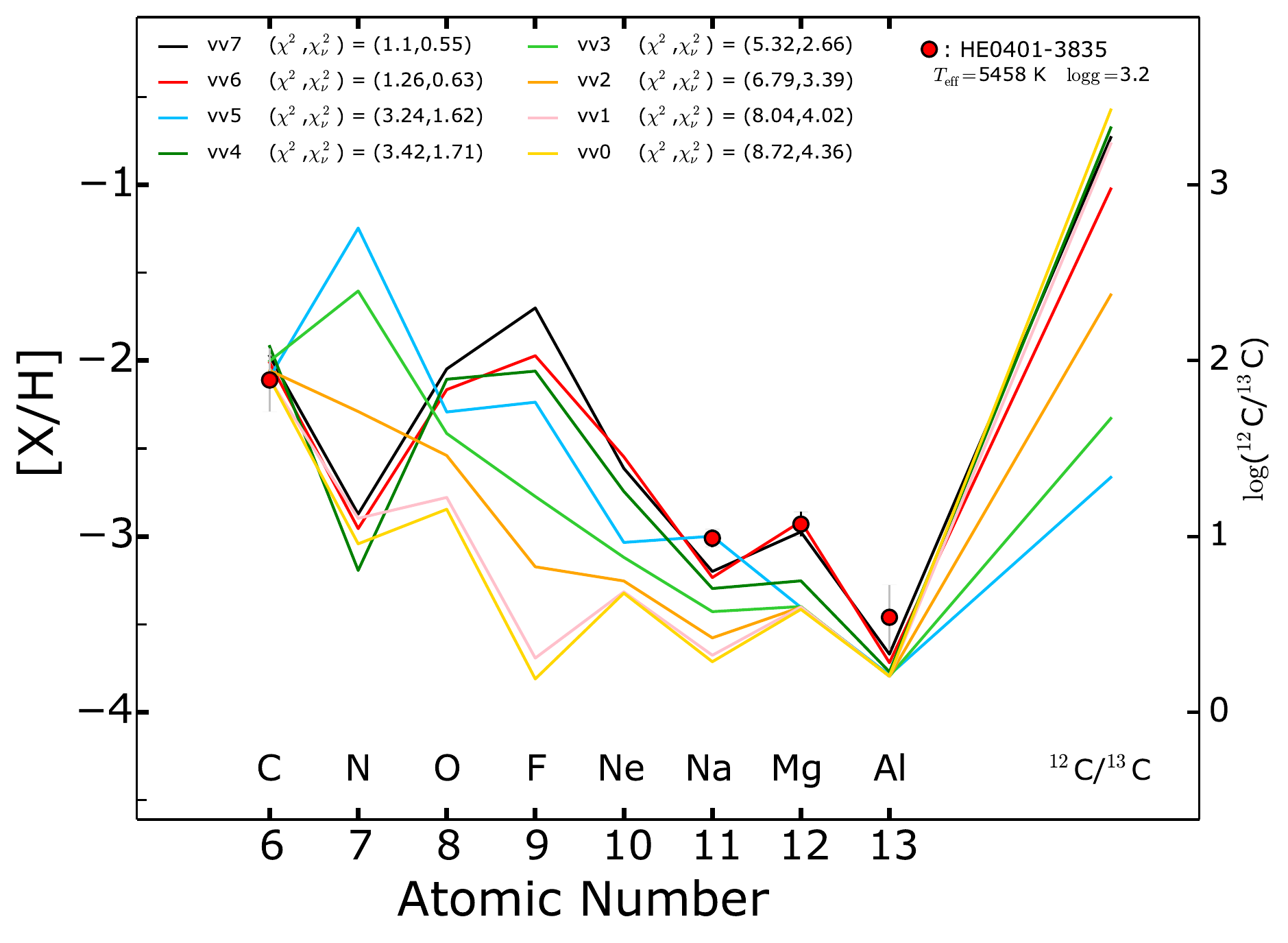}
   \end{minipage}
   \begin{minipage}[c]{.33\linewidth}
       \includegraphics[scale=0.3]{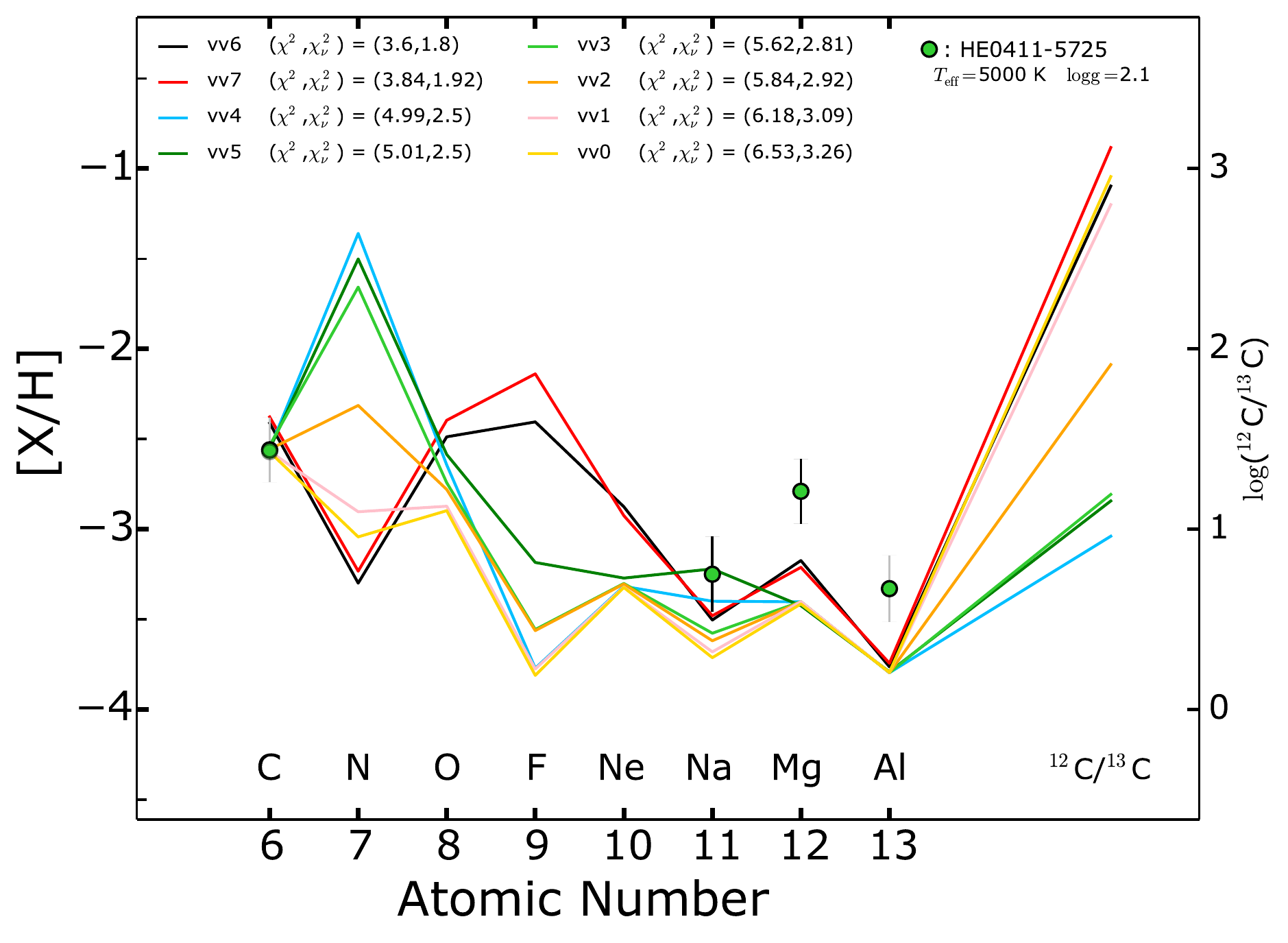}
   \end{minipage}
   \begin{minipage}[c]{.33\linewidth}
       \includegraphics[scale=0.3]{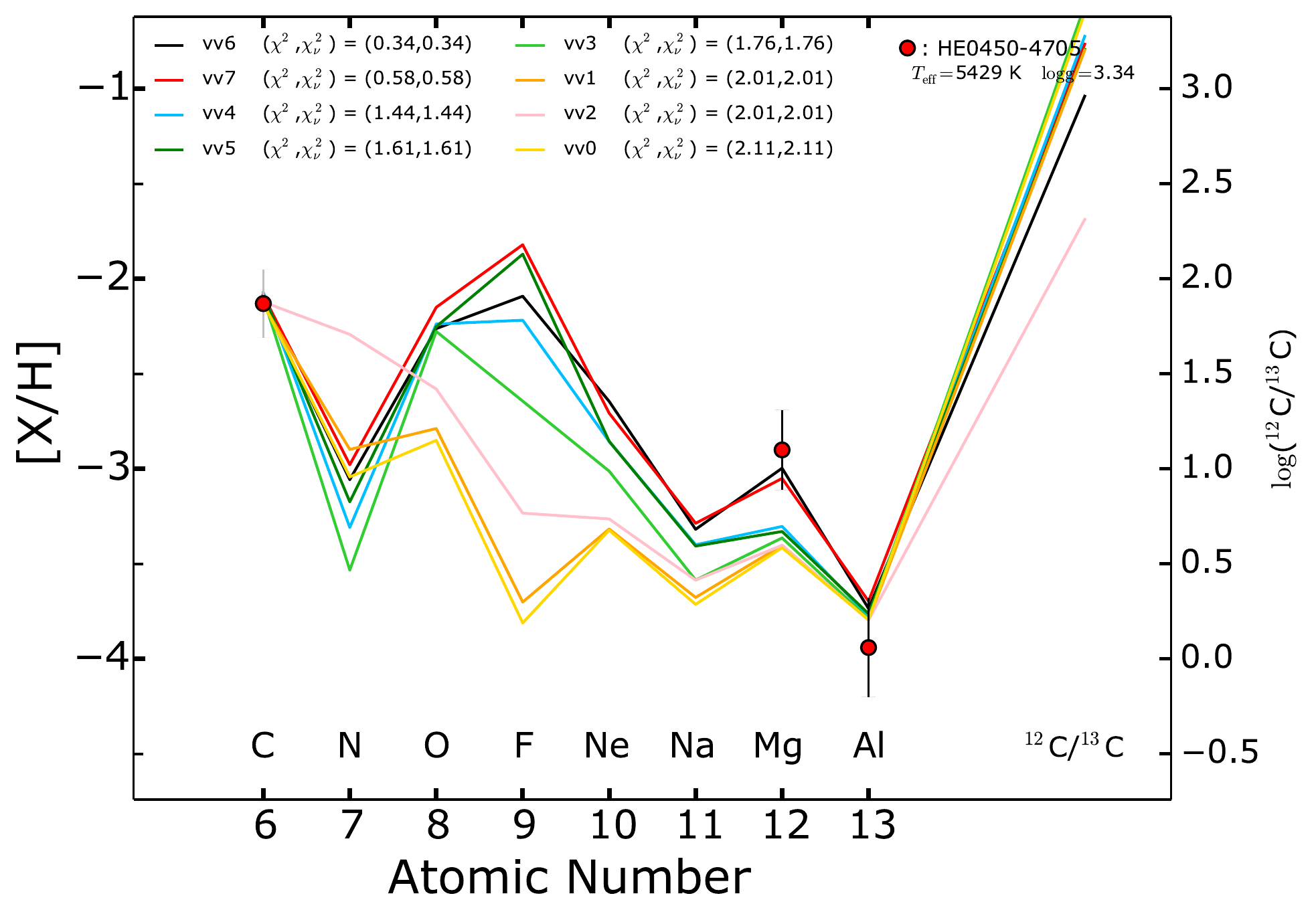}
   \end{minipage}
   \begin{minipage}[c]{.33\linewidth}
       \includegraphics[scale=0.3]{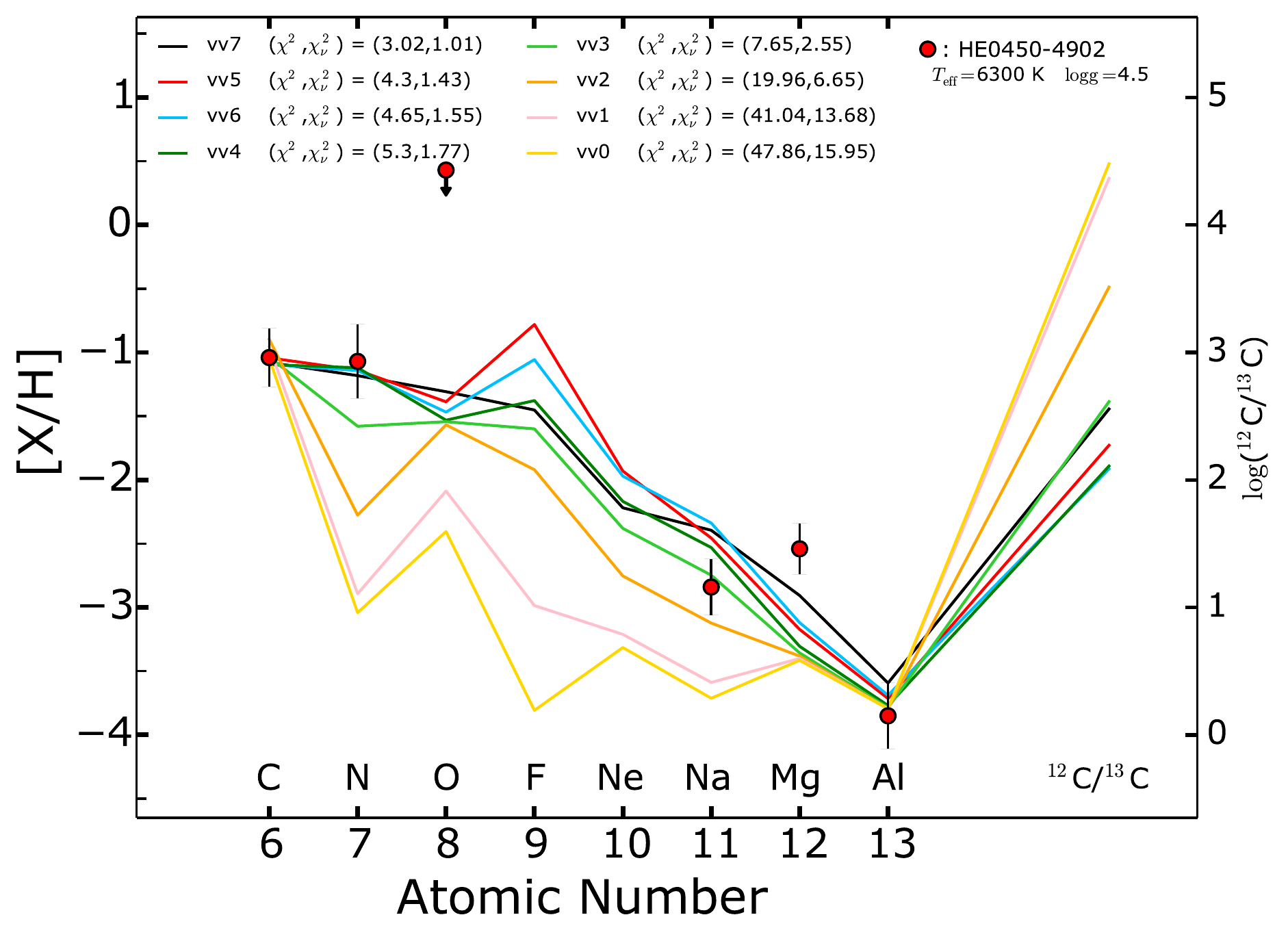}
   \end{minipage}
   \begin{minipage}[c]{.33\linewidth}
       \includegraphics[scale=0.3]{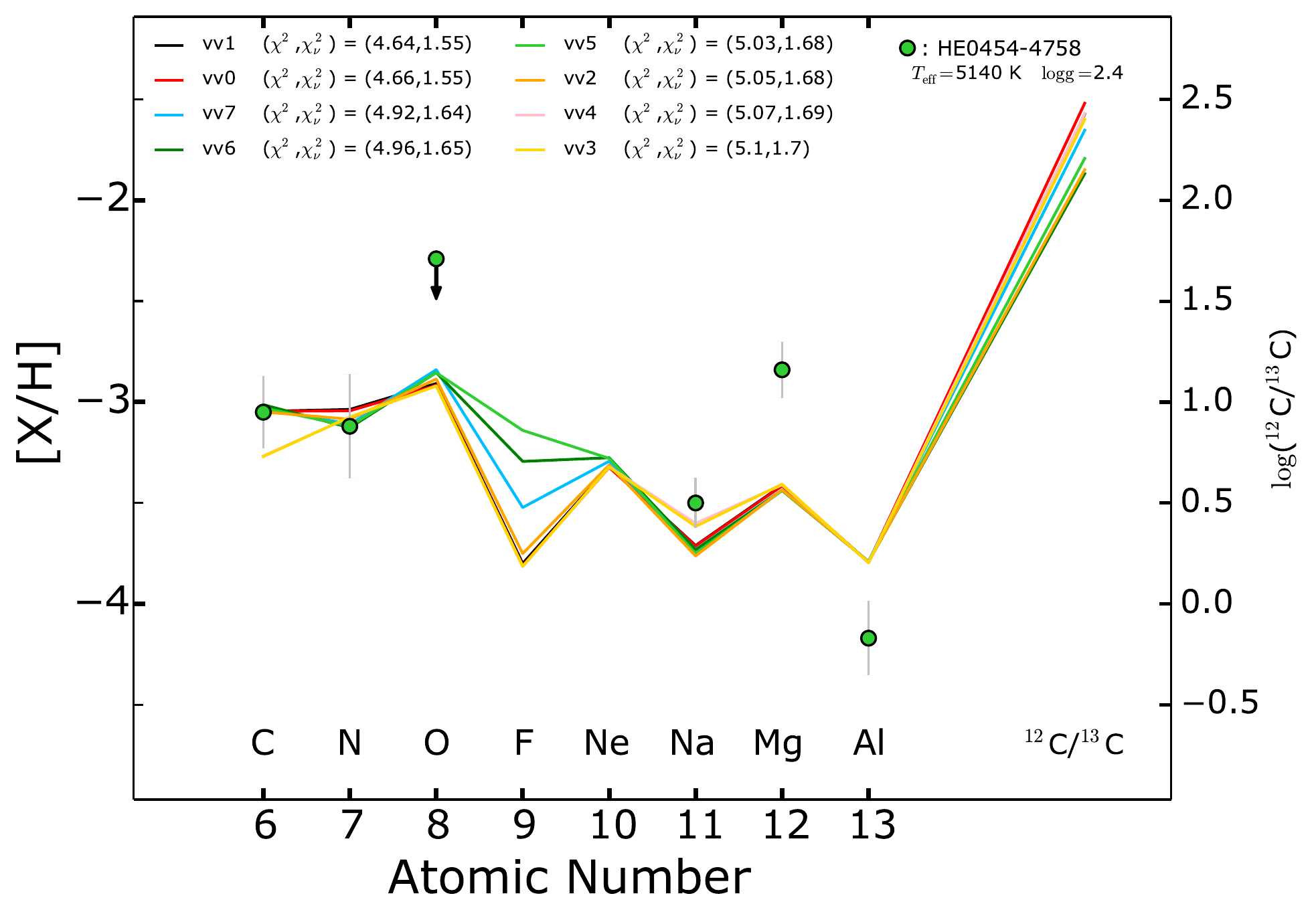}
   \end{minipage}
   \begin{minipage}[c]{.33\linewidth}
       \includegraphics[scale=0.3]{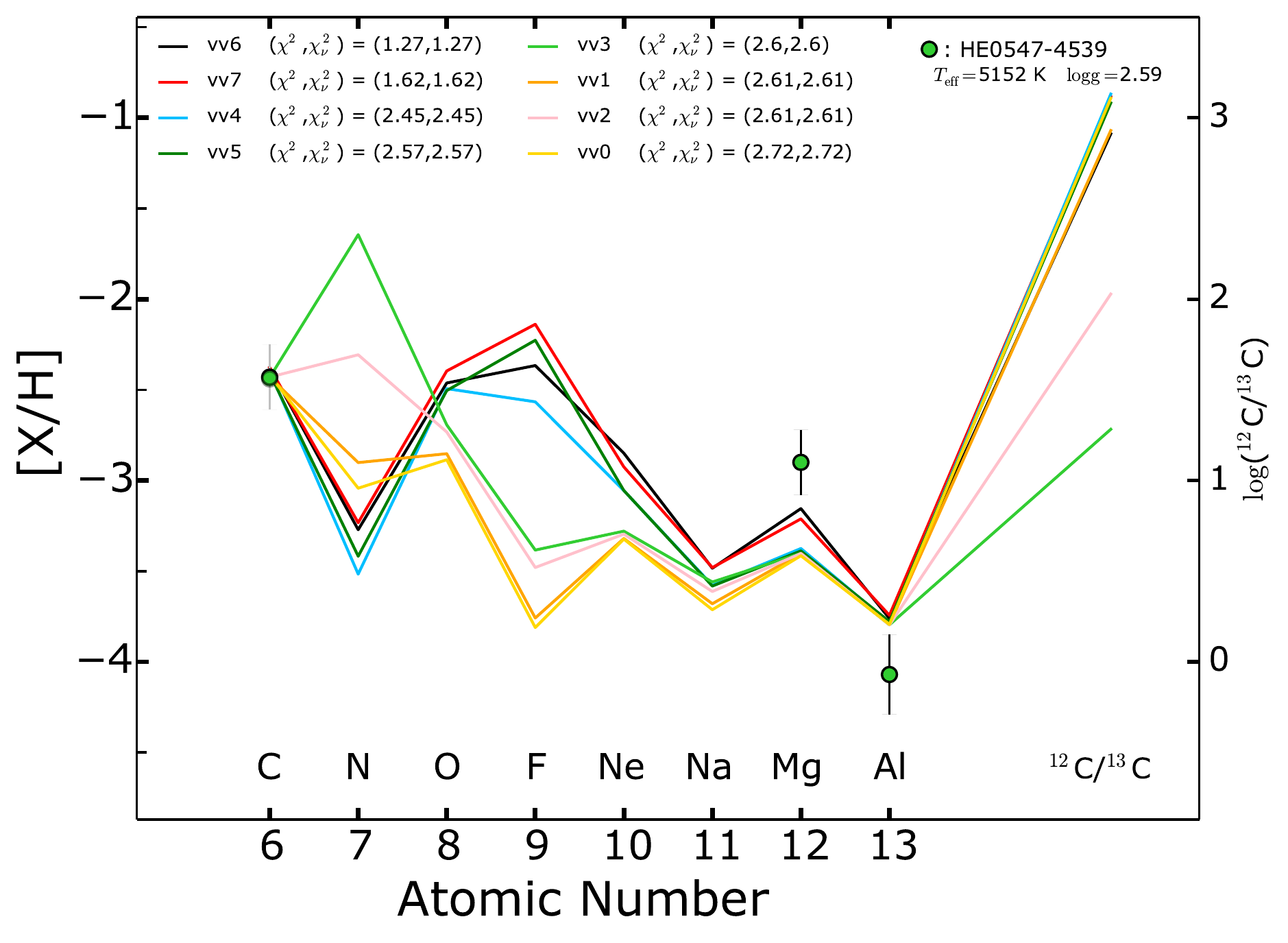}
   \end{minipage}
   \begin{minipage}[c]{.33\linewidth}
       \includegraphics[scale=0.3]{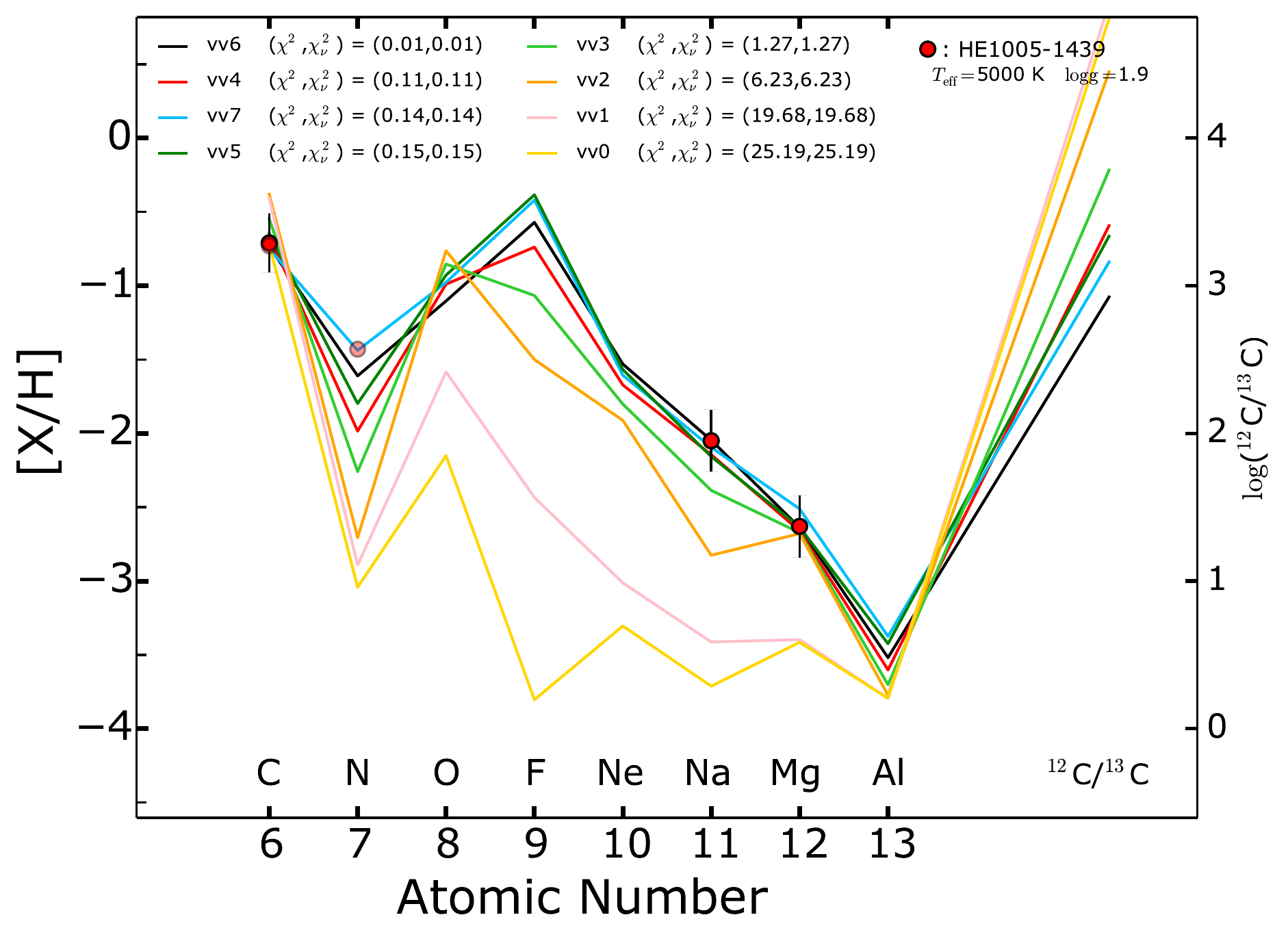}
   \end{minipage}
   \begin{minipage}[c]{.33\linewidth}
       \includegraphics[scale=0.3]{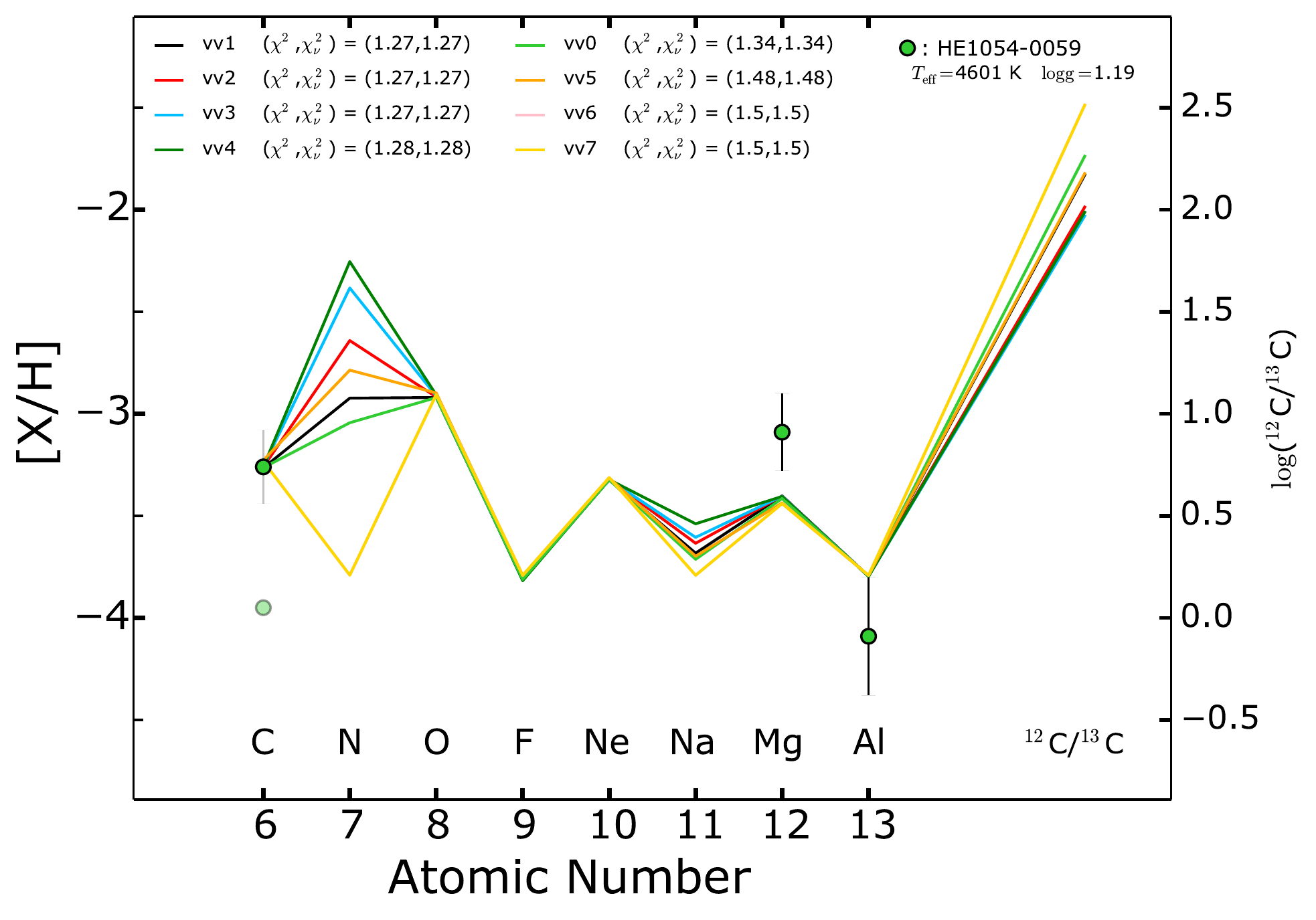}
   \end{minipage}
   \begin{minipage}[c]{.33\linewidth}
       \includegraphics[scale=0.3]{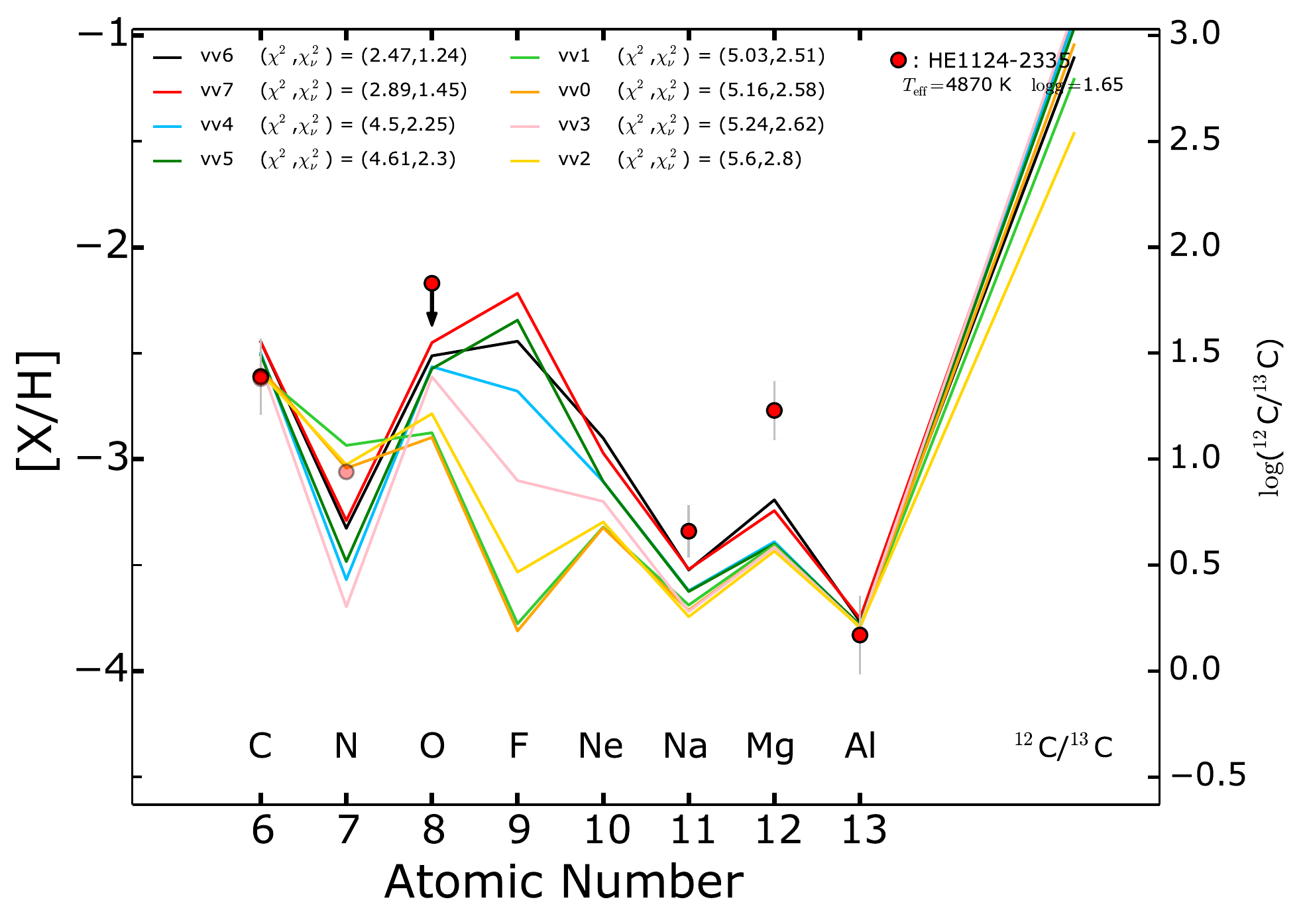}
   \end{minipage}
   \begin{minipage}[c]{.33\linewidth}
       \includegraphics[scale=0.3]{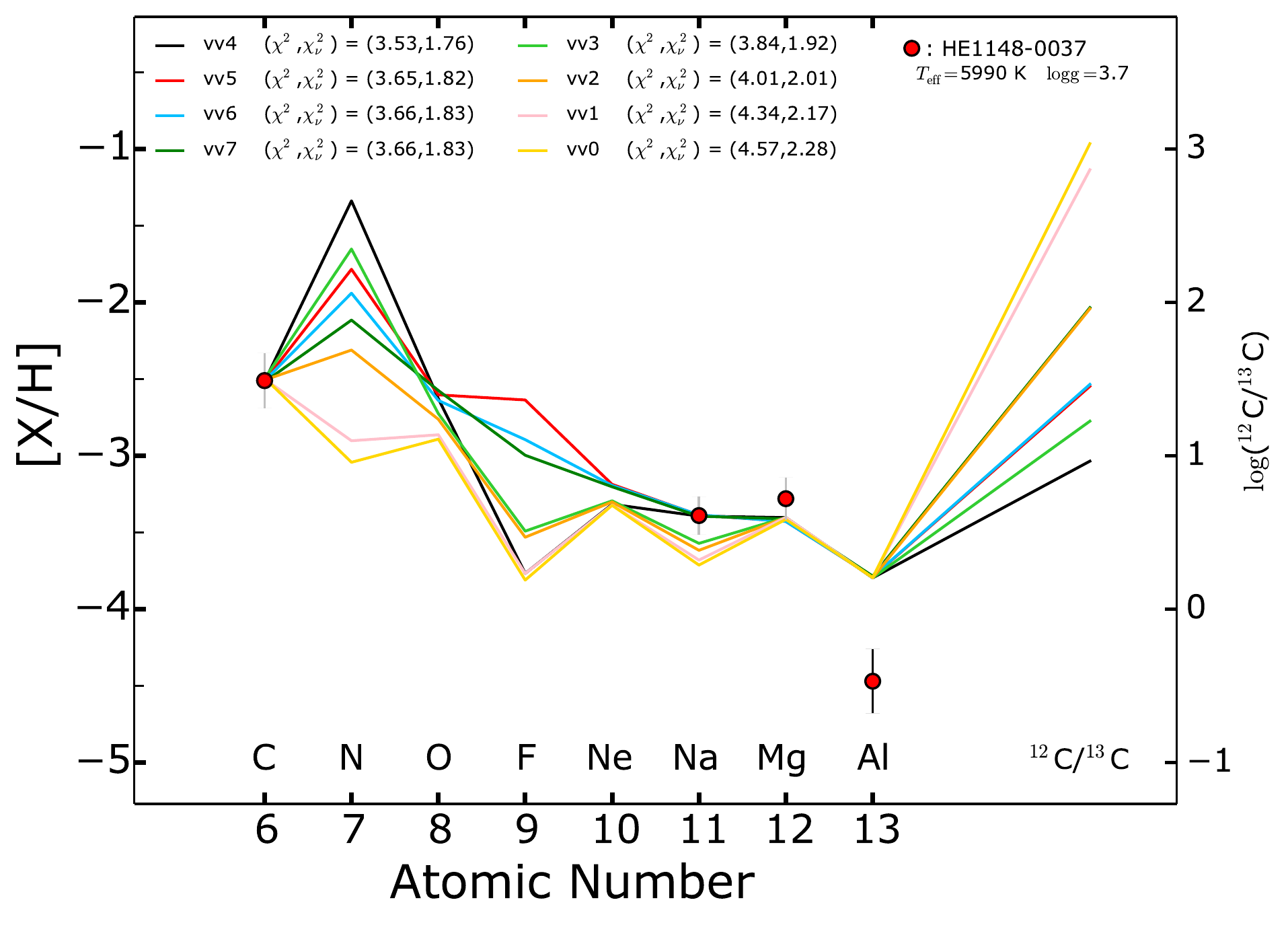}
   \end{minipage}
   \begin{minipage}[c]{.33\linewidth}
       \includegraphics[scale=0.3]{figs/XH_3best_161.pdf}
   \end{minipage}
   \begin{minipage}[c]{.33\linewidth}
       \includegraphics[scale=0.3]{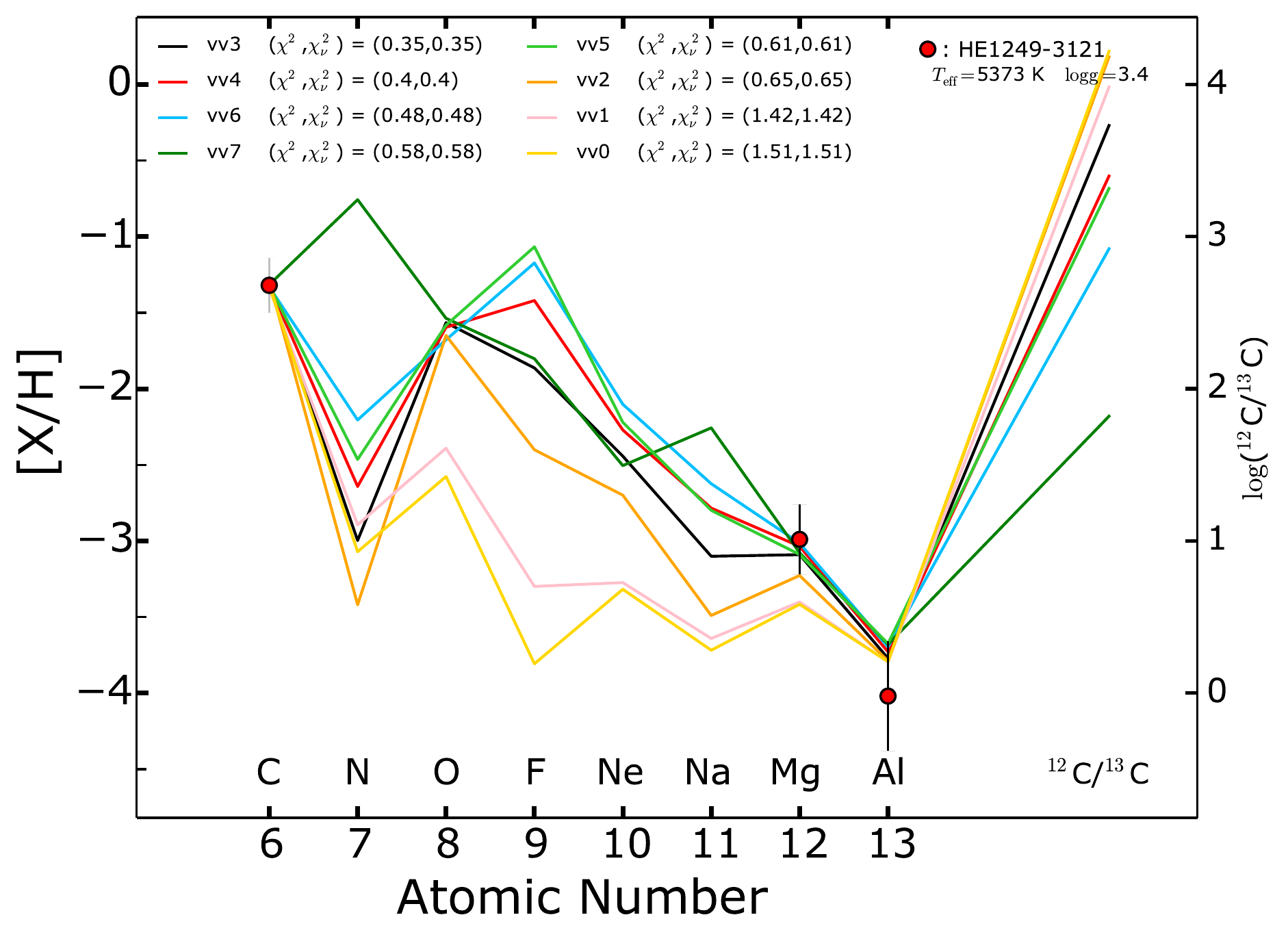}
   \end{minipage}
   \caption{Continued.}
\label{allfit1}
    \end{figure*}

   \begin{figure*}
     \ContinuedFloat 
   \centering
   \begin{minipage}[c]{.33\linewidth}
       \includegraphics[scale=0.3]{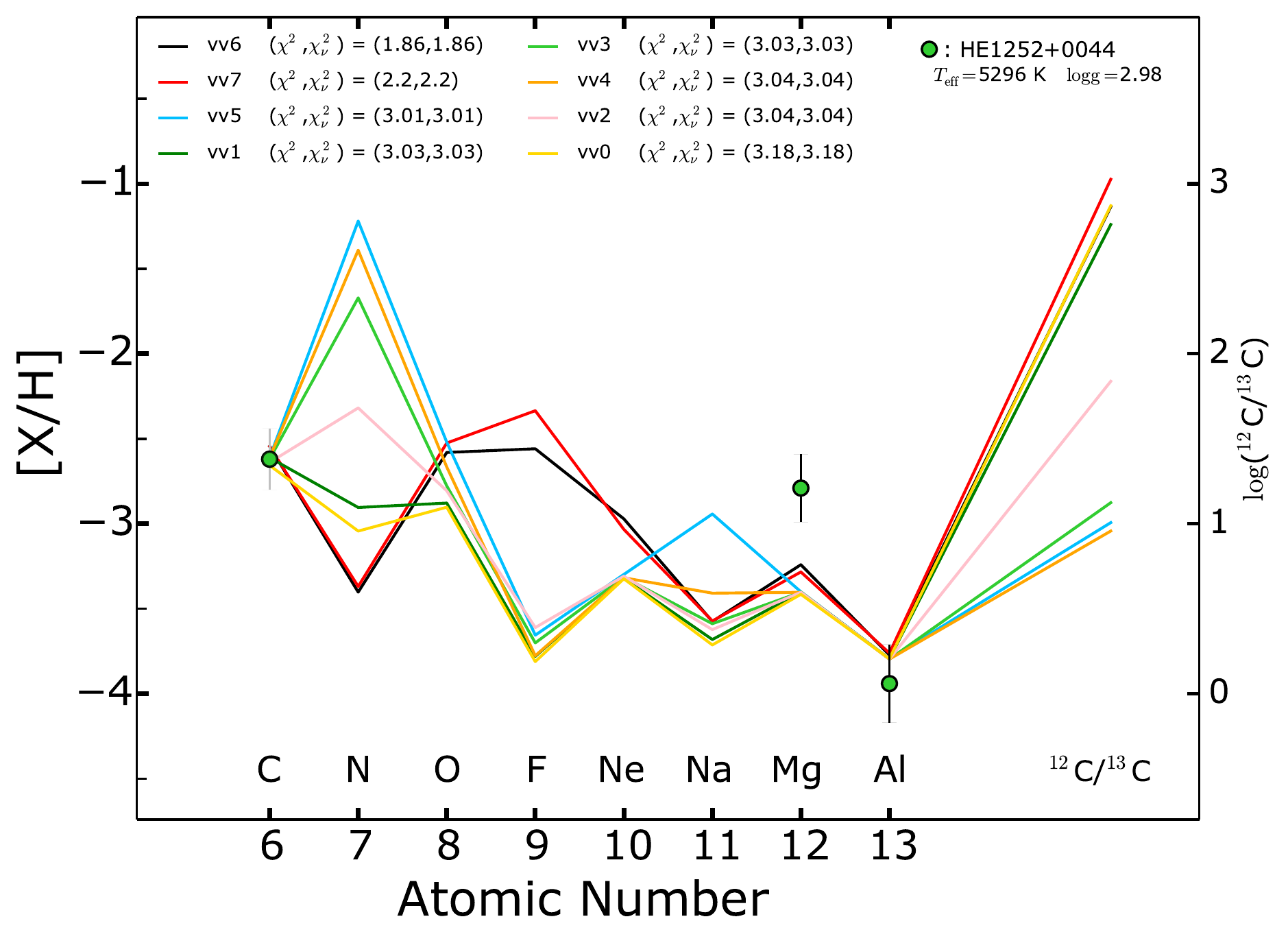}
   \end{minipage}
   \begin{minipage}[c]{.33\linewidth}
       \includegraphics[scale=0.3]{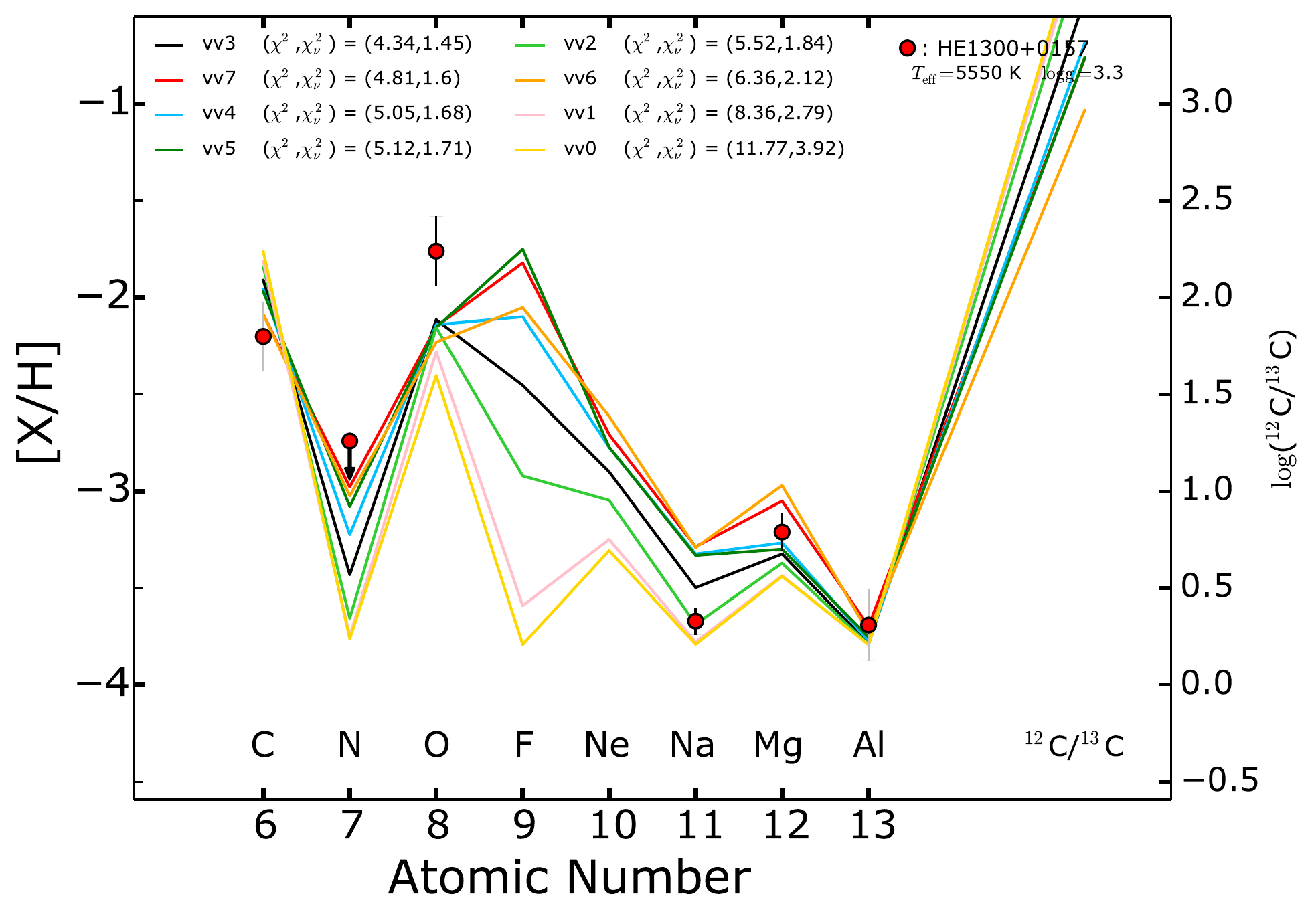}
   \end{minipage}
   \begin{minipage}[c]{.33\linewidth}
       \includegraphics[scale=0.3]{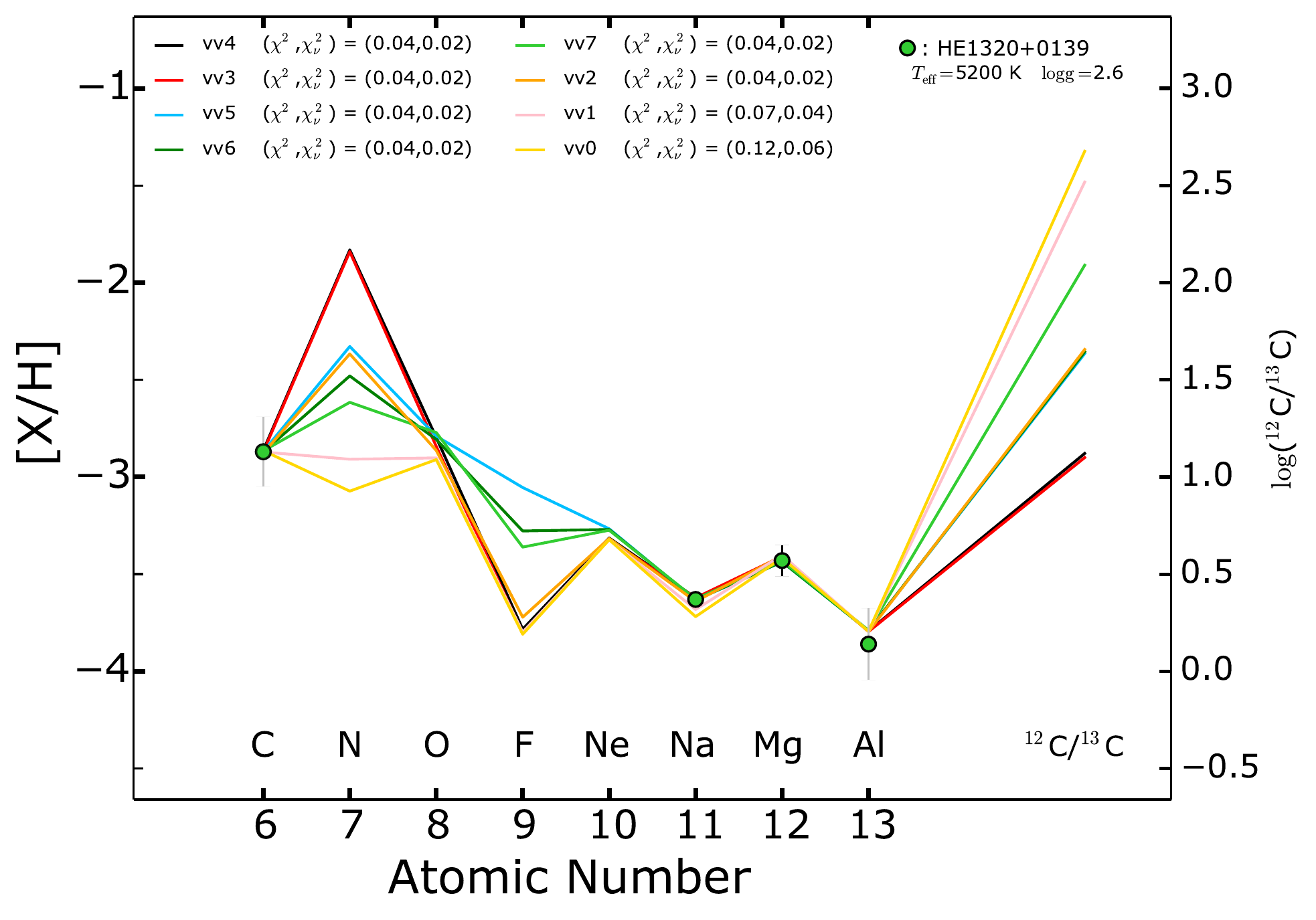}
   \end{minipage}
   \begin{minipage}[c]{.33\linewidth}
       \includegraphics[scale=0.3]{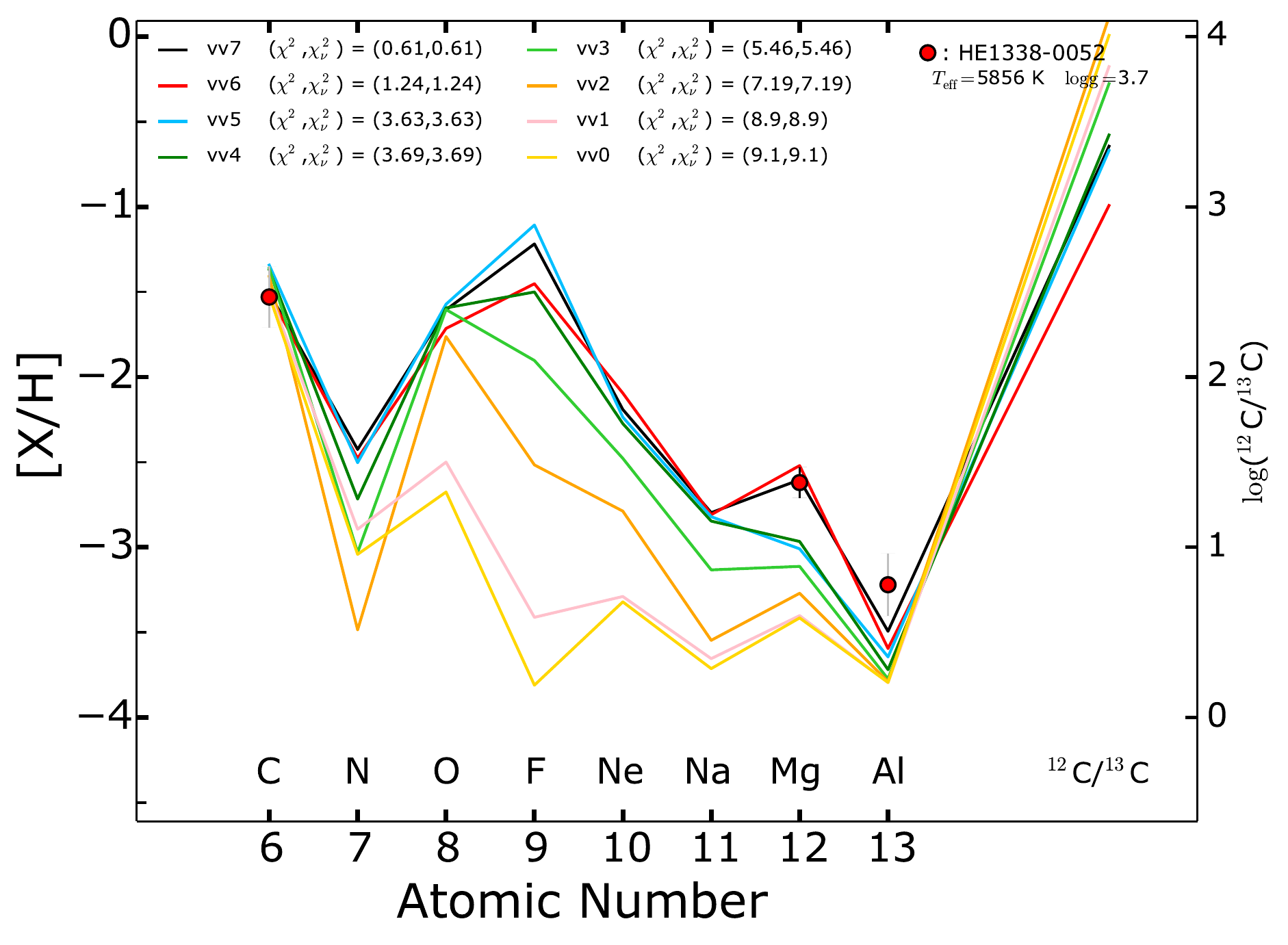}
   \end{minipage}
   \begin{minipage}[c]{.33\linewidth}
       \includegraphics[scale=0.3]{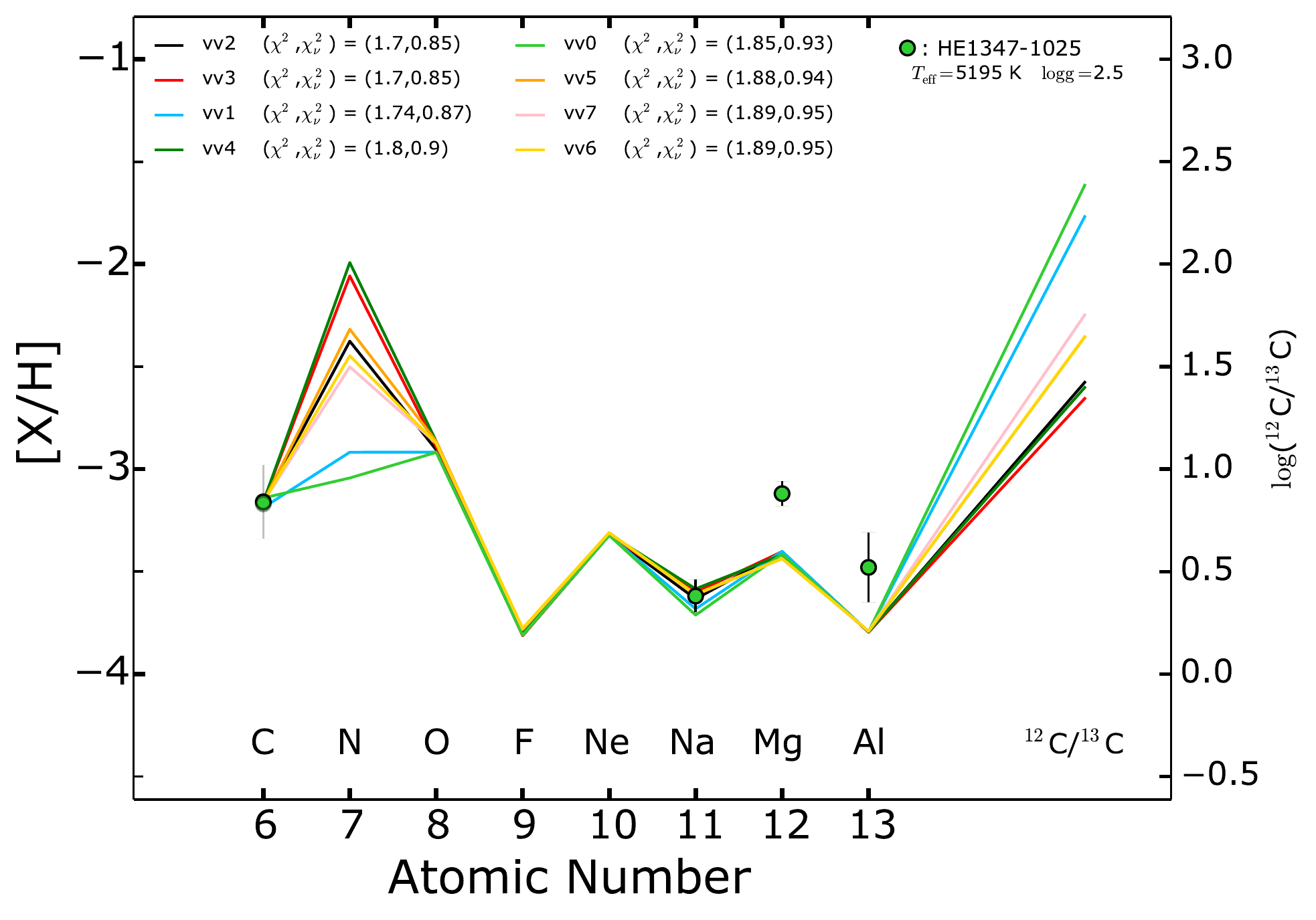}
   \end{minipage}
   \begin{minipage}[c]{.33\linewidth}
       \includegraphics[scale=0.3]{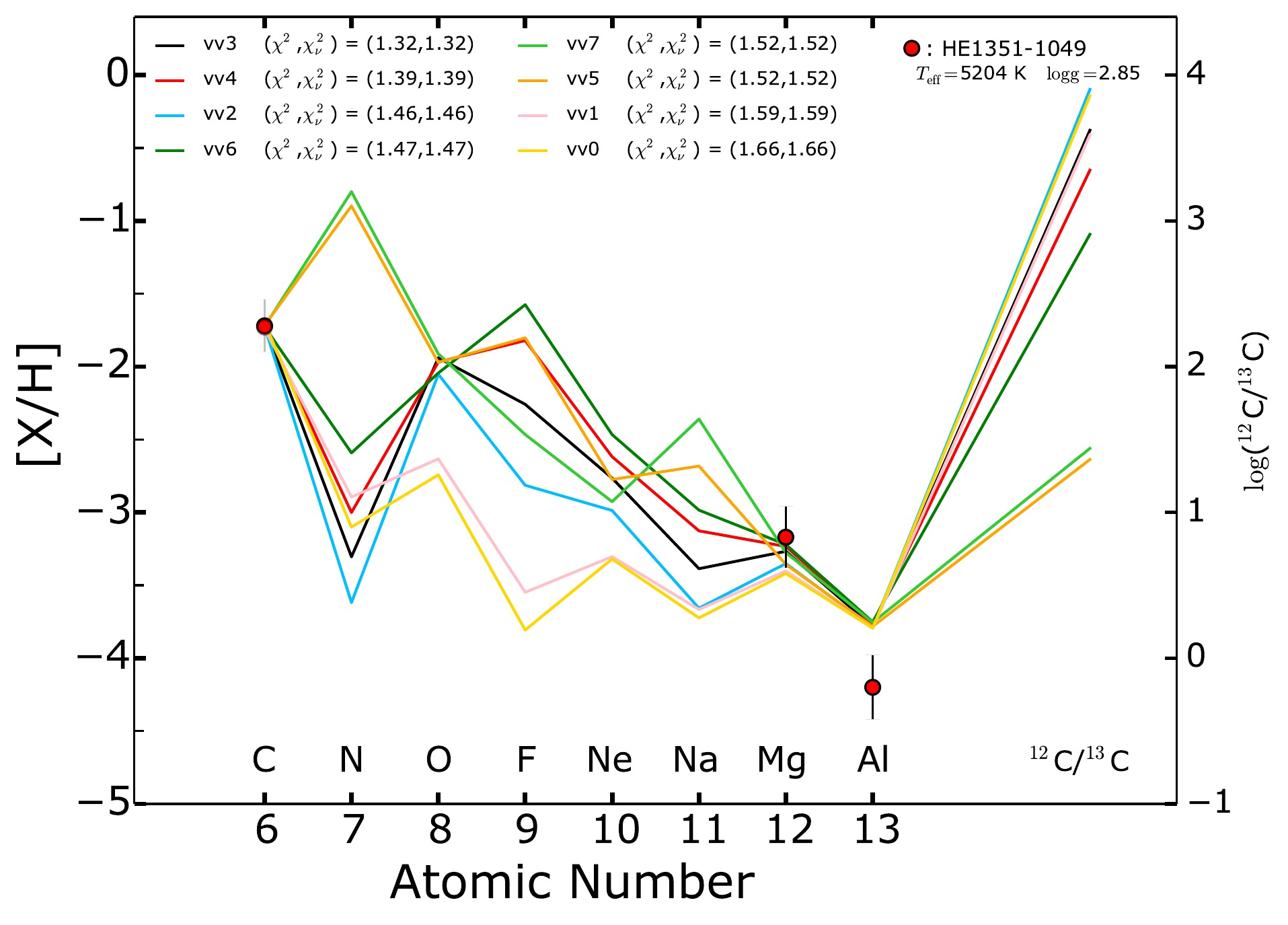}
   \end{minipage}
   \begin{minipage}[c]{.33\linewidth}
       \includegraphics[scale=0.3]{figs/XH_3best_180.pdf}
   \end{minipage}
   \begin{minipage}[c]{.33\linewidth}
       \includegraphics[scale=0.3]{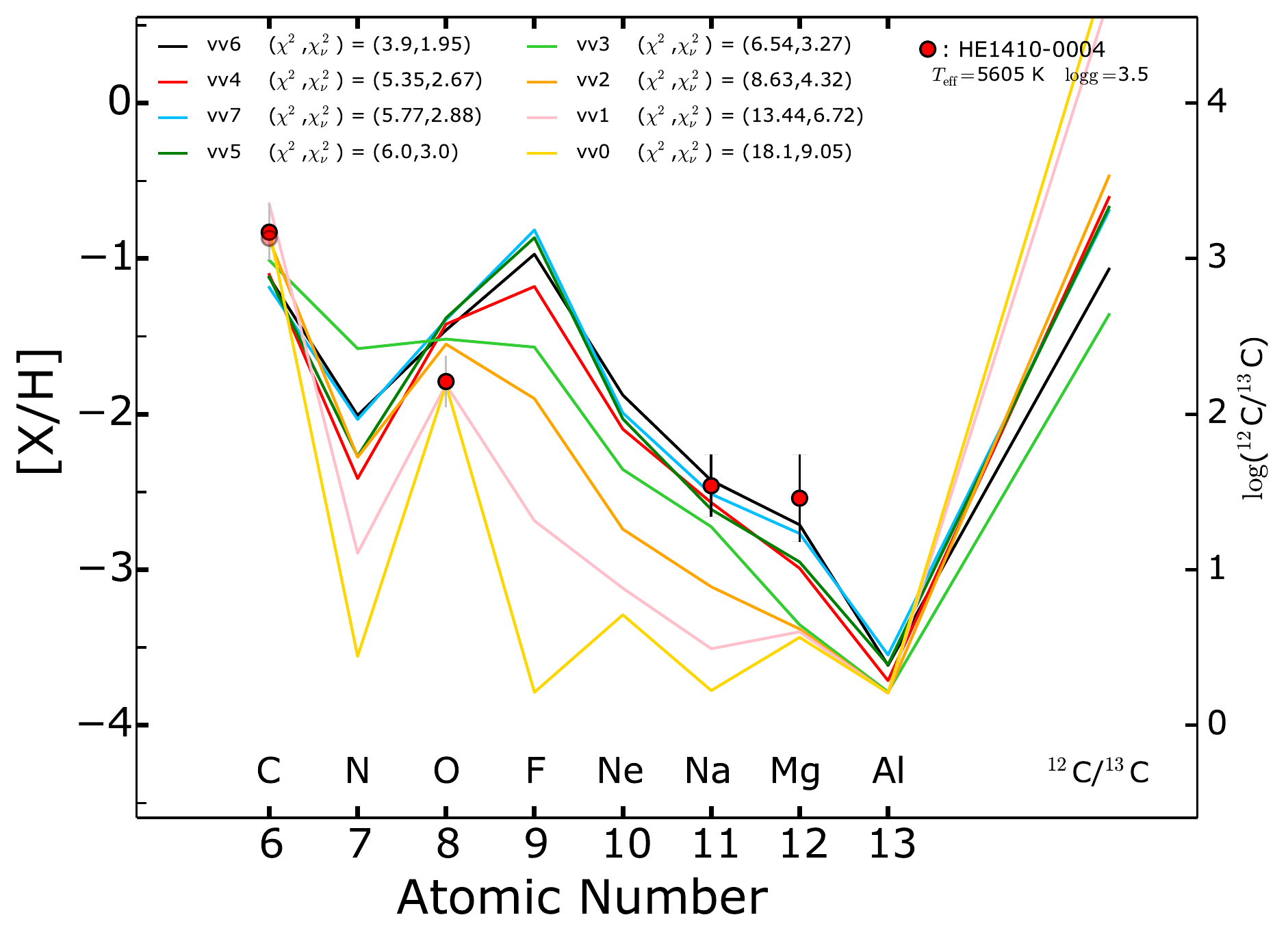}
   \end{minipage}
   \begin{minipage}[c]{.33\linewidth}
       \includegraphics[scale=0.3]{figs/XH_3best_184.pdf}
   \end{minipage}
   \begin{minipage}[c]{.33\linewidth}
       \includegraphics[scale=0.3]{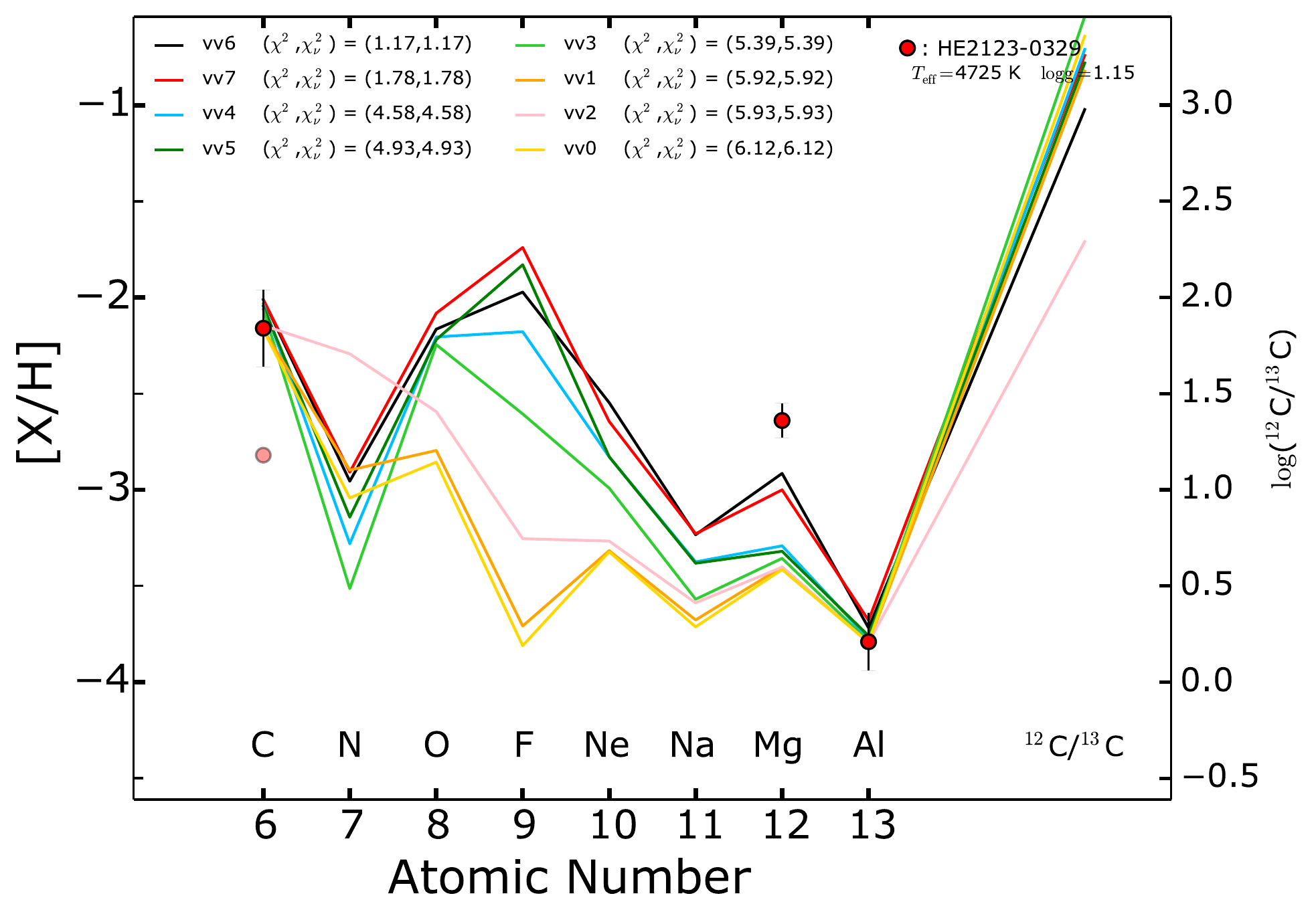}
   \end{minipage}
   \begin{minipage}[c]{.33\linewidth}
       \includegraphics[scale=0.3]{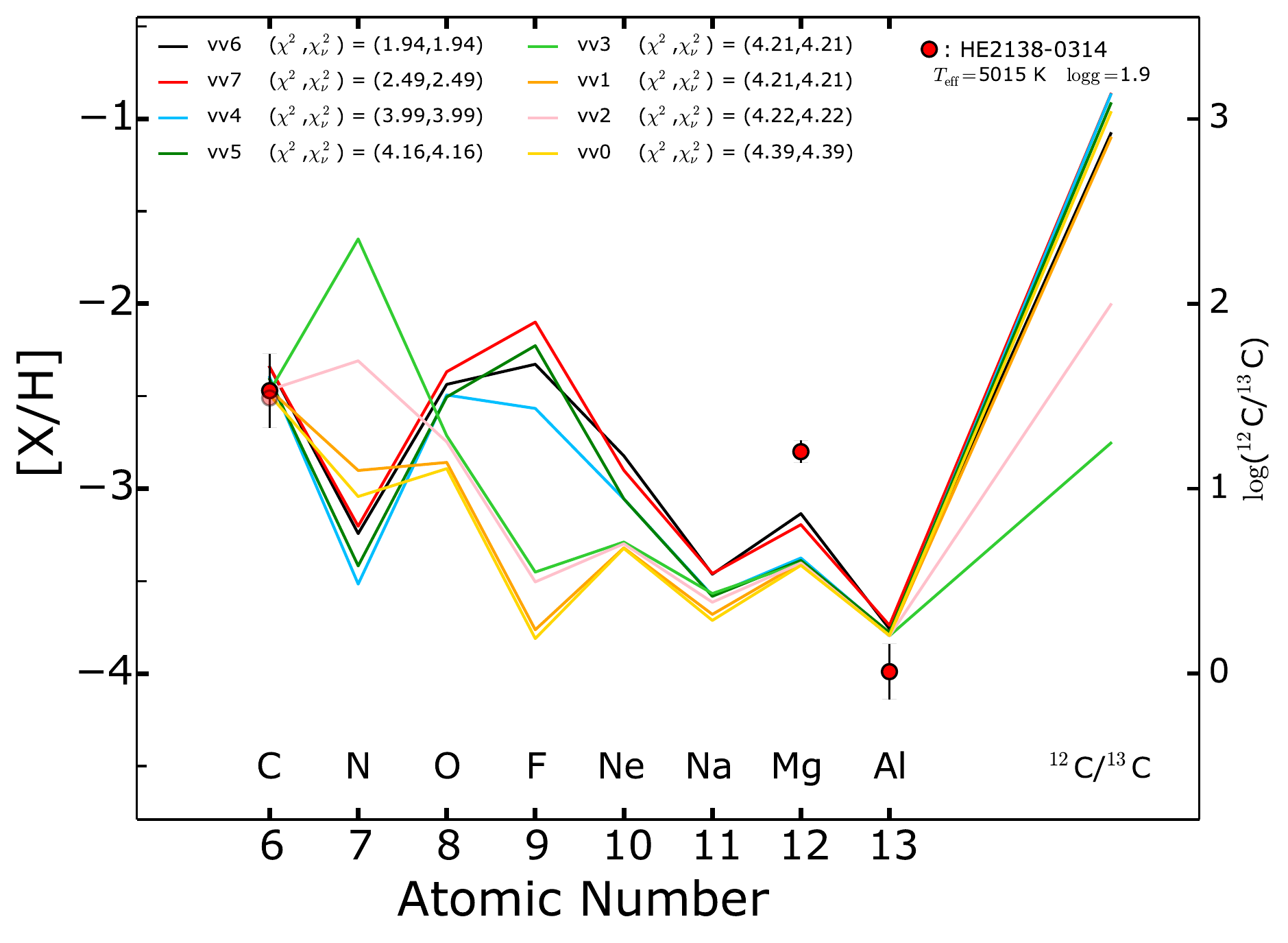}
   \end{minipage}
   \begin{minipage}[c]{.33\linewidth}
       \includegraphics[scale=0.3]{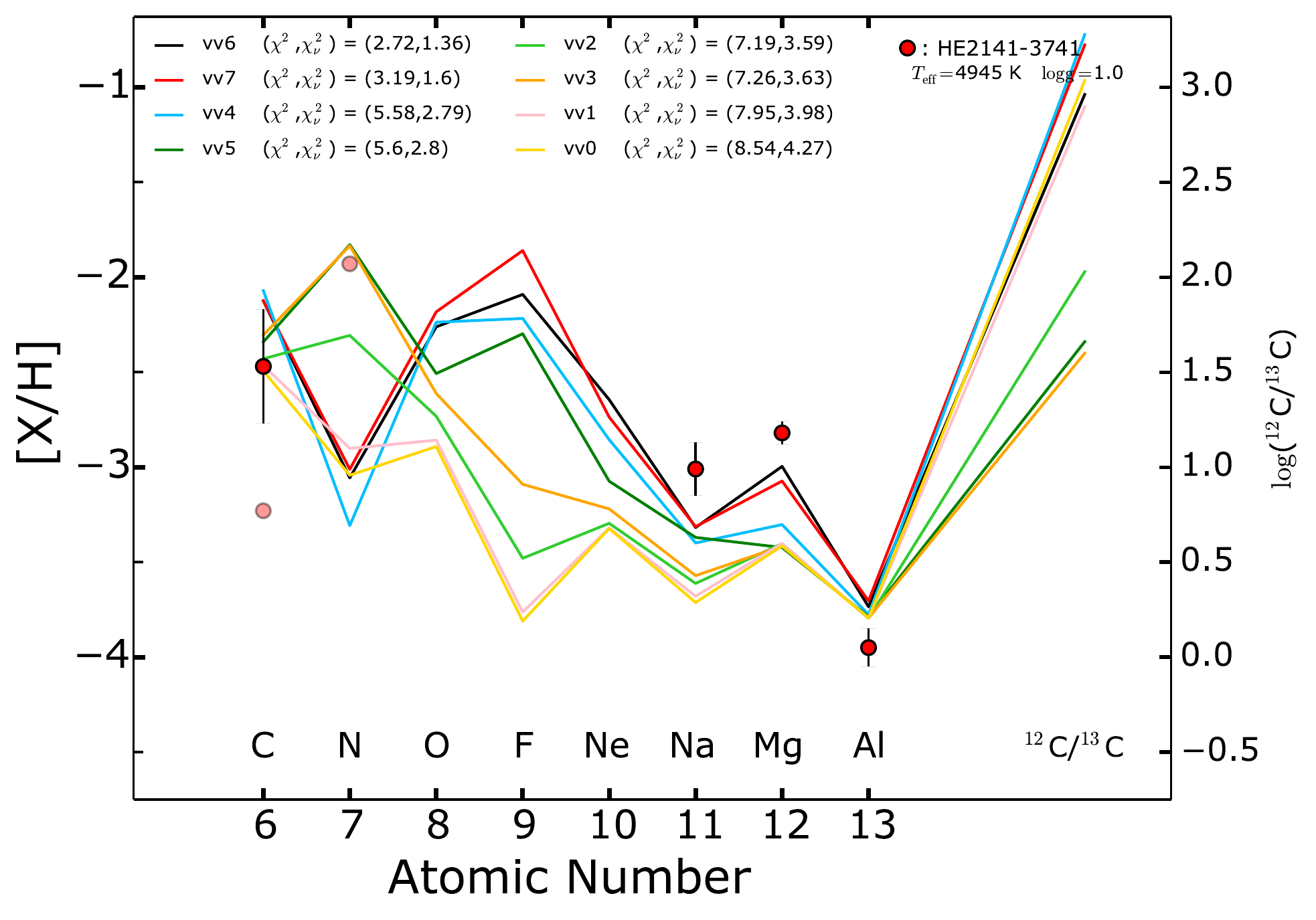}
   \end{minipage}
   \begin{minipage}[c]{.33\linewidth}
       \includegraphics[scale=0.3]{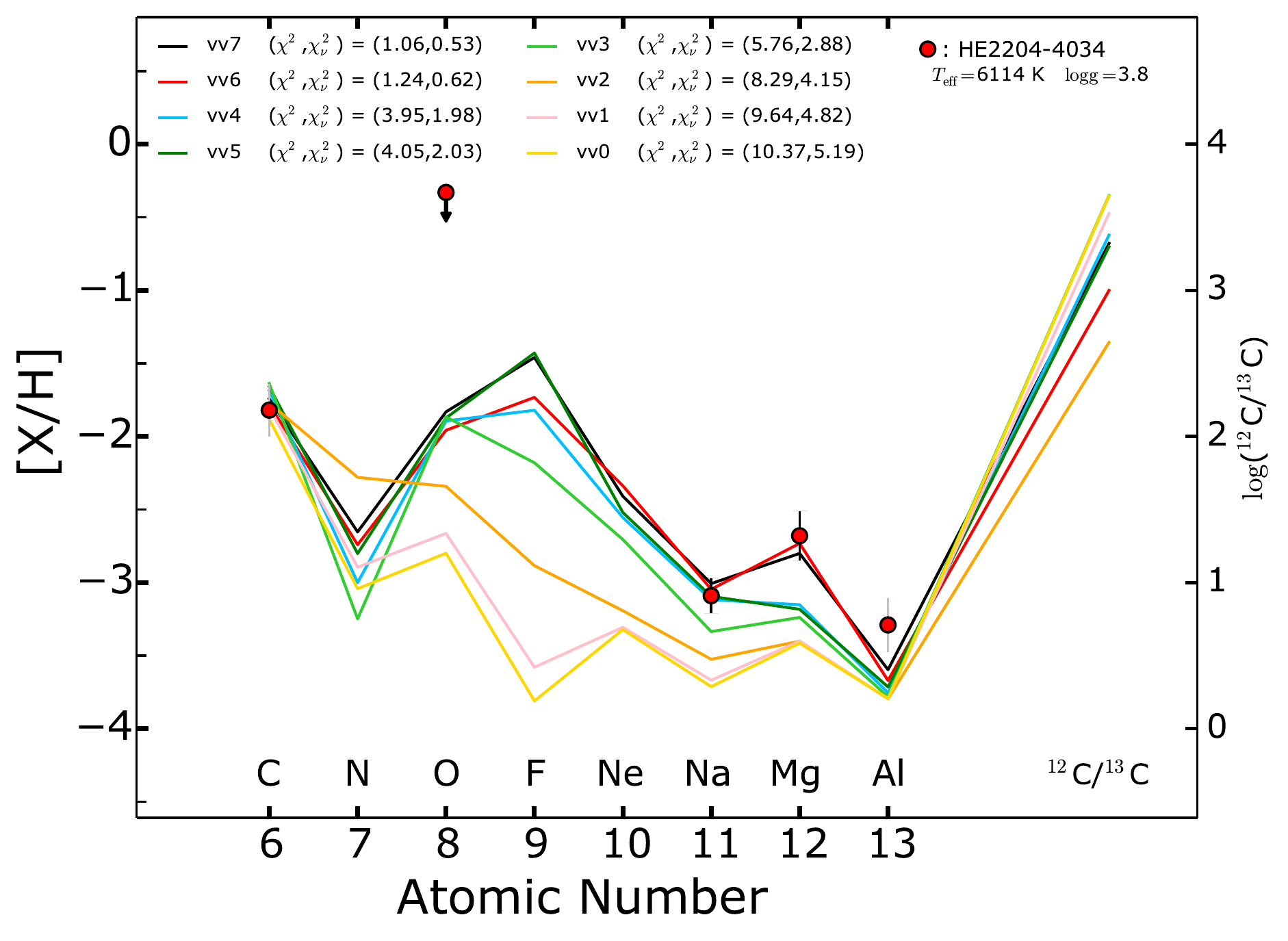}
   \end{minipage}
   \begin{minipage}[c]{.33\linewidth}
       \includegraphics[scale=0.3]{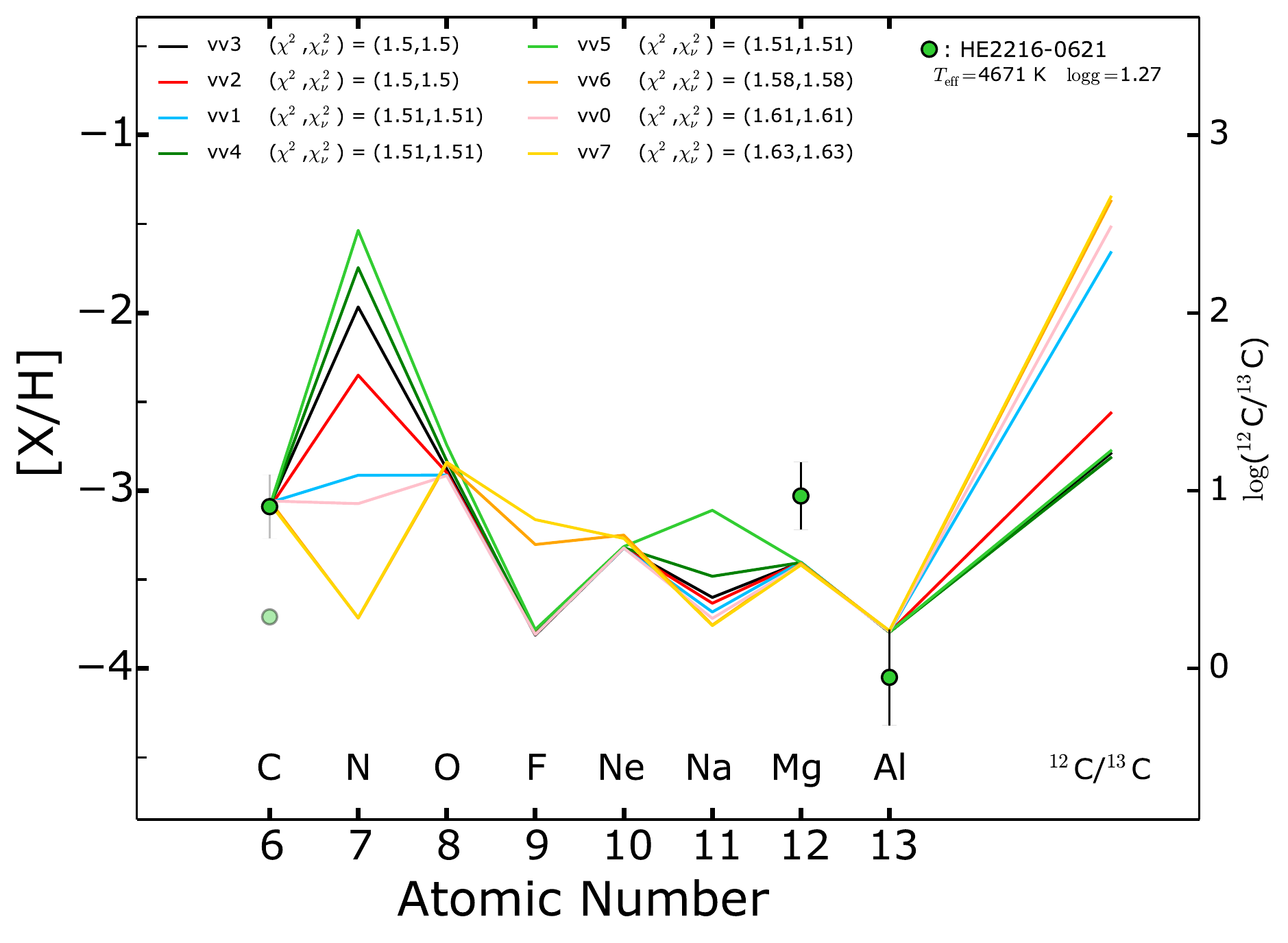}
   \end{minipage}
   \begin{minipage}[c]{.33\linewidth}
       \includegraphics[scale=0.3]{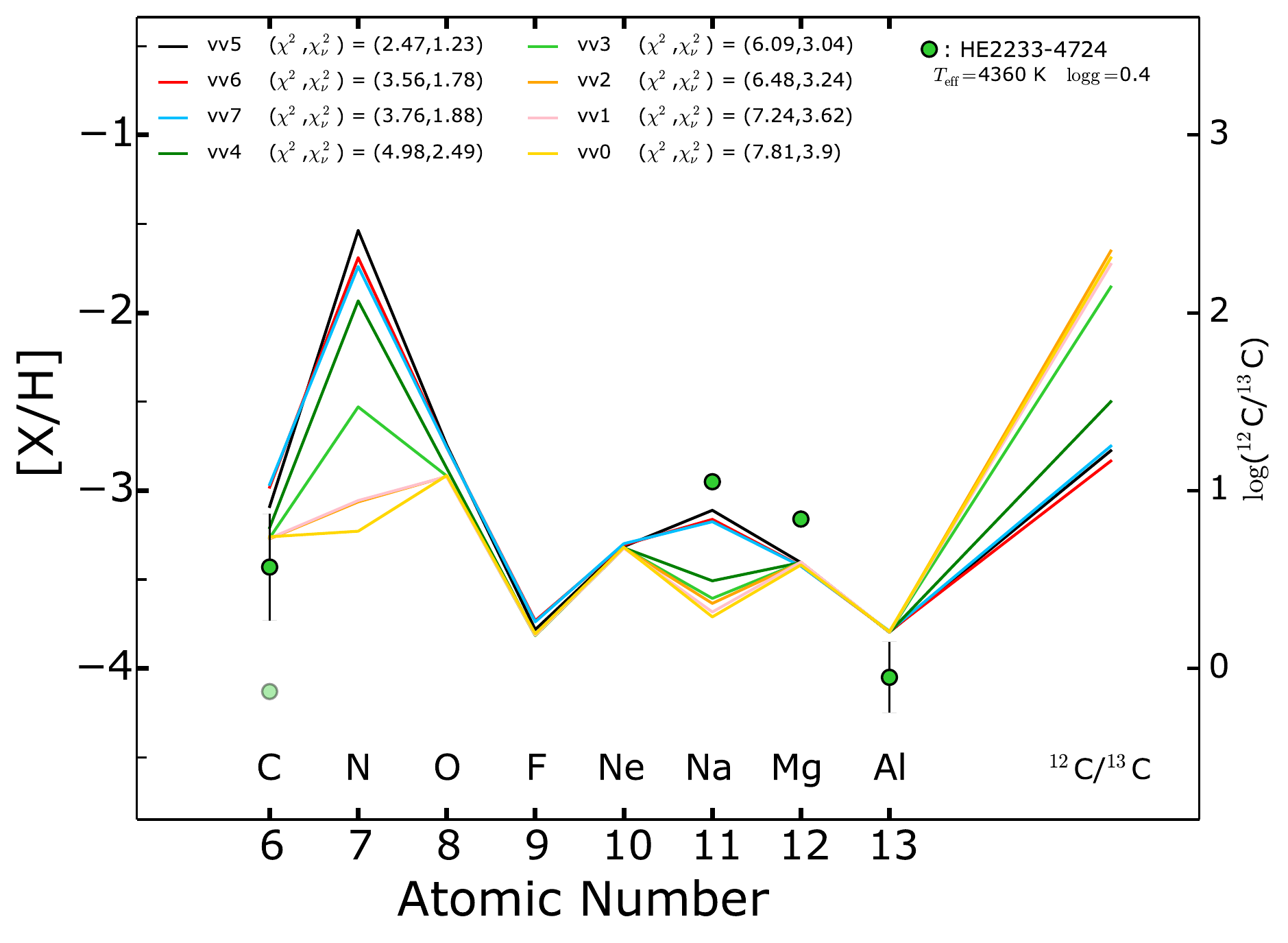}
   \end{minipage}
   \caption{Continued.}
\label{allfit1}
    \end{figure*}

   \begin{figure*}
      \ContinuedFloat
   \centering
   \begin{minipage}[c]{.33\linewidth}
       \includegraphics[scale=0.3]{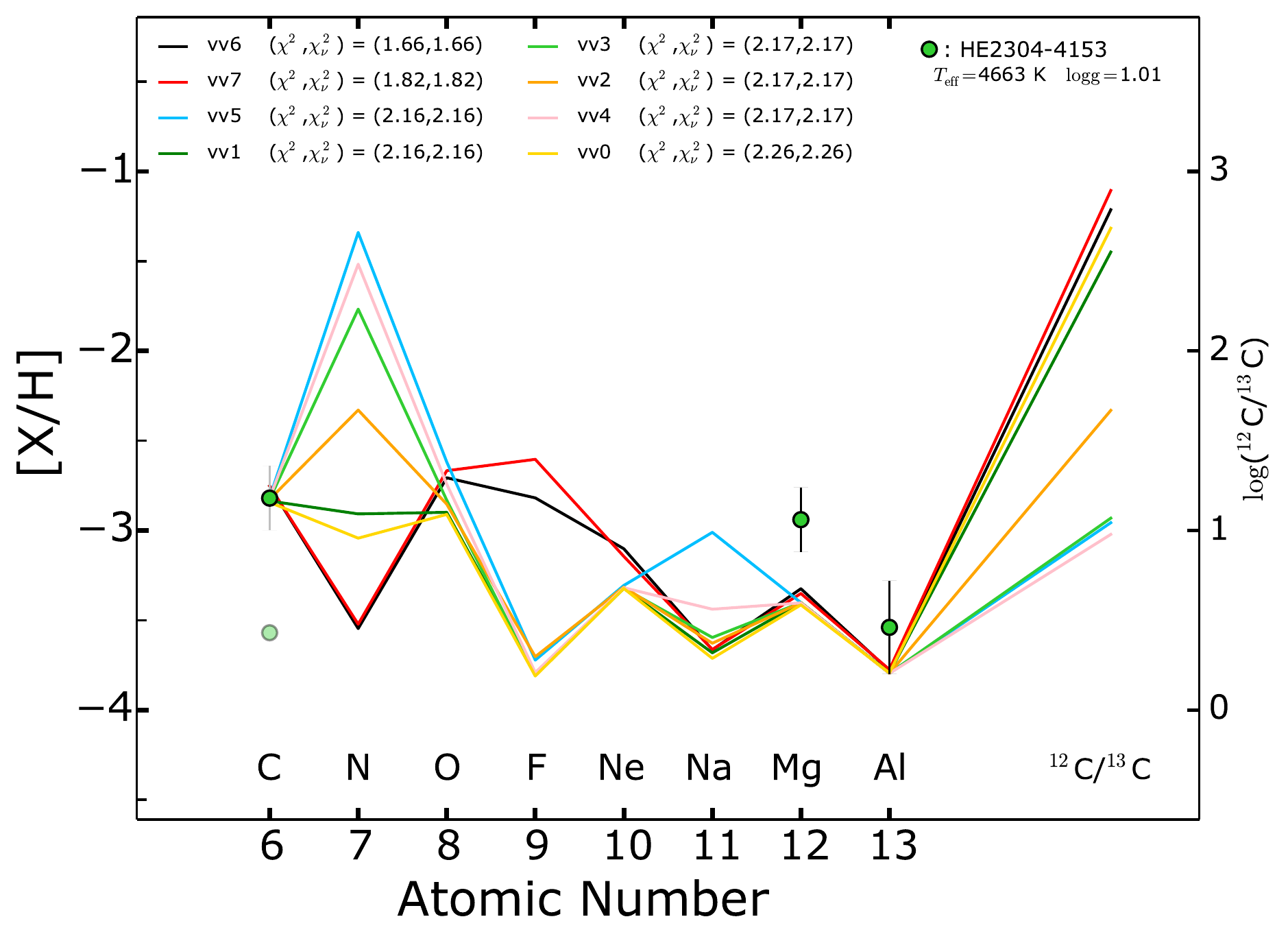}
   \end{minipage}
   \begin{minipage}[c]{.33\linewidth}
       \includegraphics[scale=0.3]{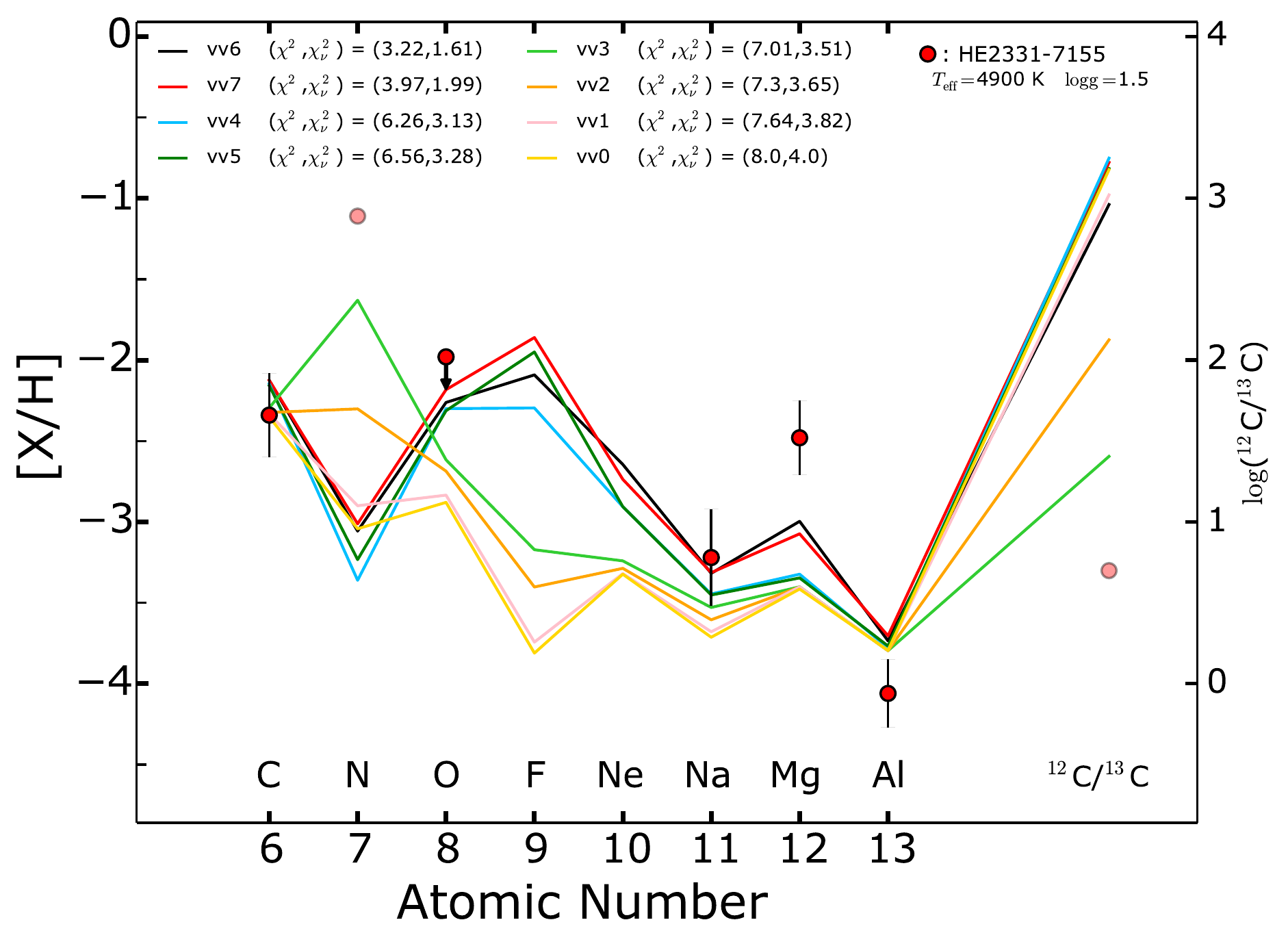}
   \end{minipage}
   \begin{minipage}[c]{.33\linewidth}
       \includegraphics[scale=0.3]{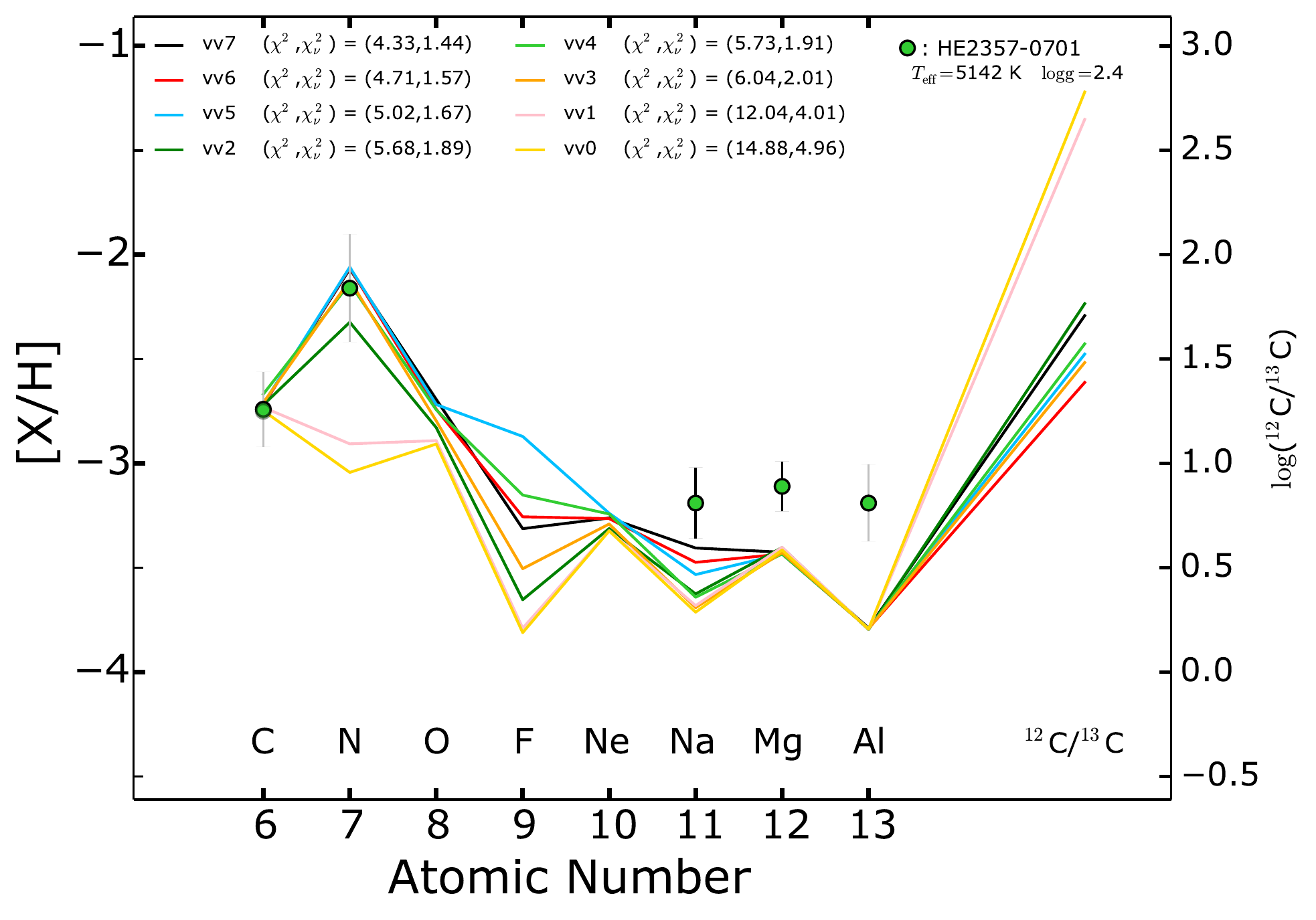}
   \end{minipage}
   \begin{minipage}[c]{.33\linewidth}
       \includegraphics[scale=0.3]{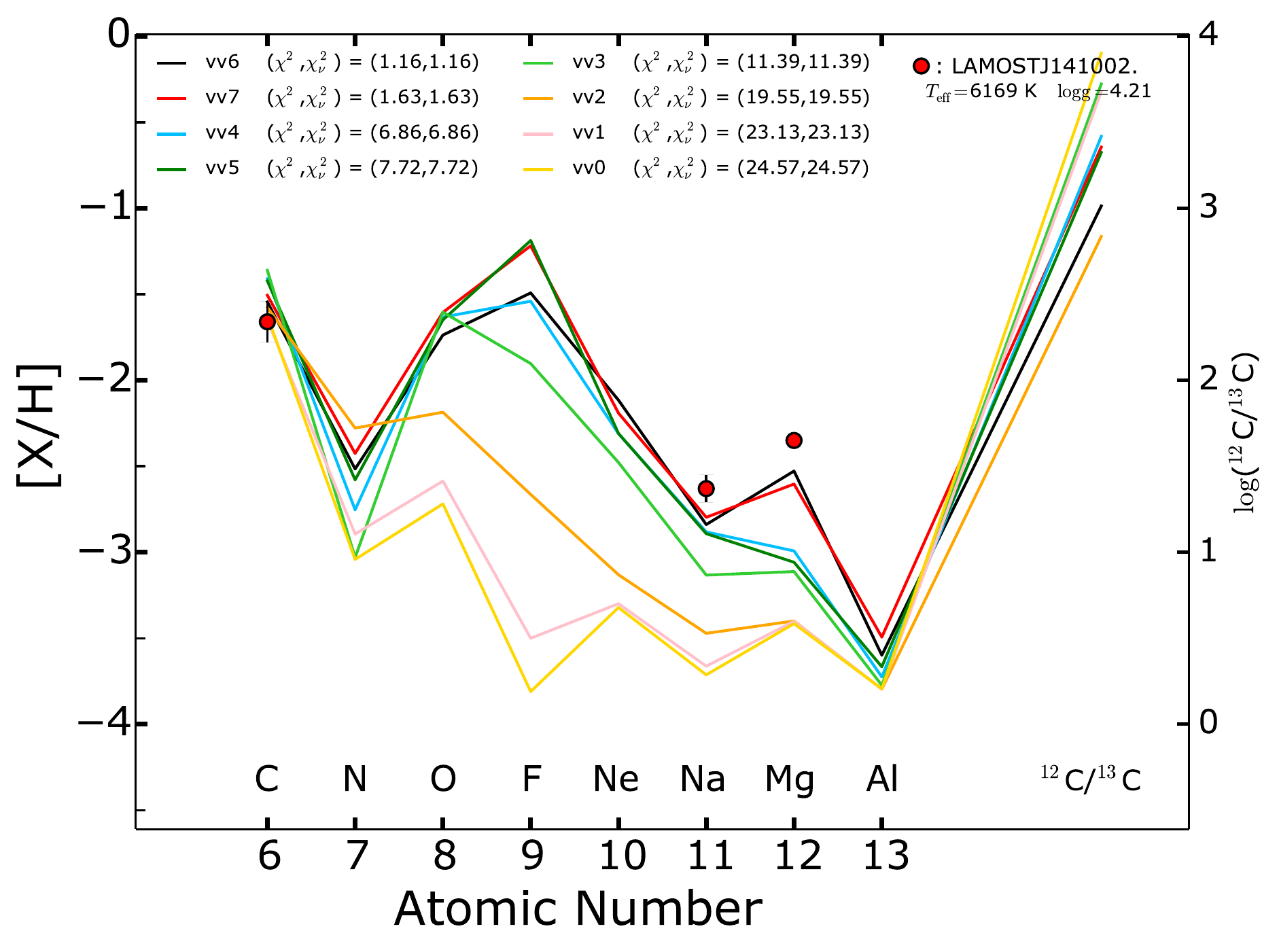}
   \end{minipage}
   \begin{minipage}[c]{.33\linewidth}
       \includegraphics[scale=0.3]{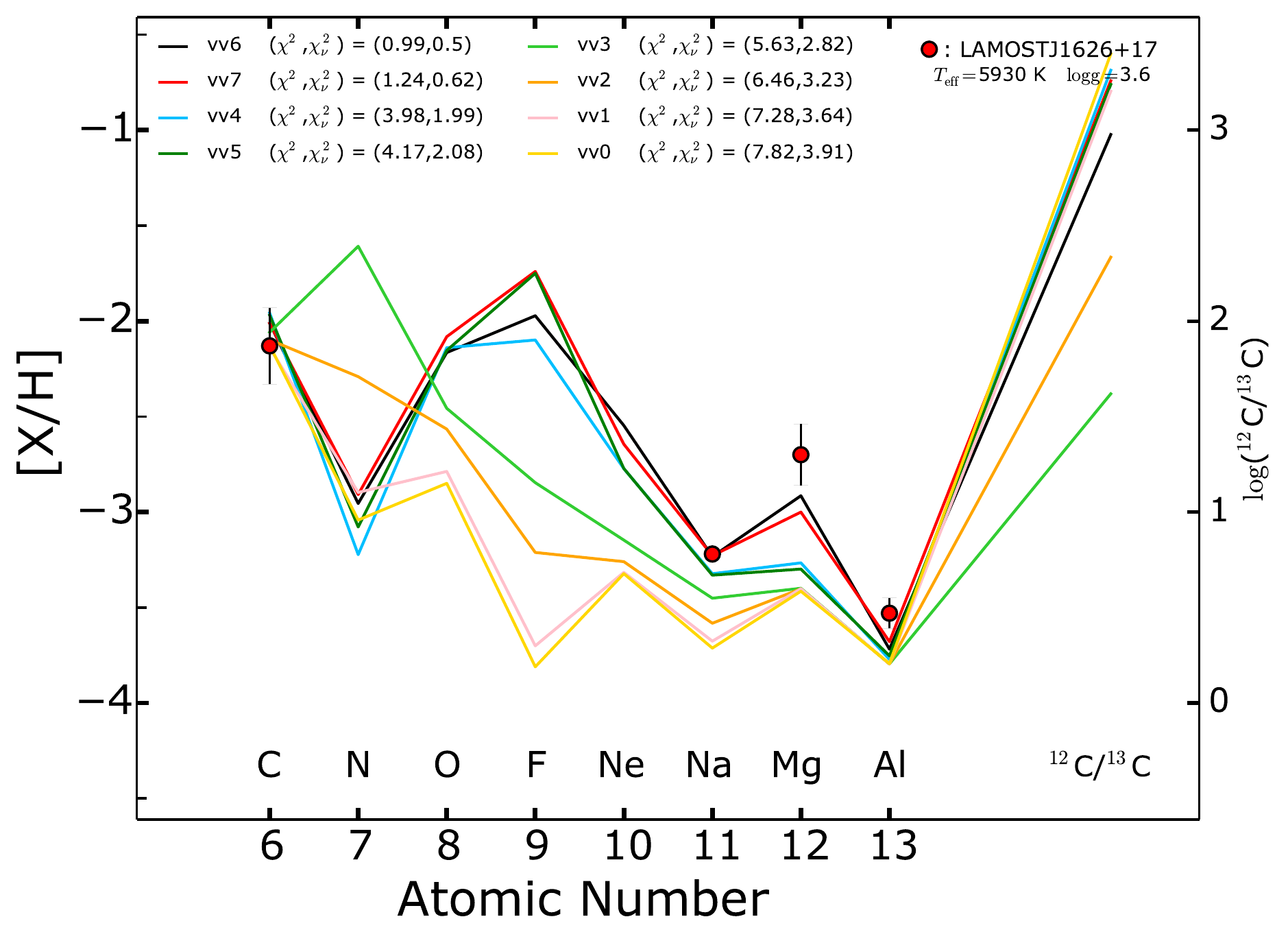}
   \end{minipage}
   \begin{minipage}[c]{.33\linewidth}
       \includegraphics[scale=0.3]{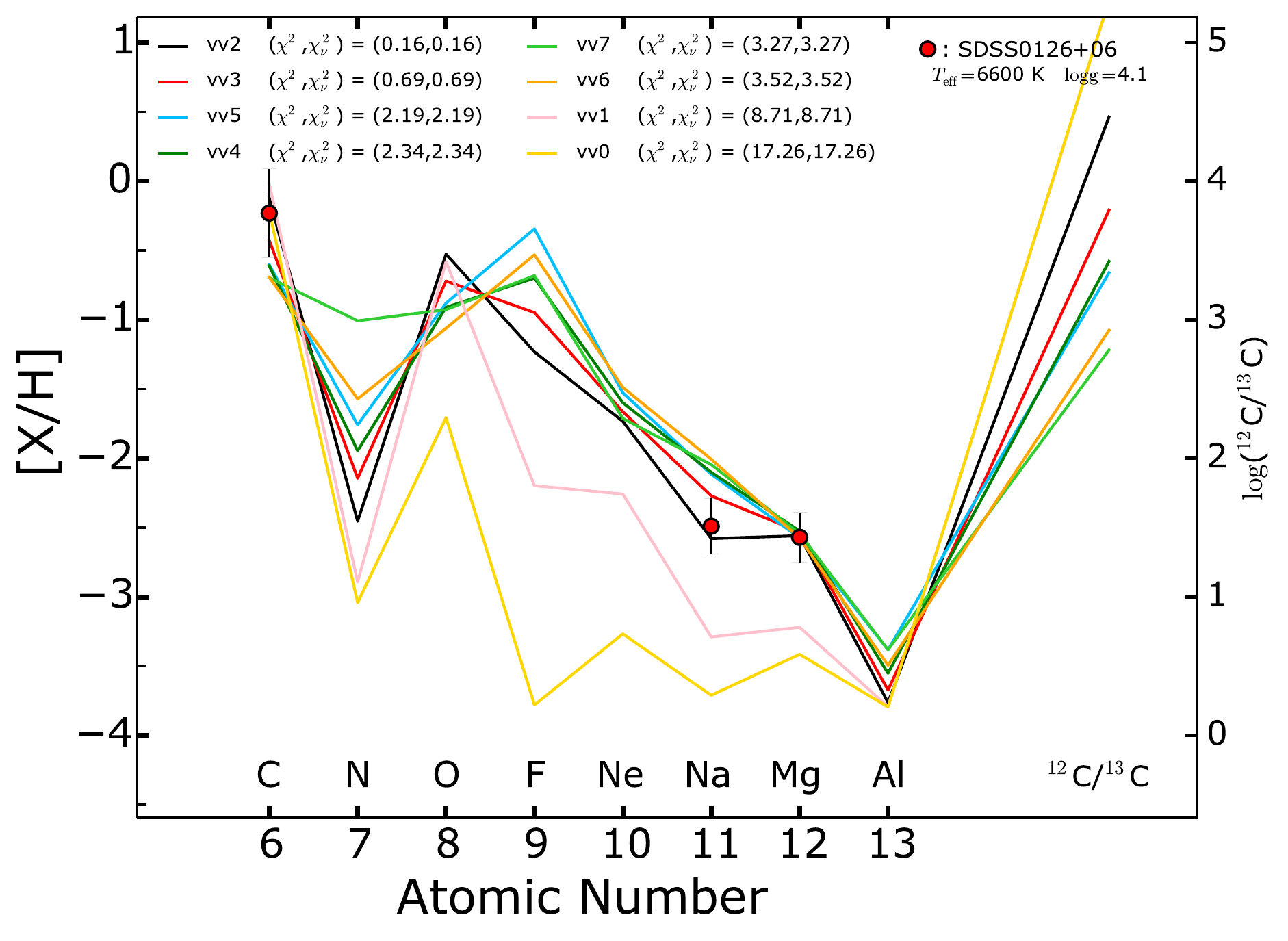}
   \end{minipage}
   \begin{minipage}[c]{.33\linewidth}
       \includegraphics[scale=0.3]{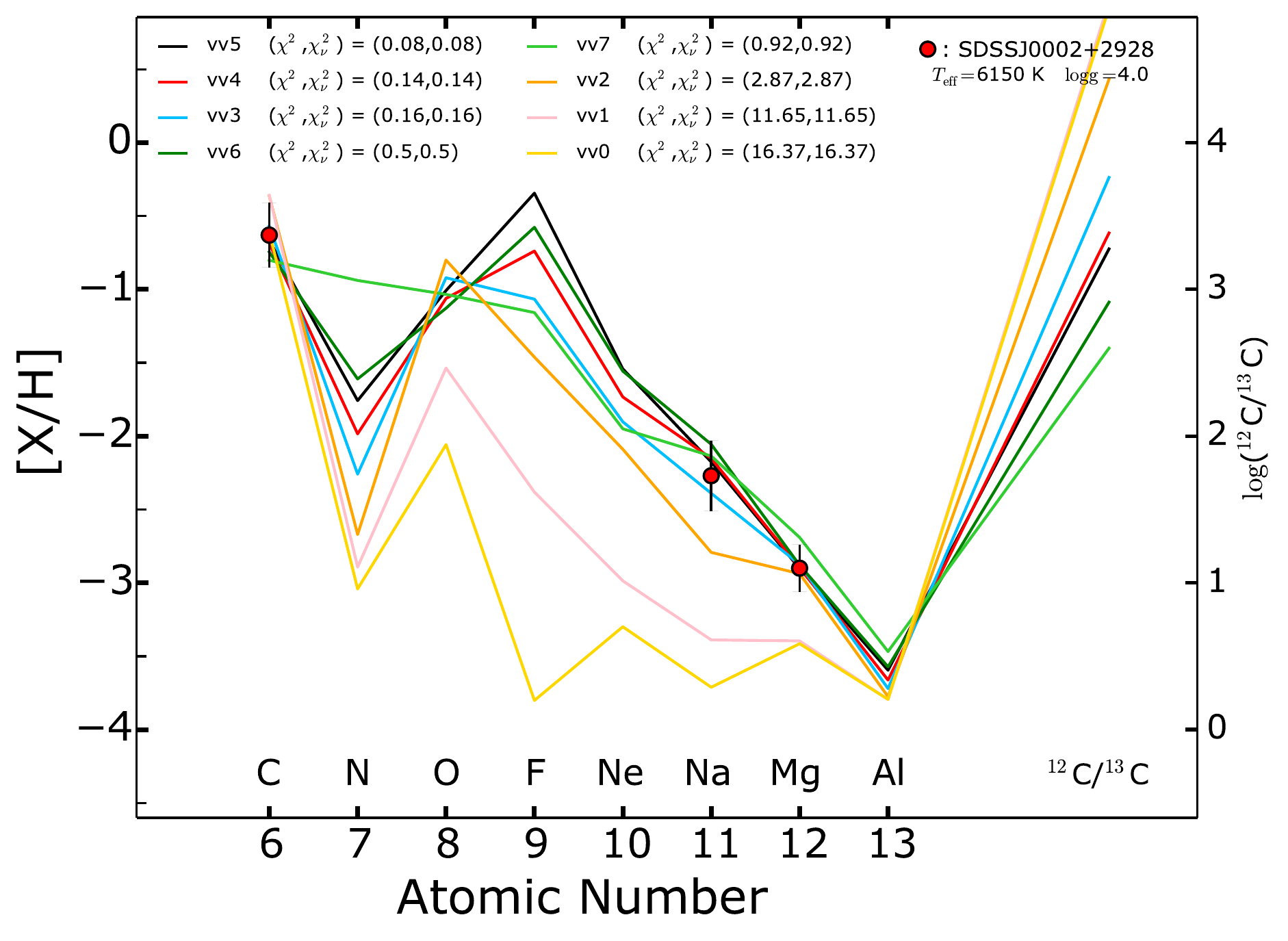}
   \end{minipage}
   \begin{minipage}[c]{.33\linewidth}
       \includegraphics[scale=0.3]{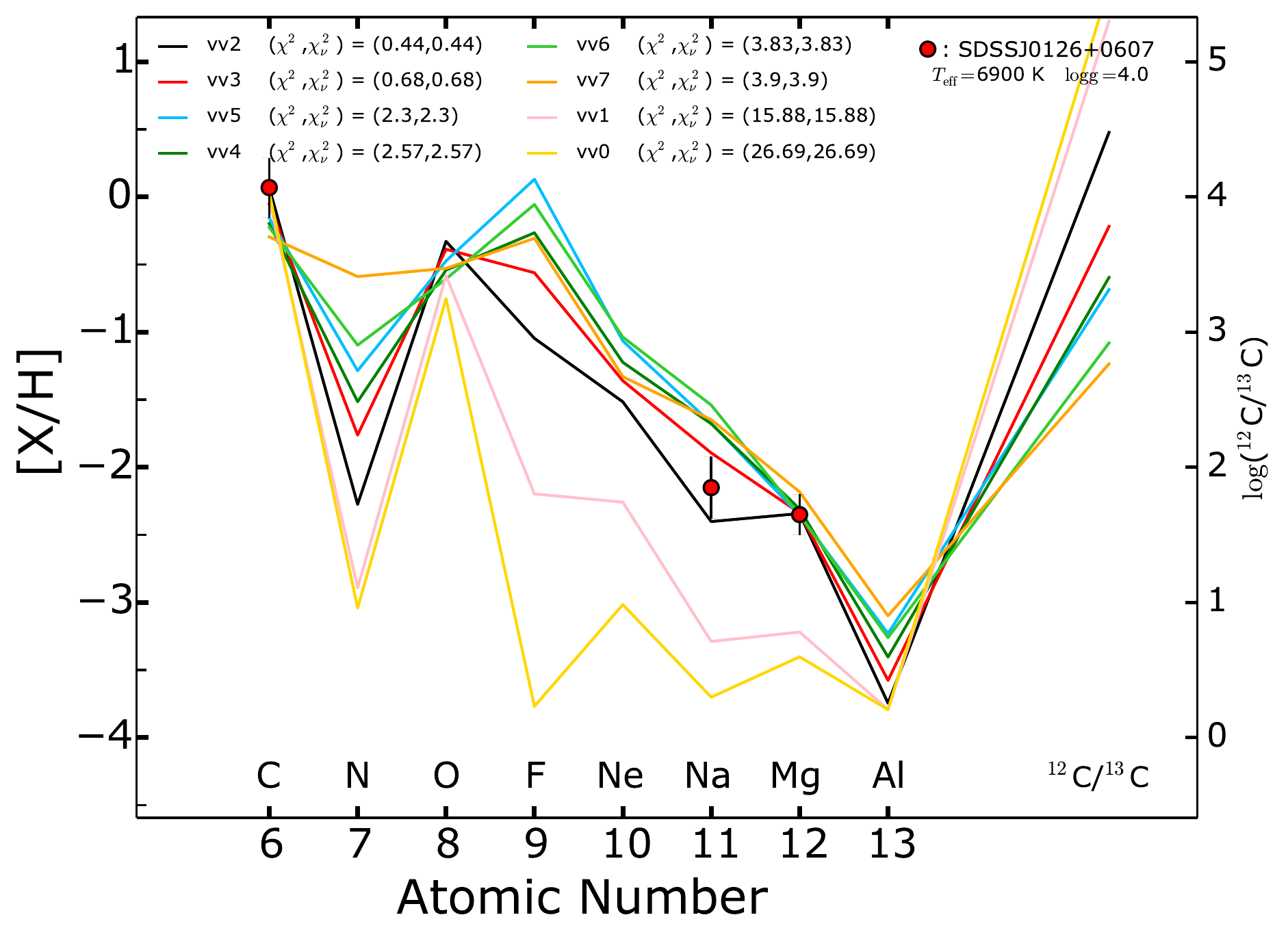}
   \end{minipage}
   \begin{minipage}[c]{.33\linewidth}
       \includegraphics[scale=0.3]{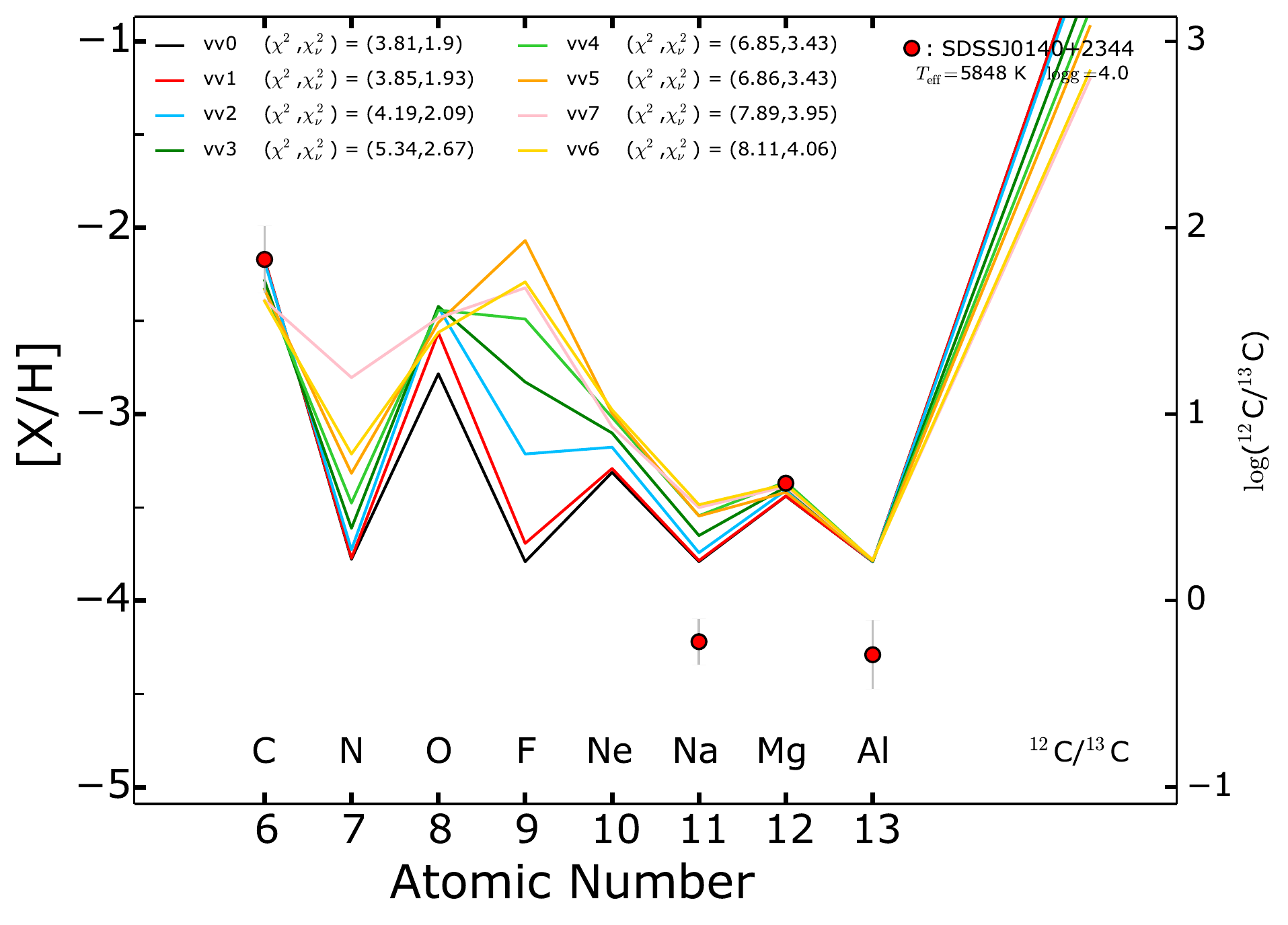}
   \end{minipage}
   \begin{minipage}[c]{.33\linewidth}
       \includegraphics[scale=0.3]{figs/XH_3best_217.pdf}
   \end{minipage}
   \begin{minipage}[c]{.33\linewidth}
       \includegraphics[scale=0.3]{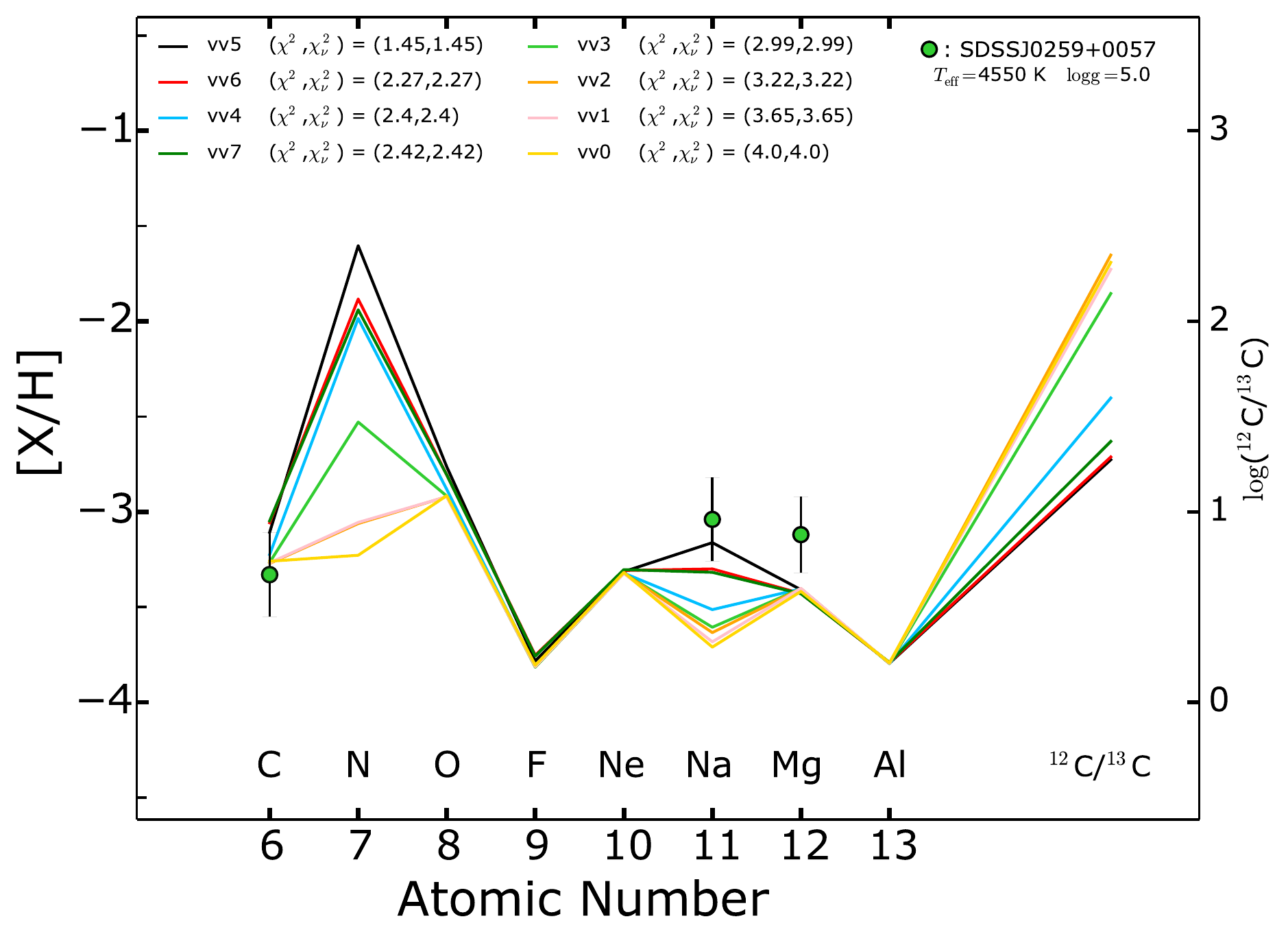}
   \end{minipage}
   \begin{minipage}[c]{.33\linewidth}
       \includegraphics[scale=0.3]{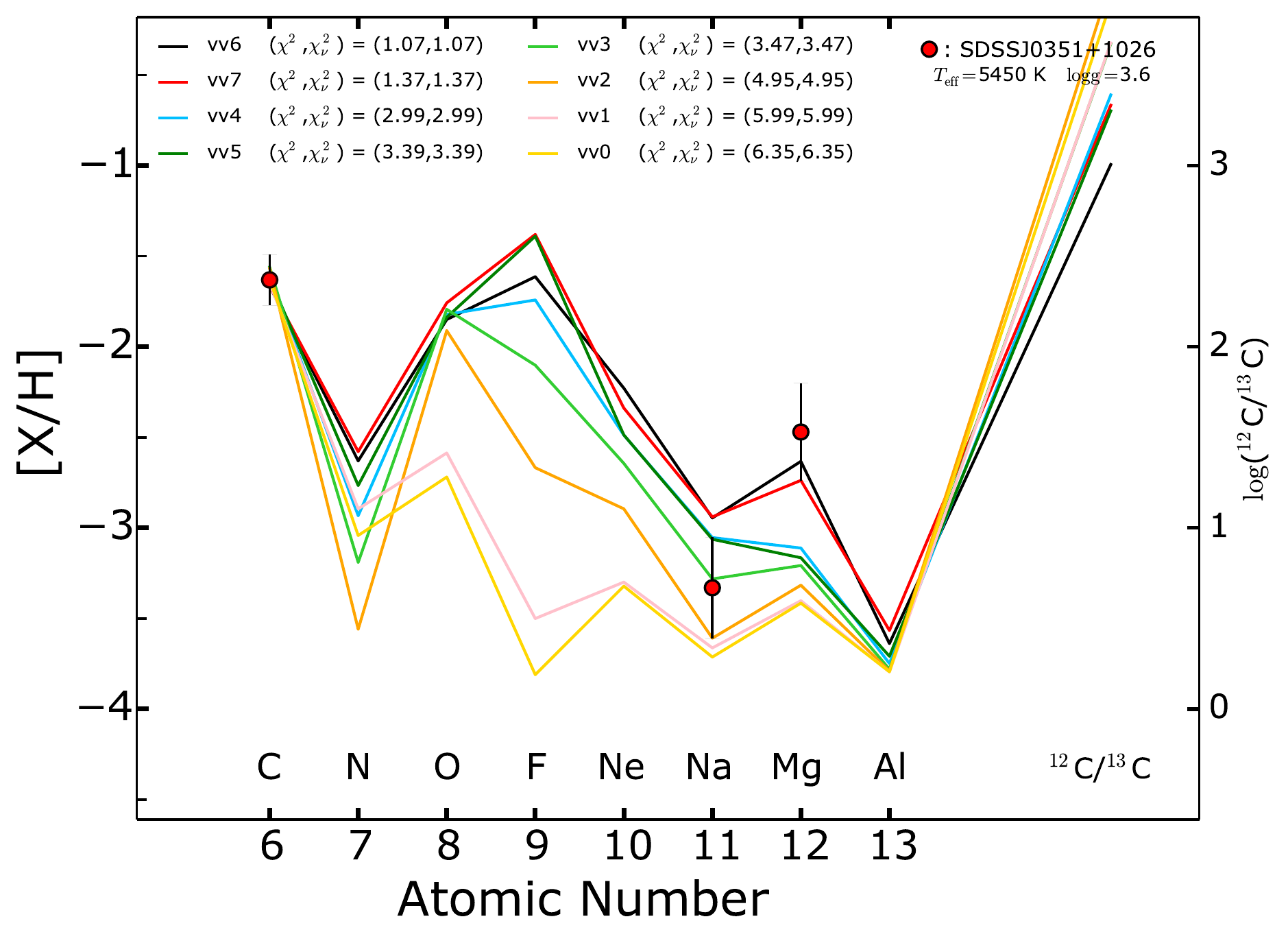}
   \end{minipage}
   \begin{minipage}[c]{.33\linewidth}
       \includegraphics[scale=0.3]{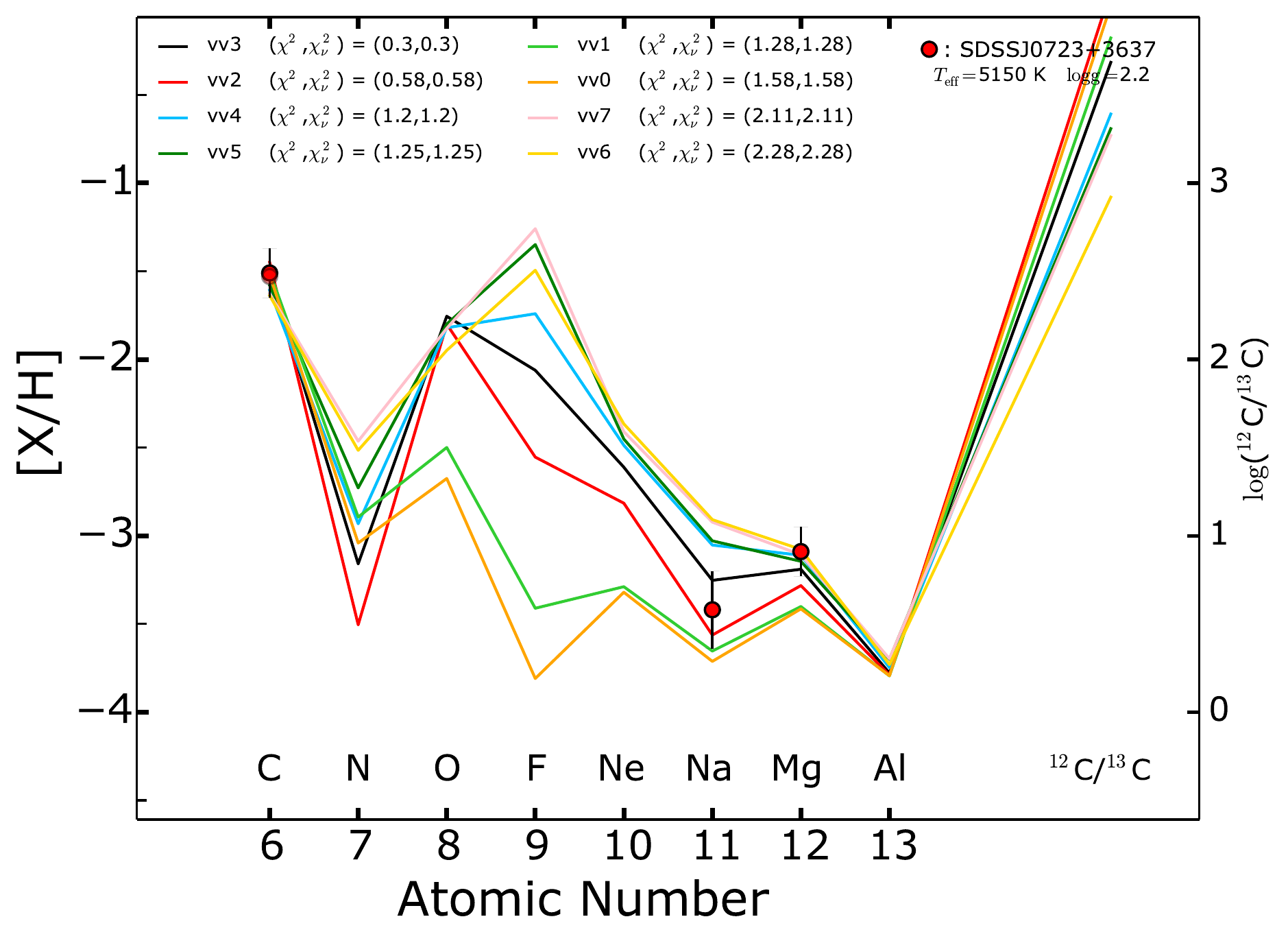}
   \end{minipage}
   \begin{minipage}[c]{.33\linewidth}
       \includegraphics[scale=0.3]{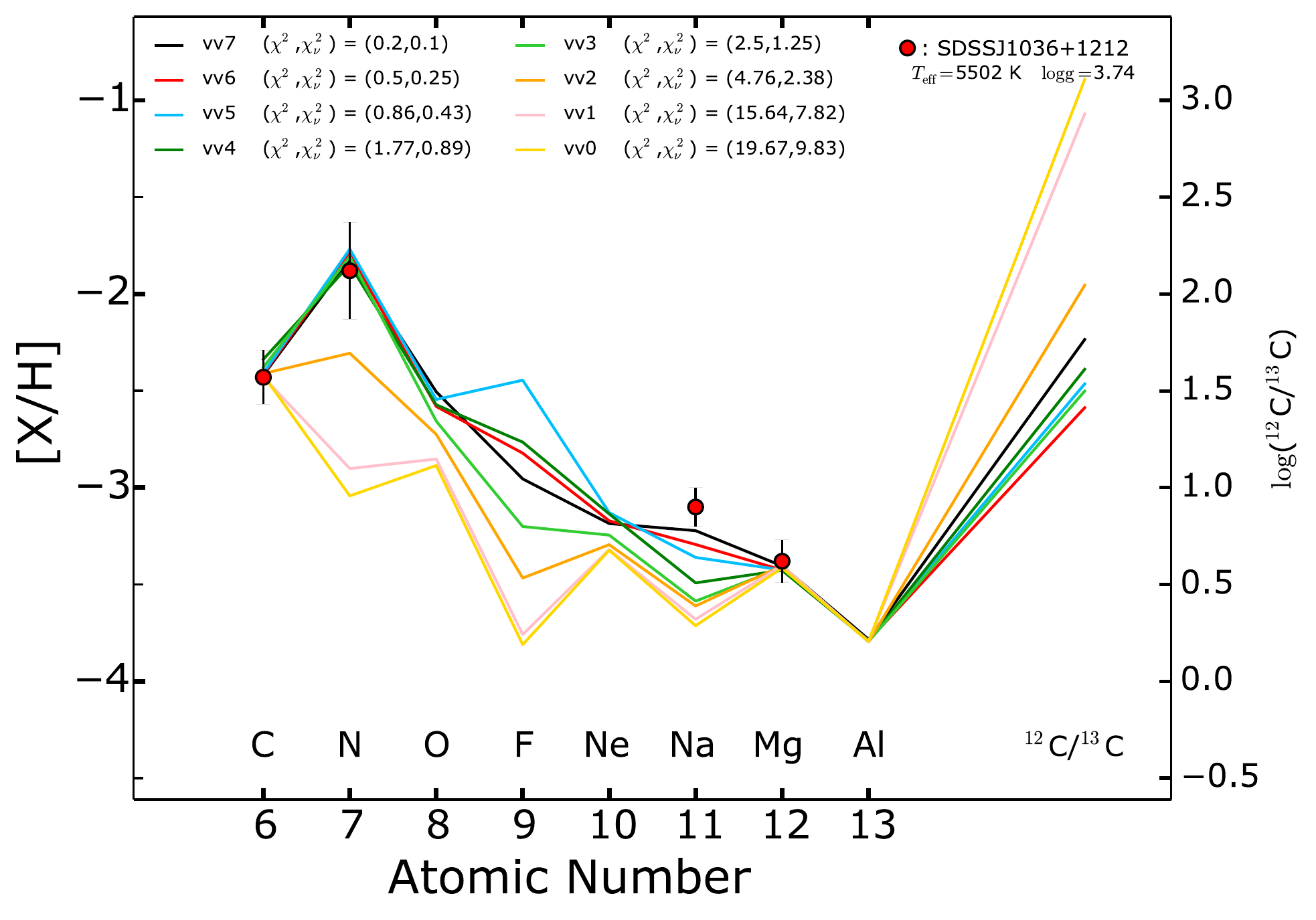}
   \end{minipage}
   \begin{minipage}[c]{.33\linewidth}
       \includegraphics[scale=0.3]{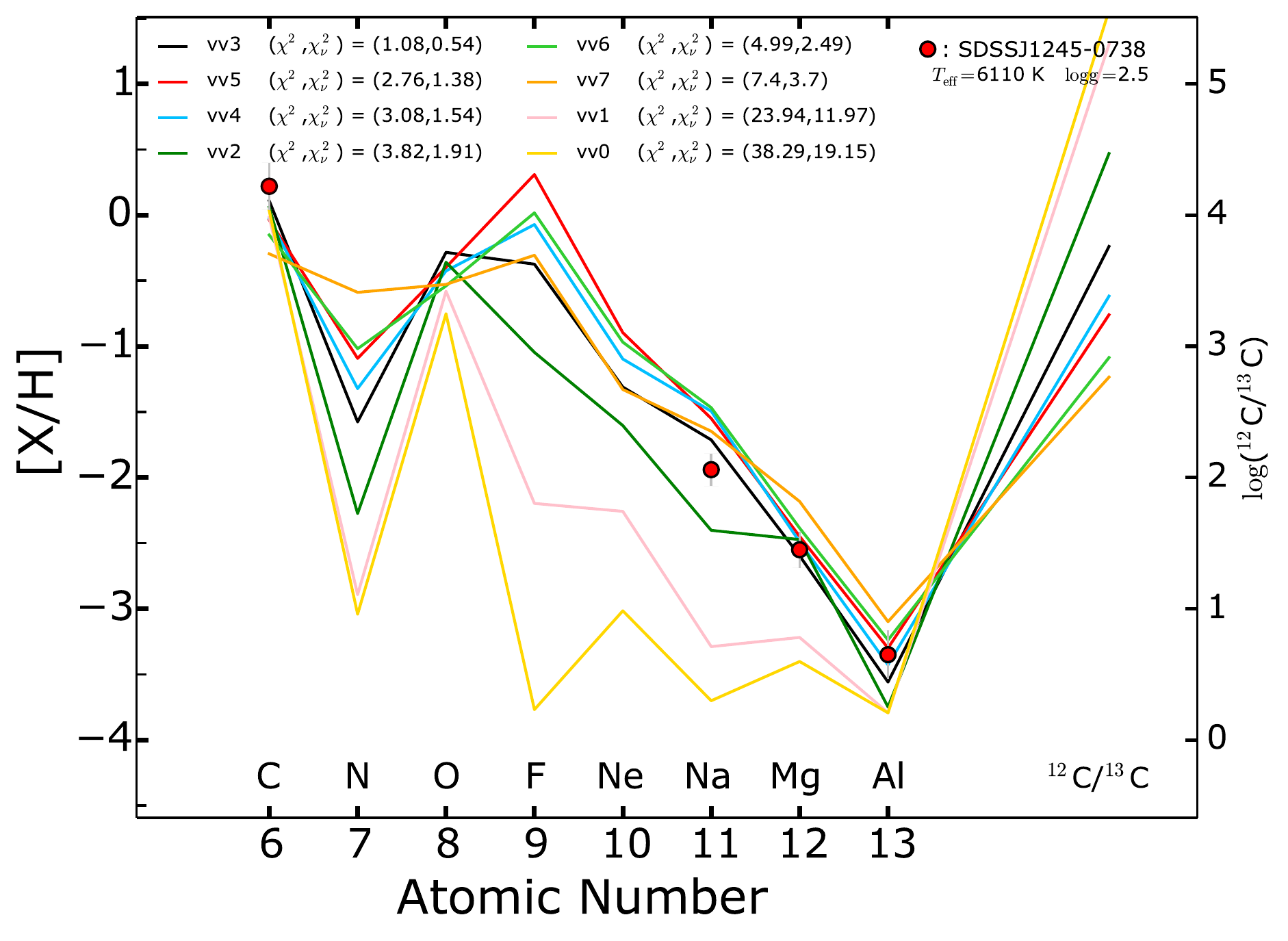}
   \end{minipage}
   \caption{Continued.}
\label{allfit1}
    \end{figure*}

   \begin{figure*}
      \ContinuedFloat
   \centering
   \begin{minipage}[c]{.33\linewidth}
       \includegraphics[scale=0.3]{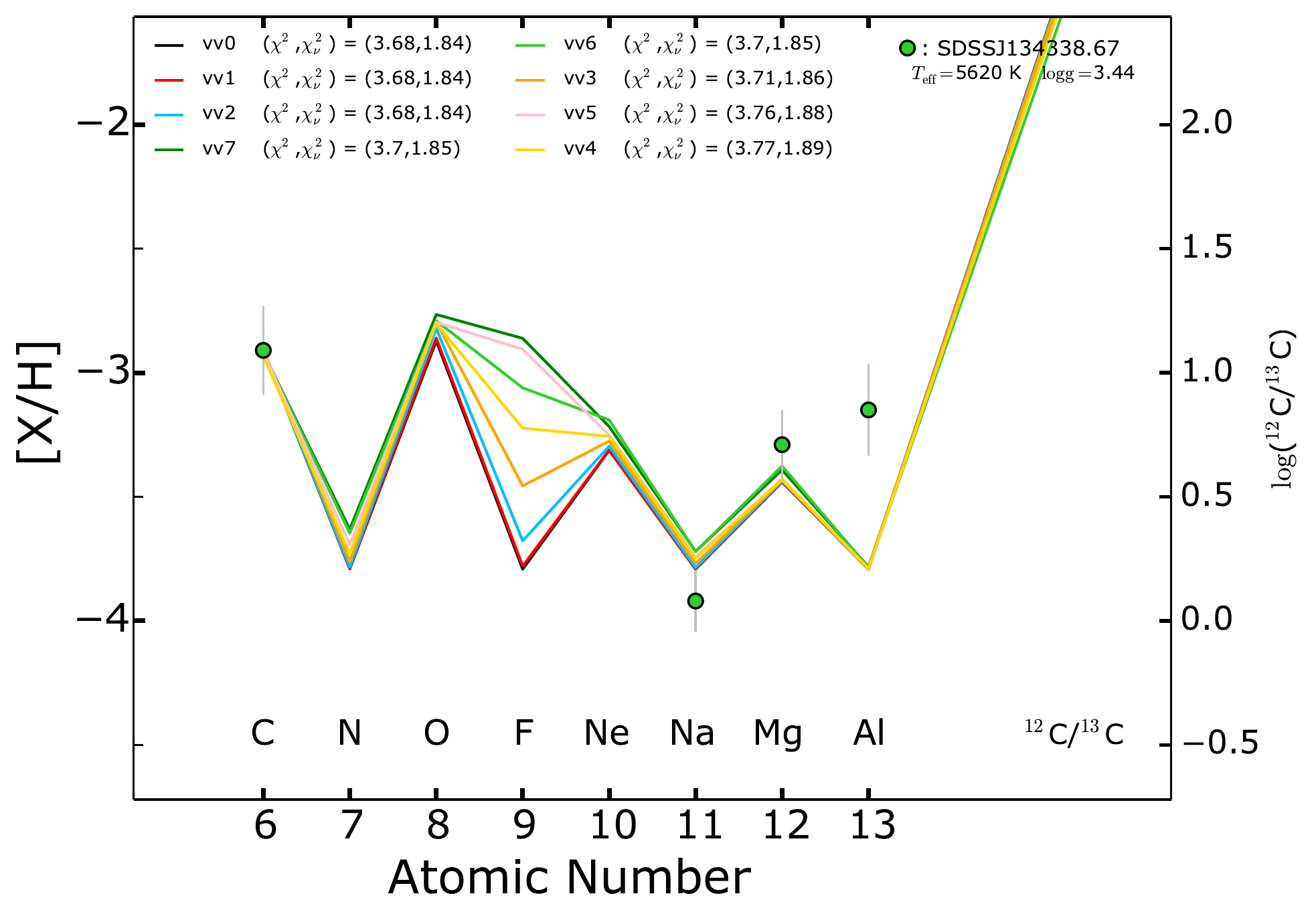}
   \end{minipage}
   \begin{minipage}[c]{.33\linewidth}
       \includegraphics[scale=0.3]{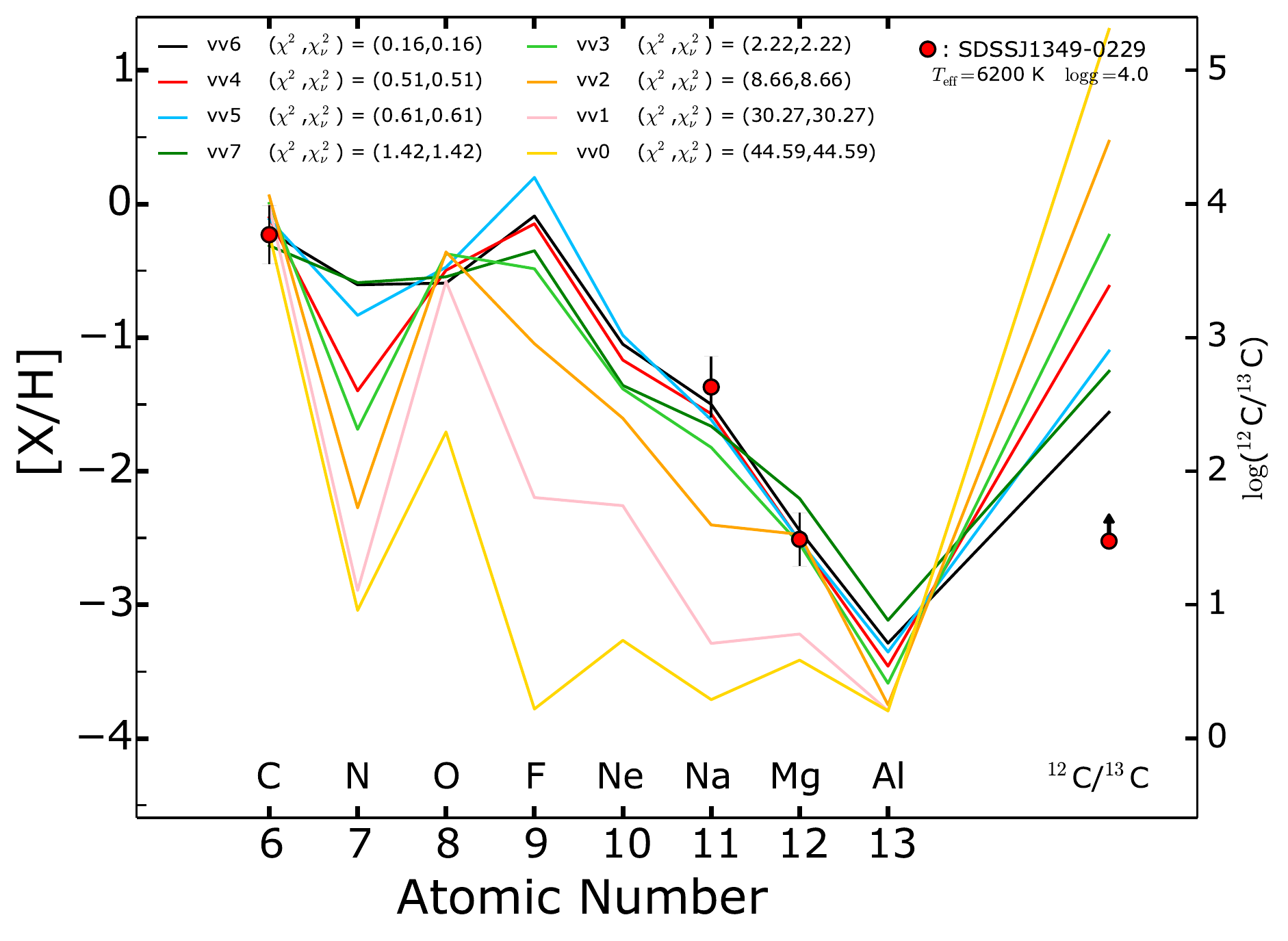}
   \end{minipage}
   \begin{minipage}[c]{.33\linewidth}
       \includegraphics[scale=0.3]{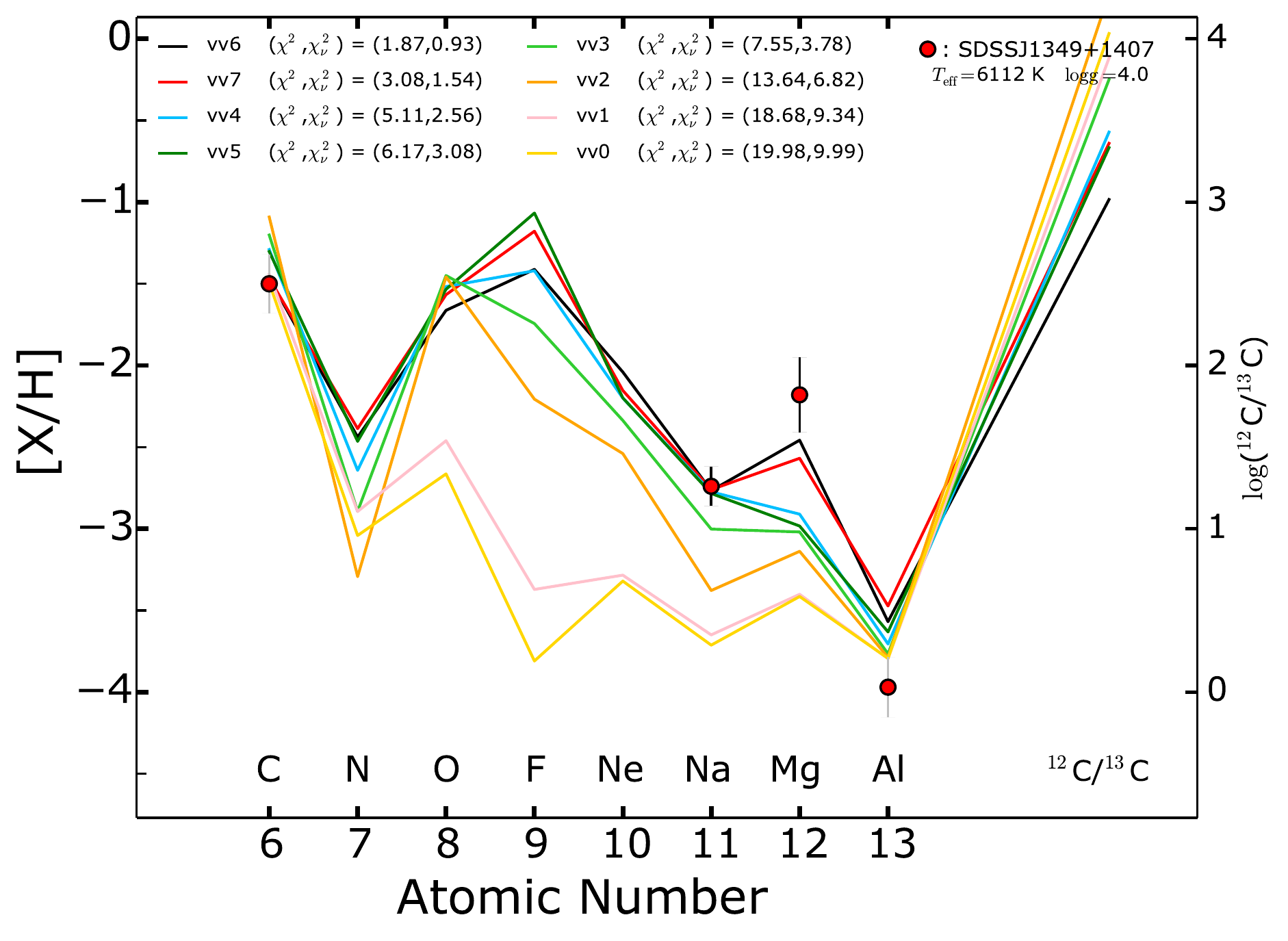}
   \end{minipage}
   \begin{minipage}[c]{.33\linewidth}
       \includegraphics[scale=0.3]{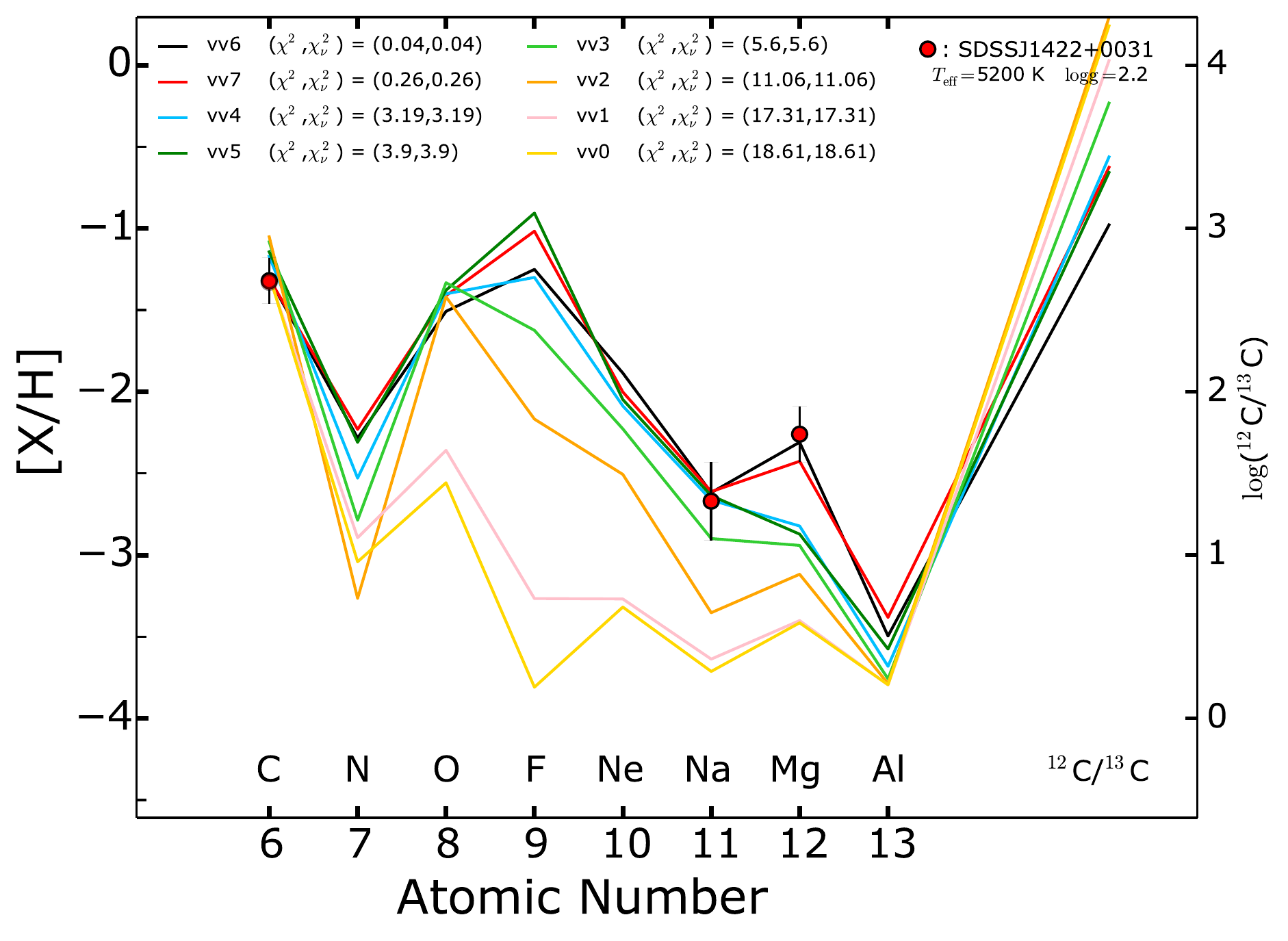}
   \end{minipage}
   \begin{minipage}[c]{.33\linewidth}
       \includegraphics[scale=0.3]{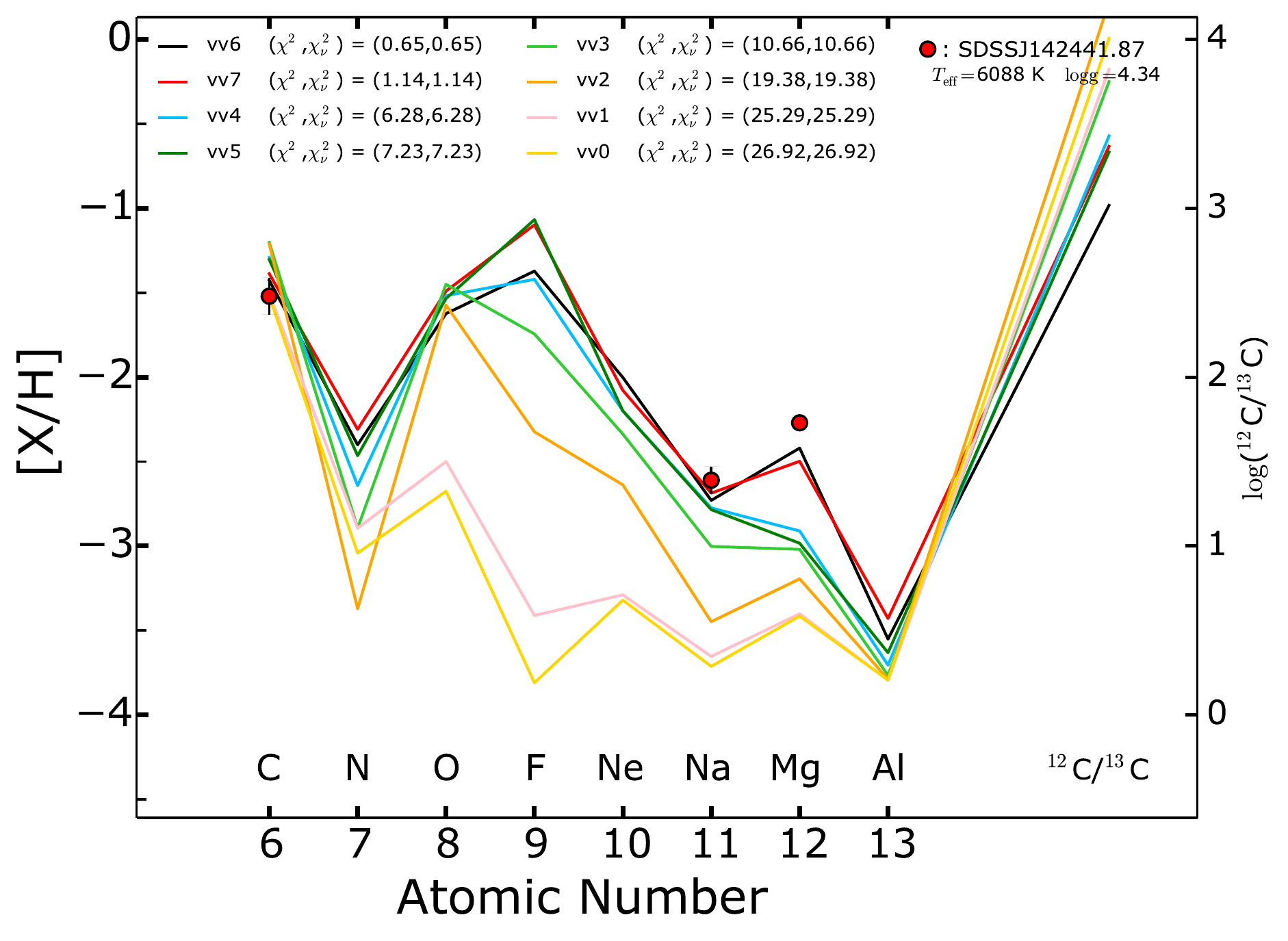}
   \end{minipage}
   \begin{minipage}[c]{.33\linewidth}
       \includegraphics[scale=0.3]{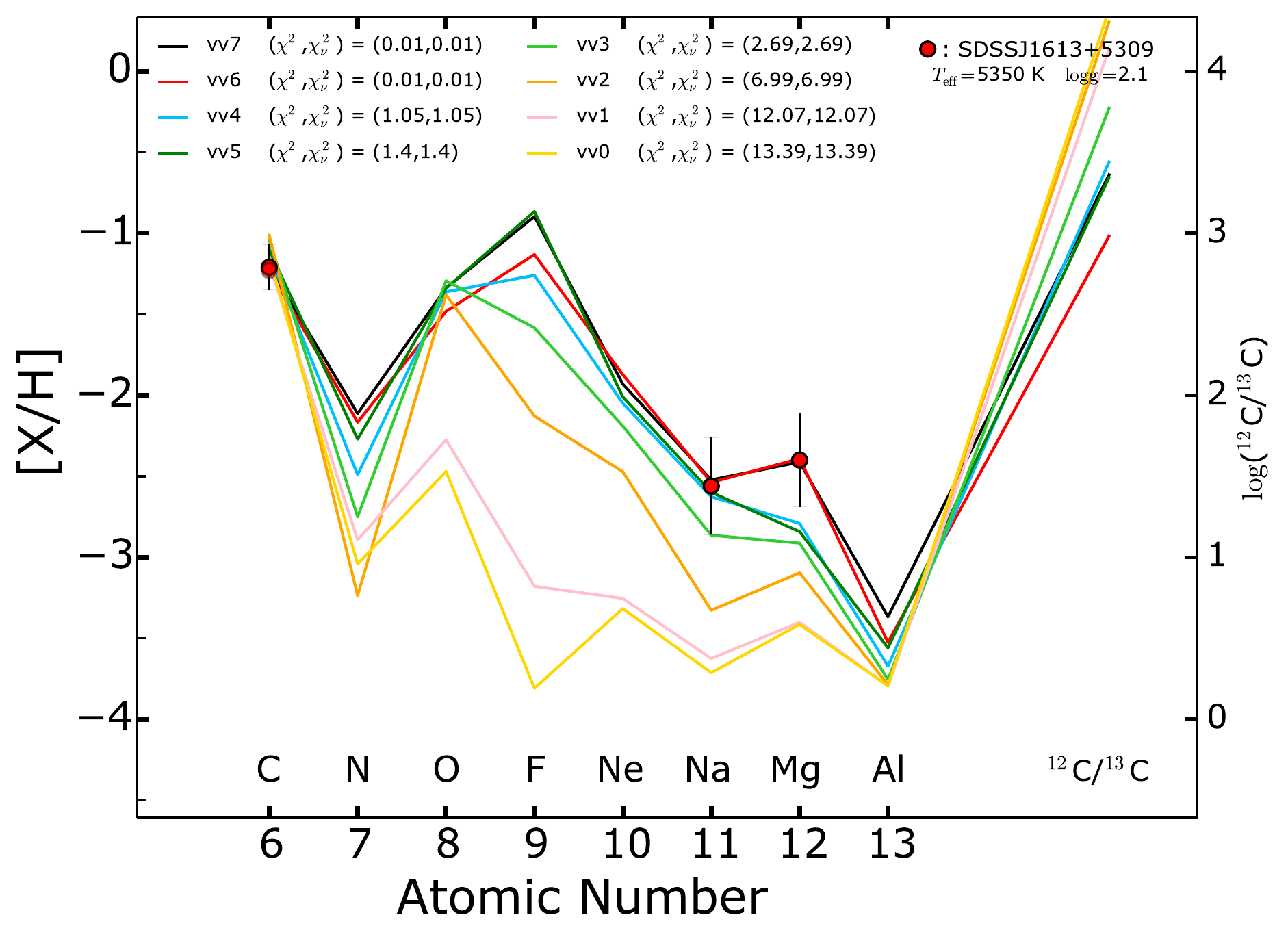}
   \end{minipage}
   \begin{minipage}[c]{.33\linewidth}
       \includegraphics[scale=0.3]{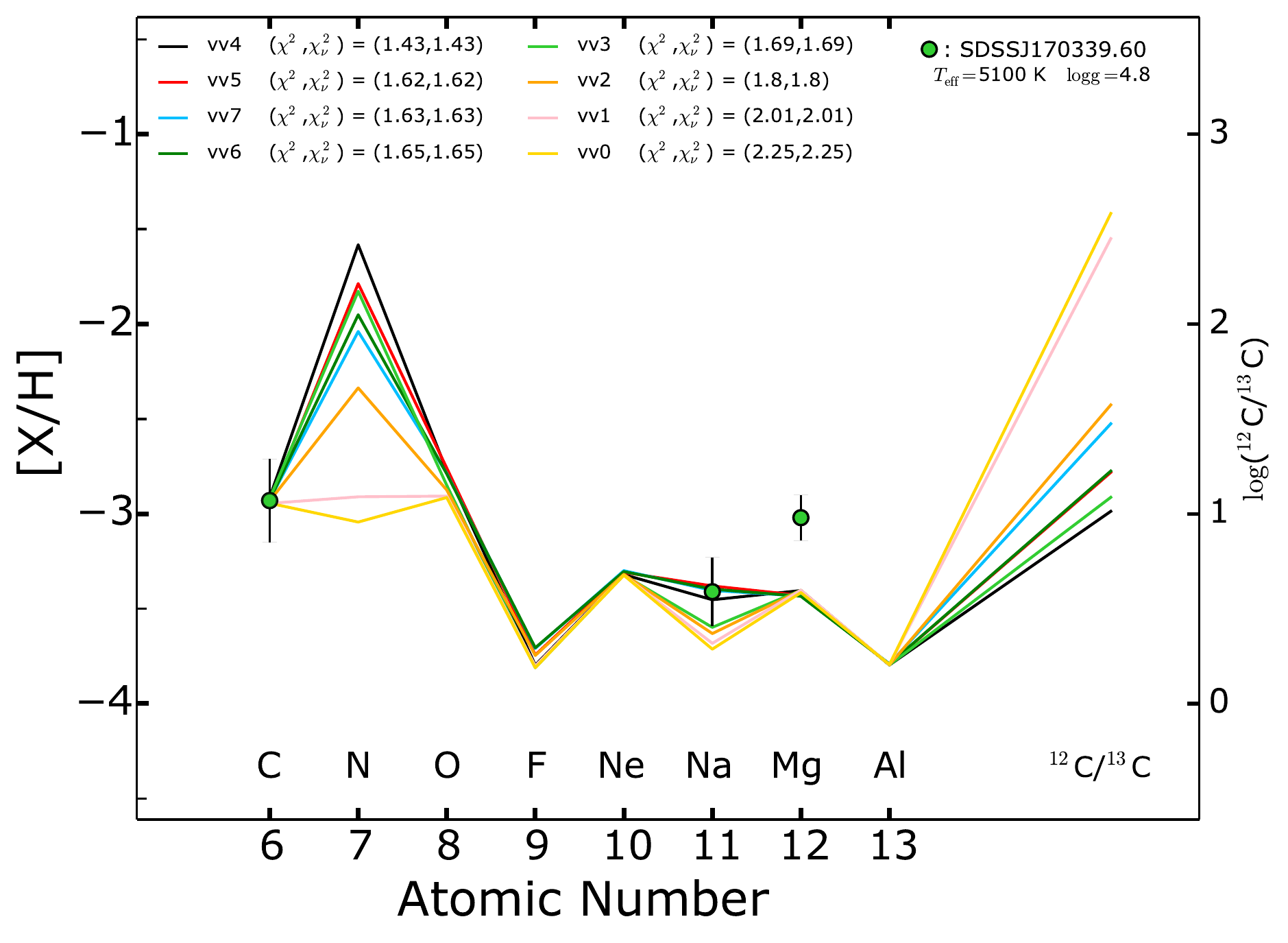}
   \end{minipage}
   \begin{minipage}[c]{.33\linewidth}
       \includegraphics[scale=0.3]{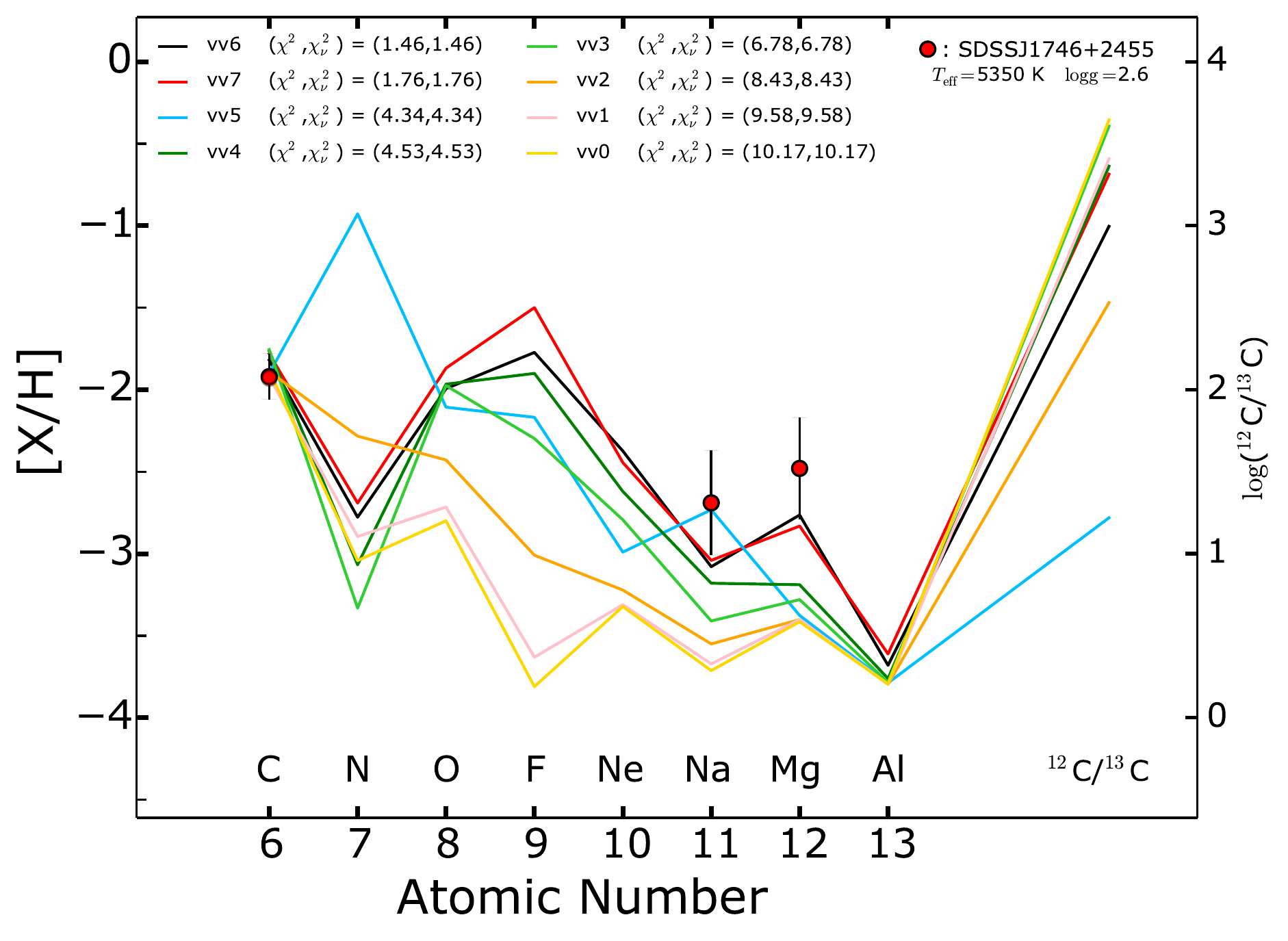}
   \end{minipage}
   \begin{minipage}[c]{.33\linewidth}
       \includegraphics[scale=0.3]{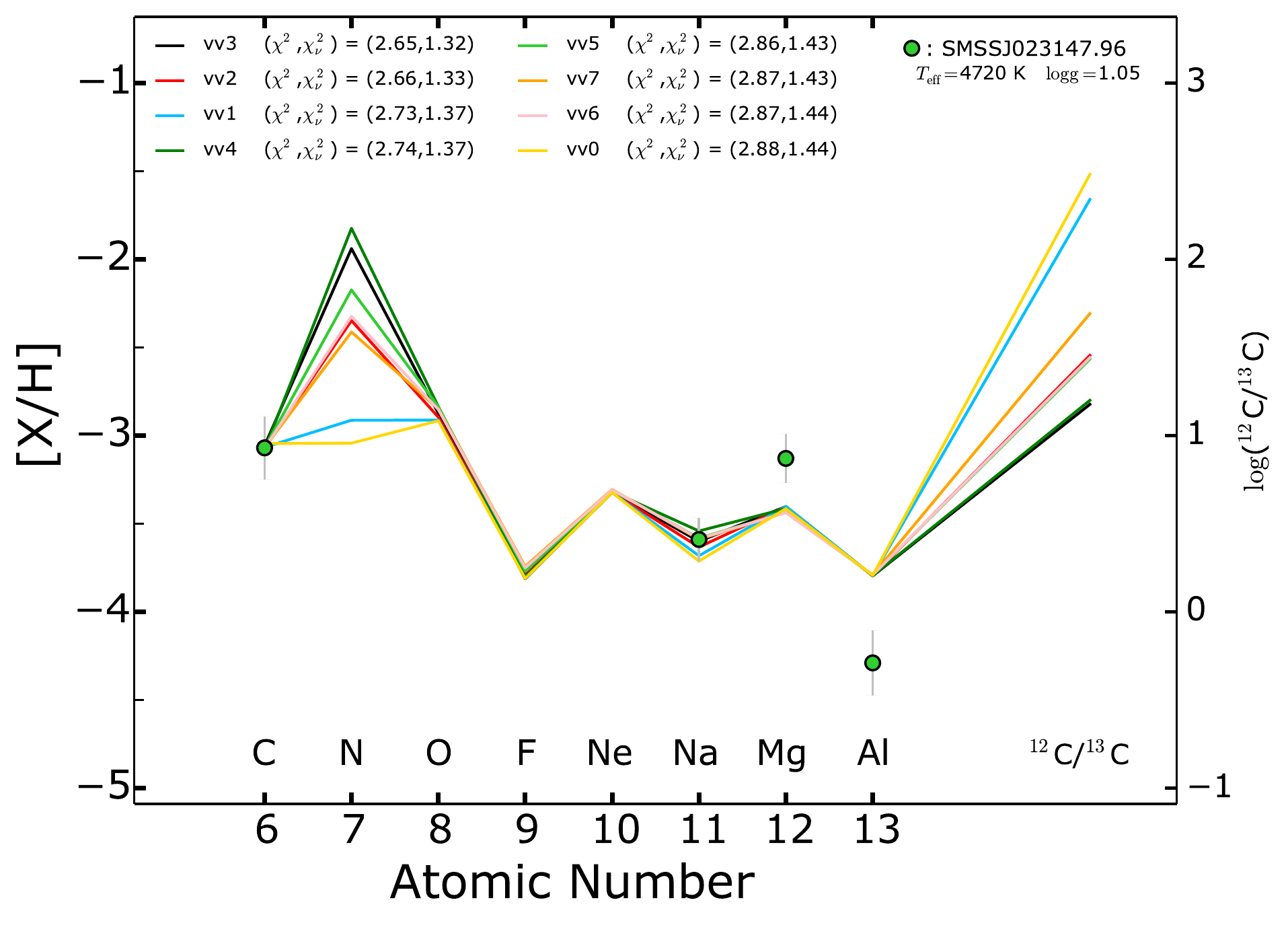}
   \end{minipage}
   \begin{minipage}[c]{.33\linewidth}
       \includegraphics[scale=0.3]{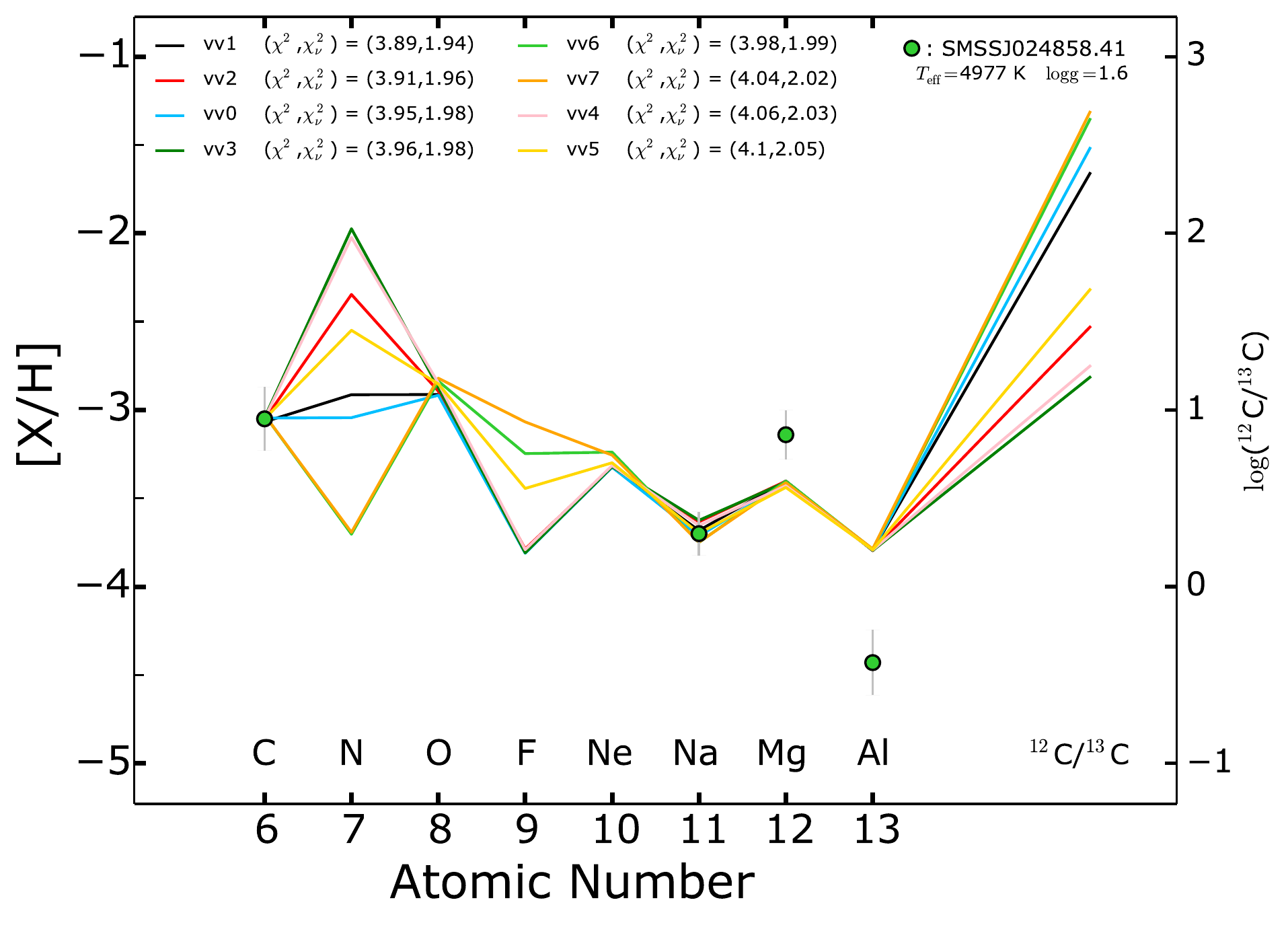}
   \end{minipage}
   \begin{minipage}[c]{.33\linewidth}
       \includegraphics[scale=0.3]{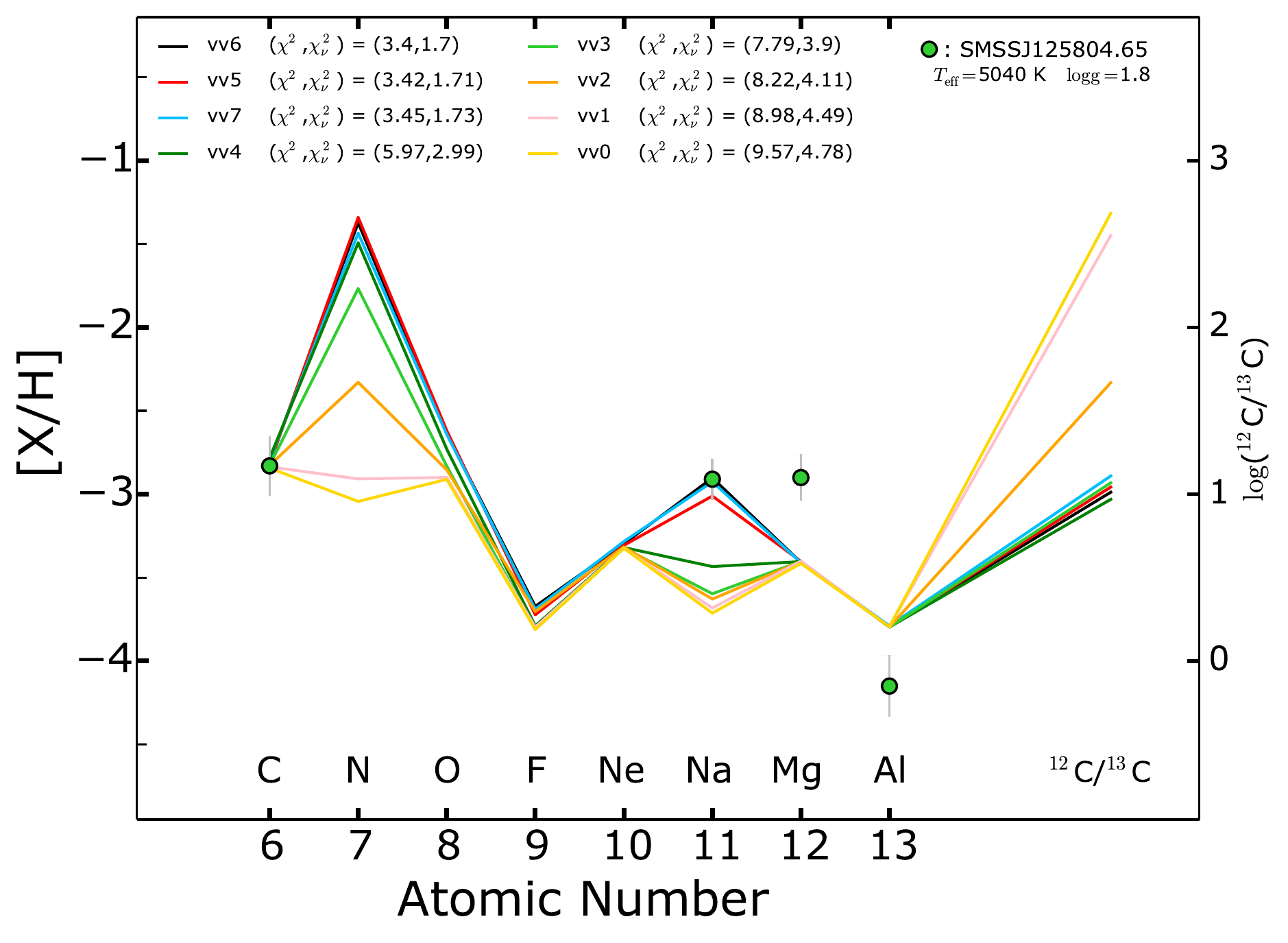}
   \end{minipage}
   \begin{minipage}[c]{.33\linewidth}
       \includegraphics[scale=0.3]{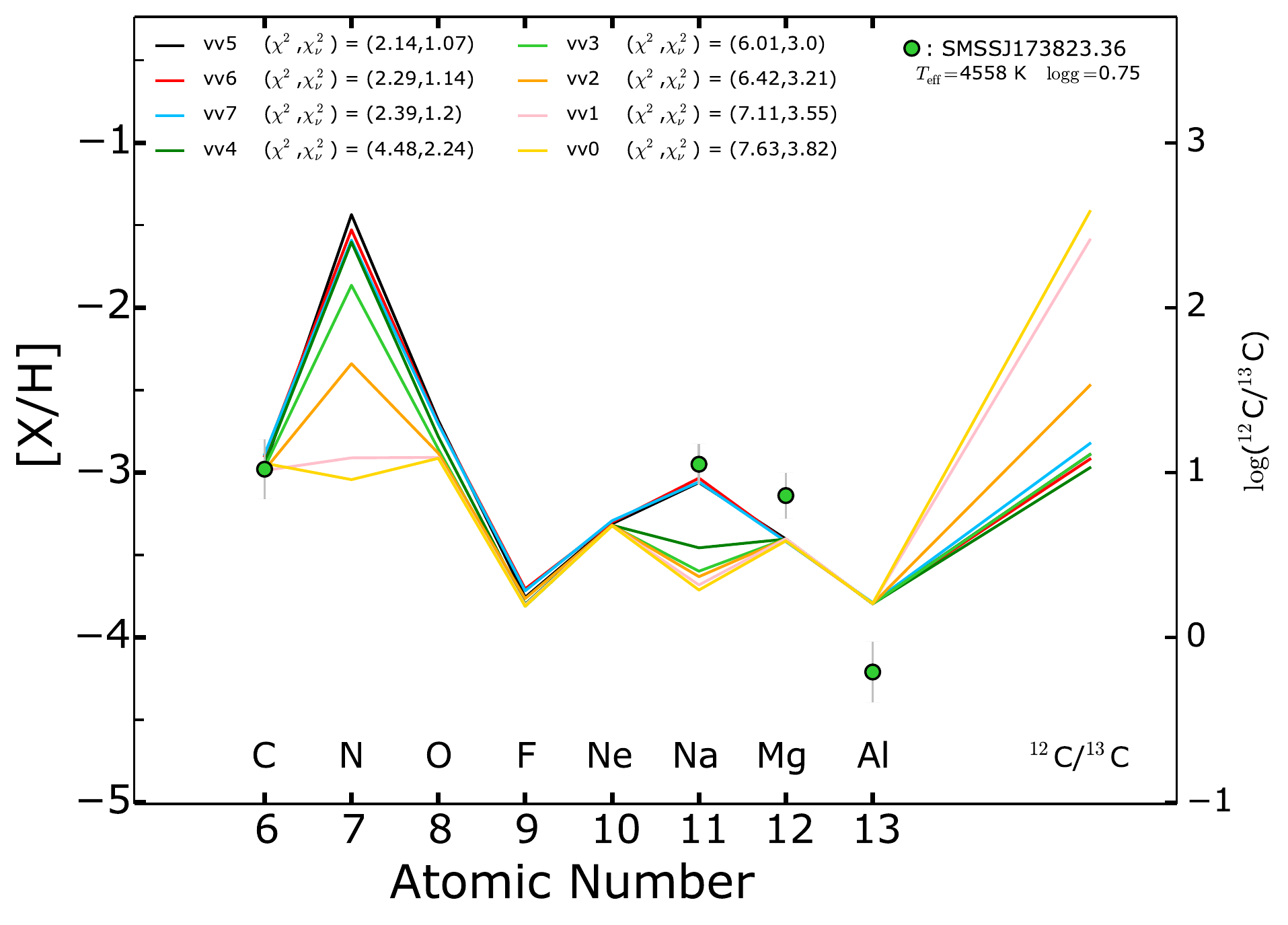}
   \end{minipage}
   \begin{minipage}[c]{.33\linewidth}
       \includegraphics[scale=0.3]{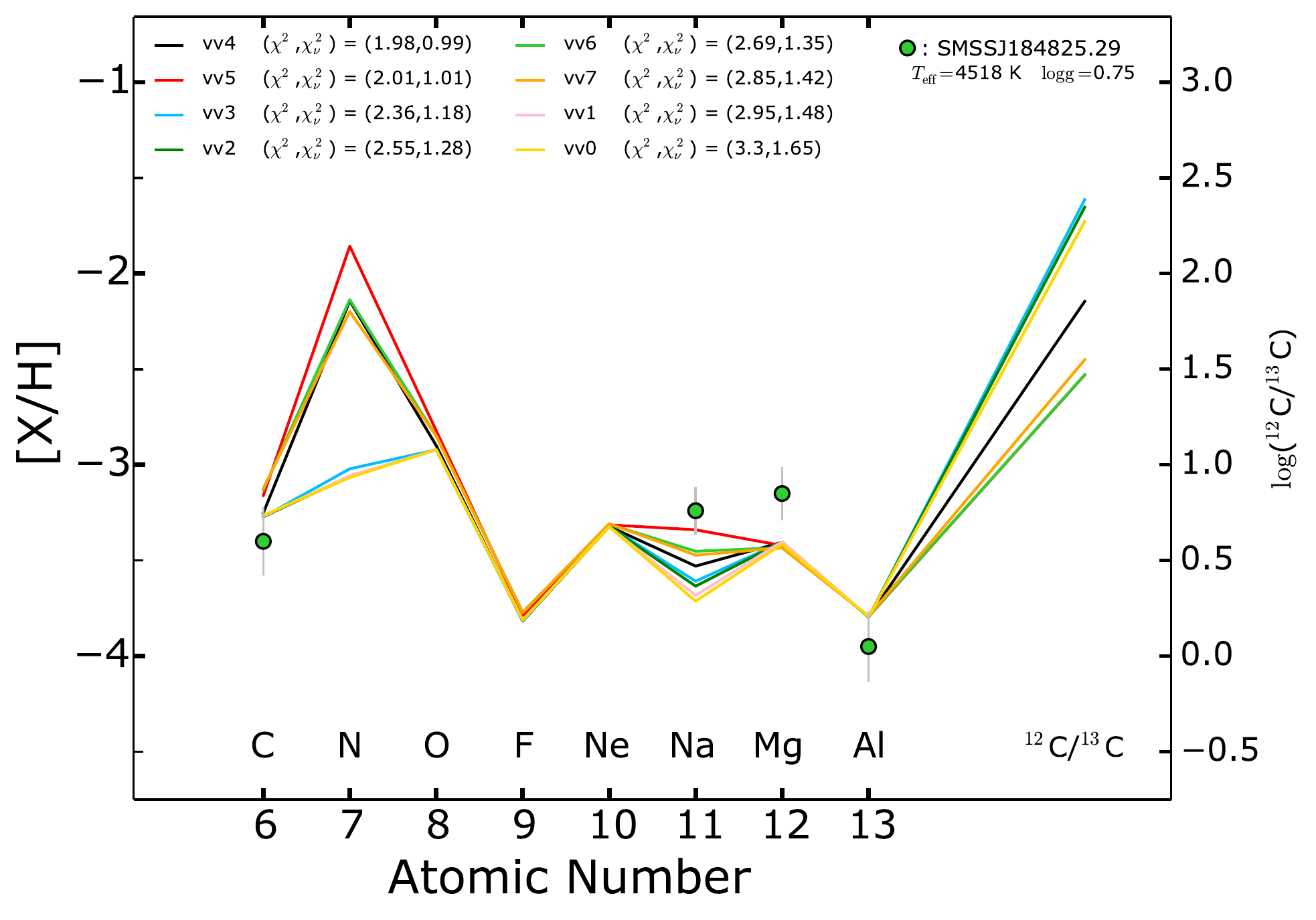}
   \end{minipage}
   \begin{minipage}[c]{.33\linewidth}
       \includegraphics[scale=0.3]{figs/XH_3best_268.pdf}
   \end{minipage}
   \begin{minipage}[c]{.33\linewidth}
       \includegraphics[scale=0.3]{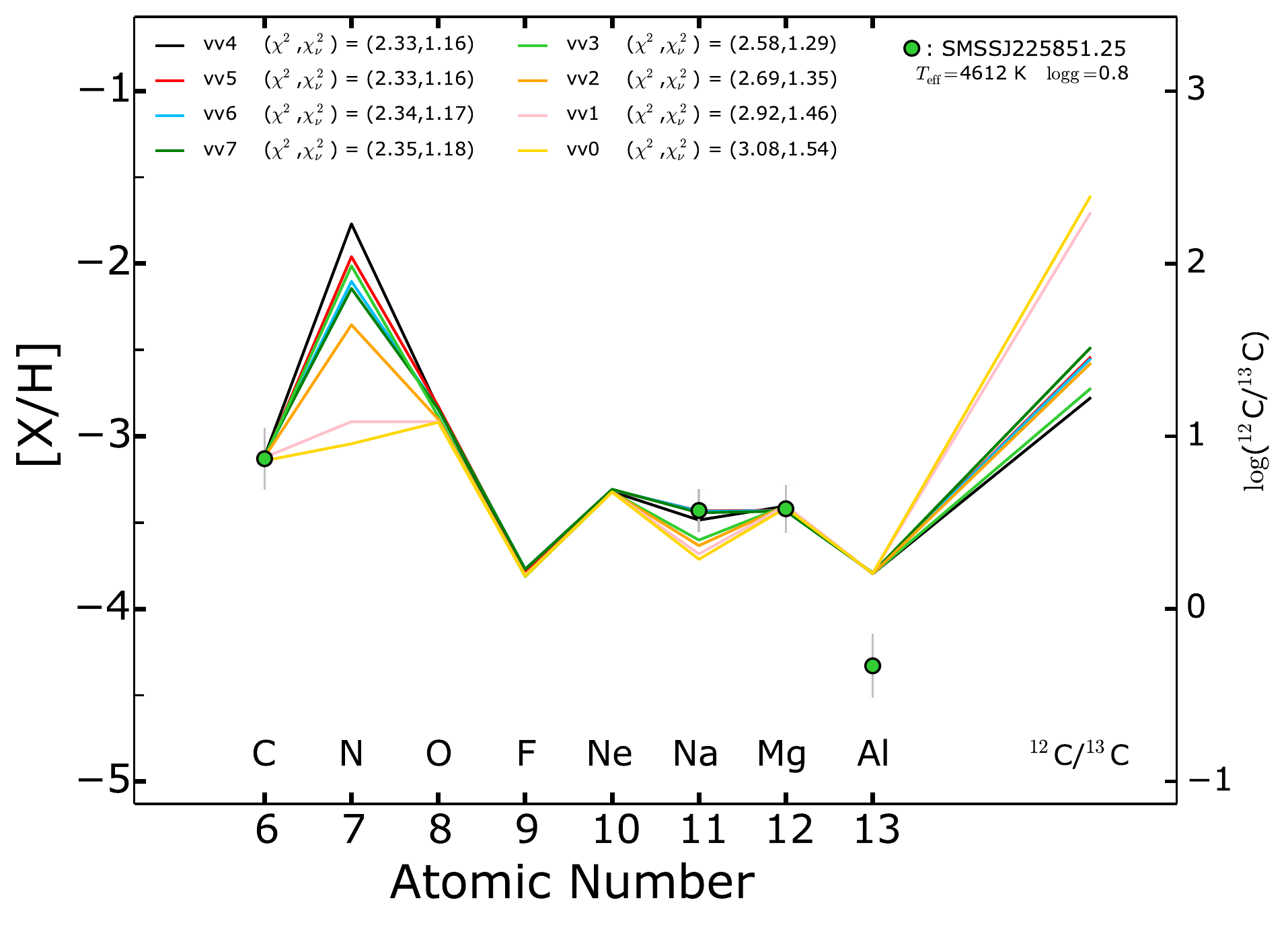}
   \end{minipage}
   \caption{Continued.} 
\label{allfit1}
    \end{figure*}

\end{document}